\documentclass[twocolumn,tighten]{aastex631}
\usepackage{multirow}
\usepackage{subfigure, apalike}
\usepackage{amssymb}
\usepackage{natbib}
\usepackage{multirow}
\usepackage{longtable}
\usepackage{geometry}
\usepackage{amsmath}
\usepackage{color}
\usepackage{times}
\usepackage{amsmath,amsfonts,amsthm,bm}
\usepackage{comment}

\newcommand{\vdag}{(v)^\dagger}
\newcommand\aastex{AAS\TeX}
\newcommand\latex{La\TeX}

\newenvironment{acknowledgmentsno}{%
  \nolinenumbers
  \section*{Acknowledgments}
}

\submitjournal{ApJS}

\begin{document}

\title{The 2025 Evaluation of Experimental Thermonuclear Reaction Rates (ETR25)}

\author[0000-0003-2381-0412]{Christian Iliadis}
\affiliation{Triangle Universities Nuclear Laboratory (TUNL), Duke University, Durham, North Carolina 27708, USA}
\affiliation{Department of Physics \& Astronomy, University of North Carolina at Chapel Hill, NC 27599-3255, USA}

\author[0000-0001-7731-580X]{Richard Longland}
\affiliation{Triangle Universities Nuclear Laboratory (TUNL), Duke University, Durham, North Carolina 27708, USA}
\affiliation{Department of Physics, North Carolina State University, Raleigh, NC 27695, USA}

\author[0000-0002-2149-5060]{Kiana Setoodehnia}
\affiliation{Triangle Universities Nuclear Laboratory (TUNL), Duke University, Durham, North Carolina 27708, USA}
\affiliation{Department of Physics, Duke University, Durham, North Carolina 27708, USA}

\author[0000-0002-1194-2920]{Caleb Marshall}
\affiliation{Triangle Universities Nuclear Laboratory (TUNL), Duke University, Durham, North Carolina 27708, USA}
\affiliation{Department of Physics \& Astronomy, University of North Carolina at Chapel Hill, NC 27599-3255, USA}

\author[0000-0002-6695-9359]{Peter Mohr}
\affiliation{HUN-REN Institute for Nuclear Research (Atomki), H-4001 Debrecen, Hungary}

\author[0000-0003-2197-0797]{Athanasios Psaltis}
\affiliation{Triangle Universities Nuclear Laboratory (TUNL), Duke University, Durham, North Carolina 27708, USA}
\affiliation{Department of Physics, North Carolina State University, Raleigh, NC 27695, USA}

\correspondingauthor{Christian Iliadis}
\email{iliadis@unc.edu}



\begin{abstract}
This work describes the formalism for estimating thermonuclear reaction rates for astrophysical applications, emphasizing modern statistical approaches such as Monte-Carlo sampling and Bayesian models. We discuss related topics including the calculation of resonance energies from nuclear Q values, indirect estimates of particle partial widths, and matching of reaction rates at elevated temperatures to statistical-model results. We have evaluated available experimental data on cross sections, resonance energies and strengths, partial widths, lifetimes, spin-parities, and spectroscopic factors. Based on these results, we have estimated numerical values of 78 experimental charged-particle thermonuclear reaction rates for target nuclei in the A = 2 to 40 mass region, for temperatures ranging from 1~MK to 10~GK. For each reaction, three rate values are provided: low, median, and high, corresponding to the 16th, 50th, and 84th percentiles, respectively, of the cumulative reaction rate probability density distribution. Additionally, we present the factor uncertainty of each rate at each temperature grid point. These results enable users to sample the reaction rate probability density in nucleosynthesis calculations, facilitating uncertainty estimates of nuclidic abundances. The rates presented here refer to their laboratory values. For use in stellar model simulations, these values need to be corrected for the effects of thermal excitations of the interacting nuclei. For each reaction, we include graphs that illustrate the fractional contributions to the overall reaction rate along with the associated uncertainty. These visuals are designed to assist both stellar modelers and nuclear experimentalists by identifying the primary sources of rate uncertainty at specific stellar temperatures. A graphical comparison with earlier Monte-Carlo rates is also provided.
\end{abstract}





\section{Introduction} \label{sec:intro}
Thermonuclear reaction rates are essential ingredients for predictive modeling of stellar structure, evolution, and explosions, as well as for models of the early universe. Since nuclear reactions generate the energy that makes stars shine and are responsible for the synthesis of the elements, spectroscopic or photometric astronomical observations cannot be explained without knowing the rates of the thermonuclear reactions. Of particular importance to stellar modelers is the convenient access to state-of-the-art evaluated rates for a large number of relevant nuclear reactions. Previous  evaluations of nuclear reaction rates have been given, e.g., by \citet{Fowler:1967tu,CF88,NACRE,Iliadis2001,Descouvemont2004,RevModPhys.83.195,NACRE2013}. Most of these reaction rates were based on information derived from nuclear physics experiments and, therefore, are distinct from those estimated from nuclear theory, e.g., Hauser-Feshbach theory \citep{RAUSCHER20001} or the nuclear shell model \citep{Herndl-1995}. 

From a historical perspective, the incorporation of Fowler’s rates \citep{Fowler:1967tu,CF88} into stellar models provided a solid nuclear physics foundation for providing reasonable estimates of nuclear energy generation and nucleosynthesis. As observations became more quantitative, a major shortcoming of the evaluations mentioned above became apparent: the reported reaction rates had no rigorous statistical meaning, in the sense that they were not derived from probability density functions. With the publication of each evaluation, the nuclear data were updated, but the reaction rates were still computed using techniques developed prior to 1988.

Thermonuclear reaction rates are highly complex quantities that are derived from a multitude of nuclear physics input that is painstakingly extracted from laboratory measurements (e.g., resonance energies and strengths, nonresonant cross sections, transfer spectroscopic factors, and so on). Modern techniques to estimate reaction rates from experimental nuclear physics input have employed Monte-Carlo sampling methods \citep{Longland:2010is} and Bayesian techniques \citep{iliadis16,odell2022}. These have a major advantage with respect to earlier results: their recommended values and associated uncertainties\footnote{Throughout this work, we use the expression ``uncertainty'' for a parameter describing the dispersion of a measured value that can, at least in principle, be described by a suitable probability density function. In contradistinction, we use the expression ``error'' when we think that a mistake has been made.} can be quantified in terms of probability densities. Some of these methods were used in the 2010 evaluation of Monte-Carlo-based experimental reaction rates \citep{ILIADIS2010b,ILIADIS2010c,ILIADIS2010d}, hereafter labeled as ``ETR10.'' 
 
We considered the available experimental data for nuclear interactions of protons and $\alpha$ particles with target nuclei in the $A$ $\le$ $40$ region, and present statistically meaningful estimates of their thermonuclear rates, including uncertainties, following the ideas presented in \citet{Longland:2010is} and \citet{iliadis16}. This information is indispensable for assessing uncertainties of abundances or nuclear energy generation in astrophysical simulations. Table~\ref{tab:overview} summarizes all reactions for which we present experimental thermonuclear rates. Our numerical results are presented on a temperature grid from $1$~MK to $10$~GK (see Appendix~\ref{sec:ratetables}). 
\startlongtable
\begin{deluxetable*}{lcr}
\tablecaption{Summary of reactions with Monte-Carlo-based thermonuclear rates.
\label{tab:overview}} 
\tablewidth{\columnwidth}
\tablecolumns{3}
\tablehead{
Reaction & $Q_{nu}$~(keV)\tablenotemark{a} & Table
}
\startdata
D(p,$\gamma$)$^3$He\tablenotemark{b}               &    $5493.428$\tablenotemark{c}          &  \ref{tab:dpg} \\
D(d,n)$^3$He\tablenotemark{b}                      &    $3268.862$\tablenotemark{c}          &  \ref{tab:ddn} \\
D(d,p)$^3$H\tablenotemark{b}                       &    $4032.664$\tablenotemark{c}          &  \ref{tab:ddp} \\
$^3$H(d,n)$^4$He\tablenotemark{b}                  &    $17589.253$\tablenotemark{c}         &  \ref{tab:tdn} \\
$^3$He(d,p)$^4$He\tablenotemark{b}                 &    $18353.054$\tablenotemark{c}         &  \ref{tab:3hedp} \\
$^3$He($^3$He,2p)$^4$He\tablenotemark{b}           &    $12859.580$\tablenotemark{c}         &  \ref{tab:3he3he} \\
$^3$He($\alpha$,$\gamma$)$^7$Be\tablenotemark{b}   &    $1586.892\pm0.070$                   &  \ref{tab:3heag} \\
$^7$Be(n,p)$^7$Li\tablenotemark{b}                 &    $1644.422\pm0.071$                   &  \ref{tab:7benp} \\
\enddata
\tablenotetext{a}{Nuclear $Q$-value, calculated using atomic masses listed in \citet{wang2021} and the total electronic binding energy estimated from Equation~(A4) of \citet{RevModPhys.75.1021}; see Section~\ref{sec:massq} and \citet{Iliadis:2019ch}.}
\tablenotetext{b}{Bayesian rate.}
\tablenotetext{c}{Uncertainty less than $5$~eV, see \citet{wang2021}.}
\tablenotetext{d}{This value is the nuclear proton separation energy used for calculating resonance energies from excitation energies. To calculate the actual $Q$ value, subtract the energy of the first excited state in $^{26}$Al ($E_x$ $=$ $228.305\pm0.013$~keV; see ENSDF).}
\tablenotetext{e}{Based on the $^{24}$Si mass excess from \citet{Puentes_2022}.}
\tablenotetext{f}{Based on the $^{36}$Ca mass excess from \citet{Surbrook-2021}.}
\end{deluxetable*}

As was the case in previous evaluations, we restrict our considerations to nondegenerate, nonrelativistic circumstances for the interacting nuclei. We also assume that the Maxwell-Boltzmann distribution holds for the relative velocities of the interacting particles. Our reaction rates are appropriate for {\it bare nuclei in the laboratory}. In other words, when used in astrophysical model simulations, the results presented here need to be corrected, when appropriate, for electron screening at elevated densities, and thermal target excitations at elevated temperatures. 

\section{Reaction rate formalism} \label{sec:rateform}
The formalism to calculate thermonuclear reaction rates can be found, e.g., in \citet{Iliadis_2015}, and has also been summarized by \citet{Longland:2010is}. We are not repeating it here, but will focus on the main results. In the following, all energies refer to the center-of-mass coordinate system, unless explicitly mentioned otherwise. 

The total laboratory thermonuclear rate (in units of cm$^3$~mol$^{-1}$~s$^{-1}$) for a reaction involving two nuclei, $0$ and $1$, in the entrance channel at a given temperature $T$ is given by 
\begin{equation}
\begin{split}
N_A\langle \sigma v \rangle_{01} = \frac{3.7318\cdot 10^{10}}{T_9^{3/2}} \sqrt{\frac{M_0 + M_1}{M_0 M_1}} \\
\times \int_0^\infty E\,\sigma(E)\,e^{-11.605\,E/T_9}\,dE 
\end{split}
\label{eq:generalrate}
\end{equation}
where the center-of-mass energy $E$ is in units of MeV, the temperature $T_9$ is in GK ($T_9 \equiv T/10^9$~K), the nuclear masses $M_i$ $=$ $m_i/m_u$ are in atomic mass units (u), and the cross section $\sigma$ is in b ($1~\mathrm{b} \equiv 10^{-24}~\mathrm{cm}^2$); $N_A$ denotes the Avogadro constant. The reaction rate is determined by the absolute magnitude and the energy dependence of the nuclear reaction cross section, $\sigma(E)$. Based on the energy-dependence of $\sigma(E)$, a number of different specialized expressions and procedures can be derived for individual contributions to the total reaction rate. 

\subsection{Nonresonant reaction rates} \label{sec:nonrates}
Nonresonant cross sections vary smoothly with energy and can be converted into the astrophysical $S$ factor, defined by
\begin{equation}
S(E) \equiv E\,e^{2\pi\eta}\,\sigma(E)	
\label{eq:sfactor}
\end{equation}
This definition removes the $1/E$ dependence of nuclear cross sections and the $s$-wave Coulomb barrier transmission probability, $e^{-2\pi\eta}$ (i.e., the Gamow factor), from the cross section. The astrophysical $S$ factor, therefore, exhibits a much weaker energy dependence. The Sommerfeld parameter, $\eta$, is numerically given by
\begin{equation}
2\pi\eta  = 0.989510\,Z_0 Z_1 \sqrt{\frac{M_0 M_1}{M_0 + M_1}\frac{1}{E}}     
\label{eq:Sommerfeld}
\end{equation}
where $Z_i$ are the charges of nuclei $0$ and $1$. When the $S$ factor depends weakly on energy, substitution of Equation~(\ref{eq:sfactor}) into Equation~(\ref{eq:generalrate}) yields an integrand whose energy dependence is dominated on the low-energy side by the penetrability through the Coulomb barrier, and on the high-energy side by the Maxwell-Boltzmann distribution. The integrand is referred to as the Gamow peak and represents the effective energy range of stellar burning for a nonresonant reaction at a given temperature and nuclear masses and charges. 

Once the $S$ factor has been estimated over an appropriate range of energies, the thermonuclear reaction rates can be computed by numerical integration of Equation~(\ref{eq:generalrate}). The energy dependence of the nonresonant $S$ factor can be estimated using microscopic nuclear models, R-matrix theory, potential models, or any other suitable method. For example, in this work we computed the $S$ factor for direct radiative capture reactions, i.e., for (p,$\gamma$) or ($\alpha$,$\gamma$), using the results of a potential model normalized to experimental data, as discussed in Section~\ref{sec:dc}.

\subsection{Isolated-resonance rates} \label{sec:broadrates}
The cross section of an isolated resonance can be described by the one-level Breit-Wigner formula. For the cross section (in units of b) of a resonance located at a center-of-mass energy $E_r$, we find
\begin{equation}
\begin{split}
\sigma_{\mathrm{BW}}(E) = 0.6566 \frac{\omega}{E} \frac{M_0+M_1}{M_0 M_1} \\
\times \frac{\Gamma_a(E)\Gamma_b(E + Q-E_f)}{(E_r - E)^2 + \Gamma(E)^2/4}	
\end{split}
\label{eq:breitwignercrosssection}
\end{equation}
where all energies and widths are in units of MeV; $\omega$ $\equiv$ $(2J+1)/[(2j_0+1)(2j_1+1)]$ is the spin factor, where $J$, $j_0$, and $j_1$ denote the spins of the resonance, projectile, and target nucleus, respectively; $Q$ is the reaction Q-value, $E_f$ is the energy of the final state in the residual nucleus, and $\Gamma_a$, $\Gamma_b$, and $\Gamma$ are the partial widths for the entrance ($a$) and exit ($b$) channel, and the total resonance width (i.e., the sum of all partial widths, $\Gamma$ $=$ $\Gamma_a$ $+$ $\Gamma_b$ $+$ $...$), respectively. In the above expression, and throughout this work, the energy-dependent partial widths denote ``observed" rather than ``formal" quantities \citep{lane58}. If the partial widths of the resonance have been measured, Equation~(\ref{eq:breitwignercrosssection}) can be substituted into Equation~(\ref{eq:generalrate}) to compute the reaction rates by numerical integration. Notice that Equation~(\ref{eq:breitwignercrosssection}) applies equally to subthreshold states, i.e., when $E_r<0$.


The Breit-Wigner cross-section for an isolated resonance, as presented in Equation~\ref{eq:breitwignercrosssection}, exhibits slight differences in its energy dependence compared to a comprehensive R-Matrix calculation of a single isolated resonance, as outlined in Section~XII of \citet{lane58}. Specifically, we employ the ``Thomas Approximation'' \citep{Thomas1951}, which estimates the level shift using a linear function. Practically, this approximation leads to a minor error when extrapolating the cross section of a broad resonance, since the energy dependence of the level shift is not fully accounted for. However, for low-energy resonances critical in astrophysically significant reactions, the level shift varies only slowly with energy. Our tests with (p,$\gamma$) and ($\alpha$,$\gamma$) resonances have indicated that the Thomas approximation introduces an error of less than 10\% in the reaction rate for a single isolated resonance.

If the partial widths are not directly known, they may be estimated using experimental nuclear structure information. The particle partial width for a given level, $\lambda$, and channel, $c$, can be written as \citep{RevModPhys.32.519}
\begin{equation}
\Gamma_{\lambda c} = 2 P_c \gamma_{\lambda c}^2 = 2  P_c \frac{\hslash^2}{\mu R^2} \theta_{\lambda c}^2
\label{eq:redtheta}
\end{equation}
where $\gamma_{\lambda c}^2$ and $\theta_{\lambda c}^2$ are the reduced width and dimensionless reduced width, respectively, $\mu$ $=$ $m_0m_1/(m_0+m_1)$ is the reduced mass, $R$ is the channel radius (see later), and $P_c$ is the penetration factor. For a single-nucleon channel, the proton or neutron partial width can be expressed as \citep{ILIADIS1997,Iliadis_2015}
\begin{align}
\Gamma_{\lambda c} = 2  P_c \frac{\hslash^2}{\mu R^2} C^2S \theta_{p c}^2    
\label{eq:partpartwidth}
\end{align}
where $C$ denotes the isospin Clebsch-Gordan coefficient (Appendix~\ref{sec:CGC}), $S$ is the nucleon spectroscopic factor, and $\theta_{p c}^2$ denotes the dimensionless single-particle reduced width. This expression has the following intuitive meaning. The probability ($\Gamma_{\lambda c}$, in energy units) of decay (or formation) of a resonance by proton or neutron emission (or absorption) is equal to the product of three probabilities: (i) the probability that the nucleons in the compound nucleus arrange themselves according to a ``target $+$ projectile'' configuration ($C^2S$), (ii) the probability that the single nucleon appears at the nuclear surface ($\theta_{p c}^2$), and (iii) the probability that the nucleon tunnels through the Coulomb and centripetal barriers ($P_c$). The partial widths of protons and $\alpha$ particles are strongly energy dependent through the penetration factor, $P_c$, which is computed from Coulomb wave functions \citep{lane58}.

Equation~(\ref{eq:partpartwidth}) has not been applied consistently, even in the recent literature. Sometimes, the factor of $2$ is replaced by a factor of $3$, or the quantity $\theta_{p c}^2$ is set equal to unity. In such cases, unnecessary errors are introduced in the calculation of the particle partial width \citep[for more information, see][]{ILIADIS1997}. In Table~\ref{tab:redwidths}, we provide for protons numerical values of $\theta_{p c}^2$ versus bombarding energy and target mass number for orbital angular momenta of $\ell$ $=$ $0$, $1$, $2$, and $3$. Another source of confusion in the literature arises from the fact that the product $C^2S$, rather than $S$, is referred to as the ``spectroscopic factor.'' Unless otherwise specified, we also adopted this terminology, as the product is the key quantity involved in Equation~(\ref{eq:partpartwidth}). 

The above expression provides a versatile way to compute a particle partial width when the spectroscopic factor has been measured in a transfer (stripping) reaction. If the value of $C^2S$ for a level of interest is unknown, but the spectroscopic factor of the mirror state (or, more generally, components of the same isospin multiplet) has been measured, Equation~(\ref{eq:partpartwidth}) can still be used to estimate the particle partial width \citep{Iliadis_1999}. See Appendix~\ref{sec:mirror} for more information. Examples for estimating proton partial widths from spectroscopic factors are given in Appendix~\ref{sec:dimwidths}.

For accurate estimates of particle partial widths according to Equation~(\ref{eq:partpartwidth}), the parameters used to derive $C^2S$, $P_c$, and $\theta_{p c}^2$ must be internally consistent. For example, many distorted-wave Born approximation (DWBA) analyses have generated the final-state wave function using a Woods-Saxon nuclear potential radius parameter of $r_0$ $=$ $1.25$~fm, a diffuseness of $a$ $=$ $0.65$~fm, and a Coulomb potential radius parameter of $r_{c0}$ $=$ $1.25$~fm. The penetration factor is commonly computed using a channel radius of $R$ $=$ $1.25 (A_0^{1/3} + A_1^{1/3})$~fm, where $A_0$ and $A_1$ denote the (integer) mass numbers of the projectile and target, respectively. Internal consistency demands that $\theta_{p c}^2$ is computed with exactly the same channel and Woods-Saxon parameters used in the estimation of $C^2S$ and $P_c$. Otherwise, an additional error is introduced in the derivation of the particle partial width. The uncertainty of indirectly derived particle partial widths will be addressed in Section~\ref{sec:uncpartwidth}.


\subsection{Narrow resonance rates} \label{sec:narrowrates}
When the partial widths entering in Equation~(\ref{eq:breitwignercrosssection}) are known, the numerical integration of Equation~(\ref{eq:generalrate}) will provide a reaction rate that takes contributions from all energy regions of the resonance into account. However, for many resonances, neither the cross section has been measured nor the partial widths estimated. Instead, all that can be measured is the integrated resonance cross section. This situation is experimentally unavoidable if the total resonance width is much smaller than the resolution of the beam and the thickness of the target. The measured cross section integral is proportional to the resonance strength, defined by
\begin{equation}
\omega\gamma \equiv \frac{2J+1}{(2j_0+1)(2j_1+1)} \frac{\Gamma_a \Gamma_b}{\Gamma}   
\label{eq:resstrength}
\end{equation}
which can be used to estimate the resonance contribution to the reaction rate. If, for a particular resonance, we can assume that the partial widths in Equation~(\ref{eq:breitwignercrosssection}) do not vary significantly with energy over the total resonance width, Equation~(\ref{eq:generalrate}) reduces to
\begin{equation}
\begin{split}
N_A \langle {\sigma v} \rangle = \frac{1.5399 \cdot 10^{11}}{T_9^{3/2}}\left( \frac{M_0 + M_1}{M_0 M_1 }\right)^{3/2} \\
\times \, \omega\gamma \, e^{-11.605\,E_r/T_9}    
\end{split}
\label{eq:narresrateexpr}
\end{equation}
where the center-of-mass resonance energy, $E_r$, and resonance strength, $\omega\gamma$, are both in units of MeV. When the astrophysically important energy range contains more than a single narrow resonance, the combined rate is given by the incoherent sum of their individual rate contributions. According to Equation~(\ref{eq:narresrateexpr}), a given narrow resonance achieves its maximum contribution to the total rate at a temperature of $T_9^{max}$ $=$ $7.737 E_r$, where the resonance energy is in units of MeV. Experimental resonance strengths will be discussed in Section~\ref{sec:wg}.

The price to pay for assuming energy-independent partial widths is that Equation~(\ref{eq:narresrateexpr}) only considers the rate contribution at the resonance energy, $E_r$. This quantity appears exponentially in the above expression, implying that its experimental value must be accurately known; otherwise, it will significantly contribute to the uncertainty in the reaction rate.

We address now the question, ``When can a resonance be considered narrow for the purpose of estimating the reaction rate?'' An experimentalist would typically consider a resonance with a total width of, say, 0.15 eV as ``narrow." However, for astrophysical purposes, caution is necessary. An example of a resonance with this total width ($0.15$~eV) is presented in Figure~10 of \citet{POWELL1999349} for the $^{24}$Mg(p,$\gamma$)$^{25}$Al reaction. At low temperatures ($\le 30$~MK), where the Gamow peak has shifted far below the resonance energy of $E_r$ $=$ $223$~keV, the contribution of the resonance tail exceeds the rate computed using the approximation of Equation~(\ref{eq:narresrateexpr}) by several orders of magnitude. In practice, the impact of this effect for a given narrow resonance is frequently lessened, because contributions from either other low-energy resonances or nonresonant processes dominate over the tail contribution of the resonance in question. Nevertheless, it is important to assess the impact of the disregarded resonance tail when using Equation~(\ref{eq:narresrateexpr}). We defined a ``narrow resonance'' as one whose total contribution to the reaction rate is accurately represented by Equation~(\ref{eq:narresrateexpr}), i.e., based solely on the knowledge of its energy, $E_r$, and strength, $\omega\gamma$.

When a total width has contributions from two channels only (e.g., the proton width, $\Gamma_p$, and $\gamma$-ray partial width, $\Gamma_{\gamma}$), at sufficiently low energies the condition $\Gamma_p$ $\ll$ $\Gamma_{\gamma}$ (or $\Gamma$ $\approx$ $\Gamma_{\gamma}$) applies, and the resonance strength becomes $\omega\gamma$ $\approx$ $\omega \Gamma_p$. Conversely, at sufficiently high energies, the proton width typically exceeds the $\gamma$-ray partial width significantly, leading to $\omega\gamma$ $\approx$ $\omega \Gamma_{\gamma}$. These approximations are frequently applicable in practice.

%
%

\section{Reaction rates and modern statistics} \label{sec:statrates}

Experimental thermonuclear reaction rates are quantities derived from measured nuclear properties (e.g., cross sections, resonance energies, strengths, and partial widths,  excitation energies, spectroscopic factors, etc.) and, therefore, are subject to measurement uncertainties. How does one then estimate the reaction rate uncertainties based on the uncertainties of all measured nuclear input quantities? One answer is analytical error propagation, a path that was first explored in the context of reaction rates by \citet{THOMPSON1999}. Their method worked reasonably well when the uncertainties in the nuclear input were small. However, this approach proved inadequate when the uncertainties of some measured quantities became significant, a common occurrance for many nuclear reactions of astrophysical interest.

When error propagation becomes analytically intractable, a modern approach necessarily resorts to Monte-Carlo techniques. This method was introduced for thermonuclear reaction rates in \citet{Longland:2010is} and proved highly successful.  The first major advantage of Monte-Carlo-based reaction rates is their robustness against the size of nuclear physics input uncertainties. The method remains effective regardless of how large these uncertainties are, as long as there is a physically meaningful probability density function for each nuclear input quantity. The second major advantage of the Monte-Carlo method is its fundamental grounding in the central limit theorem of statistics. Measured nuclear quantities can be assigned Gaussian (or lognormal) probability densities when their total uncertainty is determined by a sum (or product) of individual contributions. The central limit theorem also explains why an ensemble of reduced widths is described by a Porter-Thomas probability density distribution; see \citet{RevModPhys.81.539,Longland:2010is,Pogrebnyak2013} for details.

The Monte-Carlo method is most useful when the total rate has a significant resonant contribution. In some cases, particularly with light target nuclei, the total cross section within the astrophysically relevant energy range is determined solely by nonresonant contributions. In such cases, other techniques must be employed to estimate statistically-rigorous reaction rates, e.g., Bayesian hierarchical models (see Section~\ref{sec:bayesrates}). 

\begin{figure*}[hbt!]
\centering
\includegraphics[width=0.9\linewidth]{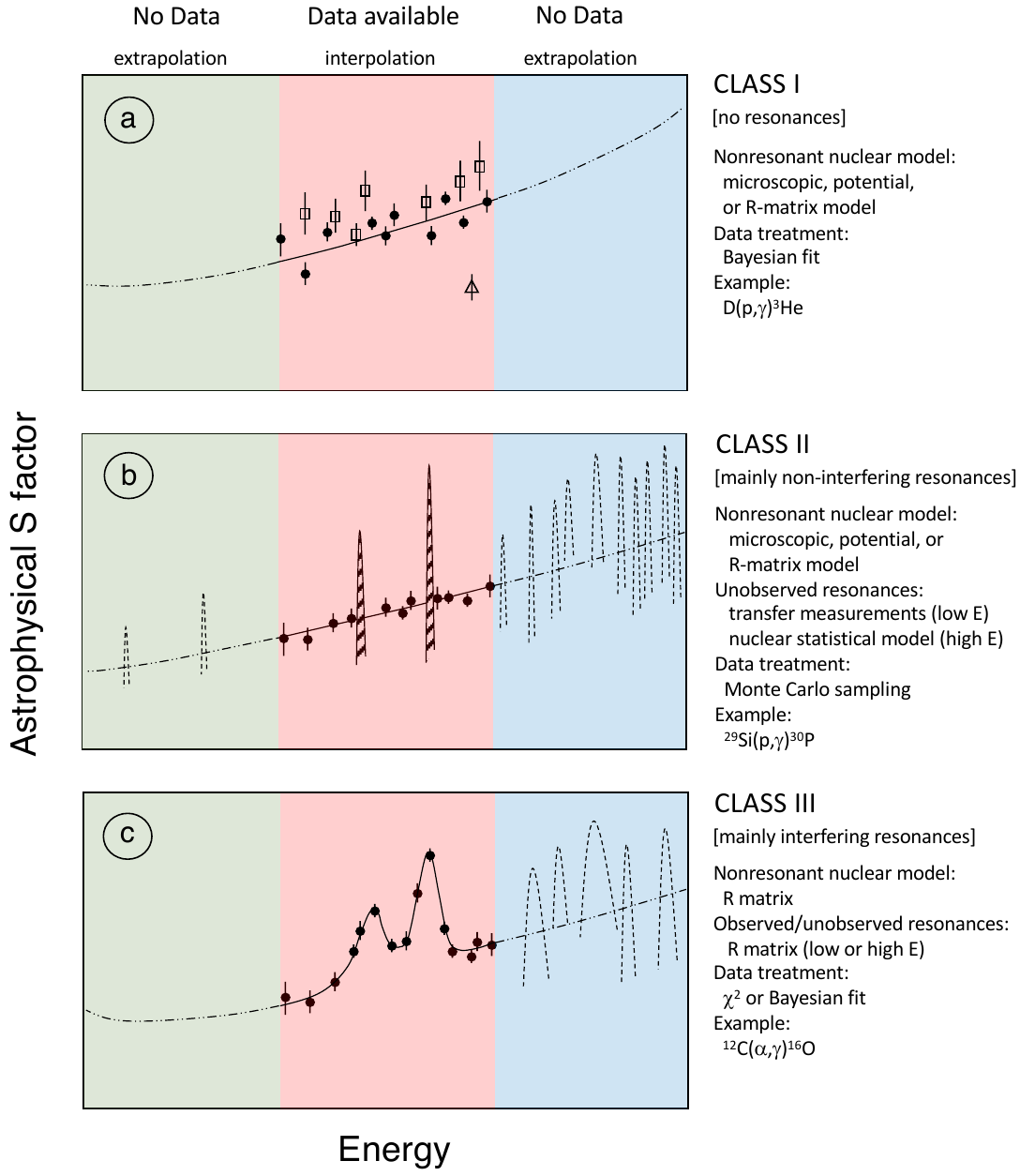}
\caption{
Idealized picture of various $S$-factor contributions to charged-particle reaction rates. In each panel, the energy range is divided into three regions. Green, Left: Directly measured $S$ factor data are not available because of the small transmission through the Coulomb barrier. Red, Middle: Region where $S$ factor data have been obtained in the laboratory. Blue, right: No data are available because of experimental limitations in the attainable beam energy. Three fundamentally different situations are encountered in practice, depending on the nature of contributions in the middle (red) region. (a) Class I reactions: nonresonant $S$ factor contributions only. (b) Class II reactions: main contributions from non-interfering (i.e., narrow) resonances. (c) Class III reactions: main contribution from interfering (i.e., broad) resonances. See text. 
}
\label{fig:idealpic}
\end{figure*}

The different statistical methods to analyze $S$ factor data, depending on the circumstances, are explained in Figure~\ref{fig:idealpic}. In each panel, the energy region is divided into three areas. In the energy range to the left (green), no direct $S$-factor data exist because of the low transmission through the Coulomb barrier. The middle (red) range contains all of the measured data. On the right-hand side (blue), no data are available, e.g., because those energies may not be attainable at low-energy accelerator facilities.

Panel (a) illustrates the scenario for what we term ``Class I reactions,'' where resonances are absent and the total rate is given by nonresonant contributions only. In this case, $S$-factor data have been measured over a wide energy range. Typically, results are available from different experiments (i.e., depicted by the solid circles and open squares in Figure~\ref{fig:idealpic}), while statistical and systematic uncertainties are separately reported for all data. The theoretical energy dependence of the $S$ factor may be calculated from a potential model, a microscopic model, or another suitable prescription. For Class I reactions, we use reaction rates based on fitting the theoretical $S$ factor to the data using Bayesian hierarchical models \citep{iliadis16}. As will be explained in Section~\ref{sec:bayesrates}, this method has inherent advantages over the traditional $\chi^2$ fitting technique. An example of the Bayesian analysis of a Class I reaction, D(p,$\gamma$)$^3$He, is provided in \citet{moscoso2021}.

Panel (b) depicts the situation for cases where the total rate is dominated by contributions from mostly non-interfering (narrow) resonances (``Class II reactions''). In the middle region, two resonances are shaded, indicating that, for those only, the resonance energies and strengths (see Equation~(\ref{eq:resstrength})) have been measured. Additional, nonresonant, contributions appear between resonances. In the (green) region on the left-hand side, additional low-energy resonances may contribute to the total rate because they correspond to known levels in the compound nucleus. If these states are populated in a proton or $\alpha$-particle transfer reaction, their experimental spectroscopic factors can be used to calculate the particle partial width according to Equation~(\ref{eq:partpartwidth}),  thereby allowing for the estimation of their contribution to the total rate. For Class II reactions, the total rate must be computed from a large body of experimental nuclear physics input: resonance energies, strengths, partial widths, nonresonant contributions, etc. We evaluated the rates of such reactions using the Monte-Carlo sampling method \citep{Longland:2010is,Iliadis2015}, as will be explained in Section~\ref{sec:mc}. An example for the Monte-Carlo analysis of a Class II reaction, $^{29}$Si(p,$\gamma$)$^{30}$P, can be found in \citet{Downen2022}. In the right-hand (blue) region, the contribution of the many unobserved higher-energy resonances must also be estimated. We calculated the rates at these high temperatures by resorting to the statistical (Hauser-Feshbach) nuclear reaction model, and by matching this contribution at a temperature where the total rate is fully determined by the available data \citep{PhysRevC.78.025805}. This will be discussed in more detail in Section~\ref{sec:tmatch}.

The most challenging situation, in terms of applying a rigorous statistical technique to the data analysis, is depicted in panel (c). Here, the total rates are dominated by interfering amplitudes from broad resonances and nonresonant processes. We refer to such cases as ``Class III reactions.'' Since the amplitudes of different contributions interfere, one cannot simply describe each resonance by the Breit-Wigner formula (Equation~(\ref{eq:breitwignercrosssection})) and add their contributions incoherently, as is done for Class II reactions. A suitable nuclear model to apply in these situations is the phenomenological R-matrix theory \citep{lane58,Descouvemont_2010}. R-matrix fits of the data have been performed so far almost exclusively using traditional (i.e., $\chi^2$) methods. A prominent example for a Class III reaction is $^{12}$C($\alpha$,$\gamma$)$^{16}$O and its ($\chi^2$) analysis was presented in \citet{RevModPhys.89.035007}. We predict that Class III reactions will be analyzed in the future by implementing R-matrix theory into Bayesian hierarchical models. This has been achieved mostly in the simplest of cases, when the rate is dominated by a single broad resonance \citep{desouza2019a,desouza2019b,PhysRevC.105.014625}; see also \citet{PhysRevC.106.055803}. For the more complex cases, such as $^{12}$C($\alpha$,$\gamma$)$^{16}$O, $^{12}$C(p,$\gamma$)$^{13}$N, or $^{14}$N(p,$\gamma$)$^{15}$O, we do not present numerical results in this work, because their reaction rates have not yet been based on rigorous probability density functions. 

The above division of reactions into three different classes, based on the complexity of the statistical analysis, is highly idealized. In particular, the meanings of ``nonresonant`` and ``broad-resonance'' contributions overlap significantly. Nevertheless, Figure~\ref{fig:idealpic} demonstrates the key differences. In the following, we focus on Bayesian- and Monte-Carlo-based techniques to estimate rigorous thermonuclear reaction rates for Class I and II reactions, respectively.

\section{Bayesian-based reaction rates} \label{sec:bayesrates}

\subsection{Overview} \label{sec:bayesoverview}

The first study to present a general Bayesian model for estimating thermonuclear reaction rates was published in \citet{iliadis16}. Bayesian thermonuclear rates have so far been estimated for most of the key Big Bang nucleosynthesis reactions \citep[see][and references therein]{Iliadis_2020}.

The Bayesian approach has several advantages over other methods. It allows for a consistent propagation of uncertainties within a well-established statistical framework. Each fitting (model) parameter can be associated with a physically motivated probability density function (e.g., normal, lognormal, Poisson, or uniform distributions), eliminating the need for any implicit approximation by Gaussian probability densities. Bayesian techniques also allow for quantifying the selection of different models used to fit the data. Perhaps most importantly, the interpretation of the results is intuitive, in the sense that all model parameters are random variables (unlike in ``frequentist'' statistics) and their values can be estimated rigorously for any given coverage probability. The price to be paid is an added layer of complexity: the analysis must resort to numerical techniques (i.e., Markov Chain Monte Carlo algorithms), and one has to ensure that the results are numerically stable. In the following, we summarize the formalism as it applies to astrophysical $S$ factors and thermonuclear reaction rates. For more information, see \citet{iliadis16}.

When Bayes' theorem is applied to data, represented by the vector $y$, and model parameters, given by the vector $\theta$, it can be written as \citep{2017bmad}
\begin{equation}
    p(\theta|y) = \frac{\mathcal{L}(y|\theta)\pi(\theta)}{\int \mathcal{L}(y|\theta)\pi(\theta)d\theta}
    \label{eq:Bayes}
\end{equation}
All factors in this expression serve as probability densities: $\mathcal{L}(y|\theta)$ is the likelihood, i.e., the probability that the data, $y$, were obtained for given values of the model parameters, $\theta$; $\pi(\theta)$ is the prior, which represents our state of knowledge about each parameter before seeing the data; the product of the likelihood and the prior defines the posterior, $p(\theta|y)$. This quantity is of primary interest because it contains the specific information we seek: the probability of obtaining the values of a specific set of model parameters given the data. The general meaning of Equation~(\ref{eq:Bayes}) is that the availability of new data updates our prior knowledge about the model parameters, and our revised knowledge is fully described by the posterior distribution. Once the posterior is calculated, it can be used to estimate model parameters for any given coverage probability, or to make predictions about physical quantities, such as cross sections or $S$ factors. The normalization factor in the denominator of Equation~(\ref{eq:Bayes}) is irrelevant when a specific Bayesian model is used for parameter estimation or value prediction, but it becomes important when performing Bayesian model selection.

\subsection{Likelihoods and priors}
Consider as a simple example an astrophysical $S$ factor subject to statistical uncertainties, $\sigma_{stat}$. If the probability density of each data point is given by a normal (i.e., Gaussian) distribution, the likelihood is given by
\begin{equation} 
    \mathcal{L}(S^{exp}|\theta)=\prod_{i=1}^N \frac{1}{\sigma_{stat,i}\sqrt{2\pi}}
    e^{-\frac{[S_i^{exp} - S(\theta)_i]^2}{2\sigma^2_{stat,i}}}
    \label{eq:likelihood}
\end{equation}
where the product runs over all data points, labeled by the index $i$. The theoretical $S$ factor, $S(\theta)_i$, is determined by a suitable nuclear reaction model (e.g., a microscopic model, a potential model, or R-matrix theory). The above likelihood represents a product of normal distributions, each with a mean of $S(\theta)_i$, and a standard deviation, $\sigma_{stat,i}$, given by the experimental statistical uncertainty of datum $i$. In symbolic notation, we can rewrite the above expression as
\begin{equation}
S_i^{exp} \sim N (S(\theta)_i, \sigma^2_{stat,i})
\label{eq:Si}
\end{equation}
where ``$N$'' denotes a normal probability density and the symbol ``$\sim$'' means ``has the probability distribution of.'' 

Each of the model parameters contained in the vector $\theta$ of Equation~(\ref{eq:Bayes}) requires a prior. Priors should be chosen to best represent the physics involved. For example, if all we know about a specific parameter, $\theta_j$, before seeing the data is that its value must lie somewhere in a region between zero and $\theta_{m}$, we can write the prior as
\begin{equation}
 \theta_j \sim U(0, \theta_{m})
 \label{eq:prior}
\end{equation}
where ``$U$'' denotes a uniform (i.e., constant) probability density between the boundaries. If the parameter in question can take values in excess of $\theta_{m}$, but with a probability inverse to its magnitude, we may choose instead a truncated normal prior
\begin{equation}
 \theta_j \sim T(0, \infty) N(0, \theta_{m}^2)
 \label{eq:prior2}
\end{equation}
which represents a normal distribution centered at zero with a standard deviation of $\theta_{m}$, multiplied by a truncation function, $T$, which suppresses samples outside of the interval defined by its arguments. When the data are of high quality, reasonable choices for the prior, say Equation~(\ref{eq:prior}) or (\ref{eq:prior2}), will have an insignificant effect on the posterior in Equation~(\ref{eq:Bayes}). We call such priors ``non-informative.'' If the prior impacts the posterior moderately or strongly, we call them ``moderately informative'' and ``highly informative'' priors, respectively. In Bayesian analyses, different reasonable choices of prior distributions should always be explored to assess their impact on the posterior inference and parameter estimation. For more information on particular choices of priors, see \citet{e19100555}. 

\subsection{Systematic uncertainties}\label{sec:sys}
Systematic uncertainties \citep{heinrich07} require careful consideration in the data analysis. We define systematic effects by the following attributes: they do not usually signal their existence by a larger fluctuation of the data; they are not reduced by combining the results from different measurements or by collecting more data; when the experiment is repeated, the presence of systematic effects may not produce different answers. In a nuclear physics experiment, systematic effects impact the overall normalization by shifting all points of a given data set into the same direction. They are correlated from data point to data point, in the sense that if one happened to know how to correct such an uncertainty for one data point, then one could calculate the correction for all other data points as well. Specifically, a reported systematic uncertainty of, say, $±5$\%, implies a systematic factor uncertainty of $f.u.$ $=$ $1.05$. The true value of the multiplicative normalization factor, $f$, is unknown at this stage.
However, we can reasonably assume that the expectation value of the normalization factor is unity, otherwise we would have corrected the data for the systematic effect.

A useful distribution for normalization factors is the lognormal probability density, which is characterized by two quantities: the lognormal location parameter, $\mu_L$, and the spread parameter, $\sigma_L$. The median value of the lognormal distribution is given by $x_{med} = e^{\mu_L}$, while the factor uncertainty, for a coverage probability of $68$\%, is $f.u.$ $=$ $e^{\sigma_L}$. Systematic effects can be included in a hierarchical Bayesian model as a highly informative, lognormal prior with a median of $x_{med} = 1.0$ (or $\mu_L$ $=$ $\ln x_{med}=0$), and a factor uncertainty given by the systematic uncertainty. In the above example, we would choose $f.u.$ $=$ $1.05$ (or $\sigma_L$ $=$ $\ln f.u.$ $=$ $\ln (1.05))$. The prior is then explicitly given by 
\begin{equation}
 \pi(f) = \frac{1}{\ln (f.u.) \sqrt{2\pi}f}e^{-\frac{[\ln f]^2}{2[\ln (f.u.)]^2}}
\label{eq:lognor}
\end{equation}
where ``$\ln$'' represents the natural logarithm. In symbolic notation, this is expressed as
\begin{equation}
f \sim LN(0, [\ln (f.u.)]^2)
\label{eq:lognor2}
\end{equation}
where ``$LN$'' denotes a lognormal probability density. The multiplicative normalization factor, $f$, is applied to the theory prediction, $S(\theta)$. For more information on this prior choice for systematic uncertainties, see \citet{iliadis16}. 

In conventional $\chi^2$ fitting, normalization factors are often treated as systematic shifts applied directly to the data. In Bayesian inference, by contrast, the reported data are considered fixed $-$ reflecting observed measurements $-$ and are not adjusted to accommodate model assumptions. However, data may still be rescaled for numerical stability, provided such transformations are properly accounted for in the likelihood. Instead, the true (but unknown) $S$ factor is multiplied by the normalization factor, $f$. This means that, during the fitting, each data set pulls on the true $S$ factor curve with a strength inversely proportional to the systematic uncertainty: a data set with a small systematic uncertainty will pull the true $S$ factor curve more strongly toward it compared to one with a large systematic uncertainty. This ``pulling'' is independent of the data set size: disregarding statistical uncertainties for a moment, a set consisting of a single datum will have the same weight in the fitting as one containing many data points, if both sets are described by the same factor uncertainty, $f.u.$  

\subsection{Additional effects}
The Bayesian model should be designed to closely reflect the data-generating processes. For example, in some cases it is apparent that the observed scatter of the measured data cannot be explained solely by the reported statistical uncertainties. This is illustrated in panel (a) of Figure~\ref{fig:idealpic}, where most of the data points indicated by the black circles miss the best-fit (black solid) line. This indicates the presence of additional sources of statistical uncertainty that were unknown to the experimenter. These have been termed ``extrinsic uncertainties'' in \citet{desouza2019b}. Since the observed scatter in the data contains information on the unreported statistical uncertainty, the Bayesian model can predict its magnitude for a given data set. When both statistical and extrinsic uncertainties are present in a measurement, the overall likelihood is given by a nested (hierarchical) expression. Using again the symbolic notation, we can replace Equation~(\ref{eq:Si}) by 
\begin{equation}
\label{eq:Si2}
 S^{\prime}_{i} \sim N(S(\theta)_{i},\sigma^2_{extr})
\end{equation}
\begin{equation}
\label{eq:Si3}
 S^{exp}_i \sim N(S^{\prime}_i,\sigma^2_{stat;i})
\end{equation}
Equations~(\ref{eq:Si2}) and (\ref{eq:Si3}) provide a hierarchical approach for constructing the overall likelihood. First, statistical uncertainties quantified by the standard deviation, $\sigma_{extr}$, of an assumed normal probability density, perturb the true (but unknown) value of the $S$ factor, $S(\theta)_i$, of a data point $i$, at a given energy to produce a value of $S^{\prime}_i$. Second, the latter value is perturbed, in turn, by the reported experimental statistical uncertainty, quantified by the standard deviation, $\sigma_{stat;i}$, of a normal probability density, to produce the observed value of $S^{exp}_i$. The above example demonstrates how any quantifiable experimental effect impacting the data can be readily incorporated into a hierarchical Bayesian model.
 

In other cases, individual data points appear outside the range suggested by the bulk of the data (``outliers''; see the open triangle in panel (a) of Figure~\ref{fig:idealpic}), or all data points of a specific experiment deviate systematically from the results of other experiments (``discrepant data''; see the open squares in panel (a) of Figure~\ref{fig:idealpic}). A robust algorithm to include such data in the analysis is presented in \citet{andreon2015bayesian}. The method treats the complete body of data as a mixture of two populations: one of supposedly correctly measured uncertainties, and another for which the reported uncertainty estimates are too optimistic. The membership to these two populations is described by  including additional parameters in the Bayesian model. The algorithm automatically identifies and reduces the weight of data points with overoptimistic uncertainties. Data points contribute more significantly to the posterior the smaller their uncertainty and the higher the probability that the reported uncertainty is accurate. Therefore, all data points are considered in the analysis, and none are discarded subjectively. The algorithm also quantifies the outlier probability of a given datum or experiment. For details, see \citet{iliadis16}.

\subsection{Markov Chain Monte Carlo}
Except in the simplest cases, the posterior of Equation~(\ref{eq:Bayes}) cannot be computed analytically, necessitating the use of numerical techniques. The implementation, since the late 1980s, of Markov Chain Monte Carlo (MCMC) methods into multi-parameter Bayesian hierarchical models is the main reason for the exponential growth in the use of Bayesian inference across many fields \citep{fienberg2006did}. The MCMC technique allowed for the computation of the posterior and arbitrary functions of the model parameters without the need for approximations. The main idea is, first, to construct a Markov chain whose stationary distribution is equal to the posterior and, second, to take a sufficiently long random walk by drawing samples from the Markov chain. Several related algorithms have been devised to solve this problem \citep[e.g., Metropolis-Hastings, Gibbs, Hamiltonian Monte Carlo, etc.; see][]{doi:10.1146/annurev-astro-082214-122339}.  

A number of software tools are publicly available and have been successfully used to estimate astrophysical $S$ factors from data. Examples are NIMBLE \citep{de_valpine_programming_2017}, JAGS \citep{Plummer2003}, DREAM \citep{terBraak:2008:DEM}, emcee \citep{2013PASP..125..306F}. All these packages require the user to define the physical model, likelihood, and priors. They also allow for the initialization, adaptation, and monitoring of the Markov chain. Running a model refers to generating random samples from the posterior distribution of all model parameters. The initial steps of the random walk (``burn-in''), before the Markov chain has reached convergence, must be discarded. When equilibrium has been achieved, a sufficient number of samples is drawn to ensure that Monte-Carlo fluctuations become negligible compared to the statistical, systematic, and extrinsic uncertainties.

The sampling returns the model parameters at each step of the Markov chain. For each set of parameters, a credible $S$ factor is obtained, from which the reaction rate at all temperatures can be found by numerical integration of Equations~(\ref{eq:generalrate}) and (\ref{eq:sfactor}). At a given temperature, the ensemble of rate values found in this manner represents the reaction rate probability density. The recommended rate is estimated by adopting the $50$th percentile of the rate probability density, while the rate uncertainty is found from the $16$th and $84$th percentiles (for a coverage probability of 68\%). See \citet{Longland:2010is} and Section~\ref{sec:mc}.

So far, astrophysical $S$ factors and reaction rates estimated using Bayesian hierarchical models have  been published for nine light-ion (Class I) reactions: D(p,$\gamma$)$^3$He, D(d,n)$^3$He, D(d,p)$^3$H, $^3$H(d,n)$^4$He, $^3$He(d,p)$^4$He, $^3$He($^3$He,2p)$^4$He, $^3$He($\alpha$,$\gamma$)$^7$Be, $^7$Be(n,p)$^7$Li, and $^{16}$O(p,$\gamma$)$^{17}$F. These are labeled by table note ``b'' in Table~\ref{tab:overview}. The rates of some of these reactions were re-calculated in the present work (see table notes in Appendix~\ref{sec:ratetables}), because the Bayesian model assumptions have evolved since the original formulation in \citet{iliadis16}, as will be explained below. However, the changes in these rates are very small ($\le 1.5$\%).

\subsection{Practical considerations} \label{sec:morebayes}
We now address a few issues of practical interest when estimating $S$ factors and thermonuclear rates using Bayesian hierarchical models.  

First, for a given reaction, the analysis should start with a collection of data from the entire relevant literature. It is important not to dismiss data simply because they were published some time ago or because they have larger uncertainties than other studies. Unless there is reason to believe that errors were made in a given publication, all published data should be taken into account in the data analysis.

Second, one should estimate the separate contributions of statistical and systematic uncertainties, when possible. This is particularly important because these enter in a very different manner in the Bayesian model, as discussed above. Sometimes, only the mean values of the $S$ factor or cross section are reported without any uncertainties whatsoever. Such results still provide useful information, but such $S$-factor data need to be implemented in a different manner in a Bayesian model. Instead of using Equation~(\ref{eq:lognor}), one can choose to scale the true (unknown) $S$ factor by a factor of $10^g$, e.g., using a mildly informative prior of 
\begin{equation}
   g \sim U(-k,+k)
   \label{eq:mildinfo}
\end{equation}
corresponding to a uniform prior between $-k$ and $+k$. In other words, the normalization factor, $10^g$, is varied by up to $k$ orders of magnitude up or down during the sampling. Such ``relative data'' provide only information on the energy dependence of the $S$ factor, but little information on its absolute normalization. This method was applied in the analysis of the D(p,$\gamma$)$^3$He reaction rate by \citet{moscoso2021} for experiments that did not report any uncertainties.

Third, a suitable physical model for the true (unknown) $S$ factor must be chosen in the data fitting. Reasonable choices are $S$ factors computed using microscopic models, potential models, or R-matrix theory, because all of these are grounded in nuclear reaction theory. In the simplest case, when microscopic-model $S$ factors are employed in the fitting, only a single physical parameter, i.e., the scale factor of the model $S$ factor used, enters into the Bayesian analysis \citep[see, e.g.,][]{gomez17}. A slightly modified strategy was pursued in \citet{moscoso2021}, who introduced two fitting parameters: a multiplicative scale factor, $a$, by which the microscopic model $S$ factor is multiplied, and an offset, $b$, according to
\begin{equation}
   S_{true} (E) = a S_{model} (E) + b
   \label{eq:modmicro}
\end{equation}
The rationale for this choice was that microscopic cross-section calculations represent model-based Hamiltonian approaches with a priori difficult-to-quantify uncertainties. Bias in the microscopic-theory $S$ factor may arise from the truncation of the set of basis states used to determine the matrix elements or the exclusion of operators in the Hamiltonian. These issues cannot be entirely disregarded, despite the fact that these type of model calculations are usually tuned to experimental scattering data and binding energies. The Bayesian fit returns estimates of the parameters $a$ and $b$, which can be used to assess the microscopic-model $S$-factor prediction (i.e., by checking how much the fit results differ from the values of $a$ $=$ $1$ and $b$ $=$ $0$).

Fourth, polynomials are sometimes employed in the data fitting, which has a number of advantages: a simple form, well-known properties, moderate flexibility of shapes, and computational ease of use. However, they also have limitations: poor interpolatory and extrapolatory properties, a poor trade-off between degree and shape, and a disregard of nuclear theory. In extreme cases, these issues may lead to numerically unstable models. The use of polynomials is permissible for several reactions occuring during primordial nucleosynthesis, when the cross section data sets fully cover the astrophysically-important energy region \citep{Mossa20,Yeh21}. In the case of the D(p,$\gamma$)$^3$He reaction, a comparison of results obtained with the two assumptions, a microscopic model prescription versus a polynomial, gave consistent results, both for the derived mean $S$ factor and the associated uncertainties \citep{moscoso2021}. However, the adoption of polynomials as physical model functions should be avoided if the $S$ factor needs to be extrapolated to energy regions devoid of data. 

Fifth, choices have to be made for the probability density functions of all likelihoods and priors in Equation~(\ref{eq:Bayes}). Choices of priors have already been discussed above. In most Bayesian hierarchical models, normal (Gaussian) likelihood functions are assumed for describing statistical uncertainties of  data points. Lognormal instead of normal likelihoods were adopted in \citet{iliadis16,gomez17}. This was justified on the grounds of the central limit theorem: since astrophysical $S$ factors are experimentally determined by products and ratios of several nuclear physics input quantities (e.g., measured signal intensities, incident beam charge, detection efficiencies, number of target nuclei, stopping powers, etc.), the probability density of the derived $S$ factor will tend toward a lognormal distribution. Furthermore, a normal density function predicts a finite probability for negative values of the random variable, which is unphysical for manifestly positive quantities, such as astrophysical $S$ factors. However, it was found that neither the $S$-factor fit nor the predicted parameters were sensitive to this choice for the reactions studied. Consequently, later publications employed normal likelihood functions \citep{desouza2019a,desouza2019b,desouza2020}. The reason for this insensitivity is that Gaussian functions closely approximate lognormal densities when the uncertainties are not too large (say, if the standard deviation is $\lesssim 10$\% of the mean value). 

Sixth, the type of numerical sampler needs to be considered carefully, depending on the complexity  and magnitude of the parameter space. Suppose, one wishes to analyze two independent data sets using a physical model $S$ factor as given by Equation~(\ref{eq:modmicro}). In this case, the data analysis will involve six parameters: two physical model parameters ($a$ and $b$), two parameters describing the extrinsic scatter in the two data sets, and two additional parameters for the normalization factors. The computation is relatively simple from a numerical point of view, and can be performed, e.g., with Gibbs samplers \citep{10.2307/2685208} that are implemented in software instruments such as JAGS \citep{Plummer2003} or NIMBLE \citep{de_valpine_programming_2017}. When the physical model becomes more complex, e.g., an R-matrix expression \citep[see][]{desouza2019a,PhysRevC.105.014625}, several of the parameters may be highly correlated, giving rise to unacceptably slow convergence of the Markov chains. In such a case, one must resort to more sophisticated samplers designed to handle the parameter correlations \citep{terBraak:2008:DEM,2013PASP..125..306F}.

\section{Monte-Carlo-based reaction rates} \label{sec:mc}

\subsection{Overview} \label{sec:mcoverview}
For the Class II reactions referred to in Section~\ref{sec:statrates}, a Monte-Carlo technique is used to estimate the reaction rates and associated uncertainties. This method was first described in detail by \citet{Longland:2010is}. We provide a short summary here, together with a discussion of improvements made since 2010. The calculations are performed using the computer code RatesMC \citep{RatesMC_2_3}, which is publicly available (Section~\ref{sec:git}) and widely used by many research groups.

The Monte-Carlo method relies on assigning probability density distributions to each input variable of the reaction rate calculation (e.g., resonance energies, partial widths, resonance strengths, and spectroscopic factors). Once these have been assigned, a random sample from each probability distribution is drawn. This forms the basis for a single reaction rate sample, which is calculated using Equations~\eqref{eq:generalrate} $-$ \eqref{eq:narresrateexpr}. The process of drawing random samples and calculating the reaction rate is repeated many times to obtain an ensemble of reaction rates. This represents the rate probability density, from which the recommended rate and uncertainties can be derived according to rigorously defined probabilities. 

The task involves assigning probability distributions to each type of input parameter, with the central limit theorem playing a significant role in guiding these choices. For example, resonance energies are usually derived either from the difference of excitation energy and particle separation energy ($Q$ value), or from an accelerator or magnet calibration. Low-energy resonances, in particular, are often determined using the former method. The central limit theorem states that the {\it sum} of many random variables is distributed normally, regardless of the form of the individual probability densities. This also implies a finite possibility for sampling a resonance with a negative energy. Such a case is treated consistently in our formalism as a subthreshold resonance, according to Equation~\eqref{eq:partpartwidth}. 

Resonance strengths, partial widths, or nonresonant $S$ factors, on the other hand, are derived from experimental yields. Conversion of a yield to a physical quantity involves the multiplication or division of several quantities (e.g., target thickness, detector efficiency, or beam current). In such cases, the central limit theorem states that the {\it product} of many random variables is distributed lognormally, regardless of the form of the individual probability densities. Since this probability density is only defined for positive values of the random variable, this choice is also consistent with the fact that these observables are manifestly positive. The lognormal distribution is given by
\begin{equation}
f(x)  = \frac{1}{\sigma\sqrt{2\pi}} \frac{1}{x} e^{-(\rm{ln}\, x-\mu)^2/(2\sigma^2)}, 0 < x < \infty\label{eq:logneq}
\end{equation}  
where $\mu$ and $\sigma$ are the lognormal location and shape parameter, respectively, and ``$\ln$'' denotes the natural logarithm. These parameters are related to the expectation (mean) value, $E[x]$, and variance, $V[x]$, by
\begin{equation}
\mu=\rm{ln}(E[x])-\frac{1}{2}\rm{ln} \rm{\Big( 1+\frac{V[x]}{E[x]^2}\Big)} 
\label{eq:mu1}
\end{equation}
\begin{equation}
\sigma=\sqrt{\rm{ln}\Big( 1+\frac{V[x]}{E[x]^2}\Big)}\label{eq:mu2}
\end{equation}
or, equivalently,
\begin{equation}
E[x] = e^{\mu+\sigma^2/2}  
\end{equation}
\begin{equation}
V[x] = e^{2\mu+\sigma^2}(e^{\sigma^2}-1)\label{eq:muinv}
\end{equation}
The median (50th percentile) of the lognormal distribution is given by $x_{med}$ $=$ $e^\mu$, while the factor uncertainty, for a coverage probability of 68\%, can be written as $f.u.$ $\equiv$ $e^{\mu+\sigma}/e^\mu$ $=$ $e^{\mu}/e^{\mu-\sigma}$ $=$ $e^\sigma$. 

Reported values and uncertainties (i.e., for a resonance strength, partial width, or nonresonant $S$ factor) can usually be equated with the expectation value, $E[x]$, and square root of the variance, $\sqrt{V[x]}$, respectively. The lognormal parameters are then found from Equations~\eqref{eq:mu1} and \eqref{eq:mu2}, allowing for random sampling according to Equation~\eqref{eq:logneq}. In some cases, reported values and uncertainties of a given quantity are presented in the literature as $x_{med}$ and $f.u.$. From these, the lognormal parameters can be determined with $\mu$ $=$ $\ln(x_{med})$ and $\sigma$ $=$ $\ln(f.u.)$.


The Monte-Carlo calculation of the reaction rate takes energy correlations explicitly into account. For example, when a narrow-resonance reaction rate is estimated from an energy and particle partial width of a resonance located near the particle threshold, and the resonance energy is sampled according to a normal probability density, the same sampled energy value must be used to estimate the particle partial width. 

In this work, we provide a tabulation of the recommended reaction rate for a given temperature, including the median value (50th percentile) of the sampled reaction-rate probability density. Additionally, we present ``low" and ``high" rates, corresponding to the 16th and 84th percentiles, respectively, which together cover a probability range of 68\%. It is important to note that these values do not represent strict lower or upper limits, as there remains a 32\% chance that the true (but unknown) reaction rate falls outside this interval. 

Reaction rates are often (but not always) distributed according to a lognormal distribution, for reasons discussed in detail by \citet{Longland:2010is}. If a rate is lognormally distributed, its median value is related to the lognormal location parameter by $x_{med}$ $=$ $e^\mu$, and the factor uncertainty of the rate will be given by $f.u.$ $=$ $e^\sigma$ (see Section \ref{sec:sys}). Alongside the reaction rates, the tables also include the values of $f.u.$ at each temperature. Unlike the tabulated values for the low, median, and high rates, which are derived from the percentiles of the actual rate probability density, the values of $f.u.$ are based on the lognormal approximation of the reaction rate density.




\subsection{Upper limits of partial widths} \label{sec:ul}
For many resonances near the particle threshold, the particle partial width is neither known from experiment nor from theory. In such cases, the particle partial width cannot be described by normal or lognormal probability density functions. Instead, a physically-motivated probability density can be derived using a fundamental assumption about the Gaussian orthogonal ensemble of random matrix theory. The dimensionless reduced width amplitude, $\theta_{\lambda c}$, appearing in Equation~\eqref{eq:redtheta}, is given by the sum of contributions from many different parts of the nucleon configuration space, with the sign and magnitude of a particular contribution being random from level to level and independent in sign and magnitude from all other parts. According to the central limit theorem, the probability density function of $\theta_{\lambda c}$ will then be approximately Gaussian, with an expectation value of zero. Consequently, the probability density function for $\theta^2_{\lambda c}$, i.e., the square of the amplitude, is given by a chi-squared distribution with one degree of freedom. 

This probability density function, also known as Porter-Thomas distribution, can be written as \citep{PorterThomas} 
\begin{equation}
  \label{eq:PT}
  f(\theta^2_{\lambda c}) = \frac{1}{ \sqrt{2 \pi \theta^2_{\lambda c} \left<\theta^2_{\lambda c} \right >}} e^{- \theta^2_{\lambda c}/ \left( 2 \left<\theta^2_{\lambda c} \right > \right )}
\end{equation}
The mean value, $\left<\theta^2_{\lambda c} \right >$, may vary with increasing excitation energy because the complexity of the compound nucleus will increase. Therefore, the quantity $\left<\theta^2_{\lambda c} \right >$ represents the local mean value, applicable to a given region of excitation energy. The above expression implies that the reduced widths for a single reaction channel, i.e., for a given nucleus and set of quantum numbers, vary by several orders of magnitude, with a higher probability for smaller values of the reduced width. 

Random sampling and implementation of Equation~\eqref{eq:PT} into the Monte-Carlo reaction rate formalism requires knowledge of $\left<\theta^2_{\lambda c} \right >$. This quantity is not predicted by random matrix theory, but values can be obtained by analyzing a large body of data. A preliminary estimate \citep{Longland:2010is} yielded values of $\langle\theta_p^2\rangle$ $=$ $0.0045$ and $\langle\theta_{\alpha}^2\rangle$ $=$ $0.010$ for protons and $\alpha$ particles, respectively, but disregarded any dependence on excitation energy or spin-parity. These results were used in \cite{ILIADIS2010b,ILIADIS2010c,ILIADIS2010d} without an assigned uncertainty. 

Subsequent work by \cite{Pogrebnyak2013} analyzed a larger data set, in the target mass range of $A$ $=$ $28$ $-$ $67$, using a maximum-likelihood method. They reported values of $\left<\theta^2_{\lambda c} \right >$, including uncertainties, as a function of target mass and charge, excitation energy, spin-parity, and orbital angular momentum. For $\alpha$ particles, we adopted a global mean value of $\langle\theta_{\alpha}^2\rangle$ $=$ $0.017$, with an uncertainty of a factor of $1.7$, consistent with the results presented in Figures~2 and 3 of \cite{Pogrebnyak2013}. For protons, the local mean values reveal a significant scatter, as displayed in their Figures~4 and 5. In the present work, we adopted a global value of $\langle\theta_{p}^2\rangle$ $=$ $0.001$, with a factor of $5$ uncertainty, which encompasses the reported local mean values. Using these numerical results, we described $\left<\theta^2_{\lambda c} \right >$ by a lognormal probability density, and sampled values of an unknown reduced width, $\theta^2_{\lambda c}$, according to Equation~\eqref{eq:PT}.

It must be emphasized that reduced widths follow a Porter–Thomas distribution only if the nuclear matrix elements have contributions from many different parts of the configuration space. This is clearly not the case for low-lying bound levels of near closed-shell character or $\alpha$-cluster states, where the matrix elements may be dominated by a few large contributions. However, such states frequently exhibit large values of $\theta^2_{\lambda c}$, and are, thus, likely to be observed in transfer reaction studies.  If a level is too weak to be observed in a transfer study, it can be safely assumed that its reduced width can be estimated by sampling from a Porter-Thomas distribution.

When an upper limit of the reduced width of a given level has been determined experimentally, we sample the Porter-Thomas distribution of Equation~\eqref{eq:PT} by truncating it at the experimental upper limit value\footnote{Versions of RatesMC between 2013 and 2022 used a rejection sampling technique that favored small values of $\Gamma_{\lambda c}$. Tests revealed that this approach artificially reduced the recommended rate contributions from upper limit resonances by approximately 20\%.}. For more details, see \cite{Longland:2010is}.

\subsection{Ambiguous spins and parities} \label{sec:jpi}
The rate contribution of a resonance that has not been directly measured can be found by estimating the involved partial widths (see Equation~\eqref{eq:breitwignercrosssection}).
If the $J^\pi$ value of the resonance is not known unambiguously, only a range of orbital angular momenta, $\ell$, can be determined. This ambiguity impacts the estimate of the particle partial width, according to Equation~\eqref{eq:redtheta}, through the $\ell$-dependence of the penetration factor, $P_c$. In such cases, we describe the ambiguous spin-parity assignment by a discrete probability distribution, where each allowed $J^\pi$ (or $\ell$) value is assigned a weight reflecting the probability of a given choice. This probability distribution is implemented in the Monte-Carlo sampling of the total rate. If such a resonance dominates the total rate, it may result in a multi-modal rather than a lognormal rate probability density. In such cases, the tabulated values of the low, median, and high rates, and the factor uncertainty, $f.u.$, need to be interpreted carefully (see Section~\ref{sec:mcoverview}).

The method of including ambiguous spin-parity assignments in the Monte-Carlo sampling of a total rate was first applied in \cite{Mohr2014}, to which the reader is referred for details.

\subsection{Correlations among measured resonance energies and strengths} \label{sec:corr}
In most experimental studies of nuclear reactions, the measured quantities are expected to be correlated. However, these correlations have not always been quantified in the literature, particularly in older work. Therefore, reasonable assumptions must be made to investigate the impact of such correlations on the total reaction rate. In the present work, we applied the procedures that were first discussed in \cite{Longland2017} and \cite{Longland2020} for resonance strength and resonance energy correlations, respectively. A brief summary is given below.

Resonance strengths and partial widths measured in a given experiment are expected to be correlated through a common normalization uncertainty. This arises from systematic uncertainties in the beam current, target composition, detector efficiency, etc. Frequently, strengths and widths are obtained relative to a well-known reference resonance. This issue will be discussed further in Section~\ref{sec:wg}. It is reasonable to assume that the systematic uncertainties in the strength or partial widths of this standard (and presumably strong) resonance dominate over the corresponding statistical ones. Furthermore, it can be assumed that the standard resonance has the smallest total uncertainty among the entire ensemble of resonances.

A correlation parameter, $\rho_j$, for resonance $j$ with measured strength, $\omega\gamma_j \pm \delta\omega\gamma_j$, can be estimated, according to \citet{Longland2017},\footnote{Equation~(18) of \cite{Longland2017} contains an error in the second equality: the correlation parameter is not equal to the ratio of factor uncertainties. Instead, the correct expression is given by Equation~(\ref{eq:correlation-rho}).} by
\begin{equation}
  \label{eq:correlation-rho}
  \rho_j = \frac{\delta\omega\gamma_r}{\omega\gamma_r}\frac{\omega\gamma_j}{\delta\omega\gamma_j} = \frac{(f.u.)_r - 1}{(f.u.)_j -1}
\end{equation}
where the subscript $r$ denotes the reference resonance and $f.u.$ represents the factor uncertainty in the resonance strength. Since the reference resonance contains the smallest fractional uncertainty of a given data set, $\rho_j$ $\ge$ $1$.

Given the set of correlation parameters, $\rho_j$, the procedure to generate correlated Monte-Carlo samples of the resonance strengths is as follows: (i) Generate normally-distributed, uncorrelated random samples, $i$, of the strength for each resonance, $j$, denoted by $y_{j,i}$. (ii) Find the correlated samples, $y'_{j,i}$, according to
\begin{equation}
  \label{eq:correlate}
  y'_{j,i} = \rho_j x_{r,i} + y_{j,i} \sqrt{1-\rho_j^2} 
\end{equation}
where $x_{r,i}$ are the samples associated with the reference resonance, $r$. Visual inspection of this expression shows that for fully correlated resonances, we find $\rho_j$ $=$ $1$ and $y'_{j,i}$ $=$ $x_{r,i}$. In other words, the correlated samples for a given resonance are identical to those of the reference resonance. Conversely, if $\rho_j$ $=$ $0$, then $y'_{j,i} = y_{j,i}$, and the uncorrelated samples remain unchanged. (iii) Calculate the correlated and lognormal samples of a given resonance strength with factor uncertainty $(f.u.)_j$ from the mean value, $\omega\gamma_{j}$, by using
\begin{equation}
  \label{eq:correlated-strength}
  \omega \gamma_{j,i} = \omega \gamma_{j} \left(f.u. \right)_j^{y'_{j,i}}
\end{equation}
This procedure also holds for partial widths when $\omega\gamma$ is replaced by $\Gamma_{\lambda}$.

Correlated resonance energies can be treated similarly, with some modifications. First, resonance energy uncertainties are described by a normal, not lognormal, probability density (see Section~\ref{sec:mcoverview}). For resonance energies, $E_j \pm \delta E_j$, Equations~\eqref{eq:correlation-rho} and \eqref{eq:correlated-strength} are replaced with
\begin{align}
  \label{eq:erho}
  \rho_j = & \frac{\delta E_r}{\delta E_j} \\
  \label{eq:ecorr}
  E_{j,i} = & E_{j} + y'_{j,i} \delta E_j
\end{align}
Second, energy correlations must be used to calculate the partial widths, as described in Section~\ref{sec:mcoverview}.

The correlations discussed above impact reaction rate uncertainties when multiple resonances contribute significantly to the total rate at a given temperature. When applied to resonance strengths, correlations can increase the rate uncertainties, in some cases by a factor of three \citep{Longland2017}, compared to the case of uncorrelated strengths. 
The impact of resonance energy correlations depends not only on how many resonances contribute to the total rate, but also on the energy location of these resonances with respect to the Gamow window. 
For example, \cite{Longland2020} found that resonance energy correlations in the $^{35}$Ar(p,$\gamma$)$^{36}$K reaction slightly decrease the rate uncertainty at $400$~MK, but increase it somewhat in the $^{39}$Ca(p,$\gamma$)$^{40}$Sc reaction at $30$~MK. This effect is generally larger  when only a few resonances contribute to the total rate and when they are all located on the same side of the Gamow peak.

The effect of resonance energy and strength correlations needs to be assessed  selectively. Resonance strengths measured and scaled to a single reference resonance are expected to be correlated, while those derived from a single-nucleon transfer measurement are not. Such quantities are sensitive to DWBA model uncertainties that are not necessarily correlated, even within a single measurement. See Section~\ref{sec:uncpartwidth} for more details. 


\section{Changes in policy compared to ETR10} \label{sec:proc}
Compared to the Monte-Carlo-based thermonuclear reaction rate evaluation by \citet{ILIADIS2010b,ILIADIS2010c,ILIADIS2010d} (ETR10), several policy changes have been implemented. These changes are discussed below.

\subsection{Masses and reaction $Q$ values}
\label{sec:massq}
The masses and $Q$ values adopted in ETR10 were sourced from the 2003 Atomic Mass Evaluation \citep[][]{AME2003}. Two changes have been made in the present work. First, the {\it atomic} masses are now based on the 2020 Atomic Mass Evaluation \citep{wang2021}. Second, because the interacting nuclei in a stellar plasma are fully ionized, {\it nuclear} instead of {\it atomic} masses are used in our reaction rate calculations. These masses are related by
\begin{equation}
m_{at}(A,Z) = m_{nu}(A,Z) + Z m_e - B_e(Z)
\label{eq:mass1}
\end{equation}
where $A$ and $Z$ denote the mass number and atomic number, respectively, $m_e$ is the electron rest mass, and $B_e(Z)$ is the total electron binding energy in the neutral atom of atomic number $Z$. We assign a positive sign to the binding energy. It can be approximated by the expression
\begin{equation}
B_e(Z) = 14.4381~Z^{2.39} + 1.55468 \times 10^{-6} Z^{5.35}~\mathrm{eV}
\label{eq:mass2}
\end{equation}
which is based on the neutral-atom electron binding energies calculated of \citet{Huang1976} using the relaxed-orbital relativistic Hartree-Fock-Slater formalism. 

As can be seen from the factors involving the projectile and target masses in Equations~(\ref{eq:generalrate}), (\ref{eq:breitwignercrosssection}), and (\ref{eq:narresrateexpr}), 
the use of nuclear instead of atomic masses causes insignificant changes to the reaction rate (typically less than 0.1\%). However, this distinction becomes important when reaction $Q$ values are involved in the calculation. The $Q$ values calculated from nuclear and atomic masses are related by
\begin{equation}
Q_{nu} = Q_{at} + \left ( \sum_i B_e^i - \sum_f B_e^f \right )
\label{eq:mass3}
\end{equation}
where $\sum B_e^i$ and $\sum B_e^f$ are the sum of the total electron binding energies before and after the nuclear reaction, respectively.

For example, based on Equation~(\ref{eq:mass2}), the total electron binding energy for Ar and K is 14.4~keV and 16.4~keV, respectively. This implies that the atomic $Q$ value exceeds the nuclear counterpart by about $2$~keV. As a result, when the $Q$ value is derived from nuclear masses, the resonance energy $-$ determined by both the energy of the excited state and the $Q$ value $-$ increases by $2$~keV. This energy shift is significant because it enters exponentially in the expression, Equation~(\ref{eq:narresrateexpr}), for the narrow-resonance reaction rate. For example, the $^{36}$Ar(p,$\gamma$)$^{37}$K reaction rate decreases by $\approx 40$\% near $70$~MK, when the nuclear instead of atomic $Q$ value is adopted. 

For this reason, many of our calculated center-of-mass resonance energies differ from the values adopted in ETR10 by a few kilo electron volts. For more information on atomic versus nuclear $Q$ values, see \citet{Iliadis:2019ch}. 

\subsection{Excitation energies, $J^\pi$ values, and resonance energies}\label{sec:energiesJpi}
In ETR10, level energies and $J^\pi$ values predominantly relied on the evaluations of \citet{Endt1990,Endt1998}. In the present study, we primarily adopted the information for these quantities from the Evaluated Nuclear Structure Data File (ENSDF).\footnote{See: \url{https://www.nndc.bnl.gov/ensdf/.}} Notably, some nuclides in the ENSDF compilation have not been updated for over a decade. In such instances, we deemed it essential to take into account results published after the last update of ENSDF. The references, for a given reaction, are provided in the corresponding RatesMC input file (see Section~\ref{sec:git}).

We have also modified the determination of resonance energies. The work of \citet{Endt1990,Endt1998} often cited resonance energies directly measured from the energy location of thick- or thin-target yield curves. The procedure adopted in ETR10 was then to calculate center-of-mass resonance energies either from excitation energies and $Q$ values, or directly from laboratory resonance energies, opting for the method that provided the smaller uncertainty. In the present work, center-of-mass resonance energies, $E_r$, are always calculated using the expression
\begin{equation}
E_r = E_x - Q_{nu}
\label{eq:mass4}
\end{equation}
where $E_r$, $E_x$, and $Q_{nu}$ are the center-of-mass resonance energy, excitation energy, and the nuclear $Q$ value, respectively. 

When possible, we disregarded resonance energies obtained from yield curves because they are influenced by laboratory electron screening. In contrast, values derived from Equation~(\ref{eq:mass4}) are not impacted by this issue assuming that the excitation energies were derived from $\gamma$-ray spectroscopy. For the $^{27}$Al(p,$\gamma$)$^{28}$Si resonance at a laboratory energy of $992$~keV (Table~\ref{tab:wg}), the screened energy (i.e., the one measured in the laboratory) is about $1.7$~keV lower than the unscreened value (i.e., the one calculated from Equation~(\ref{eq:mass4})). Again, it is important to take an energy shift of this magnitude into account because the reaction rate depends exponentially on the resonance energy (see Equation~(\ref{eq:narresrateexpr})). Laboratory electron screening is discussed in more detail in Appendix~\ref{sec:screening}.

\subsection{$S$ factor parameterizations}
Nonresonant reaction rates were discussed in Section~\ref{sec:nonrates}. The procedure adopted in ETR10 was to approximate the experimental $S$ factor by a polynomial 
\begin{equation}
\label{eq:sfac}
S(E) \approx S(0) + S^\prime(0)E + \frac{1}{2} S^{\prime\prime}(0) E^2
\end{equation}
where the primes indicate derivatives with respect to the center-of-mass energy, $E$. The nonresonant reaction rate can then be conveniently calculated using an analytical expression (see Equations~(6) -- (8) in \citet{Longland:2010is}). Furthermore, this expression was multiplied by a cutoff factor, given by \citep{Fowler1975}
\begin{equation}
\label{eq:cutoff}
f_{\mathrm{cutoff}} = e^{-(T_9/T_{9,\mathrm{cutoff}})^2}
\end{equation}
where $T_{9,\mathrm{cutoff}}$ corresponds to the temperature at which the $S$-factor expansion of Equation~(\ref{eq:sfac}) becomes inaccurate. This scaling factor was chosen to ensure a smoothly diminishing nonresonant contribution at higher temperatures, but it has no physical significance. The cutoff temperature was found from an energy cutoff, which was typically  set equal to the energy of the ``first strong and not-too-narrow resonance'' \citep{Fowler1975}.

In the present work, we did not adopt any of these procedures. Instead, we estimated experimental nonresonant $S$ factors by fitting or normalizing the results of nuclear reaction models to data, and then  numerically integrating Equation~(\ref{eq:generalrate}). To this end, the experimental nonresonant astrophysical $S$ factor is entered in tabular form in the RatesMC input file.

Numerical tests were performed to assess the error previously introduced by the use of Equations~(\ref{eq:sfac}) and (\ref{eq:cutoff}). We found that this error is generally less than 10\% in the ideal case of a constant $S$ factor (i.e., $S'(0)$ $=$ $0$ and $S''(0)$ $=$ $0$), provided that the cutoff temperature is chosen appropriately. This error is frequently less than the estimated uncertainty in a typical nonresonant reaction rate at low temperatures. Nevertheless, numerical integration of the $S$ factor, as performed in the present work, provides several advantages: (i) It allows for the accurate determination of the nonresonant rate, even in cases where the $S$ factor may be poorly approximated by Equation~(\ref{eq:sfac}); (ii) The maximum energy limit in the numerical integration can be defined unambiguously rather than by relying on a nonphysical expression such as Equation~(\ref{eq:cutoff}); (iii) It is straightforward to estimate the nonresonant rate uncertainty based on the energy-dependent uncertainty of the nonresonant S-factor.

\section{Experimental resonance strengths} \label{sec:wg}
It was mentioned in Section~\ref{sec:narrowrates} that the strength of a narrow resonance, defined by Equation~(\ref{eq:resstrength}), is proportional to the area under the narrow-resonance cross section curve. For most narrow resonances, it is this area, rather than the cross section, that is determined in laboratory experiments. This is a fortunate circumstance, because the energies and strengths are the only parameters required to compute the narrow-resonance reaction rate, as can be seen from Equation~(\ref{eq:narresrateexpr}). We focus in this section on directly-measured resonances, as opposed to the indirect estimation of resonance strengths from spectroscopic factors or asymptotic normalization coefficients (see Section~\ref{sec:broadrates}).

A detailed discussion of how to extract a resonance strength from the measured resonance yield of emitted reaction products is given in \citet{Iliadis_2015}. Most experimental resonance strengths have been determined from the plateau height of the thick-target yield assuming an infinitely thick target, $Y^{max}_{\Delta E \rightarrow \infty}$, by using
\begin{equation}
\omega\gamma = \frac{2 \epsilon_r}{\lambda^2_r} Y^{max}_{\Delta E \rightarrow \infty}
\label{eq:yield_1}
\end{equation}
where $\epsilon_r$ is the stopping power, $\lambda_r$ is the de Broglie wavelength of the projectile, and $Y^{max}_{\Delta E \rightarrow \infty}$ denotes the plateau yield for an infinitely thick target. The kinematic quantities, $\epsilon_r$ and $\lambda_r$, are both evaluated at the resonance energy and refer to the center-of-mass system. Therefore, the quantity $\omega\gamma$ in Equation~(\ref{eq:yield_1}) is the center-of-mass resonance strength. 

The maximum yield, $Y_{max}$, is estimated from the number of emitted reaction products, the number of incident projectiles, the detector efficiency, the branching ratios, and the angular correlation effects. It must also be corrected for the finite target thickness used in the measurement. 

When the target consists of a compound, $X_aY_b$, with $n_X$ active nuclei and $n_Y$ inactive nuclei per square centimeter, the quantity $\epsilon_r$ must be replaced by the effective stopping power, defined as
\begin{equation}
\epsilon_{\mathrm{eff}} \equiv \epsilon_X + \frac{n_Y}{n_X} \epsilon_Y
\label{eq:yield_2}
\end{equation}
where $n_Y/n_X$ $=$ $b/a$. Note that $\epsilon_{\mathrm{eff}}$ is not the same as the total stopping power of a compound \citep{Iliadis_2015}.

While Equation~(\ref{eq:yield_1}) is useful for determining relative values of the resonance strength, it may not be applicable to estimate absolute resonance strengths. The main issues are: 

(i) The dependence of $\omega\gamma$ on the stopping power, $\epsilon_r$, which is usually adopted from tabulations\footnote{See: \url{https://physics.nist.gov/PhysRefData/Star/Text/PSTAR-t.html}.} or codes, such as SRIM\footnote{See: \url{http://www.srim.org}.}. It is particularly difficult to estimate an uncertainty for these compiled stopping power values. An impression can be obtained from Table~4.3 in \citet{Iliadis_2015}, where estimated stopping power uncertainties for protons and $\alpha$ particles in selected absorbers range from $4$ $-$ $8$\%. Nevertheless, much smaller uncertainties are sometimes adopted in the literature for determining resonance strengths, without any obvious justification. 

(ii) In most cases, targets consist of compounds, and stopping powers are estimated using Bragg's (additivity) rule \citep{bragg1905}, which is implicitly adopted in the above definition of the effective stopping power. Bragg's rule states that the stopping power of a compound can be approximated by taking a weighted average of the stopping powers of its constituent elements, where the weights are usually based on the number of atoms of each element in the compound. The reliability of Bragg's rule is limited because the energy loss of an incident ion to the electrons in the absorber (i.e., the target) depends on the detailed atomic (electronic) structure. Consequently, differences in electron bonding between elemental materials and compounds can lead to inaccuracies in Bragg's rule. The inaccuracies can amount to up to 20\%  \citep{ZIEGLER20101818}, depending on the identity and energy of the incident projectile, as well as on the nature of the absorber. 

(iii) Corrections for incident beam spread, straggling, and the total resonance width have to be performed carefully to estimate $Y^{max}_{\Delta E \rightarrow \infty}$ in Equation~(\ref{eq:yield_1}) from the measured yield.

The experimental resonance strength can also be found from
\begin{equation}
\omega\gamma = 2 \frac{A_Y}{n \lambda^2_r}
\label{eq:yield_3}
\end{equation}
where $A_Y$ denotes the area under the resonance yield curve (i.e., the measured yield versus energy). The strength estimated using Equation~(\ref{eq:yield_3}) is independent of beam resolution, straggling, target thickness, stopping power, and resonance width, but requires knowledge of $n$, the  number of (active) target nuclei per square centimeter. The latter quantity can be determined from the measured yield curve by using $n$ $=$ $\Delta E$/$\epsilon_r$, with $\Delta E$ denoting the center-of-mass target thickness in energy units. However, this introduces again a dependence of $\omega\gamma$ on the stopping power.

More reliable, absolute resonance strength values can be determined when the measurement is performed relative to Rutherford scattering \citep[see, e.g.,][]{TRAUTVETTER197537}. 
Some absolute resonance strengths have been determined by measuring Rutherford scattering {\it simultaneously} with the reaction of interest \citep{POWELL1998263,PhysRevC.65.064609}. The results obtained with this technique are independent of the properties of the target (stopping power, stoichiometry, uniformity) and the beam (current integration, straggling). Therefore, they are more reliable than those depending on approximate stopping powers and stoichiometries.

Table~\ref{tab:wg} provides a set of measured absolute resonance strengths. Most of these values were determined relative to Rutherford scattering. Some of the results have been obtained in inverse-kinematics experiments, again relative to the Rutherford scattering yields. We have adopted these values as standards for normalizing relative resonance strengths (i.e., those obtained from tabulated stopping powers and assumed stoichiometries). A few of the listed values were not measured directly, but were estimated using the absolute value of a resonance strength involving a different isotope of the same element, together with the resonance strength ratio reported by \citet{ENGELBERTINK196612}. The latter ratio was obtained from the ratio of the areas under the respective thin-target yield curves and, therefore, is independent of stopping powers, stoichiometries, beam resolution, straggling, target thickness, or total resonance width.

We address now the confusion in the literature regarding the definition of resonance strength. First, the modern definition is given by Equation~(\ref{eq:resstrength}). In the older literature, the strength is defined by $S$ $\equiv$  $(2j_0+1)(2j_1+1)$ $\omega\gamma$ $=$ $(2J+1)\Gamma_a\Gamma_b/\Gamma$, where $j_0$, $j_1$, and $J$ denote the spin of the projectile, target, and resonance, respectively. The historical definition of the strength, $S$, is sometimes confused with the modern definition, $\omega\gamma$. Second, we already mentioned that all kinematic quantities in Equations~(\ref{eq:yield_1}) and (\ref{eq:yield_3}) are given in the center-of-mass system. Since stopping power tabulations, or the code SRIM, provide laboratory values, some authors prefer to express the center-of-mass resonance strength, $\omega\gamma$, in terms of the laboratory stopping power, according to
\begin{equation}
\omega\gamma = \frac{2}{\lambda^2_r} \frac{M_1}{M_0 + M_1} \epsilon_r^{lab} Y^{max}_{\Delta E \rightarrow \infty}
\label{eq:yield_4}
\end{equation}
where the kinematic factor, $K$ $\equiv$ $M_1/(M_0 + M_1)$ takes into account the conversion of the stopping power from the laboratory to the center-of-mass frame. If the center-of-mass de Broglie wavelength is expressed in terms of the laboratory resonance energy, we find
\begin{equation}
\frac{\lambda_r^2}{2} = \left ( \frac{M_0+M_1}{M_1}  \right)^2 \frac{(\pi \hbar)^2}{M_0 E_r^{lab}}   
\label{eq:yield_5}
\end{equation}
The square of the kinematic factor arises, first, from the definition of the reduced mass, and, second, from the laboratory-to-center-of-mass conversion of the resonance energy. From Equations~(\ref{eq:yield_1}), (\ref{eq:yield_4}), and (\ref{eq:yield_5}), we find
\begin{equation}
\omega\gamma = \left ( \frac{M_1}{M_0+M_1}  \right)^3 \frac{M_0 E_r^{lab}}{(\pi \hbar)^2} \epsilon_r^{lab} Y^{max}_{\Delta E \rightarrow \infty}
\label{eq:yield_6}
\end{equation}
In the older literature, Equation~(\ref{eq:yield_6}) without the cube of the kinematic factor is sometimes used to define a {\it laboratory} resonance strength. However, the reported values have been mistakenly interpreted as center-of-mass strengths. This issue is discussed in \citet{PhysRevC.88.038801} for the $^{20}$Ne(p,$\gamma$)$^{21}$Na reaction, where $K^3$ $=$ $0.86$. In other instances, $K$ instead of $K^3$ appears mistakenly in the expression of the center-of-mass strength \citep[see, e.g.,][]{PhysRevLett.95.031101}.

Table~\ref{tab:wg} also includes a subset of standard resonance strengths based on the work of Sargood and collaborators \citep{PAINE1979,ANDERSON1980154,SARGOOD198261}. Their strengths, which they refer to as ``laboratory'' values, are defined similarly to Equation~(\ref{eq:yield_6}), but with a kinematic factor of $K^2$ instead of $K^3$. This has caused considerable confusion, because their definition of ``laboratory'' strength differs from that in the older literature, as discussed above. In other words, their reported results must be multiplied by $K$ $=$ $M_1/(M_0+M_1)$ to arrive at the center-of-mass resonance strength, $\omega\gamma$. We have performed this correction for all values listed in Table~\ref{tab:wg}.
\startlongtable
\begin{deluxetable*}{lrlrl}
\tablecaption{Strength standards of directly measured resonances adopted in the present reaction rate evaluation.
\label{tab:wg}} 
\tablewidth{\columnwidth}
\tablecolumns{3}
\tablehead{
Reaction                            & $E_r^{lab}$~(keV)\tablenotemark{a} &     $\omega\gamma_{c.m.}$~(eV)\tablenotemark{b}   &   Uncertainty  & Reference 
}
\startdata
$^{14}$N(p,$\gamma$)$^{15}$O        &	278       &  $(1.26 \pm 0.03)\times 10^{-2}$   &   2.4\%  &  \tablenotemark{c}        \\
$^{18}$O(p,$\gamma$)$^{19}$F     	&	151       &  $(9.77 \pm 0.35)\times 10^{-4}$   &   3.6\%  &  \tablenotemark{d}        \\	
$^{18}$O(p,$\alpha$)$^{15}$N     	&	151       &  $(1.67 \pm 0.12)\times 10^{-1}$   &   7.2\%  &  \tablenotemark{e}        \\	
$^{22}$Ne(p,$\gamma$)$^{23}$Na      &	479       &  $(5.06 \pm 0.53)\times 10^{-1}$   &   11\%   &  \tablenotemark{f}        \\	
$^{23}$Na(p,$\gamma$)$^{24}$Mg     	&	512       &  $(8.75 \pm 1.20)\times 10^{-2}$   &   14\%   &  \tablenotemark{g}        \\	
$^{23}$Na(p,$\alpha$)$^{20}$Ne     	&	338       &  $(7.16 \pm 0.29)\times 10^{-2}$   &   4.1\%  &  \tablenotemark{h}        \\	
$^{24}$Mg(p,$\gamma$)$^{25}$Al     	&	223       &  $(1.27 \pm 0.09)\times 10^{-2}$   &   7.1\%  &  \tablenotemark{i}        \\	
                                   	&	419       &  $(4.16 \pm 0.26)\times 10^{-2}$   &   6.3\%  &  \tablenotemark{j}        \\
\enddata   
\tablenotetext{a}{Resonance energy in the laboratory system.}
\tablenotetext{b}{Resonance strength in the center-of-mass system.}
\tablenotetext{c}{From Table~IV of \citet{Daigle16}.}
\tablenotetext{d}{Weighted average of values listed in Table V of \citet{PhysRevC.104.025802}.}
\tablenotetext{e}{From \citet{1995ZPhyA.351..453B}.}
\tablenotetext{f}{Weighted average of \citet{Endt1990,PhysRevC.92.035805,PhysRevC.102.035801}.}
\tablenotetext{g}{From \citet{PAINE1979}.}
\tablenotetext{h}{From \citet{PhysRevC.65.064609}.}
\tablenotetext{i}{From \citet{POWELL1999349}.}
\tablenotetext{j}{From \citet{POWELL1998263}.}
\tablenotetext{k}{Weighted average of \citet{TRAUTVETTER197537,ENGEL2005491}.}
\tablenotetext{l}{From \citet{ANDERSON1980154}.}
\tablenotetext{m}{From \citet{PhysRevC.59.1790}.}
\tablenotetext{n}{From Table~1 of \citet{SARGOOD198261}, estimated using the absolute value of a resonance strength involving a different isotope of the same element and the resonance strength ratio reported by \citet{ENGELBERTINK196612}.}
\end{deluxetable*}
%

\section{Indirectly estimated particle partial widths} \label{sec:uncpartwidth}
Unobserved low-energy resonances (i.e., those in the left-hand region of Figure~\ref{fig:idealpic}) may completely dominate the total reaction rates at low temperatures. Therefore, the estimation of their contribution by indirect means is of crucial importance. We discuss now the overall uncertainty that can be assigned to a particle partial width, $\Gamma_{\lambda c}$, that is estimated indirectly, using Equation~(\ref{eq:partpartwidth}), when the spectroscopic factor, $C^2S$, is extracted from particle transfer data using Distorted-Wave Born Approximation (DWBA) theory \citep{GLENDENNING198324}. Most of the discussion in this section is related to single-particle transfer reactions and proton partial widths. We only briefly comment on $\alpha$-particle transfer reactions.

The uncertainty of $C^2S$ will include contributions from both experimental parameters (e.g., counting statistics and the cross section normalization) and theoretical model approximations (e.g., parameters of the optical model for the entrance and exit channel, parameters determining the final-state radial wave function, finite range and non-locality effects). In addition, except for subthreshold resonances, the levels for which we wish to estimate the partial width according to Equation~(\ref{eq:partpartwidth}) are unbound. In this case, the radial form factor no longer decays exponentially, but oscillates with a constant amplitude for large distances, causing difficulties in the numerical integration of the DWBA matrix elements. Some DWBA codes, e.g., the version of DWUCK4 extended by J. Comfort, account for this circumstance \citep{PhysRevC.2.782,Cooper_1982}. With other codes, such as FRESCO \citep{THOMPSON1988167}, calculations are typically performed by initially assuming that the unbound level is weakly bound. The $C^2S$ value is then estimated by extrapolating the theoretical cross section across the particle threshold to the measured excitation energy. The latter procedure is not rigorous and may contribute to the overall uncertainty. For weakly populated levels, additional sources of uncertainty need to be considered. For example, contributions from multi-step (i.e., coupled-channel) processes and compound-nucleus formation can be evaluated quantitatively, but the reliability of such estimated corrections has not always been fully assessed. 

Another source of uncertainty originates from specific choices for the total ($j$) and orbital ($\ell$) angular momenta, and the principal quantum number ($n$), of the transferred particle. For odd-A target nuclei, the calculated DWBA cross section depends on the transferred orbital angular momentum, which may have contributions from two values, $\ell$ and $\ell + 2$. If the higher component dominates the stripping data, it is especially difficult to extract the $C^2S$ value of the lower component from a two-component fit, even though the latter may dominate the total partial width. Often, both $j$ $=$ $\ell$ $-$ $1/2$ and $j$ $=$ $\ell$ $+$ $1/2$ can be added vectorially to the target spin to form the final-state spin (even if the latter is unambigiously known). Which $j$ value to assume in the analysis is usually guided by shell-model arguments, which may be questionable, especially for unbound states. Similarly, it is not always obvious to decide which value of $n$ to assume in the analysis. For example, in the middle of the $sd$ shell, $1p$ hole states and $2p$ particle states may occur at about the same excitation energy, leading to the expectation that their configurations might mix.

A systematic evaluation of experimental proton and neutron spectroscopic factors in the $A$ $=$ $21$ $-$ $44$ region has been reported by \citet{Endt1977}. The overall uncertainty of the experimental values was assessed by calculating ratios of $C^2S$ values for pairs of stripping reactions populating the same level, e.g., (d,n) and ($^3$He,d) for proton transfer, and pairs of reactions exciting mirror states, e.g., (d,p) and ($^3$He,d); see also Appendix~\ref{sec:mirror}. Deviations from unity in these ratios provide an estimate for the experimental uncertainty of individual measurements. \citet{Endt1977} concluded that ``for strong transitions, an error of about $25$\% should be assigned to measured spectroscopic factors.'' Considering the values listed in Tables I $-$ X of \citet{Endt1977}, a ``strong transition'' refers to a $C^2S$ value exceeding $0.1$. \citet{Endt1977} only considered bound states. The above result agrees with the uncertainty of $40\pm24$\% reported by \citet{THOMPSON1999}, who analyzed neutron spectroscopic factors in the $sd$ shell for strong transitions to both bound and unbound levels up to $2.5$~MeV excitation.  

\begin{figure}[hbt!]
\centering
\includegraphics[width=1.0\linewidth]{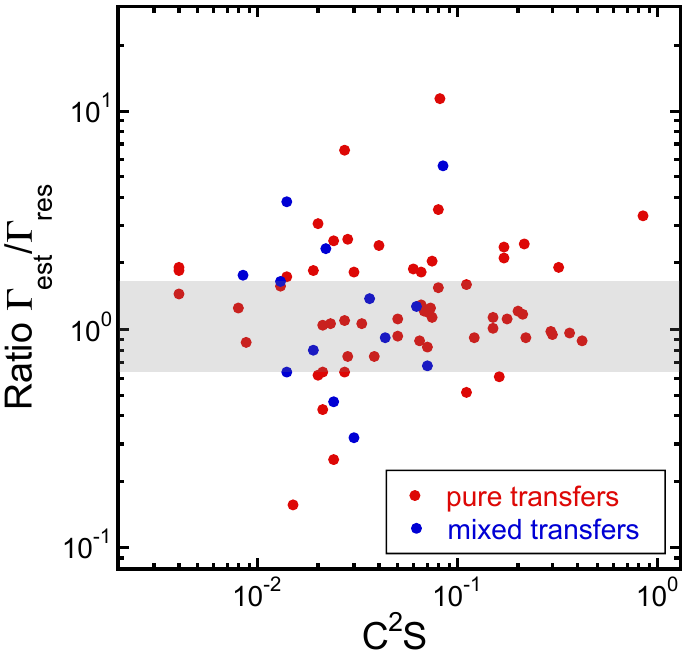}
\caption{
Ratio of proton partial widths estimated from Equation~(\ref{eq:partpartwidth}) to those measured in resonance reaction studies versus magnitude of the transfer spectroscopic factor, $C^2S$. The displayed data, courtesy of Art Champagne, form the basis of Figure~7 in \citet{PhysRevC.70.045802}. The red and blue circles refer to pure and mixed transitions, respectively. The gray-shaded area indicates a factor of $1.6$ uncertainty band. See text.  
}
\label{fig:comp}
\end{figure}

In Section~\ref{sec:broadrates}, we pointed out that spectroscopic factors represent only intermediate steps in the calculation of the quantity of primary astrophysical interest, i.e., the particle partial width, $\Gamma_{\lambda c}$. Therefore, a different kind of test was performed in \citet{PhysRevC.70.045802}, who compared proton partial widths estimated from proton-transfer spectroscopic factors (using Equation~(\ref{eq:partpartwidth})) with those extracted from resonance reactions studies (i.e., elastic scattering or resonance strength measurements). This information was simultaneously available for $72$ levels in $A$ $=$ $21$ $-$ $41$ compound nuclei. The test included only unbound levels of known spin-parity with center-of-mass resonance energies below $2$~MeV. In addition, their sample was limited to levels with total widths of $\Gamma$ $<$ $20$~keV, and excluded mixed-$\ell$ transitions, for which the strength of the lower $\ell$ component could not be extracted from the measured angular distribution. The data are displayed in Figure~\ref{fig:comp}. The red and blue circles denote pure and mixed transitions, respectively. For the whole set, the ratios deviate on average from unity by a factor of $1.6$ (gray band), with no significant difference in the scatter of the red or blue circles. We suspect that the average uncertainty of a proton partial width   estimated indirectly using a $C^2S$ value is less than a factor of $1.6$, because the above comparison disregarded the uncertainties of the partial widths measured in the resonance reactions. We also emphasize that the results presented in Figure~\ref{fig:comp} are subject to selection bias, because it is easier in an experiment to populate a single-particle state (with a large $C^2S$ value) than one with a complex nucleon configuration (implying a small $C^2S$ value). In particular, Figure~\ref{fig:comp} provides no information on uncertainties when the $C^2S$ value is smaller than, say, $0.01$. Additional systematic studies of this kind are highly desirable. 

In the absence of more information, we adopted the following uncertainties for proton (or neutron) partial widths estimated from reported experimental spectroscopic factors, depending on the magnitude of $C^2S$: we assumed an uncertainty factor of $1.6$, $2$, and $3$ for $C^2S$ $>$ $0.1$, $0.1$ $\ge$ $C^2S$ $\ge$ $0.01$, and $C^2S$ $<$ $0.01$, respectively. When additional information was available, e.g., when authors carefully assessed the impact of the systematic effects mentioned above and obtained smaller uncertainties, we generally adopted their reported values. For example, \cite{Marshall2020} performed a Markov Chain Monte Carlo investigation of the $^{70}$Zn(d,$^{3}$He)$^{69}$Cu proton pick-up reaction to estimate the uncertainties of spectroscopic factors derived from a DWBA analysis. They found uncertainties ranging from 35\% to 108\%, depending on the state populated. Additional examples of using statistical methods to quantify uncertainties arising from optical model potentials can be found in \cite{King2018}, \cite{Flavigny2018}, and \cite{Lovell2019}. 

The above uncertainty factors apply to proton (or neutron) partial widths. The uncertainties are generally larger when estimating $\alpha$-particle partial widths from $\alpha$-particle spectroscopic factors. Because systematic studies comparing $\alpha$-particle transfer and capture are scarce, we determined the uncertainties in our estimated $\alpha$-particle partial widths on a case-by-case basis. We emphasize again that, for either nucleons or $\alpha$-particles, the same potential parameters used in the extraction of $C^2S$ values from transfer data must also be employed in the calculation of the partial widths \citep{Fortune2003}. See Appendix~\ref{sec:dimwidths}. 

\section{Direct radiative capture}\label{sec:dc}
For reactions of type (p,$\gamma$) or ($\alpha$,$\gamma$), the direct radiative capture process contributes to the nonresonant $S$ factor \citep{Christy1961,ROLFS1973,Kim1987,Krausmann1996}. We computed this contribution using a single-particle potential model, together with experimental spectroscopic factors. 

The potential model assumes a single-step process, where the projectile is directly captured, without formation of a compound nucleus, into a final bound state with the emission of a photon. The electric dipole (E1) part of the total direct capture cross section usually dominates over the E2 and M1 contributions, except in cases involving $\alpha$-particle capture on self-conjugate (i.e., $N$ $=$ $Z$) target nuclei, where the E1 contribution is isospin-suppressed. 

The E1 contribution to the theoretical cross section (in $\mu$b) for capture from an initial scattering state with orbital angular momentum $\ell_i$ to a final bound state with orbital angular momentum $\ell_{f}$ and principal quantum number $n$ (i.e., the number of wave-function nodes), is given by \citep{ROLFS1973}
\begin{equation}
\label{eq:sp1}
\begin{split}
\sigma^{DC}(E1, n, \ell_{i}, \ell_{f})  & = 0.0716 \mu^{\frac{3}{2}}\left (\frac{Z_{0}}
         {M_{0}} - \frac{Z_{1}}{M_{1}} \right)^{2} \\
        & \hspace{-2cm} \times \frac{E_{\gamma}^{3}}{E^{\frac{3}{2}}} \frac{(2J_{f}+1)(2\ell_{i}+1)}
         {(2j_{0}+1)(2j_{1}+1)(2\ell_{f}+1)}  \\
        & \hspace{-2cm} \times (\ell_{i} 0 1 0 | \ell_{f} 0)^{2} R_{n \ell_{i} 1 \ell_{f}}^{2} 
\end{split}
\end{equation}
and
\begin{equation}
\label{eq:sp2}
    R_{n \ell_{i} 1 \ell_{f}} =
         \int_{0}^{\infty} u_{s}(r) \mathcal{O}_{E1}(r) u_{b}(r) dr 
\end{equation}
where $\mu$ is the reduced mass, $Z_0$, $Z_1$ and $M_0$, $M_1$ are the charges and nuclear masses (in $u$), respectively, of projectile and target; $j_{0}$, $j_{1}$, $J_{f}$ are the spins of projectile, target and final state, respectively; $E$ and $E_\gamma$ are the center-of-mass energy and the energy of the emitted $\gamma$ ray, respectively; $\mathcal{O}_{E1}$ is the radial part of the E1 multipole operator; and $u_{s}$ and $u_{b}$ are the radial wave functions of the initial scattering state and final bound state, respectively.

The radial bound-state wave function, with $u_b(r=0)$ $=$ $0$ and $\int_0^\infty{u_b^2 dr}$ $=$ $1$, was generated using a potential consisting of Woods-Saxon, angular momentum, and Coulomb terms, given by
\begin{equation}
    V(r) = \frac{-V_0}{1 + e^{(r-R_{WS})/a}} + \frac{\hbar^2}{2\mu}\frac{\ell_f(\ell_f + 1)}{r^2} + V_C(r) 
\end{equation}
where $R_{WS}$ $=$ $r_0 A_1^{1/3}$ and $a$ are the Woods-Saxon potential radius\footnote{Alternative formulations for the Woods–Saxon potential radius exist, such as: $R_{WS}$ $=$ $r_0 (A_0^{1/3} + A_1^{1/3})$ or $R_{WS}$ $=$ $r_0 (A_0 + A_1)^{1/3}$. } and  diffuseness, respectively; $A_1$ is the mass number of the target nucleus; $V_C$ corresponds to a uniformly charged sphere of radius $R_{WS}$. The well depth, $V_0$, was chosen to reproduce the binding energy of the final state. The calculation of the radial scattering wave function will be discussed at the end of this section.

The radial integration in Equation~(\ref{eq:sp2}) was performed to $500$~fm, as for some weakly-bound states close to the particle threshold, the integrand has a maximum located far beyond the nuclear radius at the lowest center-of-mass energies explored here. For the same reason, and defining $\rho$ $\equiv$ $k_\gamma r$, where $k_\gamma$ $=$ $E_\gamma /\hbar c$ is the wave number of the emitted photon, we used the exact expression for the radial part of the E1 operator \citep{BAILEY1967502}
\begin{equation}
\label{eq:e1op}
    \mathcal{O}_{E1} = \frac{3r}{\rho^3} \left[ (\rho^2 - 2) \sin \rho + 2 \rho \cos \rho    \right]
\end{equation}
instead of its long-wavelength approximation, $\mathcal{O}_{E1}$ $\approx$ $r$, which only applies if $\rho$ $\ll 1$.

The cross section computed using the single-particle potential model described above must be multiplied by the spectroscopic factor to account for the fractional parentage of the initial and final states. When the direct capture to a specific final state can proceed via several values of orbital angular momenta, $\ell_i$ and $\ell_f$, the cross section is given by an incoherent sum, 
\begin{equation}\label{eq:dc}
\sigma^{DC}_{total} = \sum_{k} \sum_{\ell_{i} \ell_f} C^{2}S_{k \ell_f}
               \sigma^{DC}_{k}(n, \ell_{i}, \ell_{f})
\end{equation}
with $S$ and $C$ denoting the spectroscopic factor and isospin Clebsch-Gordan coefficient (not to be confused with the ANC discussed below), respectively. The index $k$ runs over all bound states of the final nucleus.

The calculated single-particle cross section, $\sigma^{DC}_{sp}$, will depend strongly on the adopted choice of the Woods-Saxon potential radius parameter, $r_0$, and diffuseness, $a$. Similar to the discussion in Section~\ref{sec:broadrates}, this means that the bound-state wave functions entering in the calculation of the direct capture cross section (see Equations~(\ref{eq:sp1}) and (\ref{eq:sp2})) and the DWBA differential cross section (when the spectroscopic factor is extracted from stripping data) must be generated using the same Woods-Saxon potential parameters to avoid systematic bias.

When the direct capture proceeds at low bombarding energy to a weakly-bound final state, the integrand in Equation~(\ref{eq:sp2}) will peak outside the nuclear radius. For such a peripheral reaction, the single-particle radial bound-state wave function is asymptotically given by \citep{Akram2001}
\begin{equation}\label{eq:whitt}
u_{b,\ell_f}(r) \rightarrow b_{\ell_f} W_{-\eta, \ell_f+1/2}(2 \kappa r)
\end{equation}
where $b_{\ell_f}$ is the single-particle asymptotic normalization coefficient (spANC) and $W$ is the Whittaker function \citep{Hebbard1963}; $\kappa$ is the bound-state wave number, with $\kappa^2$ = $2 \mu E_b / \hbar^2$, where $\mu$ is the reduced mass and $E_b$ $=$ $Q - E_x$ is the binding energy of the final state; $\eta$ $=$ $ e Z_0 Z_1 \mu / (\kappa \hbar^2 )$ is the bound-state Coulomb parameter. 

In a microscopic nuclear model, the capture cross section can be described in terms of the overlap function, $I_B$, and a many-body wave function for the relative motion. Assuming a single-particle model, the radial dependence of the overlap function, which represents the projection of the bound final state onto the product bound-state wave functions of the target and projectile, can be approximated by
\begin{equation}\label{eq:overlap}
I_B(r) \approx \sqrt{C^2 S_{\ell_f}} u_{b,\ell_f}(r)
\end{equation}
At large distances between target and projectile, the complicated many-body effects will diminish, and the radial dependence of the overlap function becomes asymptotically
\begin{equation}\label{eq:anc}
I_B(r) \rightarrow C_{\ell_f} W_{-\eta, \ell_f+1/2}(2 \kappa r)
\end{equation}
where $C_{\ell_f}$ is the asymptotic normalization coefficient (ANC). Comparison of Equations~(\ref{eq:whitt}) $-$ (\ref{eq:anc}) yields the relationship between $C^2 S_{\ell_f}$, $C_{\ell_f}$, and $b_{\ell_f}$ \citep{Akram2001}
\begin{equation}
\label{eq:relation}
C^2 S_{\ell_f} = \frac{C^2_{\ell_f}}{b^2_{\ell_f}}
\end{equation}
In this approach, $C_{\ell_f}^2$ appears as an observable quantity, while both $C^2 S_{n \ell_f}$ and $b_{\ell_f}^2$ are derived quantities that depend strongly on the parameters of the adopted single-particle potential model. In other words, for peripheral reactions, substitution of Equation~(\ref{eq:relation}) into Equation~(\ref{eq:dc}), and adopting experimental ANCs obtained from transfer reactions, will yield a direct capture cross section that is relatively insensitive to the Woods-Saxon potential parameters. For a comparison of the two methods just discussed, see \citet{PhysRevC.66.055804}. 

The last ingredient needed to compute the direct capture $S$ factor is the radial scattering wave function (see Equation~(\ref{eq:sp2})). Beyond the range of the nuclear potential, it is given by
\begin{equation}
\label{eq:phase}
    u_s(r) =  F_{\ell_i}(r) \cos \delta_{\ell_i} + G_{\ell_i}(r) \sin \delta_{\ell_i}
\end{equation}
where $\delta_{\ell_i}$ is the nuclear phase shift, and $F_{\ell_i}$ and $G_{\ell_i}$ are the regular and irregular Coulomb wave functions, respectively. At low bombarding energies, the scattering phase shifts for the charged-particle reactions of interest in the present work are small and have usually not been measured. Therefore, we have assumed $\delta_{\ell_i}$ $\approx$ $0$, or $u_s(r)$ $\approx$ $F_{\ell_i}(r)$, corresponding to the assumption that the scattering potential is zero. In the past, hard-sphere scattering phase shifts have been adopted in many low-energy direct capture calculations. This issue was discussed by \citet{iliadis2004}, who showed that the adoption of a zero-energy instead of a hard-sphere nuclear scattering potential better reproduces the measured direct capture cross section, especially for deeply-bound states.

\begin{figure}[hbt!]
\centering
\includegraphics[width=1.0\linewidth]{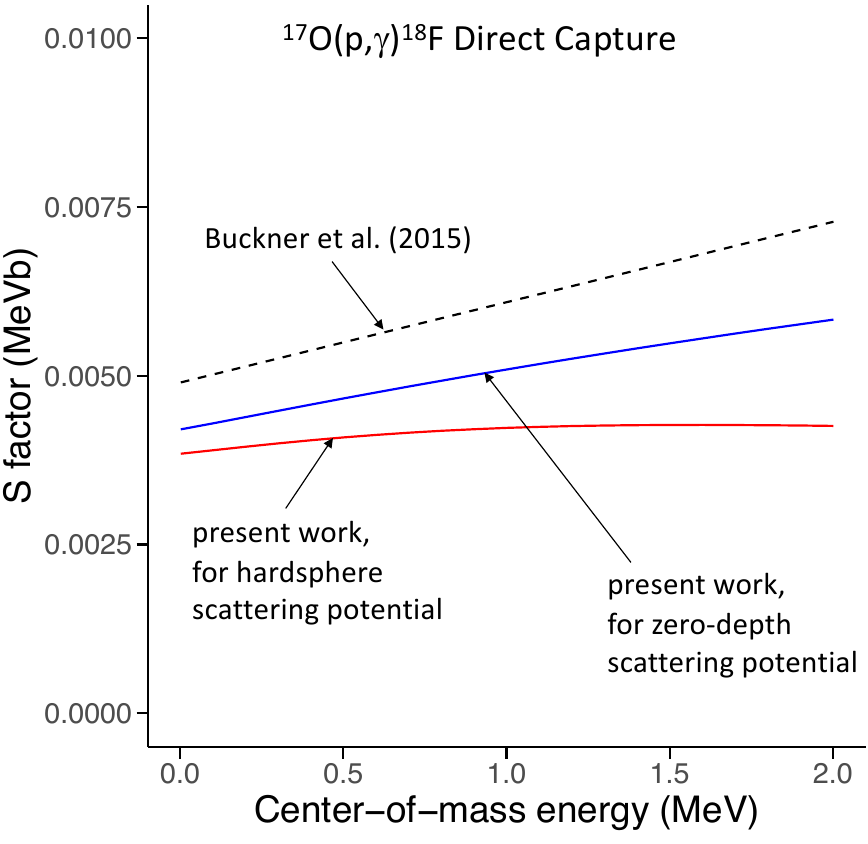}
\caption{
Direct capture $S$ factor in $^{17}$O(p,$\gamma$)$^{18}$F, summed over transitions to $21$ levels in $^{18}$F. (Blue line) $S$ factor calculated from Equation~(\ref{eq:dc}), assuming a zero-depth nuclear potential for the calculation of the radial scattering wave function, $u_s$; (Red line) $S$ factor calculated from Equation~(\ref{eq:dc}), assuming a hard-sphere scattering potential; (Dashed black line) From \citet{buckner2015}. Note that the blue and black lines agree within $\approx 20$\%. See text for details.
}
\label{fig:17opg}
\end{figure}

To test the method, Figure~\ref{fig:17opg} presents the total direct capture $S$ factor in the $^{17}$O(p,$\gamma$)$^{18}$F reaction, incorporating all contributions from bound final states in $^{18}$F. The blue and red lines are derived from Equation~(\ref{eq:dc}), assuming zero-depth and hard-sphere scattering potentials, respectively. The  experimental spectroscopic factors were adopted from the transfer measurements by \citet{Polsky1969,Landre1989}. The dashed black line represents the estimate from \citet{buckner2015} using an independent approach. It involved the direct measurement of the $^{17}$O(p,$\gamma$)$^{18}$F cross section for the strongest transitions at low bombarding energies (below approximately 300~keV) and the extraction of the direct components by fitting the measured partial yields. It can be seen that the dashed black and solid blue lines agree within $\approx$ $20$\%, which supports the method employed in this study for estimating the total direct capture $S$ factor using single-particle potential model calculations weighted by experimental spectroscopic factors.

The uncertainties assigned to direct capture cross sections are based on a combination of available experimental data and our judgment. All direct-capture uncertainties were determined on a case-by-case basis: in favorable cases, when the results of multiple transfer measurements were consistent, we adopted a 30\% uncertainty; in contrast, when data are sparse or of questionable quality, we increased the uncertainty to factors between 3 and 5 (see also the discussion in Section~\ref{sec:uncpartwidth}).  

We will now comment on interference effects between resonant and nonresonant (direct) capture. In principle, such effects need to be considered, depending on the angular momenta involved, when estimating the total cross section from a measured differential one. However, the impact of such interference effects on the total reaction rates is often negligible because the vector coupling coefficients in the interference expression (see, e.g., Equation~(A.38) in \citet{ROLFS1973}) are frequently zero. For example, the Clebsch-Gordan coefficient, $(\ell_R 0 \ell_i 0 | k 0)$, entering the expression for $k$ $=$ $0$ (i.e., the term impacting the total cross section) vanishes unless the orbital angular momenta of the resonant and direct capture, $\ell_R$ and $\ell_i$, respectively, are equal. Similar arguments apply to the coefficient $(L_R 1 L_D -1 | k 0)$, where $L_R$ and $L_D$ denote the $\gamma$-ray multipolarities of the resonant and direct capture process, respectively. As a result, interference effects in the total cross section have been ignored for all reactions evaluated in this work.

We already mentioned that usually the E1 contribution dominates the total direct capture cross section. However, in the case of $\alpha$-particle capture on self-conjugated target nuclei (i.e., $N$ $=$ $Z$), the E1 component becomes small owing to the factor $(Z_0/M_0 - Z_1/M_1)^2$ in Equation~\ref{eq:sp1}. In such cases, the total direct capture cross section is usually dominated by the E2 component. We estimated this contribution along similar lines to what we discussed for the E1 component above. The main differences are two. 

First, it is important to accurately account for the number of nodes, $N$, in the bound-state wave function of the $\alpha$-particle. This value, along with the orbital angular momentum, $L_f$, is determined by the Talmi-Moshinsky relation \citep{MOSHINSKY1959104}
\begin{equation} 
\label{eq:TM} 
   2N + L_f = \sum_k ({2n_k + \ell_{f,k}}) 
\end{equation}
Here, $n_k$ and $\ell_{f,k}$ represent the principal quantum number and orbital angular momentum, respectively, of the nucleons involved in the transfer. This equation applies to the three-dimensional harmonic oscillator and ensures the conservation of total oscillator quanta across different coordinate transformations (e.g., from single-particle to relative and center-of-mass coordinates of the $\alpha$-particle cluster). Consider the example of an $\alpha$-particle transfer to low-lying bound states of $^{28}$Si from $^{24}$Mg. For bound states of positive parity, a typical $(sd)^4$ configuration suggests that $2N$ $+$ $L$ $=$ $8$, while for configurations where the transferred nucleons are in a $(sd)^3(fp)$ arrangement, $2N$ $+$ $L$ $=$ $9$.

Second, experimental $\alpha$-particle spectroscopic factors, often derived from reactions such as ($^6$Li,d) or ($^7$Li,t), exhibit substantially greater uncertainties compared to those of single-particle spectroscopic factors discussed in Section~\ref{sec:uncpartwidth}. This variation can be substantiated by comparing values obtained from independent Distorted Wave Born Approximation (DWBA) analyses conducted at various bombarding energies. A primary source of uncertainty stems from the choice of optical-model potentials used for both the incoming (e.g., $^{6}$Li) and outgoing (e.g., d) scattering channels. In our study, we have assigned a conservative uncertainty factor of $5$ to the estimated E2 direct capture component for a transition to a given bound state. As previously emphasized, to reduce bias, it is crucial to consistently use the same bound state potential parameters and number of nodes, $N$, in the direct capture calculations as were used in the DWBA analysis of the $\alpha$-transfer data.

\section{Rate extrapolation to high temperatures} \label{sec:tmatch}
Previously, we discussed how to estimate statistically meaningful reaction rates based on available information about the nuclear physics input. However, for a number of reasons, any experiment will have an associated cutoff at some maximum bombarding energy. This cutoff may be dictated by the highest energy attainable by a particle accelerator. Alternatively, measurements may be terminated at an energy where data analysis becomes intractable, possibly because of strongly overlapping resonances that obscure the resonance structure. This implies that the total reaction rate can be computed directly with the methods discussed so far only up to a limiting temperature. Therefore, a method of extrapolation is required to estimate the total rate all the way up to a temperature of $10$~GK.

The process of extrapolating a total reaction rate to higher temperatures can be broadly divided into two tasks. First, the lowest temperature at which the available experimental information is no longer sufficient to calculate the reaction rate reliably needs to be estimated. Second, based on this temperature limit, the extrapolation must be performed. In the present work, we follow the procedure detailed in \cite{Newton2008}. At any given temperature, an Effective Thermonuclear Energy Range (ETER) is determined, which is based on the experimental distribution of fractional resonant-rate contributions. 

Specifically, the ETER for a given temperature is obtained by the following steps: (i) The cumulative distribution of fractional resonant rates is computed, which resembles a step function; (ii) The 50th percentile (i.e., the median) of the cumulative distribution is identified with the energy location, $E(T^{\text{ETER}})$, of the ETER; (iii) The 8th and 92nd percentiles of the cumulative distribution define the energy width, $\Delta E(T^{\text{ETER}})$, of the ETER. This energy region has generally a significantly different location and width compared to the Gamow peak. See \cite{Newton2008} for details. 

As the temperature increases, the ETER moves to a higher energy and becomes broader. A limiting temperature, $T^{\text{ETER}}_{\text{match}}$, is eventually reached, where the upper boundary of the ETER coincides with the maximum bombarding energy, $E^{\text{exp}}_{\text{max}}$, for which experimental information is available. This  matching temperature is given by the condition
\begin{equation}
  \label{eq:matchingT}
  E(T^{\text{ETER}}_{\text{match}}) + \Delta E(T^{\text{ETER}}_{\text{match}}) = E^{\text{exp}}_{\text{max}} 
\end{equation}
For temperatures beyond $T^{\text{ETER}}_{\text{match}}$, the total rate must be found by extrapolation.

To perform the rate extrapolation, we adopted theoretical estimates of reaction rates based on the statistical (Hauser–Feshbach) model of nuclear reactions. These have been calculated using TALYS, which is a modern nuclear reaction code that includes many state-of-the-art nuclear models to cover the main reaction mechanisms \citep{TALYS2023}. In particular, TALYS was updated to estimate nuclear reaction rates of relevance to astrophysics. The uncertainties of the TALYS predictions have mainly two origins: (i) the description of the reaction mechanism, i.e., the model of formation and de-excitation of the compound nucleus, including a possible pre-equilibrium and direct capture contribution; and (ii) the evaluation of the nuclear quantities entering the calculation of the transmission coefficients for each entrance and exit channel. For more information, see \cite{Sallaska}.

\begin{figure}[hbt!]
\centering
\includegraphics[width=1.0\linewidth]{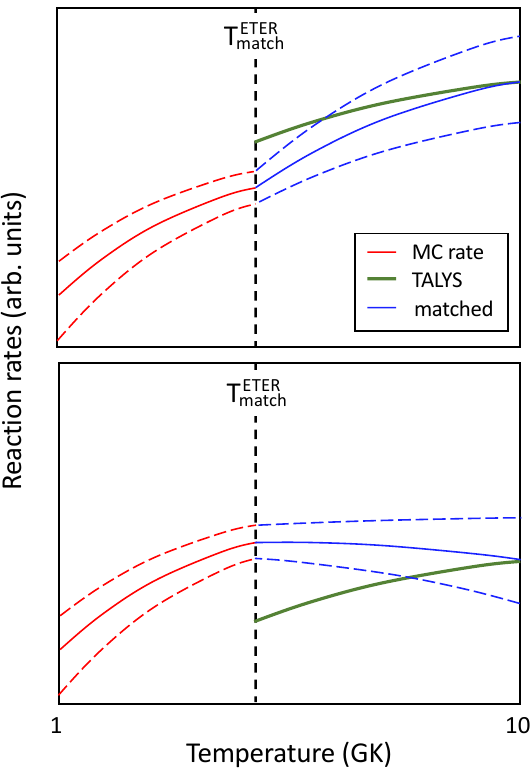}
\caption{Matching of experimental Monte-Carlo rates (red) to TALYS results (green) at high temperatures. The top and bottom panels depict the situation where the TALYS rates at the matching temperature, $T^{\text{ETER}}_{\text{match}}$ (vertical dashed line), are larger and smaller, respectively, than the median experimental rate. In either case, the TALYS prediction at the highest temperature, 10~GK, is adopted at face value, with an assumed uncertainty of a factor of 10. The matched rates and their uncertainties (blue) beyond $T^{\text{ETER}}_{\text{match}}$ are found from connecting the experimental rates at $T^{\text{ETER}}_{\text{match}}$ to the TALYS results at 10~GK, according to Equation~(\ref{eq:match1}). All rates depicted here refer to laboratory rates (i.e., assuming that the target is in its ground state).
}
\label{fig:TMatchExample}
\end{figure}

As a final step, experimental Monte-Carlo-based rates must be matched to the TALYS results at the matching temperature, $T^{\text{ETER}}_{\text{match}}$. We adopted the procedure depicted in Figure~\ref{fig:TMatchExample}. The top and bottom panels correspond to situations where the estimated TALYS rate is larger and smaller, respectively, than the median experimental rate at $T^{\text{ETER}}_{\text{match}}$ (vertical dashed line). In either case, the TALYS rates at the highest temperature, $10$~GK, is adopted at face value and the matched recommended rate between $T^{\text{ETER}}_{\text{match}}$ and $10$~GK is found from the expression
\begin{align}
    \langle \sigma v \rangle_{\text{rec}} =  \langle \sigma v \rangle^{\text{TALYS}} \times \nonumber \\  
    \left[ 
       \left(
       \frac{10 - T_9}{10 - T^{\text{ETER}}_{\text{match}}}
       \right)
       \frac{\langle \sigma v \rangle^{\text{ETER}}_{\text{match}}}{\langle \sigma v \rangle^{\text{TALYS}}_{\text{match}}}
        + 
       \left(
       \frac{T_9 - T^{\text{ETER}}_{\text{match}}}{10 - T^{\text{ETER}}_{\text{match}}}  
       \right)
       \right] 
     \label{eq:match1}
\end{align}
where $\langle \sigma v \rangle^{\text{TALYS}}$ is the temperature-dependent TALYS rate, and $\langle \sigma v \rangle^{\text{ETER}}_{\text{match}}$ and $\langle \sigma v \rangle^{\text{TALYS}}_{\text{match}}$ are the experimental and TALYS rates at the matching temperature, $T^{\text{ETER}}_{\text{match}}$, respectively, in units of GK.

The uncertainties of the matched rate are found from a similar procedure. We assumed a factor 10 uncertainty for the TALYS rate at 10~GK. The TALYS high and low bounds are then connected to the high and low experimental rates, respectively, at $T^{\text{ETER}}_{\text{match}}$ using an expression similar to Equation~(\ref{eq:match1}). 

The procedure for matching reaction rates at elevated temperatures in the present evaluation differs from that employed in ETR10 \citep{ILIADIS2010b}. These differences are discussed in detail in Appendix~\ref{sec:compTab}.

%
%

\section{Use of present reaction rates in nucleosynthesis studies} \label{sec:network}
A number of recent studies have examined how uncertainties in thermonuclear reaction rates affect nucleosynthesis predictions obtained from reaction network simulations. Examples include sensitivity studies of Big Bang nucleosynthesis \citep{Iliadis_2020}, classical novae \citep{Wallace_2025,Psaltis_2025}, presolar grain isotopic signatures \citep{Iliadis_2018,downen2022b,Ward_2025}, and abundance anomalies in globular cluster stars \citep{Dermigny_2017}. These investigations share two principal goals: (1) to determine realistic final abundance yields and quantify their uncertainties arising from thermonuclear reaction–rate uncertainties, and (2) to identify which reaction(s) contribute most strongly to the final abundance uncertainty of a given nuclide.

In contemporary sensitivity studies, Monte-Carlo methods are used to vary all reaction rates simultaneously in each network run. Forward and reverse rates associated with the same reaction are not sampled independently; instead, their variations are correlated according to the reciprocity theorem. Previous work has shown that thermonuclear reaction rates are well described by lognormal probability density functions \citep{Longland:2010is}; see also Section~\ref{sec:mcoverview}. Accordingly, a sampled rate, $x$, for a given reaction $j$ at temperature $T$ is drawn from
\begin{equation}
\label{eq:ln}
f[x(T)_j] = \frac{1}{\sigma \sqrt{2\pi}} \frac{1}{x(T)_j}
e^{-\frac{[\ln x(T)_j - \mu(T)_j]^2}{2 \sigma(T)_j^2}},
\end{equation}
where the lognormal parameters $\mu$ and $\sigma$ represent the location and spread, respectively.

Sampling is performed using \citep{Longland2012}
\begin{equation}
\label{eq:sample}
x(T)_{kj} = x(T)_{\mathrm{med},j}\,\bigl[f.u.(T)\bigr]_j^{p(T)_{kj}},
\end{equation}
where $x_{\mathrm{med}}$ is the median rate, and $f.u.$ is the tabulated factor uncertainty. The variation exponent $p_{kj}$ is drawn from a standard normal distribution (mean = 0, standard deviation = 1). Importantly, $p_{kj}$ is sampled once per reaction per network run and remains fixed for all temperatures in that run. Repeating the simulation for $n$ Monte-Carlo samples produces an ensemble of final abundance yields. For each nuclide, the 50th percentile defines the median abundance $X_{\mathrm{med}}$, while the 16th and 84th percentiles give $X_{\mathrm{low}}$ and $X_{\mathrm{high}}$. The corresponding abundance factor uncertainty is $\Delta f$ $=$ $\sqrt{X_{\mathrm{high}}/X_{\mathrm{low}}}$ corresponding to a $68\%$ confidence interval.

The influence of a reaction’s rate uncertainty on the abundance of a given nuclide is assessed by correlating its variation exponent $p_{kj}$ with the resulting final abundance $X_f$ across the ensemble of Monte-Carlo runs. Earlier work used Pearson’s linear correlation coefficient $r$ \citep[e.g.,][]{Coc_2014}, or Spearman’s rank–order correlation coefficient $r_s$ \citep{Iliadis2015}. However, correlations encountered in nucleosynthesis simulations are often neither linear nor monotonic, limiting the usefulness of $r$ and $r_s$.

A more robust alternative is the Mutual Information (MI) metric from information theory \citep{linfoot1957,coverthomas}. MI measures how much knowing one random variable (here, $p_{kj}$) reduces the uncertainty in another (the final abundance $X_f$). For random variables $Y$ and $Z$ with values $\{y_i\}$ and $\{z_j\}$, respectively,

\begin{equation}
\label{eq:mi}
MI = \sum_y \sum_z P(y,z)
\log \left[ \frac{P(y,z)}{P(y)\,P(z)} \right],
\end{equation}

where $P(y)$ and $P(z)$ are the marginal distributions and $P(y,z)$ is the joint probability distribution. By definition, $MI = 0$ if and only if the variables are statistically independent. MI therefore provides a model–agnostic measure of reaction importance.

The laboratory reaction rates evaluated in this work cover a significant fraction of charged–particle reactions relevant to nucleosynthesis in the Big Bang, classical novae, low–mass and AGB stars, and other astrophysical environments. Before use in nucleosynthesis simulations, these rates must be corrected for the thermal population of target excited states at elevated temperatures.

However, a reaction network may also include reactions for which no Monte-Carlo reaction rates are currently available. For such cases (e.g., Class~III reactions; see Section~\ref{sec:statrates}), we recommend that users approximate the rate factor uncertainty by $f.u.$ $=$ $\sqrt{\mathrm{High}/\mathrm{Low}}$ (Section~\ref{sec:explan}), assuming the rate is lognormally distributed. Here, ``High'' and ``Low'' denote literature estimates of the upper and lower rate bounds. Unlike the statistically rigorous uncertainties derived in the present work, these values are only approximations and may be adjusted at the discretion of the user when incorporating non–Monte Carlo literature rates into network calculations.

\section{Summary} \label{sec:summary}
In this work, we have presented a comprehensive analysis of nuclear reaction rates using modern statistical approaches, including Bayesian inference and Monte-Carlo methods. We discussed the fundamental reaction rate formalisms and outlined the improvements these methods offer in estimating reaction rates more accurately.

Our results highlight the advantages of Bayesian-based reaction rates, particularly in incorporating prior knowledge and systematically accounting for uncertainties. Additionally, we demonstrated how Monte-Carlo-based reaction rates provide a robust framework for propagating uncertainties and capturing correlations among nuclear parameters.

Experimental resonance strengths and indirectly estimated partial widths were examined in detail, emphasizing their impact on reaction rate calculations. Furthermore, we discussed direct radiative capture and the extrapolation of reaction rates to high-temperature environments relevant for astrophysical scenarios.

As the main product of this study, we provide in Appendix~\ref{sec:ratetables} a dataset of 78 reaction rates tabulated on a specified temperature grid. For each reaction, we present the fractional contributions to the total rate along with the total rate uncertainties. These experimentally derived rates serve as a valuable resource for nuclear astrophysics applications. We also provide, in Appendix~\ref{sec:compTab}, a graphical comparison of our new results with previously evaluated Monte-Carlo rates.

Importantly, we emphasize that our rates are based on experimental data rather than theoretical models. Additionally, our rates are laboratory-based and must be corrected for thermal target excitations before they can be applied in stellar evolution and explosion codes.

To facilitate transparency and reproducibility, all input files and the RatesMC computational code used in this study have been made publicly available on GitHub. This enables researchers to verify our findings, apply our methods to new data, and further improve reaction rate evaluations in the future.

These findings underscore the necessity of applying modern statistical techniques in nuclear reaction rate evaluations to achieve more reliable and precise results. Future work could focus on refining uncertainty quantification, expanding the dataset of measured resonances, and integrating additional constraints from theoretical models and experimental observations.

This study provides a framework for improving nuclear reaction rate predictions, which is crucial for applications in nuclear astrophysics, stellar nucleosynthesis, and thermonuclear processes in astrophysical environments.

\begin{acknowledgmentsno}
We would like to thank Stephane Goriely for providing the TALYS results used in the present work for estimating reaction rates at elevated temperatures. The comments of Art Champagne and Robert Janssens are highly appreciated. This work was supported in part by the DOE, Office of Science, Office of Nuclear Physics, under grants DE-FG02-97ER41041 (UNC), DE-FG02-97ER41042 (NCSU), and DE-FG02-97ER41033 (TUNL); PM was supported by NKFIH (K134197).
\end{acknowledgmentsno}

\clearpage
\appendix
\addcontentsline{toc}{section}{Appendix} 

\twocolumngrid

\section{Estimation of proton partial widths}\label{sec:dimwidths}
In this section, we explain how to estimate proton partial widths from spectroscopic factors using Equation~(\ref{eq:partpartwidth}). The penetration factors, $P_c$, can be readily computed from Coulomb wave functions \citep[see, e.g., the Appendix in][]{lane58}. To assist in these calculations, we provide in Table~\ref{tab:redwidths} numerically computed values of $\theta_{pc}^2$ for protons on a grid of energy and orbital angular momentum. These values were obtained for a channel radius of $R$ $=$ $1.25 (A_0^{1/3} + A_1^{1/3})$~fm, where $A_0$ and $A_1$ are the (integer) mass numbers of projectile and target, respectively, and optical model final-state parameters of $r_0$ $=$ $1.25$~fm, $a$ $=$ $0.65$~fm, and $r_{c0}$ $=$ $1.25$~fm. When using the $\theta_{pc}^2$ values listed in Table~\ref{tab:redwidths}, the penetration factor, $P_c$, should be computed at the same channel radius, $R$, and the spectroscopic factor $C^2S$ should be extracted from the DWBA analysis of stripping data using the same values of the optical model final-state parameters. See Section~\ref{sec:broadrates} for additional information.

As an example, consider the $s$-wave resonance at $E_r^{c.m.}$ $=$ $34.7$~keV in the $^{22}$Ne(p,$\gamma$)$^{23}$Na reaction, corresponding to the level at $E_x$ $=$ $8827.9$~keV ($1/2^+$) in the $^{23}$Na compound nucleus. A spectroscopic factor of $C^2S$ $=$ $0.020$ was measured by \citet{PhysRevC.65.015801} using the $^{22}$Ne($^3$He,d)$^{23}$Na proton transfer reaction. For $\ell$ $=$ $0$, we find a penetration factor of $3.98 \times 10^{-20}$, and from Table~\ref{tab:redwidths}, we obtain an interpolated value of $\theta_{pc}^2$ $=$ $0.62$. The Wigner limit, $\hslash^2/(\mu R^2)$, amounts to $1.918 \times 10^{6}$~eV. The resulting proton partial width from Equation~(\ref{eq:partpartwidth}) is $\Gamma_p$ $=$ $1.89 \times 10^{-15}$~eV. Because, in this case, the proton width is so small (and $\omega$ $=$ $1$), we can assume that $\omega\gamma$ $\approx$ $\Gamma_p$.
\startlongtable
\begin{deluxetable*}{lccccccccccc}
\tablecaption{Proton dimensionless single-particle reduced widths, $\theta_{pc}^2$. \tablenotemark{a}
\label{tab:redwidths}} 
\tablewidth{\columnwidth}
\tabletypesize{\footnotesize}
\tablehead{
E (keV)\tablenotemark{b}:         & 0    & 100  & 200  & 300  & 400 & 500 & 600 & 700 & 800 & 900 & 1000 \\
Mass~A \tablenotemark{c}               &      &      &      &      &     &     &     &     &     &     &     
}
\startdata
$\ell$ $=$ $0$ \tablenotemark{d}  &      &      &      &      &     &     &     &     &     &     &    \\
16               & 0.61 & 0.59 & 0.58 & 0.55 & 0.53 & 0.50 & 0.45 & 0.42 & 0.41 & 0.40 & 0.40    \\
22               & 0.62 & 0.61 & 0.61 & 0.60 & 0.59 & 0.58 & 0.56 & 0.55 & 0.52 & 0.50 & 0.48    \\
31               & 0.58 & 0.58 & 0.58 & 0.58 & 0.58 & 0.57 & 0.57 & 0.57 & 0.57 & 0.57 & 0.57    \\
40               & 0.52 & 0.52 & 0.52 & 0.53 & 0.53 & 0.53 & 0.54 & 0.54 & 0.54 & 0.54 & 0.55    \\
50               & 0.48 & 0.49 & 0.49 & 0.49 & 0.49 & 0.50 & 0.50 & 0.50 & 0.51 & 0.51 & 0.51    \\
$\ell$ $=$ $1$ \tablenotemark{d}  &      &      &      &      &     &     &     &     &     &     &    \\
16               & 0.81 & 0.78 & 0.77 & 0.75 & 0.74 & 0.73 & 0.72 & 0.71 & 0.71 & 0.71 & 0.71    \\
\enddata
\tablenotetext{a}{Calculated from Equation~(9) of \citet{ILIADIS1997}. The parameters used to compute the values listed here (a channel radius parameter of $1.25$~fm, and Woods-Saxon parameters of $r_0$ $=$ $1.25$~fm, $a$ $=$ $0.65$~fm, and $r_{c0}$ $=$ $1.25$~fm) differ somewhat from those reported in \citet{ILIADIS1997}. These values were adopted because they are commonly encountered in the literature.}
\tablenotetext{b}{Center-of-mass energy.}
\tablenotetext{c}{Target mass number.}
\tablenotetext{d}{The proton is assumed to transfer into the $2s$, $2p$, $1d$, or $1f$ single-particle orbits.}
\end{deluxetable*}

\twocolumngrid

\section{Isospin Clebsch-Gordan coefficients}\label{sec:CGC}
Results from transfer experiments report spectroscopic factors either in the form of $C^2S$ or as $S$ (see Section~\ref{sec:broadrates}). The isospin Clebsch-Gordan coefficient, $C$, takes into account that in an actual measurement the transferred nucleon is either a proton or neutron. For a proton or neutron stripping reaction, e.g., ($^3$He,d), (d,p), etc., this quantity is defined by \citep{brussaard1977}
\begin{equation}
   C = \left( T_i T_{zi} \tfrac{1}{2} t_z | T_f T_{zf} \right)
   \label{eq:iso}
\end{equation}
The symbols $T_i$ and $T_f$ denote the isospin of the initial (i.e., of the target) and final state, respectively; $T_{zi}$ and $T_{zf}$ are their respective $z$ components, with $T_z$ $=$ $(N - Z)/2$, where $N$ and $Z$ are the neutron and proton number, respectively; $t_z$ is the isospin of the transferred nucleon ($t_z$ $=$ $-1/2$ for a proton and $t_z$ $=$ $+1/2$ for a neutron). Values of $C^2$ are given in Table~\ref{tab:c2} for proton and neutron stripping reactions. 

For example, for both proton and neutron transfer to a $N$ $=$ $Z$ target nucleus (e.g., $^{20}$Ne, $^{26}$Al,...) we find $C^2$ $=$ $1$.
\begin{deluxetable}{l | cc}
\tablecaption{Square of the isospin Clebsch-Gordan coefficient, $C^2$, for single-particle stripping reactions.\tablenotemark{a}
\label{tab:c2}} 
\tablewidth{\columnwidth}
\tablehead{
         & proton    & neutron \\   
         & transfer  & transfer
}
\startdata
$T_f$ $=$ $T_i$ $-$ $\frac{1}{2}$       & $\frac{T_i + T_{zi}}{ 2T_i+1 }$      &     $\frac{T_i - T_{zi}}{ 2T_i+1 }$     \\
$T_f$ $=$ $T_i$ $+$ $\frac{1}{2}$       & $\frac{T_i - T_{zi} + 1}{2T_i+1}$    &     $\frac{T_i + T_{zi} + 1}{2T_i+1}$     \\
\enddata
\tablenotetext{a}{The symbols $T_i$ and $T_f$ denote the isospin of the initial (i.e., of the target) and final state, respectively; $T_{zi}$ and $T_{zf}$ are their respective $z$ components, with $T_z$ $=$ $(N - Z)/2$, where $N$ and $Z$ are the neutron and proton number, respectively.}
\end{deluxetable}

\section{Common mistakes}\label{sec:mistakes}

\subsection{Spectroscopic factors from analog states}\label{sec:mirror}
Since nuclear forces are approximately charge symmetric, structure information for proton-rich nuclei can be adopted from the corresponding mirror states in the neutron-rich nuclei, and vice versa. This procedure requires reliable mirror state correspondences. When a particle partial width of a resonance needs to be estimated using Equation~(\ref{eq:partpartwidth}) from a spectroscopic factor, $S_a$, but the latter quantity has not been measured, one can take advantage of the relationship $S_a$ $\approx$ $S_b$, where $S_b$ is the experimental spectroscopic factor of the corresponding mirror state (Section~\ref{sec:uncpartwidth}). This procedure is sometimes applied erroneously, as is explained below. 

The spectroscopic factor of a stripping reaction is given in the second-quantization formalism by \citep{brussaard1977} 
\begin{equation}
   S = \frac{\left< A+1; J_f T_f ||| a_{n \ell j}^\dagger ||| A; J_i T_i \right>^2}{(2 J_f + 1)(2 T_f + 1)}
   \label{eq:secquant}
\end{equation}
where the final and initial states are labeled by their respective mass numbers ($A$), spins ($J$), and isospins ($T$); $a_{n \ell j}^\dagger$ denotes the operator for the creation of a nucleon in the shell-model orbit ($n$,$\ell$,$j$); and the triple bars in the matrix element stand for reduction in both coordinate and isospin space. It can be seen that the spectroscopic factor represents an overlap integral between an initial state (the target consisting of $A$ nucleons and a single transferred nucleon) and a final state (consisting of $A+1$ nucleons). Therefore, when the spectroscopic factor of a mirror level, $S_b$, is adopted instead of $S_a$ in calculating the particle partial width, it is important to ensure that $S_b$ was measured in the {\it mirror reaction.}

\begin{figure}
\includegraphics[width=0.8\linewidth]{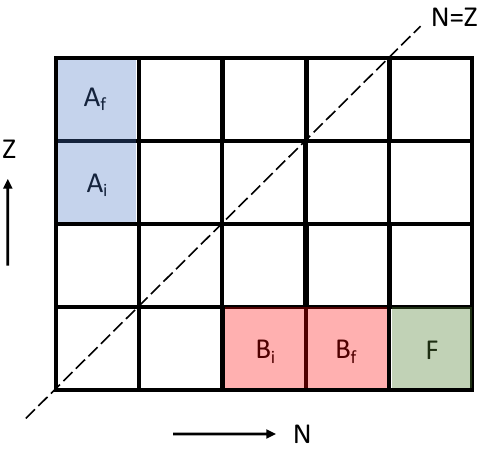}
\caption{
Sketch of the nuclidic chart displaying the number of protons, $Z$, versus the number of neutrons, $N$. The $N$ $=$ $Z$ line is depicted as a dashed line. See text.  
}
\label{fig:iso}
\end{figure}

Part of the nuclidic chart is sketched in Figure~\ref{fig:iso}. Suppose one would like to estimate the proton partial width of a resonance in the $A_i$(p,$\gamma$)$A_f$ reaction, but the required $S_a$ value is unknown. Instead, assuming $S_a$ $\approx$ $S_b$, the known value of $S_b$, measured in the neutron stripping reaction, $B_i$ $\rightarrow$ $B_f$, e.g., (d,p), can be adopted for this purpose. Notice that, in this case, the ground states of the target nuclei, $A_i$ and $B_i$, are mirror levels, as are the final levels in the nuclei $A_f$ and $B_f$.   

Sometimes, spectroscopic factors measured in pick-up reactions are erroneously used 
to calculate particle partial widths according to Equation~(\ref{eq:partpartwidth}). Consider the neutron pick-up reaction $F$ $\rightarrow$ $B_f$, e.g., (p,d), (d,t) or ($^3$He,$\alpha$). Although $A_f$ and $B_f$ are mirror nuclei, $F$ and $A_i$ are not and, therefore, a measured spectroscopic factor in a neutron pick-up reaction, $F$ $\rightarrow$ $B_f$, has no relationship to the spectroscopic factor of the $A_i$(p,$\gamma$)$A_f$ reaction, despite the fact that two levels in the nuclei $A_f$ and $B_f$ may be mirror states.  

In addition to charge symmetry, nuclear forces are also approximately charge independent. This implies that, instead of adopting nuclear structure information from only mirror states, we can also take advantage of the information across an isospin-multiplet. For an application of this method, see \citet{Iliadis_1999}.

\subsection{Laboratory electron screening}\label{sec:screening}
We mentioned in Section~\ref{sec:intro} that the results listed in this work refer to {\it laboratory} reaction rates. For use in stellar model calculations, these need to be corrected for a number of effects, including the thermal excitation of nuclear levels and electron screening. The former depends on the stellar temperature, while the latter, more precisely referred to as {\it stellar plasma screening}, depends on both temperature and density. For a review, see \citet{aliottaLanganke}.

Nuclear reaction measurements are also subject to screening because the reactions proceed in the presence of electrons. Before the laboratory reaction rates can be computed from the nuclear physics input, it is important to apply the necessary corrections for {\it laboratory electron screening}. Two situations need to be distinguished. The first refers to laboratory screening for a nonresonant reaction cross section, which is particularly noticeable at relatively small bombarding energies. Typically, the measured cross section (or $S$ factor) is fit using an appropriate function that includes a screening correction factor to extract the ``bare-nucleus'' cross section. The latter is then used to obtain the laboratory reaction rates, as discussed, e.g., in Section~\ref{sec:bayesrates}. This factor can be approximated by
\begin{equation}
f_{nr} \approx e^{\pi \eta(E) \frac{|U_e|}{E}}
\label{eq:screening}
\end{equation}
The quantity $|U_e|$ denotes the electron screening potential and is a fit parameter. For an example, the reader is referred to Figure~4 in \citet{desouza2019b}, showing the $S$-factor of the $^3$He(d,p)$^4$He reaction, with different fits representing different screening potentials depending on the identities of the target and projectile.

Notice that the screened cross section, at the same bombarding energy, is always larger than the unscreened one because the presence of the electrons reduces the thickness of the effective barrier that the projectile needs to tunnel through. The above procedure is well established, despite the fact that theory has difficulties in predicting the magnitude of the laboratory screening potential \citep{aliottaLanganke}. 

Unfortunately, Equation~(\ref{eq:screening}) is frequently applied to the strengths of narrow resonances, 
as first suggested by \citet{assenbaum87}. Several of the reported ``corrected'' resonance strengths exceed the measured values by up to 25\%, depending on the reaction and resonance energy \citep{PhysRevC.94.055804,PhysRevLett.117.142502,2012PhLB..707...60S,Sergi2015}. This issue has been discussed in detail by \citet{iliadis_screen}, who showed that: (i) it is incorrect to apply the nonresonant screening correction factor of Equation~(\ref{eq:screening}) to narrow resonances, and (ii) the actual differences between screened and unscreened resonance strengths are negligible ($<0.2$\%). 

The reason is that, for a narrow resonance, the presence of electrons causes two effects that nearly compensate each other. The first is the narrowing of the effective barrier, similar to the case of a nonresonant cross section, which increases the magnitude of the screened resonance strength. The second is the lowering of the energy of the resonance level in the compound nucleus, which implies that the energy at which the reaction proceeds is lower than the bare-nucleus resonance energy (Section~\ref{sec:energiesJpi}). This second effect decreases the magnitude of the screened resonance strength. Therefore, unless it can be demonstrated otherwise, no correction for laboratory electron screening is needed when considering the strengths of narrow resonances. For details, see \citet{iliadis_screen}.



\section{Online resources}\label{sec:git}
A majority of the charged particle induced thermonuclear reaction rates presented within this work (with the exception of the rates calculated using Bayesian methods) are calculated using the RatesMC computer code released under the GNU General Public License v3.0. Installation instructions for RatesMC, which contain information on the necessary software dependencies, can be found within a git repository hosted on GitHub: \url{https://github.com/rlongland/RatesMC}. A ``Training'' folder is also included that provides an example input file and guides users through a comprehensive rate calculation. The release version of the code used for the calculations presented in this work is \dataset[available online]{https://doi.org/10.5281/zenodo.17516448}~\citep{RatesMC_2_3}.

We also provide the RatesMC input files for all Monte-Carlo rates presented in Section~\ref{sec:ratetables}, as well as our machine-readable, tabulated reaction rates in
\dataset[ETR25]{https://doi.org/10.5281/zenodo.17610211} \citep{Input_output_files}. 

\section{Evaluated thermonuclear reaction rates}\label{sec:ratetables}
\subsection{General aspects}\label{sec:gen}
For the vast majority of reactions listed in Table~\ref{tab:overview}, the thermonuclear rates, $N_A \left < \sigma v \right >$, have been evaluated in the present work by taking into account the latest available information on the nuclear physics input. Exceptions are those rates that have been estimated using Bayesian models (Section~\ref{sec:bayesrates}). They are marked in the table with a footnote ``b'' and have been adopted, with small modifications, from the recent literature. 

Results are provided in the following tables and figures. The tables provide numerical values of the total reaction rates. Detailed information on the meaning of the table columns, and the interpretation of the figures, is provided below. The rates presented here refer to their {\it laboratory} values. This means that for use in stellar model simulations, the values provided here need to be corrected for the effects of thermal excitations of the interacting nuclei. 

The reaction rates (except for the Bayesian ones) have been computed using the code RatesMC (see Section~\ref{sec:mcoverview}), using 50,000 rate samples for each reaction. The complete input files to RatesMC for each reaction can be accessed via Github (see Section~\ref{sec:git}). These files contain the comprehensive nuclear physics input needed to perform the rate calculations.

The Low, Median, and High reaction rate values listed in the tables below are directly related to the 16th, 50th, and 84th percentiles, respectively, of the actual rate probability density function. However, the rate factor uncertainty, $f.u.$, is computed under the assumption that the rate probability density has a lognormal shape. In this case, the factor uncertainty is related to the Low, Median, and High values as explained below. However, if the actual Monte-Carlo probability density of the total rate deviates noticeably from a lognormal shape, $f.u.$ will no longer be related to the Low, Median, and High values of the total rate. 

\citet{ILIADIS2010b} provided figures of rate probability densities and numerical values of the Anderson-Darling statistic for assessing how closely the actual rate probability density follows a lognormal shape. We do not provide such details here, but the information can be requested from the authors. The second figure provided below for each reaction can be used to assess approximately the shape of the actual rate probability density at a given temperature. When the rate probability density has a lognormal shape, the normalized rate uncertainty (displayed on a logarithmic  scale of the ordinate) will be symmetric about unity.

To illustrate how to interpret the figures, consider Figures~\ref{fig:si28pg1} and~\ref{fig:si28pg2}. The former displays the fractional contributions to the total $^{28}$Si(p,$\gamma$)$^{29}$P reaction rate, while the latter depicts the uncertainty of the total reaction rate versus temperature. At low temperatures (below $\approx$100~MK), the reaction is dominated by direct capture (DC). As the temperature increases, this contribution gradually decreases, while the narrow resonance at $E_r^{c.m.}$ $=$ $358$~keV becomes dominant in the range $T$ $\approx$ $0.1$ $–$ $2$~GK. Notably, Figure~\ref{fig:si28pg2} shows a dip in the total rate uncertainty near $\approx$100~MK, where the direct capture and resonance contributions become comparable. In this regime, the independent uncertainties of each component combine quadratically, leading to a smaller overall uncertainty. At higher temperatures (above $\approx$2~GK), isolated resonances at $E_r^{c.m.}$ $=$ $1595$ and $2992$~keV begin to contribute significantly. The dotted line denotes the summed contribution from even higher-lying, individually measured resonances. It is also worth noting that the broad resonance at $E_r^{c.m.}$ $=$ $1595$~keV influences the total rate even at low temperatures ($T$ $\lesssim$ $100$~MK) through its low-energy tail.

\clearpage
\onecolumngrid
\subsection{Explanation of tables and figures}\label{sec:explan}
%


\clearpage
\subsection{Tables and figures of experimental reaction rates}\label{app:reacRatesTab}

\clearpage
\startlongtable
\begin{deluxetable*}{lcccc | lcccc}
\tablecaption{Total laboratory reaction rates for D(p,$\gamma$)$^3$He.}
\label{tab:dpg} 
\tablewidth{\columnwidth}
\tabletypesize{\footnotesize}
\tablecolumns{10}
\tablehead{
T (GK) & Low & Median & High & $f.u.$ & T (GK) & Low & Median & High & $f.u.$
}
\startdata
0.001 &   1.399E-11 &   1.454E-11 &   1.508E-11 &   1.038E+00 &   0.140 &   1.205E+01 &   1.232E+01 &   1.261E+01 &   1.023E+00  \\   
0.002 &   1.933E-08 &   2.007E-08 &   2.080E-08 &   1.038E+00 &   0.150 &   1.396E+01 &   1.428E+01 &   1.460E+01 &   1.023E+00  \\   
0.003 &   6.257E-07 &   6.494E-07 &   6.728E-07 &   1.037E+00 &   0.160 &   1.599E+01 &   1.635E+01 &   1.672E+01 &   1.022E+00  \\   
0.004 &   5.533E-06 &   5.741E-06 &   5.946E-06 &   1.037E+00 &   0.180 &   2.036E+01 &   2.081E+01 &   2.127E+01 &   1.022E+00  \\   
0.005 &   2.588E-05 &   2.684E-05 &   2.779E-05 &   1.037E+00 &   0.200 &   2.511E+01 &   2.566E+01 &   2.622E+01 &   1.022E+00  \\   
0.006 &   8.359E-05 &   8.666E-05 &   8.969E-05 &   1.036E+00 &   0.250 &   3.844E+01 &   3.926E+01 &   4.010E+01 &   1.021E+00  \\   
0.007 &   2.124E-04 &   2.202E-04 &   2.278E-04 &   1.036E+00 &   0.300 &   5.353E+01 &   5.466E+01 &   5.582E+01 &   1.021E+00  \\   
0.008 &   4.577E-04 &   4.743E-04 &   4.906E-04 &   1.036E+00 &   0.350 &   7.006E+01 &   7.155E+01 &   7.306E+01 &   1.021E+00  \\
\enddata
\tablecomments{
{\it Reaction rate:} \citet{moscoso2021}. \\
{\it $S$ factor:} See column 2 in Table~1 and Appendix~A in \citet{moscoso2021}. \\
{\it Normalization:} \citet{War63, Ma97, Sch95, Cas02, Tisma:2019ug, Mossa20, Turkat21}. \\
{\it High-temperature rates (in parentheses):} \citet{coc15}. \\
{\it Previous rates:} \citet{iliadis16, Yeh21, Pis21}. \\
{\it Other:} Bayesian rate.
}
\end{deluxetable*}

\clearpage

\begin{figure*}[hbt!]
\centering
\includegraphics[width=0.5\linewidth]{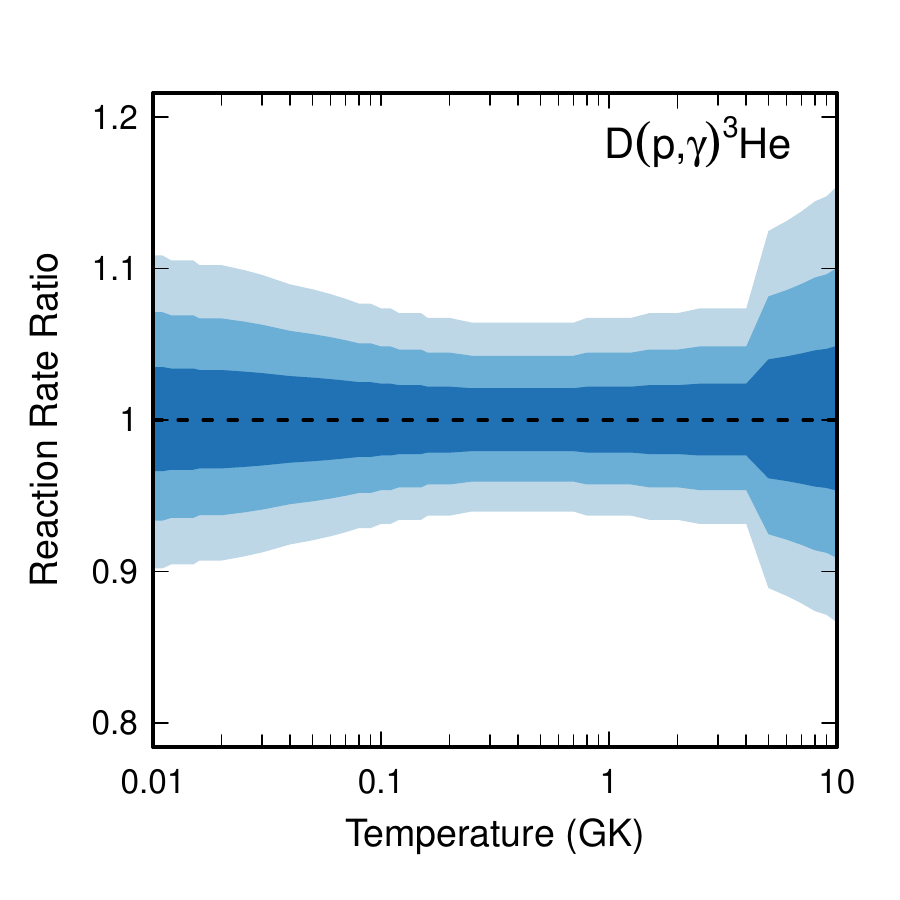}
\caption{
Reaction rate uncertainties versus temperature. The three different shades refer to coverage probabilities of 68\%, 90\%, and 98\%.}
\label{fig:dpg2}
\end{figure*}

\clearpage

\startlongtable
\begin{deluxetable*}{lcccc | lcccc}
\tablecaption{Total laboratory reaction rates for D(d,n)$^3$He.}
\label{tab:ddn} 
\tablewidth{\columnwidth}
\tabletypesize{\footnotesize}
\tablecolumns{10}
\tablehead{
T (GK) & Low & Median & High & $f.u.$ & T (GK) & Low & Median & High & $f.u.$
}
\startdata
0.001 &   1.308E-08 &   1.323E-08 &   1.336E-08 &   1.011E+00  &   0.140 &   5.579E+05 &   5.642E+05 &   5.700E+05 &   1.011E+00  \\
0.002 &   5.431E-05 &   5.493E-05 &   5.549E-05 &   1.011E+00  &   0.150 &   6.506E+05 &   6.580E+05 &   6.647E+05 &   1.011E+00  \\
0.003 &   2.993E-03 &   3.027E-03 &   3.058E-03 &   1.011E+00  &   0.160 &   7.488E+05 &   7.573E+05 &   7.650E+05 &   1.011E+00  \\
0.004 &   3.697E-02 &   3.739E-02 &   3.778E-02 &   1.011E+00  &   0.180 &   9.599E+05 &   9.708E+05 &   9.808E+05 &   1.011E+00  \\
0.005 &   2.191E-01 &   2.216E-01 &   2.239E-01 &   1.011E+00  &   0.200 &   1.189E+06 &   1.202E+06 &   1.215E+06 &   1.011E+00  \\
0.006 &   8.466E-01 &   8.562E-01 &   8.650E-01 &   1.011E+00  &   0.250 &   1.824E+06 &   1.845E+06 &   1.864E+06 &   1.011E+00  \\
0.007 &   2.482E+00 &   2.510E+00 &   2.536E+00 &   1.011E+00  &   0.300 &   2.529E+06 &   2.557E+06 &   2.583E+06 &   1.011E+00  \\
0.008 &   6.010E+00 &   6.078E+00 &   6.140E+00 &   1.011E+00  &   0.350 &   3.283E+06 &   3.321E+06 &   3.355E+06 &   1.011E+00  \\
\enddata
\tablecomments{
{\it Reaction rate:} \citet{gomez17}. \\
{\it $S$ factor:} See Table~2 and Figure~1 in \citet{gomez17}. \\
{\it Normalization:} see Table~2 of \citet{gomez17}. \\
{\it High-temperature rates (in parentheses):} \citet{coc15}. \\
{\it Previous rates:} \citet{coc15,Pis21}. \\
{\it Other:}
The rates listed in the table have been recalculated using a modified Bayesian model compared to \citet{gomez17}, as discussed in Section~\ref{sec:bayesrates}.
}
\end{deluxetable*}

\clearpage

\begin{figure*}[hbt!]
\centering
\includegraphics[width=0.5\linewidth]{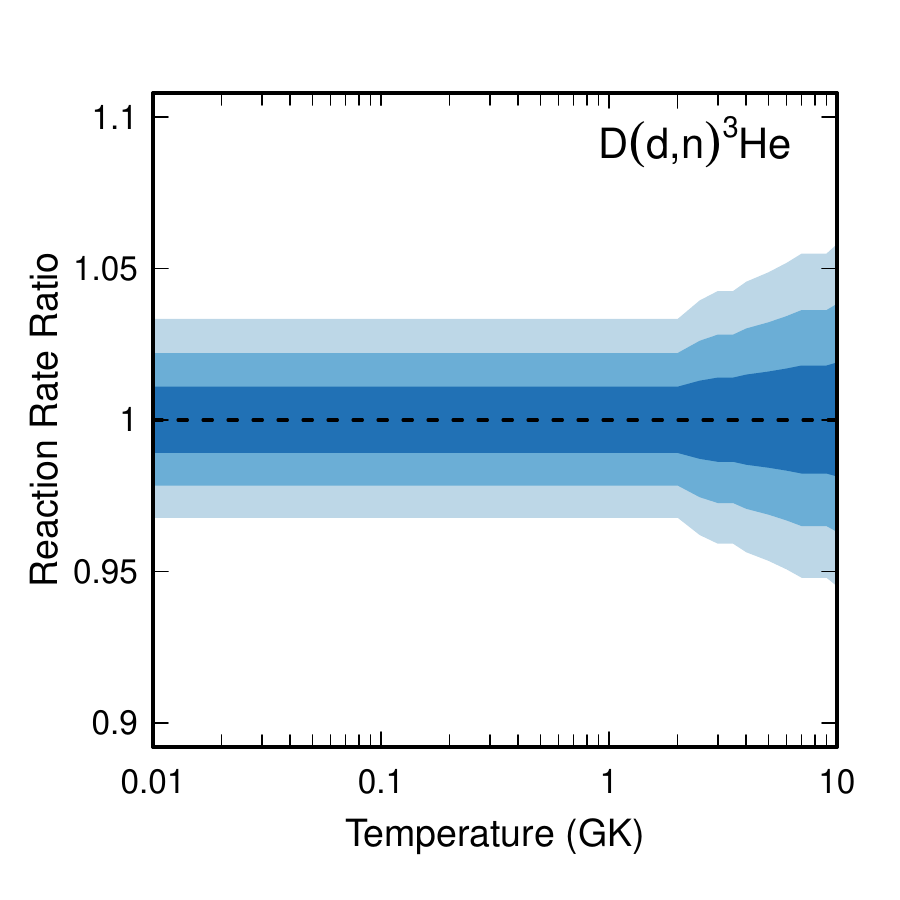}
\caption{
Reaction rate uncertainties versus temperature. The three different shades refer to coverage probabilities of 68\%, 90\%, and 98\%.}
\label{fig:ddn2}
\end{figure*}

\clearpage

\startlongtable
\begin{deluxetable*}{lcccc | lcccc}
\tablecaption{Total laboratory reaction rates for D(d,p)$^3$H.}
\label{tab:ddp} 
\tablewidth{\columnwidth}
\tabletypesize{\footnotesize}
\tablecolumns{10}
\tablehead{
T (GK) & Low & Median & High & $f.u.$ & T (GK) & Low & Median & High & $f.u.$
}
\startdata
0.001 &   1.344E-08 &   1.361E-08 &   1.381E-08 &   1.013E+00  &  0.140 &   5.257E+05 &   5.325E+05 &   5.401E+05 &   1.013E+00  \\
0.002 &   5.572E-05 &   5.644E-05 &   5.724E-05 &   1.013E+00  &  0.150 &   6.110E+05 &   6.189E+05 &   6.277E+05 &   1.013E+00  \\
0.003 &   3.065E-03 &   3.104E-03 &   3.149E-03 &   1.013E+00  &  0.160 &   7.009E+05 &   7.099E+05 &   7.201E+05 &   1.013E+00  \\
0.004 &   3.779E-02 &   3.828E-02 &   3.883E-02 &   1.013E+00  &  0.180 &   8.930E+05 &   9.046E+05 &   9.175E+05 &   1.013E+00  \\
0.005 &   2.236E-01 &   2.265E-01 &   2.298E-01 &   1.013E+00  &  0.200 &   1.100E+06 &   1.114E+06 &   1.130E+06 &   1.013E+00  \\
0.006 &   8.629E-01 &   8.740E-01 &   8.865E-01 &   1.013E+00  &  0.250 &   1.667E+06 &   1.688E+06 &   1.713E+06 &   1.013E+00  \\
0.007 &   2.526E+00 &   2.559E+00 &   2.595E+00 &   1.013E+00  &  0.300 &   2.288E+06 &   2.317E+06 &   2.350E+06 &   1.013E+00  \\
0.008 &   6.109E+00 &   6.188E+00 &   6.276E+00 &   1.013E+00  &  0.350 &   2.945E+06 &   2.983E+06 &   3.026E+06 &   1.013E+00  \\
\enddata
\tablecomments{
{\it Reaction rate:} \citet{gomez17}. \\
{\it $S$ factor:} See Table~2 and Figure~2 in \citet{gomez17}. \\
{\it Normalization:} See Table~2 of \citet{gomez17}. \\
{\it High-temperature rates (in parentheses):} \citet{coc15}. \\
{\it Previous rates:} \citet{coc15,Pis21}. \\
{\it Other:}
The rates listed in the table have been recalculated using a modified Bayesian model compared to \citet{gomez17}, as discussed in Section~\ref{sec:bayesrates}.
}
\end{deluxetable*}

\clearpage

\begin{figure*}[hbt!]
\centering
\includegraphics[width=0.5\linewidth]{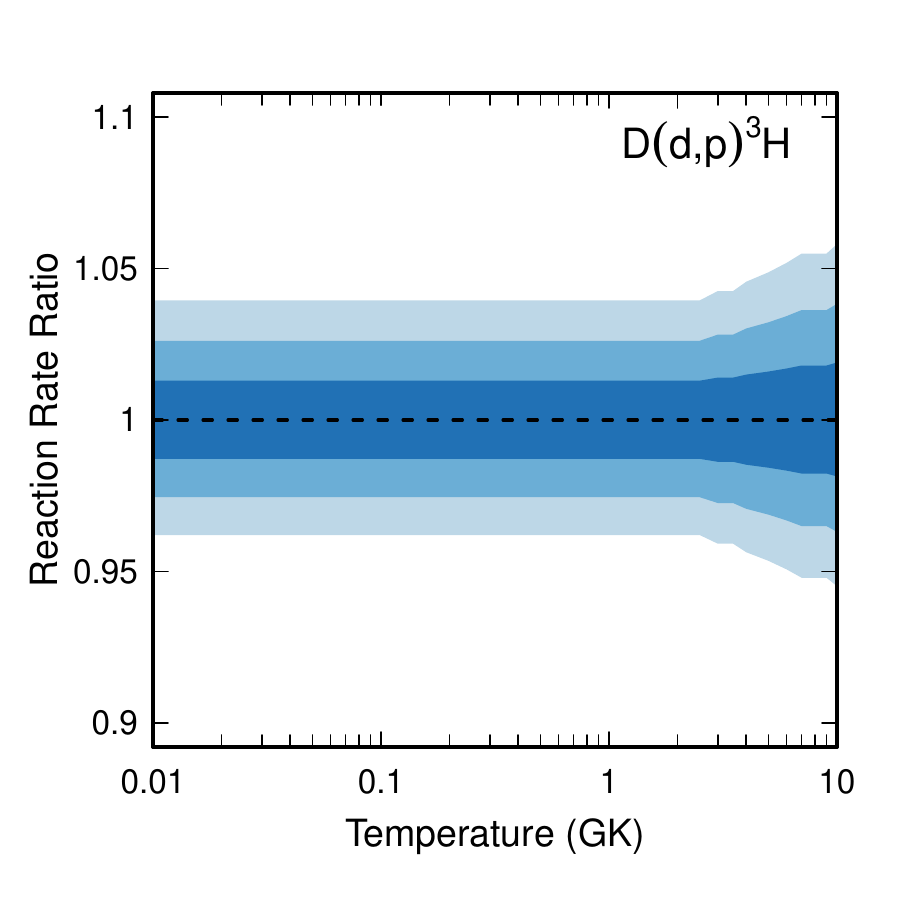}
\caption{
Reaction rate uncertainties versus temperature. The three different shades refer to coverage probabilities of 68\%, 90\%, and 98\%.}
\label{fig:ddp2}
\end{figure*}

\clearpage

\startlongtable
\begin{deluxetable*}{lcccc | lcccc}
\tablecaption{Total laboratory reaction rates for $^3$H(d,n)$^4$He.
\label{tab:tdn}} 
\tablewidth{\columnwidth}
\tabletypesize{\footnotesize}
\tablecolumns{10}
\tablehead{
T (GK) & Low & Median & High & $f.u.$ & T (GK) & Low & Median & High & $f.u.$
}
\startdata
0.001 &   1.986E-07 &   1.998E-07 &   2.010E-07 &   1.006E+00  &   0.140 &   1.047E+08 &   1.050E+08 &   1.053E+08 &   1.003E+00  \\
0.002 &   1.437E-03 &   1.445E-03 &   1.454E-03 &   1.006E+00  &   0.150 &   1.214E+08 &   1.218E+08 &   1.221E+08 &   1.003E+00  \\
0.003 &   1.040E-01 &   1.046E-01 &   1.052E-01 &   1.006E+00  &   0.160 &   1.384E+08 &   1.388E+08 &   1.392E+08 &   1.003E+00  \\
0.004 &   1.530E+00 &   1.539E+00 &   1.548E+00 &   1.006E+00  &   0.180 &   1.727E+08 &   1.732E+08 &   1.736E+08 &   1.003E+00  \\
0.005 &   1.028E+01 &   1.034E+01 &   1.040E+01 &   1.006E+00  &   0.200 &   2.064E+08 &   2.069E+08 &   2.074E+08 &   1.003E+00  \\
0.006 &   4.380E+01 &   4.405E+01 &   4.430E+01 &   1.006E+00  &   0.250 &   2.836E+08 &   2.843E+08 &   2.850E+08 &   1.002E+00  \\
0.007 &   1.389E+02 &   1.397E+02 &   1.405E+02 &   1.006E+00  &   0.300 &   3.475E+08 &   3.483E+08 &   3.492E+08 &   1.002E+00  \\
0.008 &   3.594E+02 &   3.614E+02 &   3.635E+02 &   1.006E+00  &   0.350 &   3.979E+08 &   3.988E+08 &   3.998E+08 &   1.002E+00  \\
\enddata
\tablecomments{
{\it Reaction rate:} \citet{desouza2019a}. \\
{\it $S$ factor:} See Figure~1 and Appendix in \citet{desouza2019a}. \\
{\it Observed resonances:} see Table~1 in \citet{desouza2019a}. \\
{\it Normalization:} See Figure~1 in \citet{desouza2019a}, except \citet{brown87}, who did not determine absolute cross sections. \\
{\it High-temperature rates (in parentheses):} \citet{descouvemont04}. \\
{\it Previous rates:} \citet{descouvemont04,bosch92b}. \\
{\it Other:} Bayesian rate.
}
\end{deluxetable*}

\clearpage

\begin{figure*}[hbt!]
\centering
\includegraphics[width=0.5\linewidth]{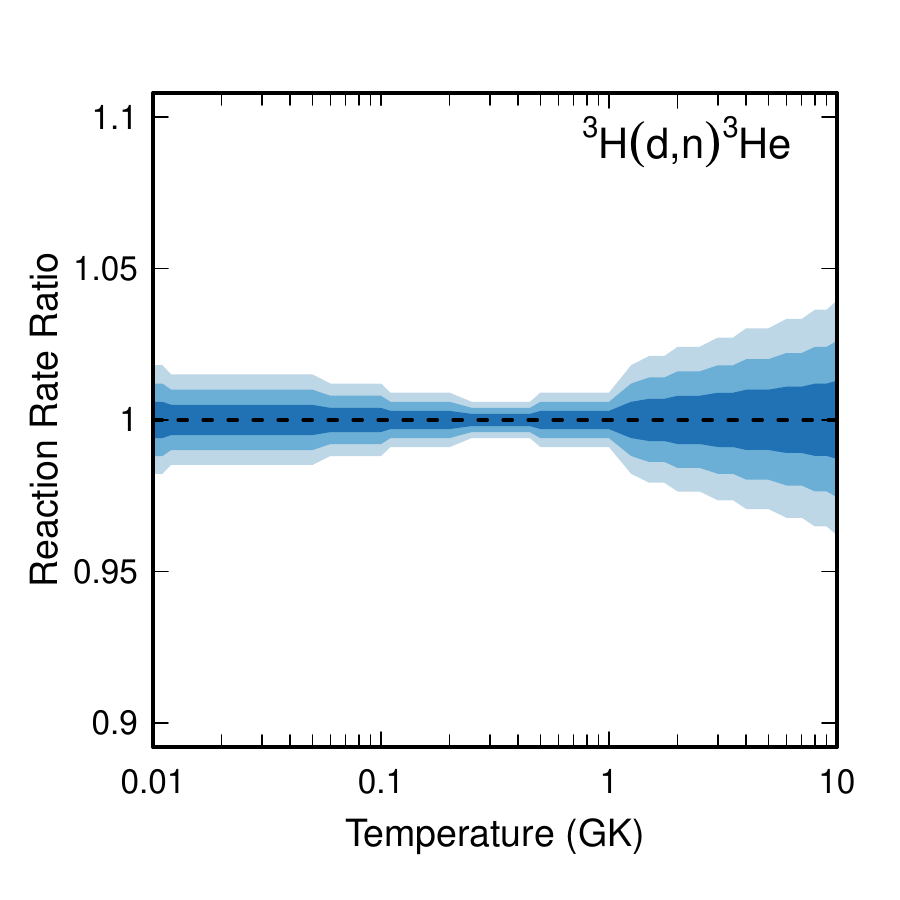}
\caption{
Reaction rate uncertainties versus temperature. The three different shades refer to coverage probabilities of 68\%, 90\%, and 98\%.}
\label{fig:tdn2}
\end{figure*}

\clearpage

\startlongtable
\begin{deluxetable*}{lcccc | lcccc}
\tablecaption{Total laboratory reaction rates for $^3$He(d,p)$^4$He.
\label{tab:3hedp}} 
\tablewidth{\columnwidth}
\tabletypesize{\footnotesize}
\tablecolumns{10}
\tablehead{
T (GK) & Low & Median & High & $f.u.$ & T (GK) & Low & Median & High & $f.u.$
}
\startdata
0.001 &   3.437E-19 &   3.496E-19 &   3.555E-19 &   1.017E+00  &  0.140 &   2.961E+05 &   3.000E+05 &   3.039E+05 &   1.013E+00  \\ 
0.002 &   5.934E-13 &   6.035E-13 &   6.138E-13 &   1.017E+00  &  0.150 &   3.967E+05 &   4.019E+05 &   4.071E+05 &   1.013E+00  \\ 
0.003 &   6.150E-10 &   6.248E-10 &   6.348E-10 &   1.016E+00  &  0.160 &   5.193E+05 &   5.255E+05 &   5.318E+05 &   1.012E+00  \\ 
0.004 &   4.848E-08 &   4.926E-08 &   5.005E-08 &   1.016E+00  &  0.180 &   8.362E+05 &   8.462E+05 &   8.564E+05 &   1.012E+00  \\ 
0.005 &   1.076E-06 &   1.093E-06 &   1.111E-06 &   1.016E+00  &  0.200 &   1.263E+06 &   1.278E+06 &   1.293E+06 &   1.012E+00  \\ 
0.006 &   1.140E-05 &   1.158E-05 &   1.176E-05 &   1.016E+00  &  0.250 &   2.894E+06 &   2.926E+06 &   2.958E+06 &   1.011E+00  \\ 
0.007 &   7.480E-05 &   7.600E-05 &   7.722E-05 &   1.016E+00  &  0.300 &   5.434E+06 &   5.493E+06 &   5.554E+06 &   1.011E+00  \\ 
0.008 &   3.525E-04 &   3.581E-04 &   3.638E-04 &   1.016E+00  &  0.350 &   8.923E+06 &   9.021E+06 &   9.120E+06 &   1.011E+00  \\
\enddata
\tablecomments{
{\it Reaction rate:} \citet{desouza2019b}. \\
{\it $S$ factor:} See Figure~4 and Appendix in \citet{desouza2019b}. \\
{\it Observed resonances:} See Table~2 in \citet{desouza2019b}. \\
{\it Normalization:} See Figure~4 and Appendix in \citet{desouza2019b}. \\
{\it High-temperature rates (in parentheses):} \citet{descouvemont04}. \\
{\it Previous rates:} \citet{descouvemont04}. \\
{\it Other:} Bayesian rate.
}
\end{deluxetable*}

\clearpage

\begin{figure*}[hbt!]
\centering
\includegraphics[width=0.5\linewidth]{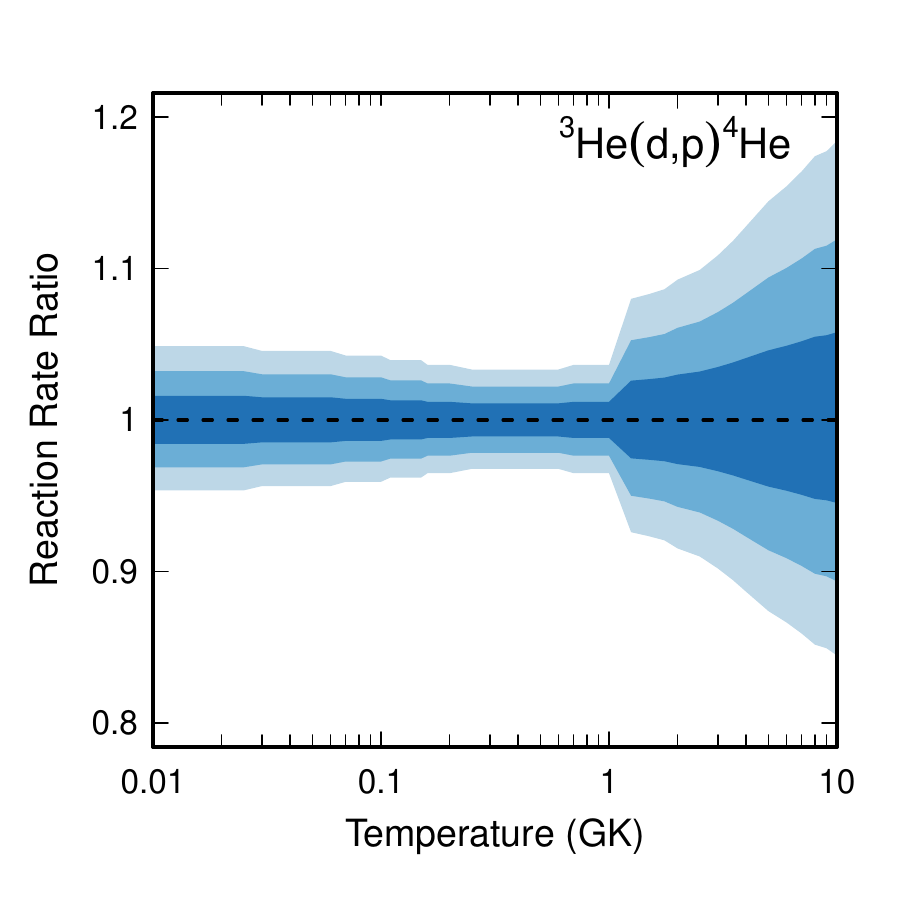}
\caption{
Reaction rate uncertainties versus temperature. The three different shades refer to coverage probabilities of 68\%, 90\%, and 98\%.}
\label{fig:he3dp2}
\end{figure*}

\clearpage

\startlongtable
\begin{deluxetable*}{lcccc | lcccc}
\tablecaption{Total laboratory reaction rates for $^3$He($^3$He,2p)$^4$He.
\label{tab:3he3he}} 
\tablewidth{\columnwidth}
\tabletypesize{\footnotesize}
\tablecolumns{10}
\tablehead{
T (GK) & Low & Median & High & $f.u.$ & T (GK) & Low & Median & High & $f.u.$
}
\startdata
 0.001 &  2.660E-41 &  2.740E-41 &  2.830E-41  &   1.031E+00  &    0.140 &  1.058E+01 &  1.084E+01 &  1.115E+01  &   1.027E+00  \\
 0.002 &  1.669E-30 &  1.719E-30 &  1.776E-30  &   1.031E+00  &    0.150 &  1.725E+01 &  1.769E+01 &  1.819E+01  &   1.027E+00  \\
 0.003 &  2.846E-25 &  2.932E-25 &  3.028E-25  &   1.031E+00  &    0.160 &  2.696E+01 &  2.764E+01 &  2.842E+01  &   1.027E+00  \\
 0.004 &  5.636E-22 &  5.804E-22 &  5.994E-22  &   1.031E+00  &    0.180 &  5.928E+01 &  6.078E+01 &  6.246E+01  &   1.026E+00  \\
 0.005 &  1.241E-19 &  1.278E-19 &  1.319E-19  &   1.031E+00  &    0.200 &  1.165E+02 &  1.195E+02 &  1.228E+02  &   1.026E+00  \\
 0.006 &  7.570E-18 &  7.794E-18 &  8.049E-18  &   1.031E+00  &    0.250 &  4.476E+02 &  4.589E+02 &  4.712E+02  &   1.026E+00  \\
 0.007 &  2.013E-16 &  2.072E-16 &  2.140E-16  &   1.031E+00  &    0.300 &  1.238E+03 &  1.270E+03 &  1.304E+03  &   1.026E+00  \\
 0.008 &  3.008E-15 &  3.096E-15 &  3.198E-15  &   1.031E+00  &    0.350 &  2.774E+03 &  2.845E+03 &  2.921E+03  &   1.026E+00  \\
\enddata
\tablecomments{
{\it Reaction rate:} \citet{iliadis16}. \\
{\it $S$ factor:} See Table~2 and Appendix~C in \citet{iliadis16}. \\
{\it Normalization:} See Table~2 of \citet{iliadis16}. \\
{\it High-temperature rates (in parentheses):} \citet{ANGULO1999}. \\
{\it Previous rates:} \citet{RevModPhys.83.195}. \\
{\it Other:}
The rates listed in the table have been recalculated using a modified Bayesian model compared to \citet{iliadis16}, as discussed in Section~\ref{sec:bayesrates}.\\
}
\end{deluxetable*}

\clearpage

\begin{figure*}[hbt!]
\centering
\includegraphics[width=0.5\linewidth]{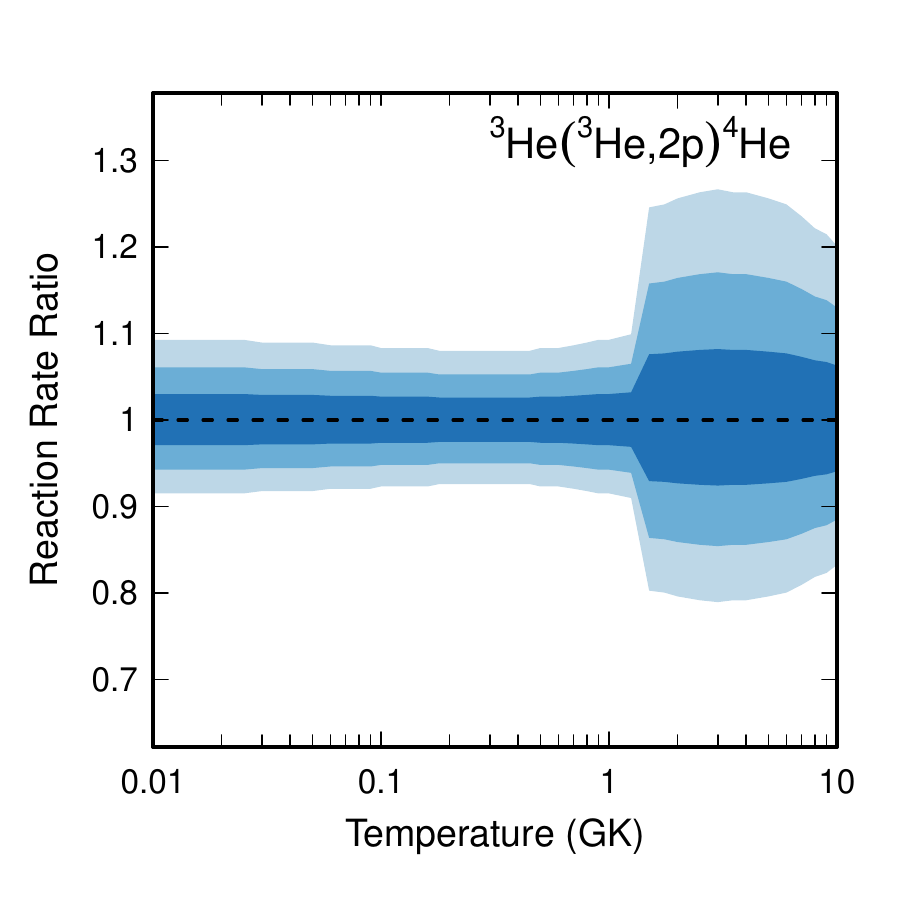}
\caption{
Reaction rate uncertainties versus temperature. The three different shades refer to coverage probabilities of 68\%, 90\%, and 98\%.}
\label{fig:he3he32}
\end{figure*}

\clearpage

\startlongtable
\begin{deluxetable*}{lcccc | lcccc}
\tablecaption{Total laboratory reaction rates for $^3$He($\alpha$,$\gamma$)$^7$Be.
\label{tab:3heag}} 
\tablewidth{\columnwidth}
\tabletypesize{\footnotesize}
\tablecolumns{10}
\tablehead{
T (GK) & Low & Median & High & $f.u.$ & T (GK) & Low & Median & High & $f.u.$
}
\startdata
 0.001 &  1.148E-47 &  1.178E-47 &  1.199E-47  &   1.021E+00  &   0.140 &  3.811E-04 &  3.911E-04 &  3.982E-04  &   1.021E+00  \\
 0.002 &  2.240E-36 &  2.299E-36 &  2.341E-36  &   1.021E+00  &   0.150 &  6.360E-04 &  6.527E-04 &  6.646E-04  &   1.021E+00  \\
 0.003 &  6.633E-31 &  6.808E-31 &  6.932E-31  &   1.021E+00  &   0.160 &  1.015E-03 &  1.041E-03 &  1.060E-03  &   1.021E+00  \\
 0.004 &  1.861E-27 &  1.910E-27 &  1.945E-27  &   1.021E+00  &   0.180 &  2.314E-03 &  2.375E-03 &  2.418E-03  &   1.021E+00  \\
 0.005 &  5.251E-25 &  5.389E-25 &  5.487E-25  &   1.021E+00  &   0.200 &  4.692E-03 &  4.816E-03 &  4.903E-03  &   1.021E+00  \\
 0.006 &  3.872E-23 &  3.974E-23 &  4.046E-23  &   1.021E+00  &   0.250 &  1.916E-02 &  1.967E-02 &  2.003E-02  &   1.021E+00  \\
 0.007 &  1.198E-21 &  1.229E-21 &  1.252E-21  &   1.021E+00  &   0.300 &  5.549E-02 &  5.695E-02 &  5.799E-02  &   1.021E+00  \\
 0.008 &  2.028E-20 &  2.082E-20 &  2.119E-20  &   1.021E+00  &   0.350 &  1.287E-01 &  1.321E-01 &  1.345E-01  &   1.021E+00  \\
\enddata
\tablecomments{
{\it Reaction rate:} \citet{iliadis16}. \\
{\it $S$ factor:} See Table~3 and Appendix~C in \citet{iliadis16}. \\
{\it Normalization:} See Table~3 of \citet{iliadis16}. \\
{\it High-temperature rates (in parentheses):} \citet{kontos2013}. \\
{\it Previous rates:} \citet{kontos2013,deboer2014,Zhang2020}. \\
{\it Other:}
For three energies (126.5~keV, 147.7~keV, and 168.9~keV), the statistical $S$ factor uncertainties listed in Table~2 of \citet{COSTANTINI2008144} are erroneous, as pointed out in \citet{ZERKIN201831} (EXFOR). The erroneous values were used by \citet{iliadis16}. The Bayesian rate listed above was obtained using the corrected values.
}
\end{deluxetable*}

\clearpage

\begin{figure*}[hbt!]
\centering
\includegraphics[width=0.5\linewidth]{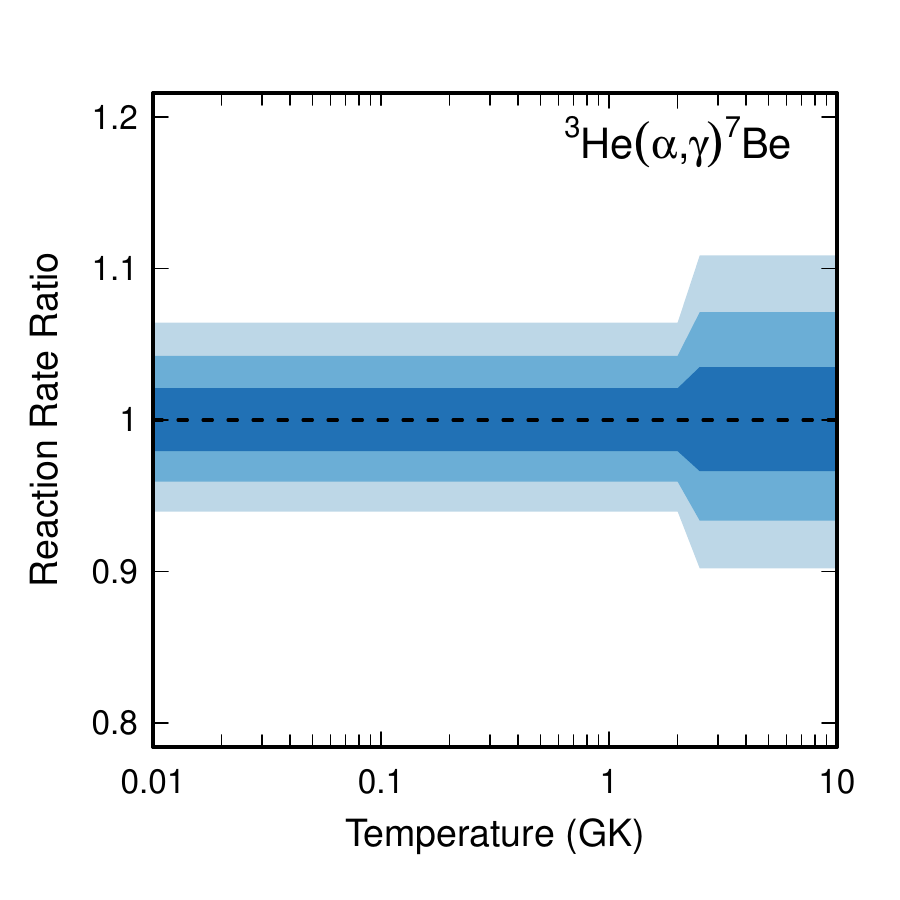}
\caption{
Reaction rate uncertainties versus temperature. The three different shades refer to coverage probabilities of 68\%, 90\%, and 98\%.}
\label{fig:he3ag2}
\end{figure*}

\clearpage

\startlongtable
\begin{deluxetable*}{lcccc | lcccc}
\tablecaption{Total laboratory reaction rates for $^7$Be(n,p)$^7$Li.
\label{tab:7benp}} 
\tablewidth{\columnwidth}
\tabletypesize{\footnotesize}
\tablecolumns{10}
\tablehead{
T (GK) & Low & Median & High & $f.u.$ & T (GK) & Low & Median & High & $f.u.$
}
\startdata
0.001 &   5.076E+09 &   5.156E+09 &   5.237E+09 &   1.015E+00  &  0.140 &   2.892E+09 &   2.944E+09 &   2.997E+09 &   1.018E+00  \\
0.002 &   4.948E+09 &   5.025E+09 &   5.103E+09 &   1.015E+00  &  0.150 &   2.839E+09 &   2.890E+09 &   2.943E+09 &   1.018E+00  \\
0.003 &   4.853E+09 &   4.928E+09 &   5.004E+09 &   1.015E+00  &  0.160 &   2.788E+09 &   2.839E+09 &   2.891E+09 &   1.018E+00  \\
0.004 &   4.775E+09 &   4.848E+09 &   4.922E+09 &   1.015E+00  &  0.180 &   2.697E+09 &   2.747E+09 &   2.798E+09 &   1.018E+00  \\
0.005 &   4.709E+09 &   4.780E+09 &   4.853E+09 &   1.015E+00  &  0.200 &   2.613E+09 &   2.662E+09 &   2.712E+09 &   1.018E+00  \\
0.006 &   4.650E+09 &   4.721E+09 &   4.793E+09 &   1.015E+00  &  0.250 &   2.439E+09 &   2.486E+09 &   2.533E+09 &   1.019E+00  \\
0.007 &   4.597E+09 &   4.666E+09 &   4.737E+09 &   1.015E+00  &  0.300 &   2.298E+09 &   2.342E+09 &   2.388E+09 &   1.019E+00  \\
0.008 &   4.548E+09 &   4.617E+09 &   4.687E+09 &   1.015E+00  &  0.350 &   2.182E+09 &   2.225E+09 &   2.268E+09 &   1.019E+00  \\
\enddata
\tablecomments{
{\it Reaction rate:} \citet{desouza2020}. \\
{\it $S$ factor:} See Figure~3 and Appendix in \citet{desouza2020}. \\
{\it Normalization:} See Figure~2 and Section~2 in \citet{desouza2020}. \\
{\it High-temperature rates (in parentheses):} \citet{descouvemont04}. \\
{\it Previous rates:} \citet{descouvemont04,dam18}. \\
{\it Other:}
Bayesian rate. Although we adopt at the highest temperatures ($T$ $\geq$ $1.25$~GK) the values of \citet{descouvemont04}, their rate uncertainties are unrealistically small over almost their entire temperature range. Therefore, we adopted a more realistic estimate of 3\% rate uncertainty for $T$ $=$ $1.25$~GK $-$ $8.0$~GK and their published rate uncertainties for $T$ $=$ $9.0$ $-$ $10.0$~GK.
}
\end{deluxetable*}

\clearpage

\begin{figure*}[hbt!]
\centering
\includegraphics[width=0.5\linewidth]{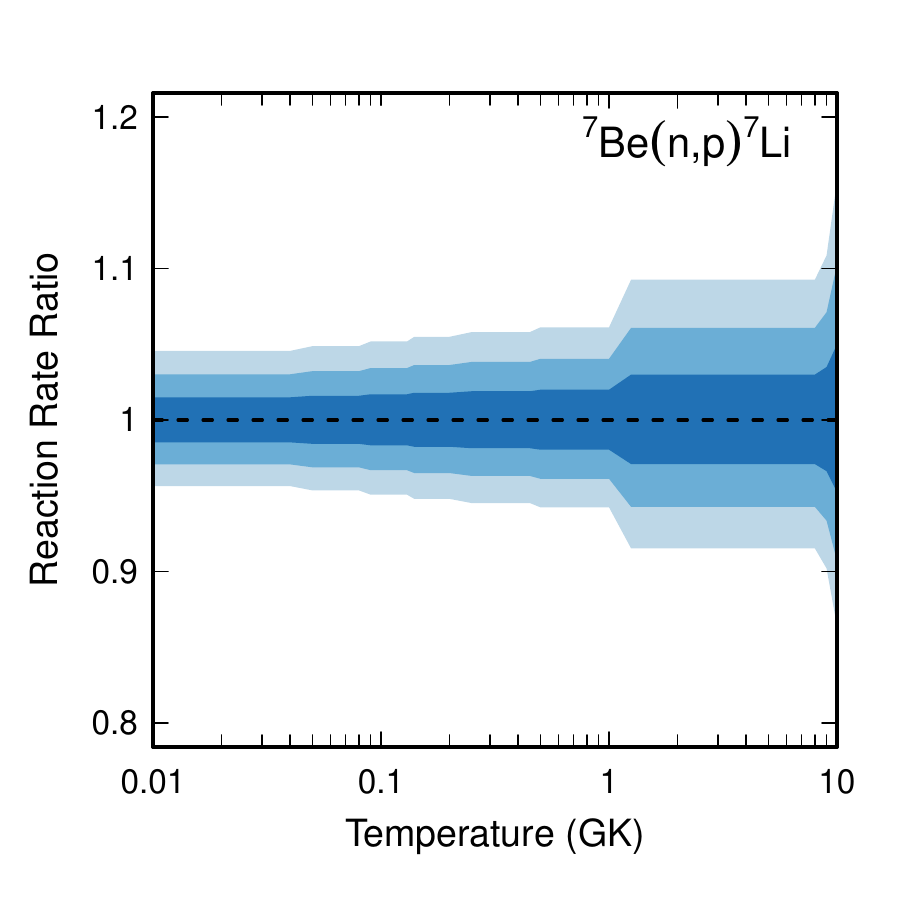}
\caption{
Reaction rate uncertainties versus temperature. The three different shades refer to coverage probabilities of 68\%, 90\%, and 98\%.}
\label{fig:be7np2}
\end{figure*}

\clearpage

\startlongtable
\begin{deluxetable*}{lcccc | lcccc}
\tablecaption{Total laboratory reaction rates for $^7$Be($\alpha$,$\gamma$)$^{11}$C.}
\label{tab:be7ag} 
\tablewidth{\columnwidth}
\tabletypesize{\footnotesize}
\tablecolumns{10}
\tablehead{
T (GK) & Low & Median & High & $f.u.$ & T (GK) & Low & Median & High & $f.u.$
}
\startdata
 0.001  &  5.381E-92  &  1.598E-91  &        4.752E-91  &  2.999E+00  &  0.140  &  6.715E-11  &  3.514E-10  &        3.050E-09  &  6.483E+00   \\ 
 0.002  &  2.125E-71  &  6.311E-71  &        1.877E-70  &  2.999E+00  &  0.150  &  1.700E-10  &  8.697E-10  &        7.487E-09  &  6.394E+00   \\ 
 0.003  &  2.115E-61  &  6.281E-61  &        1.868E-60  &  2.999E+00  &  0.160  &  3.981E-10  &  1.989E-09  &        1.698E-08  &  6.307E+00   \\ 
 0.004  &  4.301E-55  &  1.277E-54  &        3.798E-54  &  2.999E+00  &  0.180  &  1.823E-09  &  8.600E-09  &        7.174E-08  &  6.095E+00   \\ 
 0.005  &  1.322E-50  &  3.925E-50  &        1.168E-49  &  2.999E+00  &  0.200  &  7.567E-09  &  3.103E-08  &        2.475E-07  &  5.593E+00   \\ 
 0.006  &  3.505E-47  &  1.041E-46  &        3.095E-46  &  2.999E+00  &  0.250  &  5.869E-07  &  8.779E-07  &        3.365E-06  &  2.900E+00   \\ 
 0.007  &  1.475E-42  &  1.268E-41  &        1.220E-40  &  8.680E+00  &  0.300  &  2.794E-05  &  3.314E-05  &        4.808E-05  &  1.694E+00   \\ 
 0.008  &  2.505E-40  &  2.141E-39  &        2.060E-38  &  8.639E+00  &  0.350  &  4.692E-04  &  5.397E-04  &        6.434E-04  &  1.347E+00   \\
\enddata
\tablecomments{
{\it Observed resonances:} \citet{Wiescher1983,Hardie1984,Freer2012,Psaltis2022a,Psaltis2022b}. \\
{\it Unobserved resonances:} \citet{Hartos2018} for the sub-subthreshold resonance. \\
{\it High-temperature rates (in parentheses):} No matching to statistical model rates is needed over the listed temperature range. \\
{\it Previous rates:} \citet{Xu2013, Psaltis2022b}. \\
{\it Other:} 
Prescription of \citet{Hartos2018} was used for $E_r^{c.m.}$ $=$ $-44$~keV;
$\Gamma_{\alpha}$ and $\Gamma_p$ were adopted from \citet{Freer2012} for resonances with $E_r^{c.m.}$ $>$ $1500$~keV; values of $\Gamma_\gamma$ were assumed to be $1$~eV, with a factor of 10 uncertainty when no value was provided in the literature.
}
\end{deluxetable*}

\clearpage

\begin{figure*}[hbt!]
\centering
\includegraphics[width=0.5\linewidth]{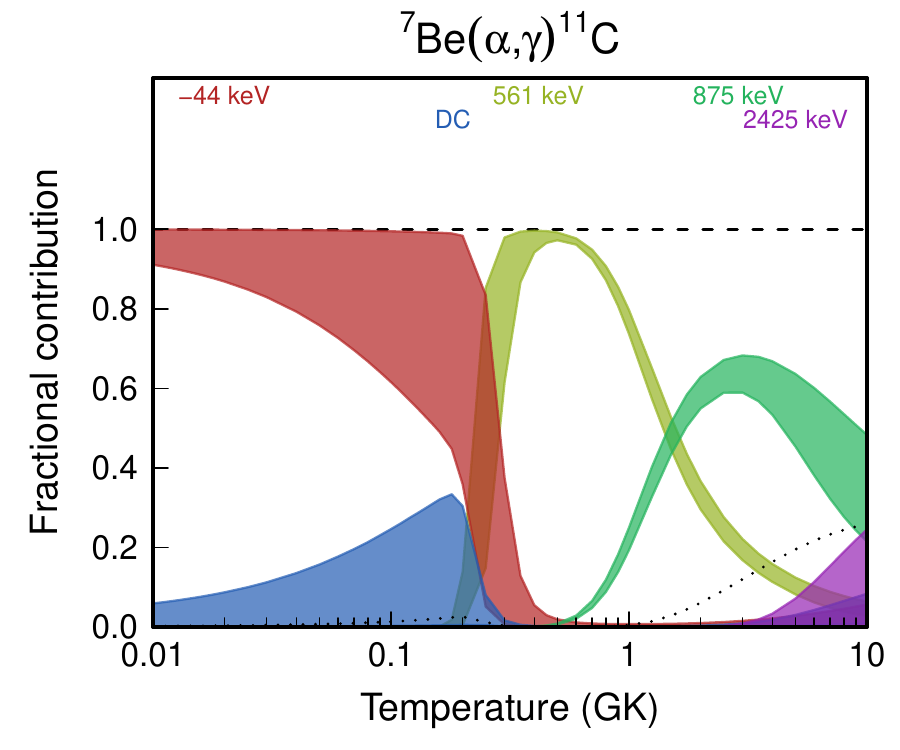}
\caption{
Fractional contributions to the total rate. ``DC'' refers to direct radiative capture. Resonance energies are given in the center-of-mass frame. 
}
\label{fig:be7ag1}
\end{figure*}
\begin{figure*}[hbt!]
\centering
\includegraphics[width=0.5\linewidth]{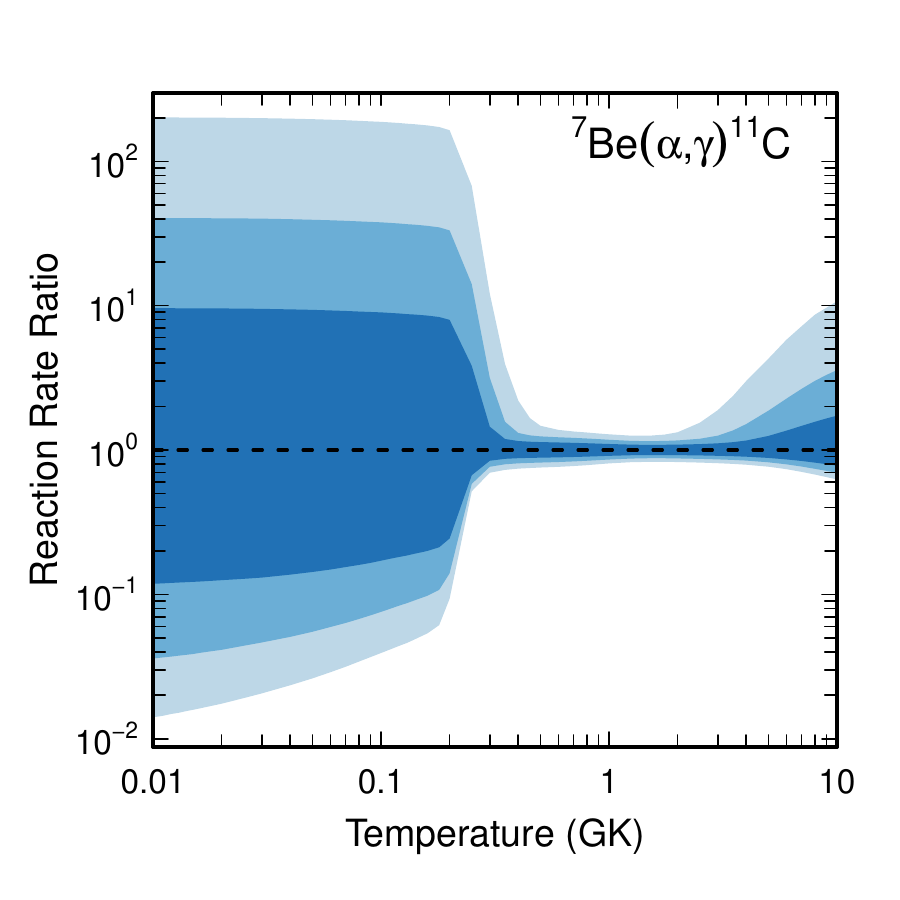}
\caption{
Reaction rate uncertainties versus temperature. The three different shades refer to coverage probabilities of 68\%, 90\%, and 98\%.}
\label{fig:be7ag2}
\end{figure*}

\clearpage

\startlongtable
\begin{deluxetable*}{lcccc | lcccc}
\tablecaption{Total laboratory reaction rates for $^{14}$C(p,$\gamma$)$^{15}$N.}
\label{tab:c14pg} 
\tablewidth{\columnwidth}
\tabletypesize{\footnotesize}
\tablecolumns{10}
\tablehead{
T (GK) & Low & Median & High & $f.u.$ & T (GK) & Low & Median & High & $f.u.$
}
\startdata
 0.001  &     1.743E-50   &         2.489E-50   &             3.611E-50   &         1.440E+00     &  0.140  &     1.021E-03   &         1.450E-03   &             2.096E-03   &         1.434E+00      \\ 
 0.002  &     2.251E-38   &         3.214E-38   &             4.664E-38   &         1.440E+00     &  0.150  &     1.786E-03   &         2.534E-03   &             3.660E-03   &         1.432E+00      \\ 
 0.003  &     1.673E-32   &         2.388E-32   &             3.464E-32   &         1.440E+00     &  0.160  &     2.980E-03   &         4.224E-03   &             6.095E-03   &         1.431E+00      \\ 
 0.004  &     8.395E-29   &         1.198E-28   &             1.738E-28   &         1.440E+00     &  0.180  &     7.416E-03   &         1.047E-02   &             1.505E-02   &         1.425E+00      \\ 
 0.005  &     3.585E-26   &         5.117E-26   &             7.423E-26   &         1.440E+00     &  0.200  &     1.649E-02   &         2.310E-02   &             3.300E-02   &         1.415E+00      \\ 
 0.006  &     3.629E-24   &         5.179E-24   &             7.513E-24   &         1.440E+00     &  0.250  &     9.046E-02   &         1.212E-01   &             1.673E-01   &         1.361E+00      \\ 
 0.007  &     1.446E-22   &         2.064E-22   &             2.994E-22   &         1.440E+00     &  0.300  &     3.785E-01   &         4.790E-01   &             6.278E-01   &         1.291E+00      \\ 
 0.008  &     3.020E-21   &         4.309E-21   &             6.250E-21   &         1.440E+00     &  0.350  &     1.262E+00   &         1.530E+00   &             1.911E+00   &         1.237E+00      \\
\enddata
\tablecomments{
{\it Observed resonances:} \citet{GORRES1990,PhysRevC.92.044328,PhysRevC.107.064617}. \\
{\it Normalization:} None. \\
{\it Unobserved resonances:} None. \\
{\it High-temperature rates (in parentheses):} Matching to statistical model rate above $T$ $=$ $2.4$~GK. \\
{\it Previous rates:} \citet{ILIADIS2010b}. \\
{\it Other:} The direct capture $S$ factor has been calculated using experimental spectroscopic factors from \citet{BOMMER1975246,SERCELY197953,ruziev}.
}
\end{deluxetable*}

\clearpage

\begin{figure*}[hbt!]
\centering
\includegraphics[width=0.5\linewidth]{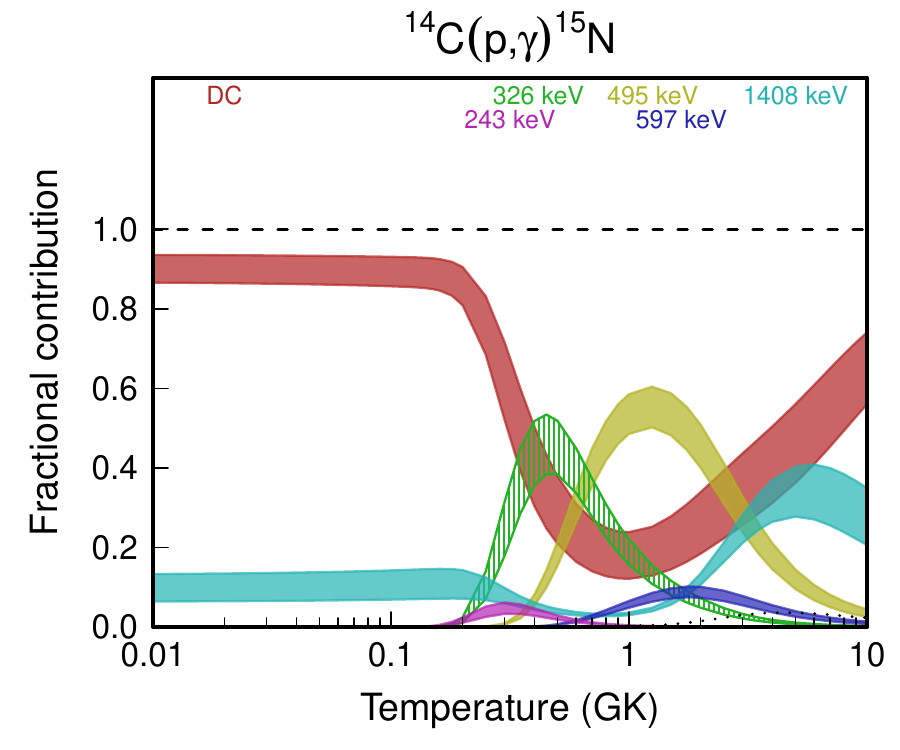}
\caption{
Fractional contributions to the total rate. ``DC'' refers to direct radiative capture. Resonance energies are given in the center-of-mass frame. 
}
\label{fig:c14pg1}
\end{figure*}
\begin{figure*}[hbt!]
\centering
\includegraphics[width=0.5\linewidth]{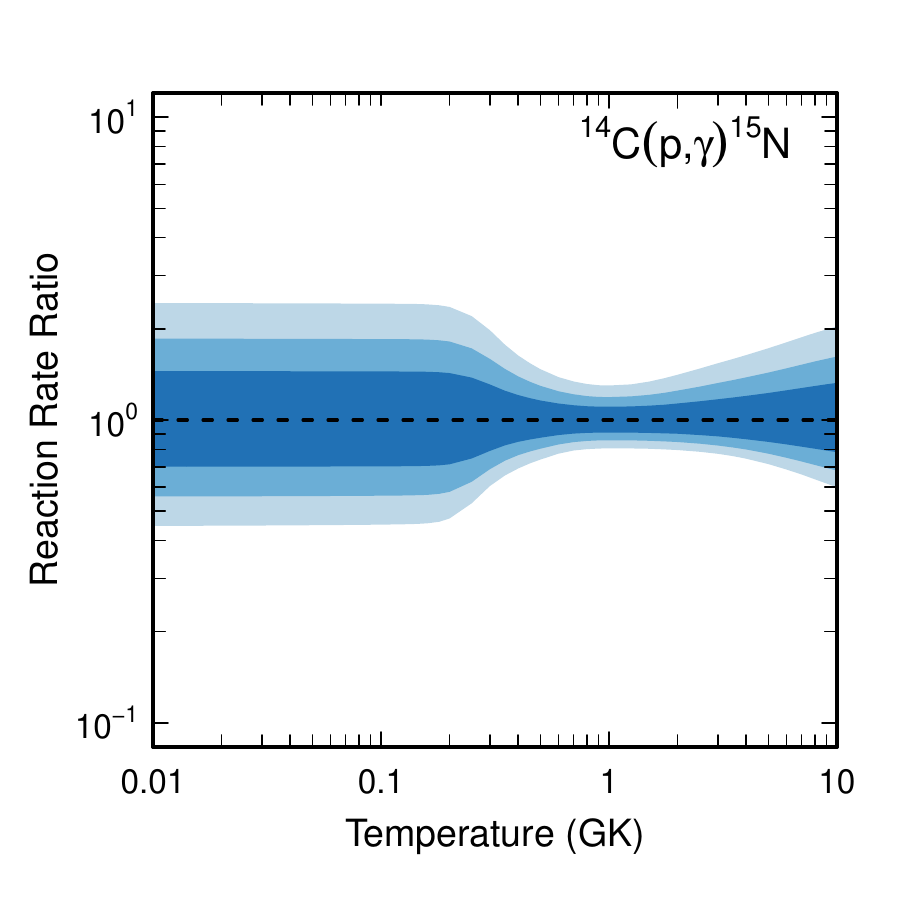}
\caption{
Reaction rate uncertainties versus temperature. The three different shades refer to coverage probabilities of 68\%, 90\%, and 98\%. 
}
\label{fig:c14pg2}
\end{figure*}

\clearpage

\startlongtable
\begin{deluxetable*}{lcccc | lcccc}
\tablecaption{Total laboratory reaction rates for $^{14}$C($\alpha$,$\gamma$)$^{18}$O.}
\label{tab:c14ag} 
\tablewidth{\columnwidth}
\tabletypesize{\footnotesize}
\tablecolumns{10}
\tablehead{
T (GK) & Low & Median & High & $f.u.$ & T (GK) & Low & Median & High & $f.u.$
}
\startdata
 0.001  &     0.000E+00   &         0.000E+00   &             0.000E+00   &         1.000E+00     &  0.140  &     4.220E-15   &         4.769E-14   &             2.317E-13   &         8.138E+00      \\ 
 0.002  &     0.000E+00   &         0.000E+00   &             0.000E+00   &         1.000E+00     &  0.150  &     1.028E-14   &         1.146E-13   &             5.561E-13   &         7.827E+00      \\ 
 0.003  &     5.185E-87   &         1.536E-86   &             4.587E-86   &         2.977E+00     &  0.160  &     2.246E-14   &         2.448E-13   &             1.186E-12   &         7.480E+00      \\ 
 0.004  &     3.838E-78   &         1.136E-77   &             3.387E-77   &         2.973E+00     &  0.180  &     8.625E-14   &         8.599E-13   &             4.137E-12   &         6.700E+00      \\ 
 0.005  &     7.869E-72   &         2.325E-71   &             6.925E-71   &         2.969E+00     &  0.200  &     2.764E-13   &         2.347E-12   &             1.110E-11   &         5.853E+00      \\ 
 0.006  &     5.153E-67   &         1.521E-66   &             4.525E-66   &         2.966E+00     &  0.250  &     3.955E-12   &         1.643E-11   &             6.519E-11   &         3.796E+00      \\ 
 0.007  &     9.573E-63   &         1.888E-62   &             4.112E-62   &         2.118E+00     &  0.300  &     1.262E-10   &         1.892E-10   &             3.539E-10   &         1.703E+00      \\ 
 0.008  &     1.362E-59   &         2.709E-59   &             5.986E-59   &         2.141E+00     &  0.350  &     9.030E-09   &         1.038E-08   &             1.198E-08   &         1.155E+00      \\
\enddata
\tablecomments{
{\it Observed resonances:} \citet{Goerres_NPA1992_c14ag,Gai_PRC1987_c14ag,Gai_PRC1992_c14ag}. \\
{\it Normalization:} None. \\
{\it Unobserved resonances:} The reduced widths were adopted from $\alpha$ transfer work \citep{Cunsolo_PRC1981_c14ag}. \\
{\it High-temperature rates (in parentheses):} Matching to statistical model rate above $T$ $=$ $1.42$~GK. \\
{\it Previous rates:} \citet{Hashimoto_APJ1986_c14ag,Buchmann_APJ1988_c14ag,Funck_APJ1989_c14ag,Lugaro_APJ2004_c14ag,ILIADIS2010b}. \\
{\it Other:} The direct capture $S$ factor (including tails of higher-lying resonances) was adopted from \citet{Goerres_NPA1992_c14ag}.
}
\end{deluxetable*}

\clearpage

\begin{figure*}[hbt!]
\centering
\includegraphics[width=0.5\linewidth]{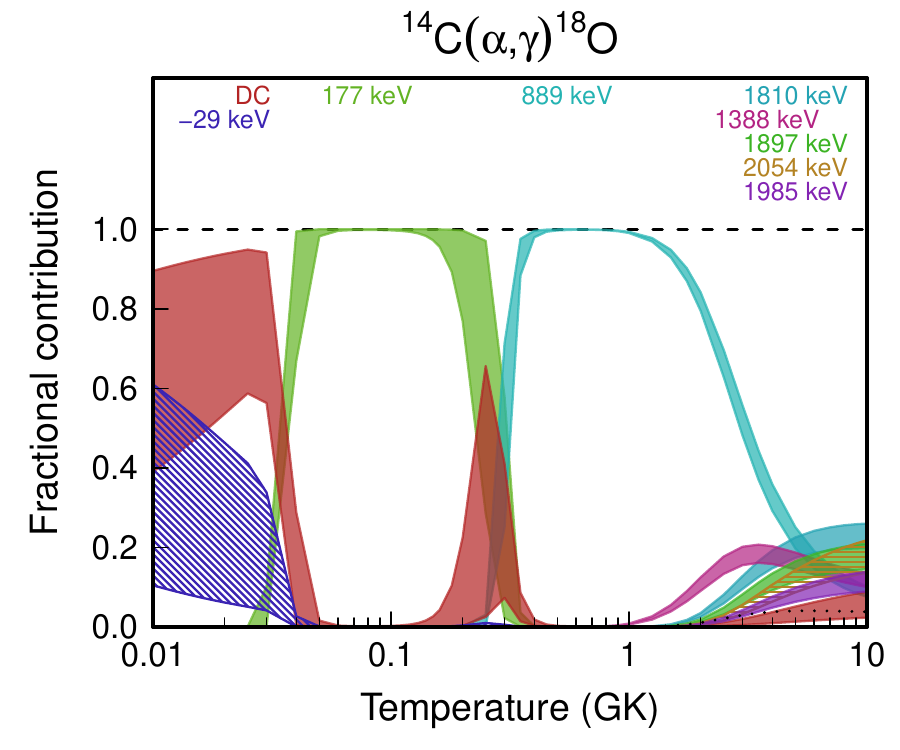}
\caption{
Fractional contributions to the total rate. ``DC'' refers to direct radiative capture. Resonance energies are given in the center-of-mass frame. 
}
\label{fig:c14ag1}
\end{figure*}
\begin{figure*}[hbt!]
\centering
\includegraphics[width=0.5\linewidth]{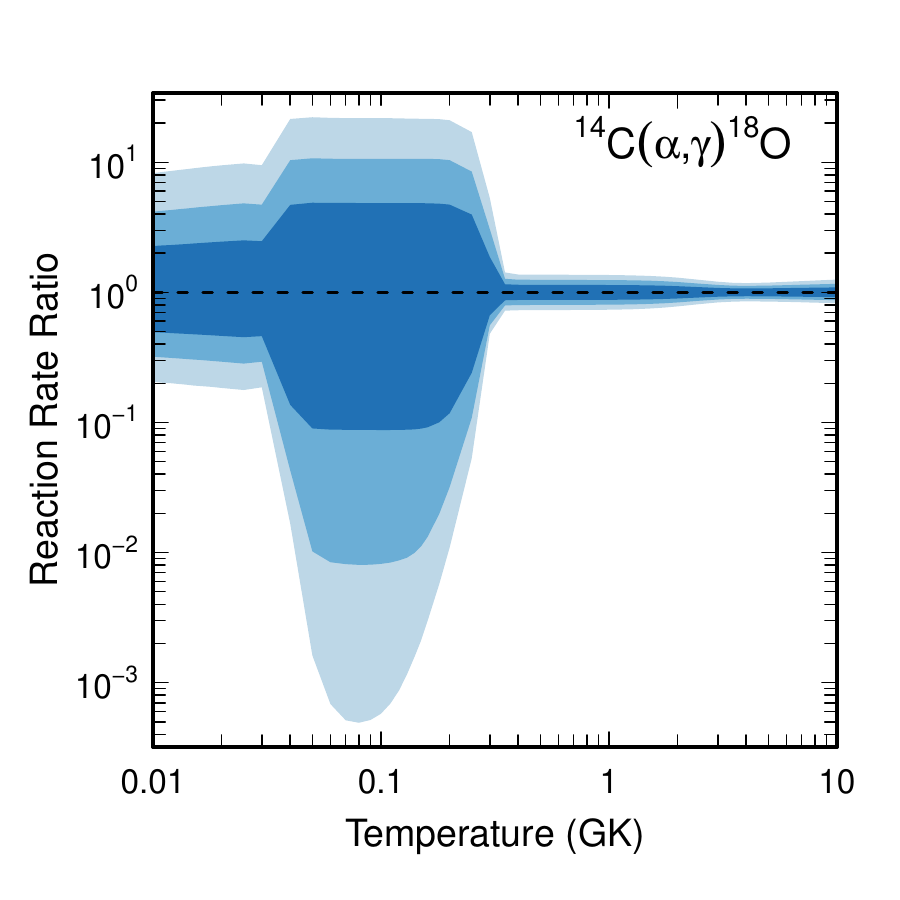}
\caption{
Reaction rate uncertainties versus temperature. The three different shades refer to coverage probabilities of 68\%, 90\%, and 98\%. 
}
\label{fig:c14ag2}
\end{figure*}

\clearpage

\startlongtable
\begin{deluxetable*}{lcccc | lcccc}
\tablecaption{Total laboratory reaction rates for $^{13}$N($\alpha$,p)$^{16}$O.}
\label{tab:n13ap} 
\tablewidth{\columnwidth}
\tabletypesize{\footnotesize}
\tablecolumns{10}
\tablehead{
T (GK) & Low & Median & High & $f.u.$ & T (GK) & Low & Median & High & $f.u.$
}
\startdata
 0.001  &  0.000E+00  &  0.000E+00  &        0.000E+00  &  1.000E+00  &  0.140  &  7.303E-13  &  1.614E-12  &        3.794E-12  &  2.275E+00   \\ 
 0.002  &  0.000E+00  &  0.000E+00  &        0.000E+00  &  1.000E+00  &  0.150  &  2.692E-12  &  5.742E-12  &        1.327E-11  &  2.219E+00   \\ 
 0.003  &  0.000E+00  &  0.000E+00  &        0.000E+00  &  1.000E+00  &  0.160  &  8.839E-12  &  1.812E-11  &        4.070E-11  &  2.152E+00   \\ 
 0.004  &  0.000E+00  &  0.000E+00  &        0.000E+00  &  1.000E+00  &  0.180  &  7.395E-11  &  1.400E-10  &        2.919E-10  &  2.003E+00   \\ 
 0.005  &  0.000E+00  &  0.000E+00  &        0.000E+00  &  1.000E+00  &  0.200  &  4.878E-10  &  8.605E-10  &        1.646E-09  &  1.865E+00   \\ 
 0.006  &  0.000E+00  &  0.000E+00  &        0.000E+00  &  1.000E+00  &  0.250  &  2.578E-08  &  4.236E-08  &        7.338E-08  &  1.710E+00   \\ 
 0.007  &  3.793E-64  &  6.139E-64  &        1.058E-63  &  1.697E+00  &  0.300  &  6.619E-07  &  1.097E-06  &        1.946E-06  &  1.744E+00   \\ 
 0.008  &  1.223E-60  &  1.986E-60  &        3.427E-60  &  1.702E+00  &  0.350  &  1.140E-05  &  1.919E-05  &        3.435E-05  &  1.766E+00   \\
\enddata
\tablecomments{
{\it Observed resonances:} \citet{Salisbury1962a, Salisbury1962b}; cross section data for $E_r^{c.m}$ $=$ $3.26$ $-$ $6.02$~MeV were taken from \citet{Jayatissa2022}.\\
{\it Unobserved resonances:} \citet{Meyer2020}, using the $\mathrm{^{13}C(^{7}Li, t)^{17}O}$ reaction for mirror states. \\
{\it Previous rates:} \citet{CF88,Meyer2020}. \\
{\it High-temperature rates (in parentheses):} No matching to statistical model rates is needed over the listed temperature range. \\
{\it Other:} Values for $\Gamma_\alpha$ were assumed to have a factor uncertainty of $2.5$; $\Gamma_p$ was assumed to have a 20\% uncertainty when no value was provided in the literature.}
\end{deluxetable*}

\clearpage

\begin{figure*}[hbt!]
\centering
\includegraphics[width=0.5\linewidth]{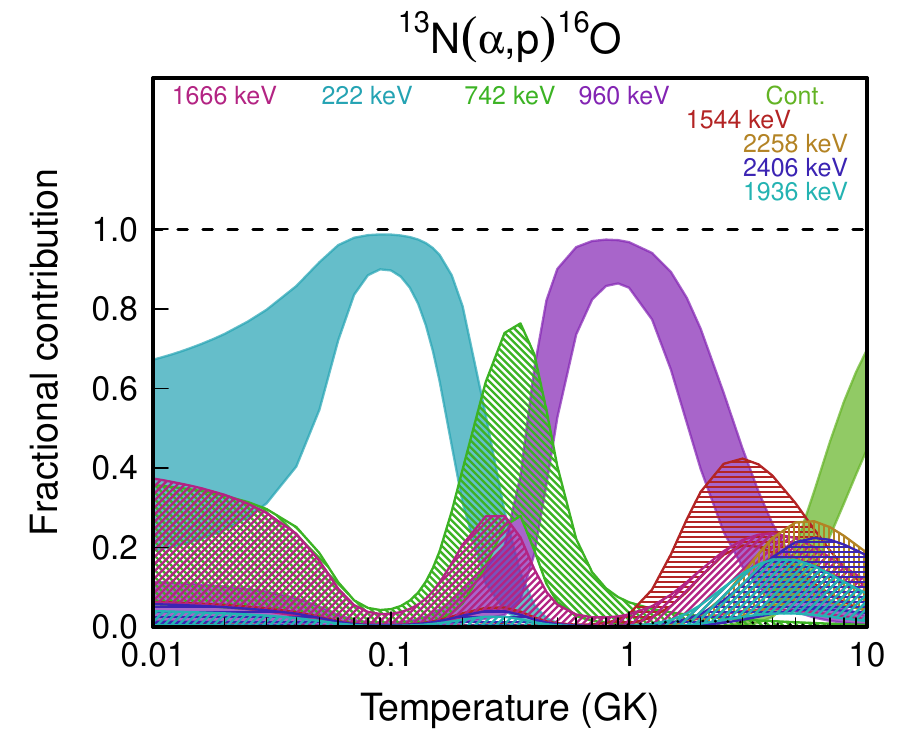}
\caption{
Fractional contributions to the total rate. ``Cont.'' refers to the continuum of higher-lying, unresolved resonances. Resonance energies are given in the center-of-mass frame. 
}
\label{fig:n13ap1}
\end{figure*}
\begin{figure*}[hbt!]
\centering
\includegraphics[width=0.5\linewidth]{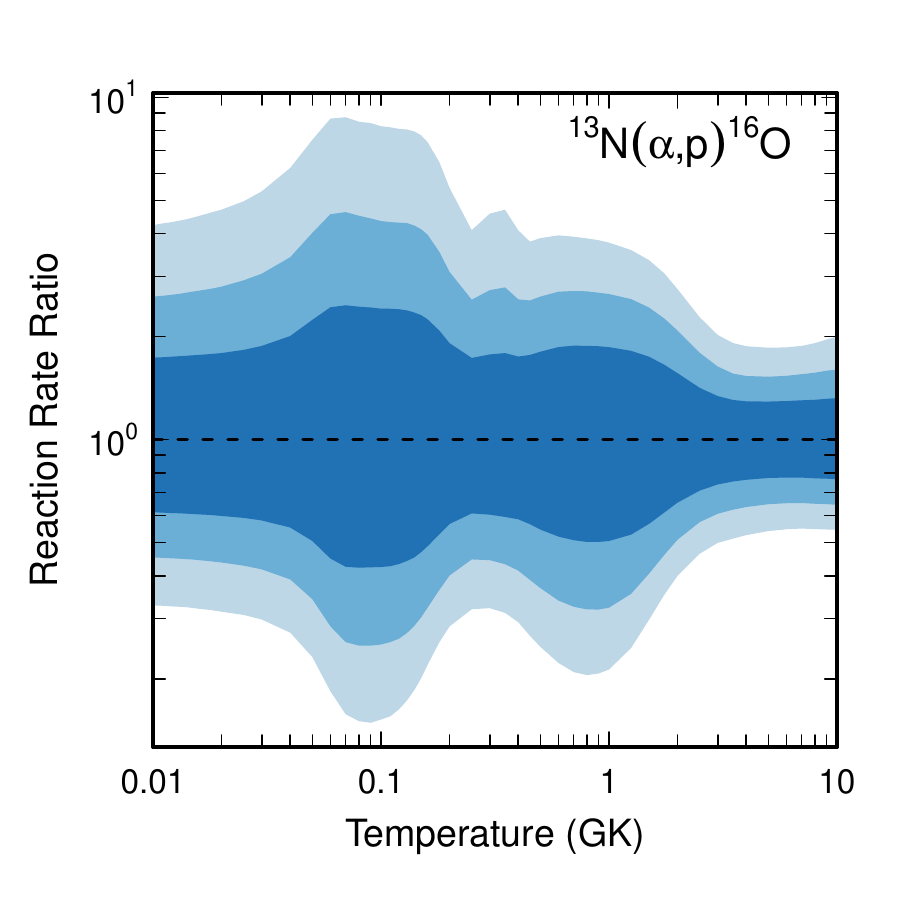}
\caption{
Reaction rate uncertainties versus temperature. The three different shades refer to coverage probabilities of 68\%, 90\%, and 98\%.}
\label{fig:n13ap2}
\end{figure*}

\clearpage

\startlongtable
\begin{deluxetable*}{lcccc | lcccc}
\tablecaption{Total laboratory reaction rates for $^{14}$N($\alpha$,$\gamma$)$^{18}$F.}
\label{tab:n14ag} 
\tablewidth{\columnwidth}
\tabletypesize{\footnotesize}
\tablecolumns{10}
\tablehead{
T (GK) & Low & Median & High & $f.u.$ & T (GK) & Low & Median & High & $f.u.$
}
\startdata
 0.001  &     0.000E+00   &         0.000E+00   &             0.000E+00   &         1.000E+00     &  0.140  &     2.286E-15   &         2.534E-15   &             2.803E-15   &         1.108E+00      \\ 
 0.002  &     0.000E+00   &         0.000E+00   &             0.000E+00   &         1.000E+00     &  0.150  &     2.446E-14   &         2.687E-14   &             2.945E-14   &         1.098E+00      \\ 
 0.003  &     4.392E-99   &         2.148E-98   &             1.057E-97   &         4.974E+00     &  0.160  &     1.933E-13   &         2.107E-13   &             2.292E-13   &         1.089E+00      \\ 
 0.004  &     3.026E-89   &         1.480E-88   &             7.285E-88   &         4.974E+00     &  0.180  &     5.961E-12   &         6.420E-12   &             6.908E-12   &         1.077E+00      \\ 
 0.005  &     3.039E-82   &         1.487E-81   &             7.318E-81   &         4.974E+00     &  0.200  &     9.095E-11   &         9.718E-11   &             1.038E-10   &         1.068E+00      \\ 
 0.006  &     6.700E-77   &         3.277E-76   &             1.613E-75   &         4.974E+00     &  0.250  &     1.161E-08   &         1.229E-08   &             1.300E-08   &         1.058E+00      \\ 
 0.007  &     1.244E-72   &         6.085E-72   &             2.995E-71   &         4.974E+00     &  0.300  &     2.787E-07   &         2.946E-07   &             3.113E-07   &         1.057E+00      \\ 
 0.008  &     1.908E-68   &         6.458E-68   &             2.079E-67   &         3.279E+00     &  0.350  &     2.598E-06   &         2.749E-06   &             2.909E-06   &         1.058E+00      \\
\enddata
\tablecomments{
{\it Observed resonances:} \citet{Parker_1968,Couch_1971,Rolfs_1973_IV,ROLFS1973257,Rolfs_1973_III,Becker_1982,Gorres_2000}. \\
{\it Normalization:} \citet{Becker_1982}. \\
{\it Unobserved resonances:} \citet{Gorodetzky_1967,Cooper_1986}. \\
{\it High-temperature rates (in parentheses):} No matching to statistical model rates is needed over the listed temperature range. \\
{\it Previous rates:} \citet{Gorres_2000}. \\
{\it Other:} For the $E_r^{c.m.}$ $=$ $238$~keV resonance, the upper-limit on $\Gamma_{\alpha}$ was deduced from the upper limit given by \citet{Couch_1971}. Since the isospin of this state is $T=1$, we assumed for $\langle\theta_{\alpha}^2\rangle$ a value of $1 \times 10^{-5}$ with a factor of 10 uncertainty.
}
\end{deluxetable*}

\clearpage

\begin{figure*}[hbt!]
\centering
\includegraphics[width=0.5\linewidth]{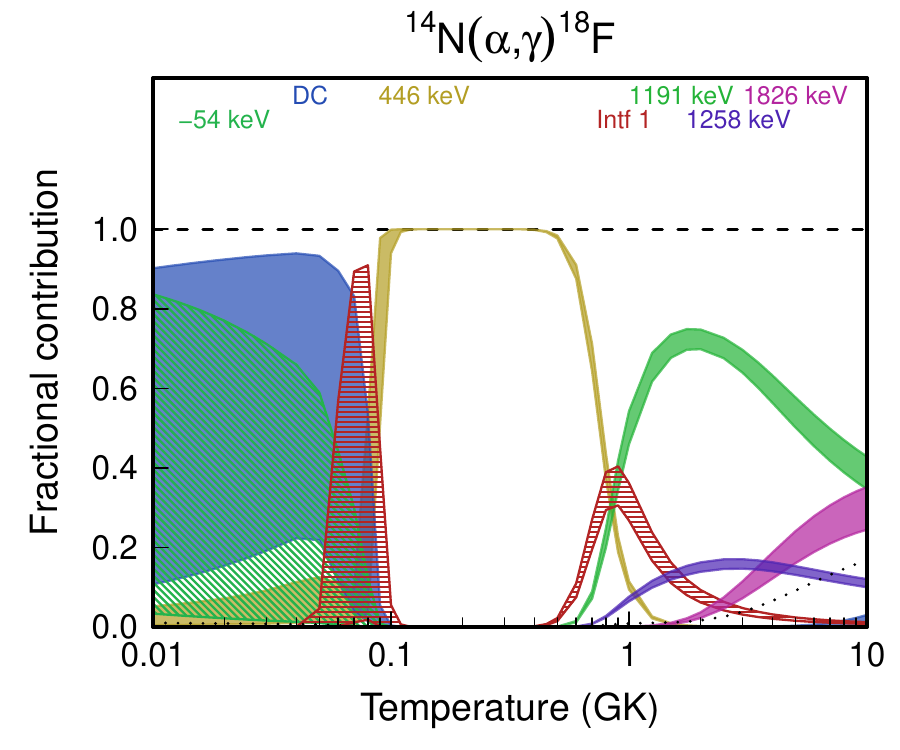}
\caption{
Fractional contributions to the total rate. ``DC'' refers to direct radiative capture. Resonance energies are given in the center-of-mass frame. ``Intf 1'' labels the combined contribution of the two interfering $4^+$ resonances at $E_r^{c.m.}$ $=$ $238$ and $884$~keV.  
}
\label{fig:n14ag1}
\end{figure*}
\begin{figure*}[hbt!]
\centering
\includegraphics[width=0.5\linewidth]{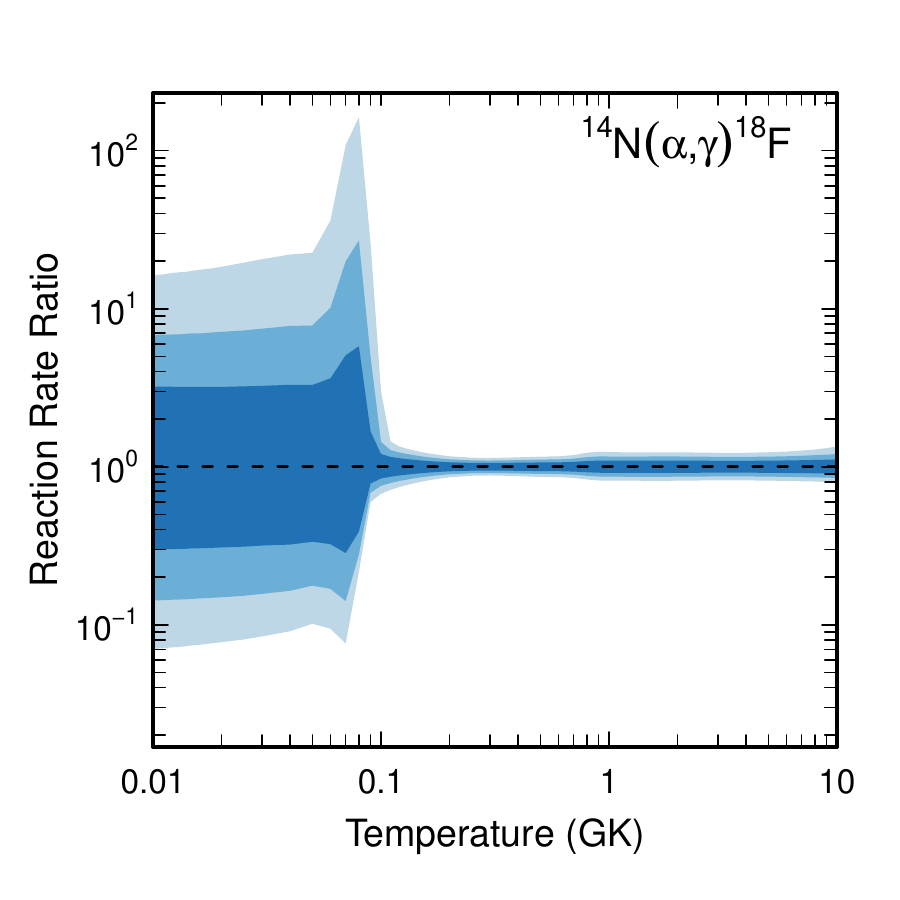}
\caption{
Reaction rate uncertainties versus temperature. The three different shades refer to coverage probabilities of 68\%, 90\%, and 98\%.}
\label{fig:n14ag2}
\end{figure*}

\clearpage

\startlongtable
\begin{deluxetable*}{lcccc | lcccc}
\tablecaption{Total laboratory reaction rates for $^{15}$N($\alpha$,$\gamma$)$^{19}$F.}
\label{tab:n15ag} 
\tablewidth{\columnwidth}
\tabletypesize{\footnotesize}
\tablecolumns{10}
\tablehead{
T (GK) & Low & Median & High & $f.u.$ & T (GK) & Low & Median & High & $f.u.$
}
\startdata
 0.001  &     0.000E+00   &         0.000E+00   &             0.000E+00   &         1.000E+00     &  0.140  &     6.370E-17   &         1.858E-16   &             5.508E-16   &         2.929E+00      \\ 
 0.002  &     0.000E+00   &         0.000E+00   &             0.000E+00   &         1.000E+00     &  0.150  &     4.568E-16   &         1.284E-15   &             3.759E-15   &         2.839E+00      \\ 
 0.003  &     8.113E-98   &         2.400E-97   &             7.179E-97   &         2.995E+00     &  0.160  &     2.703E-15   &         7.082E-15   &             2.020E-14   &         2.687E+00      \\ 
 0.004  &     6.251E-88   &         1.849E-87   &             5.531E-87   &         2.995E+00     &  0.180  &     6.690E-14   &         1.366E-13   &             3.449E-13   &         2.249E+00      \\ 
 0.005  &     6.796E-81   &         2.011E-80   &             6.014E-80   &         2.995E+00     &  0.200  &     1.197E-12   &         1.832E-12   &             3.702E-12   &         1.812E+00      \\ 
 0.006  &     1.591E-75   &         4.707E-75   &             1.408E-74   &         2.995E+00     &  0.250  &     3.539E-10   &         4.093E-10   &             5.048E-10   &         1.253E+00      \\ 
 0.007  &     3.099E-71   &         9.168E-71   &             2.742E-70   &         2.995E+00     &  0.300  &     1.934E-08   &         2.135E-08   &             2.380E-08   &         1.119E+00      \\ 
 0.008  &     2.096E-67   &         4.206E-67   &             1.054E-66   &         2.231E+00     &  0.350  &     3.765E-07   &         4.081E-07   &             4.434E-07   &         1.087E+00      \\
\enddata
\tablecomments{
{\it Observed resonances:} \citet{Smotrich_1961,Aitken_1969b,Aitken_1969,Dixon_Storey_1971b,Rogers_1972,Rogers_1972_b,Rogers_1973,Underwood_1974,Rogers_1976,Dixon_Storey_1977, Symons_1978,Wilmes_2002,diLeva_2017,Volya_2022,Fang_2024}. \\
{\it Normalization:} \citet{Aitken_1969,Dixon_Storey_1971,Wilmes_2002,diLeva_2017,Fang_2024}. \\
{\it Unobserved resonances:} \citet{1996De07,Mao-1996,Fortune2003,Fortune_2003_b,Fortune_2010}. \\
{\it High-temperature rates (in parentheses):} No matching to statistical model rates is needed over the listed temperature range. \\
{\it Previous rates:} \citet{ILIADIS2010c}. \\
{\it Other:} The resonance strength values for the $E_r^{c.m.}$ $=$ $1324$ keV resonance from \citet{Dixon_Storey_1971} and \citet{Aitken_1969} were adjusted using an updated value of the $E_r^{c.m.}$ $=$ $1258$ keV resonance in $^{14}$N($\alpha$,$\gamma$)$^{18}$F \citep{Becker_1982}.
}
\end{deluxetable*}

\clearpage

\begin{figure*}[hbt!]
\centering
\includegraphics[width=0.5\linewidth]{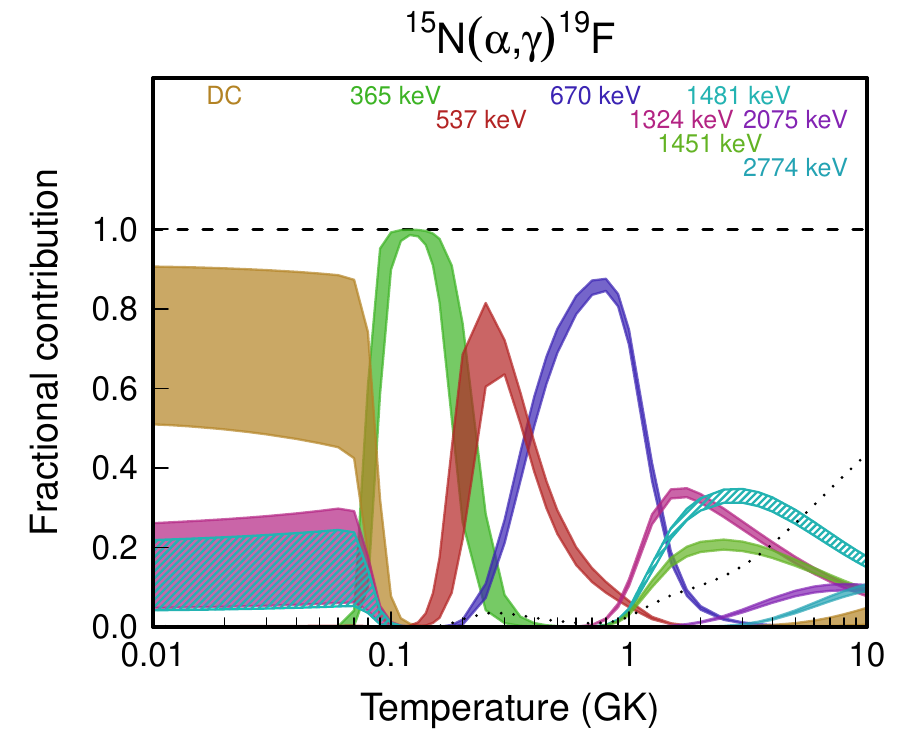}
\caption{
Fractional contributions to the total rate. ``DC'' refers to direct radiative capture. Resonance energies are given in the center-of-mass frame.  
}
\label{fig:n15ag1}
\end{figure*}
\begin{figure*}[hbt!]
\centering
\includegraphics[width=0.5\linewidth]{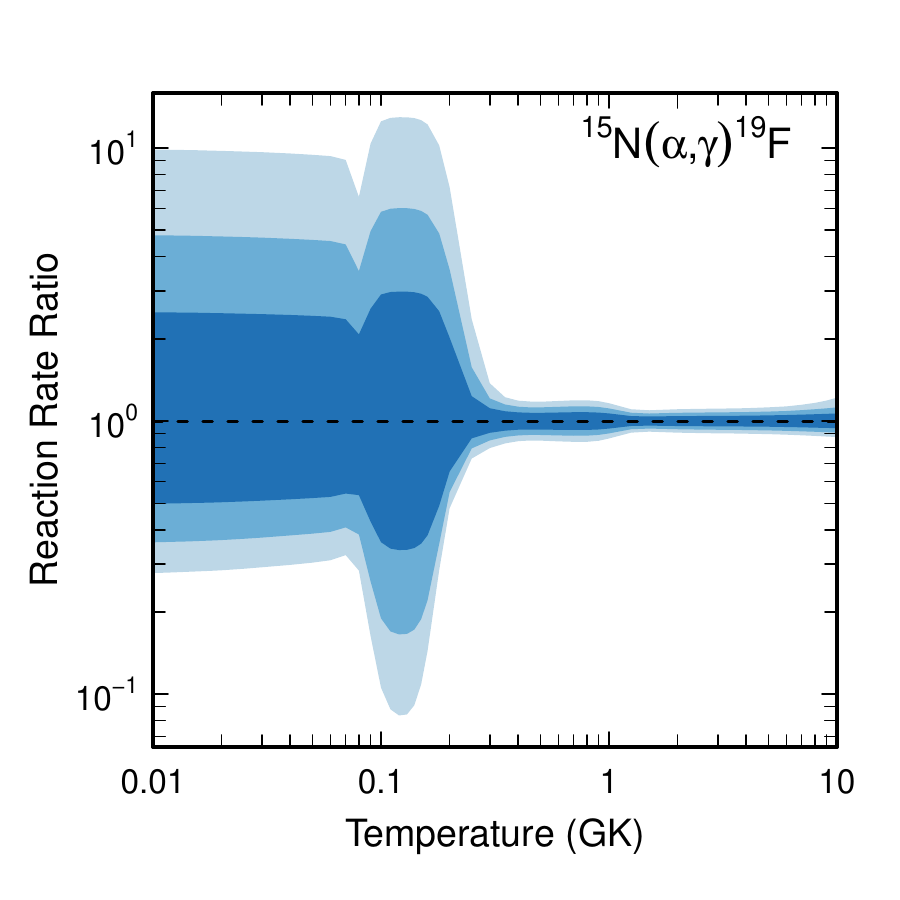}
\caption{
Reaction rate uncertainties versus temperature. The three different shades refer to coverage probabilities of 68\%, 90\%, and 98\%. 
}
\label{fig:n15ag2}
\end{figure*}

\clearpage
\startlongtable
\begin{deluxetable*}{lcccc | lcccc}
\tablecaption{Total laboratory reaction rates for $^{15}$O($\alpha$,$\gamma$)$^{19}$Ne.}
\label{tab:o15ag} 
\tablewidth{\columnwidth}
\tabletypesize{\footnotesize}
\tablecolumns{10}
\tablehead{
T (GK) & Low & Median & High & $f.u.$ & T (GK) & Low & Median & High & $f.u.$
}
\startdata
 0.001  &     0.000E+00   &         0.000E+00   &             0.000E+00   &         1.000E+00     &  0.140  &     4.481E-19   &         4.261E-18   &             4.268E-17   &         9.731E+00      \\ 
 0.002  &     0.000E+00   &         0.000E+00   &             0.000E+00   &         1.000E+00     &  0.150  &     6.614E-18   &         6.329E-17   &             6.354E-16   &         9.794E+00      \\ 
 0.003  &     0.000E+00   &         0.000E+00   &             0.000E+00   &         1.000E+00     &  0.160  &     6.994E-17   &         6.672E-16   &             6.712E-15   &         9.783E+00      \\ 
 0.004  &     5.833E-97   &         5.763E-96   &             5.723E-95   &         9.960E+00     &  0.180  &     3.548E-15   &         3.338E-14   &             3.358E-13   &         9.663E+00      \\ 
 0.005  &     2.950E-89   &         2.915E-88   &             2.895E-87   &         9.960E+00     &  0.200  &     8.223E-14   &         7.528E-13   &             7.558E-12   &         9.472E+00      \\ 
 0.006  &     2.240E-83   &         2.213E-82   &             2.198E-81   &         9.960E+00     &  0.250  &     2.416E-11   &         1.996E-10   &             1.952E-09   &         8.849E+00      \\ 
 0.007  &     1.119E-78   &         1.106E-77   &             1.098E-76   &         9.960E+00     &  0.300  &     1.101E-09   &         8.163E-09   &             7.562E-08   &         8.171E+00      \\ 
 0.008  &     1.166E-74   &         9.316E-74   &             8.464E-73   &         8.436E+00     &  0.350  &     1.751E-08   &         1.164E-07   &             1.007E-06   &         7.522E+00      \\
\enddata
\tablecomments{
{\it Observed resonances:} None. \\
{\it Normalization:} None. \\
{\it Unobserved resonances:} Spectroscopic factors used to deduce the proton partial widths were taken from \citet{1996De07}. \\
{\it High-temperature rates (in parentheses):} Matching to statistical model rate above $T$ $=$ $4.0$~GK. \\
{\it Previous rates:} \citet{1996Ha26,2009Ta09,ILIADIS2010b,2011Da24}. \\
{\it Other:} The $\gamma$-ray partial widths were scaled from those of the mirror states, except for the 850-keV resonance, where we adopted the partial width from \citet{2003Da03}. See also \citet{2011Da24}. The direct-capture $S$ factor has been calculated using experimental spectroscopic factors from \citet{1996Ma07}. For the $\alpha$-particle partial widths of resonances with $E_r^{c.m.}$ $>$ $1$~MeV, see \citet{1990Ma05,2002La29,2002Os05,2003Da03,2004Vi05} and \citet{2009Ta09}.
}
\end{deluxetable*}

\clearpage

\begin{figure*}[hbt!]
\centering
\includegraphics[width=0.5\linewidth]{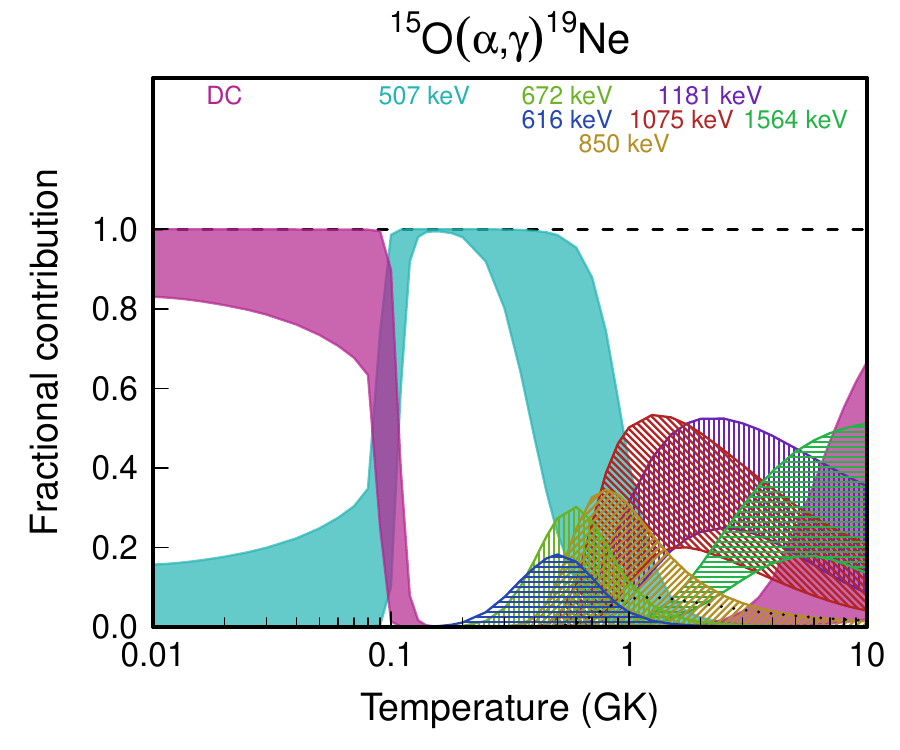}
\caption{
Fractional contributions to the total rate. ``DC'' refers to direct radiative capture. Resonance energies are given in the center-of-mass frame.  
}
\label{fig:o15ag1}
\end{figure*}
\begin{figure*}[hbt!]
\centering
\includegraphics[width=0.5\linewidth]{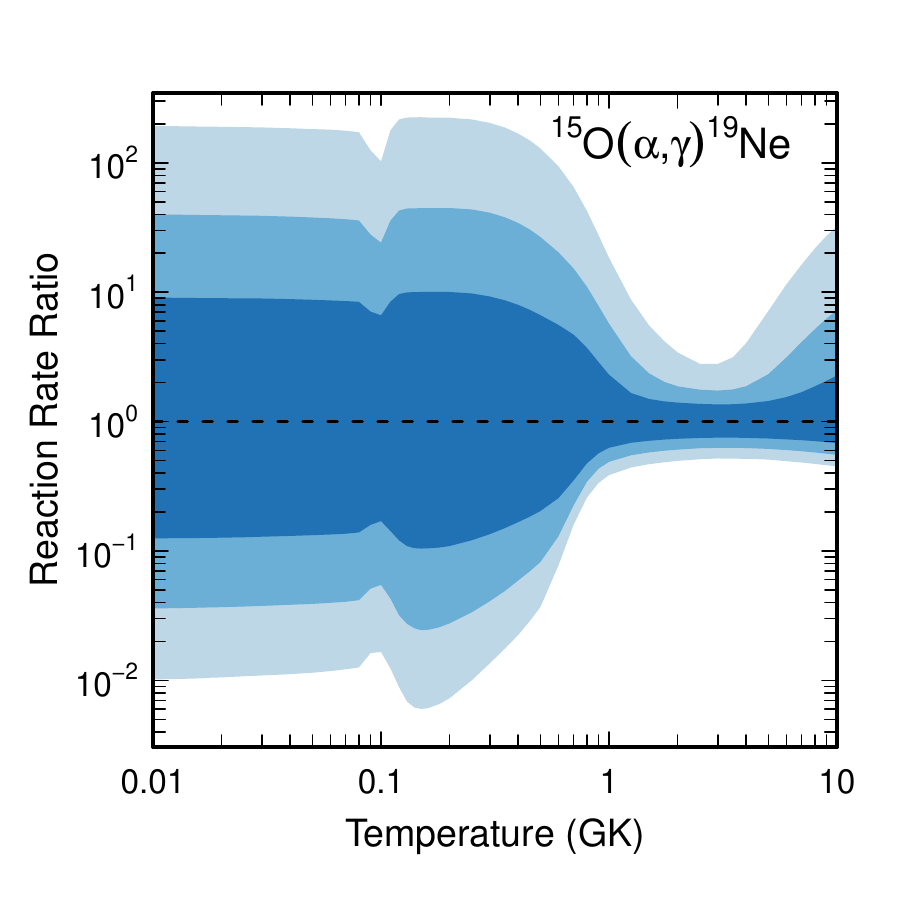}
\caption{
Reaction rate uncertainties versus temperature. The three different shades refer to coverage probabilities of 68\%, 90\%, and 98\%. 
}
\label{fig:o15ag2}
\end{figure*}

\clearpage

\startlongtable
\begin{deluxetable*}{lcccc | lcccc}
\tablecaption{Total laboratory reaction rates for $^{16}$O(p,$\gamma$)$^{17}$F.}
\label{tab:o16pg} 
\tablewidth{\columnwidth}
\tabletypesize{\footnotesize}
\tablecolumns{10}
\tablehead{
T (GK) & Low & Median & High & $f.u.$ & T (GK) & Low & Median & High & $f.u.$
}
\startdata
 0.001 &  5.229E-63 &  5.435E-63 &  5.650E-63  &  1.042E+00 &  0.140 &  4.472E-06 &  4.659E-06 &  4.851E-06  &  1.042E+00 \\ 
 0.002 &  2.935E-48 &  3.051E-48 &  3.172E-48  &  1.042E+00 &  0.150 &  8.764E-06 &  9.131E-06 &  9.507E-06  &  1.042E+00 \\ 
 0.003 &  4.164E-41 &  4.338E-41 &  4.517E-41  &  1.042E+00 &  0.160 &  1.620E-05 &  1.688E-05 &  1.757E-05  &  1.042E+00 \\ 
 0.004 &  1.343E-36 &  1.399E-36 &  1.457E-36  &  1.042E+00 &  0.180 &  4.791E-05 &  4.991E-05 &  5.197E-05  &  1.042E+00 \\ 
 0.005 &  2.154E-33 &  2.244E-33 &  2.336E-33  &  1.042E+00 &  0.200 &  1.216E-04 &  1.266E-04 &  1.319E-04  &  1.042E+00 \\ 
 0.006 &  5.980E-31 &  6.230E-31 &  6.486E-31  &  1.042E+00 &  0.250 &  7.781E-04 &  8.106E-04 &  8.440E-04  &  1.042E+00 \\ 
 0.007 &  5.332E-29 &  5.555E-29 &  5.783E-29  &  1.042E+00 &  0.300 &  3.176E-03 &  3.309E-03 &  3.446E-03  &  1.042E+00 \\ 
 0.008 &  2.162E-27 &  2.253E-27 &  2.345E-27  &  1.042E+00 &  0.350 &  9.703E-03 &  1.011E-02 &  1.053E-02  &  1.042E+00 \\
\enddata
\tablecomments{
{\it Reaction rate:} \citet{iliadis2022}. \\
{\it $S$ factor:} See Table~2 of \citet{iliadis2022}. \\
{\it Observed resonances:} The two resonances at $E_r^{c.m.}$ $=$ $2.50$ and $3.26$~MeV \citep{TILLEY_A17} are negligible for the total rate. \\
{\it Normalization:} See Table~3 of \citet{iliadis2022}. \\
{\it High-temperature rates (in parentheses)}: For temperatures of $T$ $\ge$ $4$~GK, the rates were calculated assuming a constant S-factor of $4.0\times10^{-3}$~MeVb (see Figure~2 of \citet{iliadis2008}).\\
{\it Previous rates:} \citet{iliadis2008}. \\
{\it Other:} Bayesian rate. 
}
\end{deluxetable*}

\clearpage

\begin{figure*}[hbt!]
\centering
\includegraphics[width=0.5\linewidth]{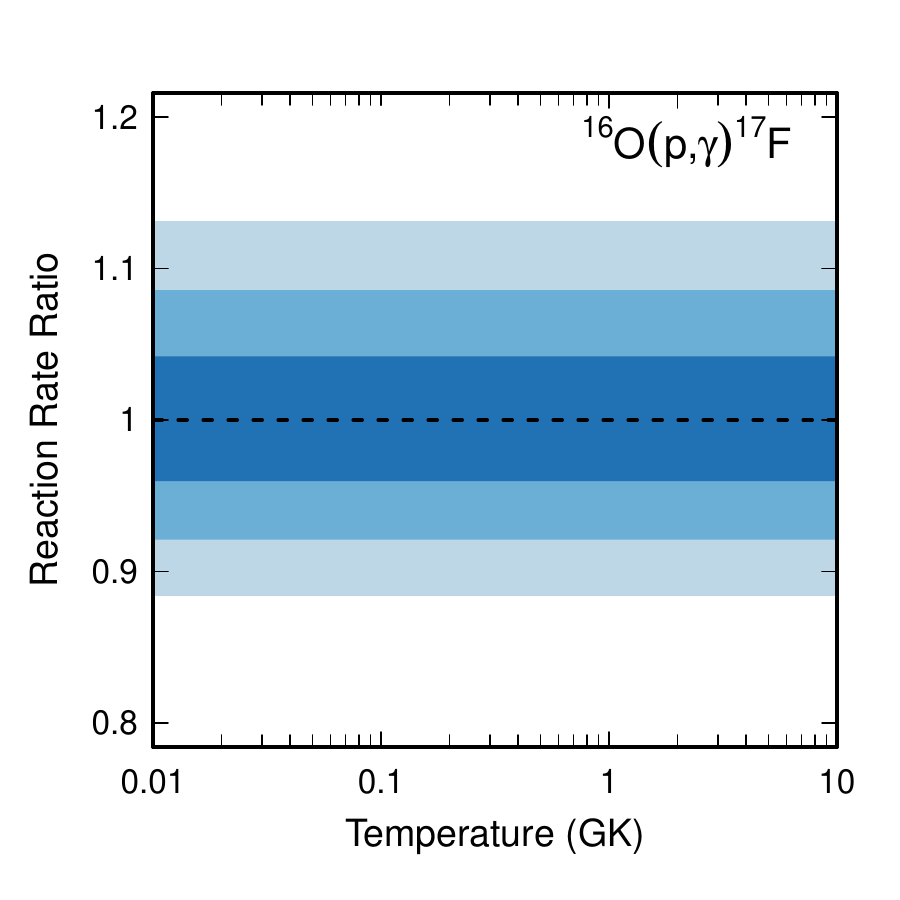}
\caption{
Reaction rate uncertainties versus temperature. The three different shades refer to coverage probabilities of 68\%, 90\%, and 98\%.}
\label{fig:o16pg2}
\end{figure*}

\clearpage

\startlongtable
\begin{deluxetable*}{lcccc | lcccc}
\tablecaption{Total laboratory reaction rates for $^{16}$O($\alpha$,$\gamma$)$^{20}$Ne.}
\label{tab:o16ag} 
\tablewidth{\columnwidth}
\tabletypesize{\footnotesize}
\tablecolumns{10}
\tablehead{
T (GK) & Low & Median & High & $f.u.$ & T (GK) & Low & Median & High & $f.u.$
}
\startdata
 0.001  &     0.000E+00   &         0.000E+00   &             0.000E+00   &         1.000E+00     &  0.140  &     3.624E-23   &         7.445E-23   &             1.587E-22   &         2.082E+00      \\ 
 0.002  &     0.000E+00   &         0.000E+00   &             0.000E+00   &         1.000E+00     &  0.150  &     1.983E-22   &         4.058E-22   &             8.628E-22   &         2.075E+00      \\ 
 0.003  &     0.000E+00   &         0.000E+00   &             0.000E+00   &         1.000E+00     &  0.160  &     9.392E-22   &         1.913E-21   &             4.057E-21   &         2.067E+00      \\ 
 0.004  &     1.024E-97   &         2.315E-97   &             5.223E-97   &         2.267E+00     &  0.180  &     1.480E-20   &         2.980E-20   &             6.281E-20   &         2.048E+00      \\ 
 0.005  &     5.537E-90   &         1.251E-89   &             2.822E-89   &         2.266E+00     &  0.200  &     1.755E-19   &         3.338E-19   &             6.808E-19   &         1.956E+00      \\ 
 0.006  &     4.414E-84   &         9.972E-84   &             2.248E-83   &         2.265E+00     &  0.250  &     3.754E-16   &         4.317E-16   &             4.989E-16   &         1.158E+00      \\ 
 0.007  &     2.289E-79   &         5.170E-79   &             1.165E-78   &         2.265E+00     &  0.300  &     2.618E-13   &         2.964E-13   &             3.357E-13   &         1.133E+00      \\ 
 0.008  &     1.920E-75   &         4.164E-75   &             9.201E-75   &         2.184E+00     &  0.350  &     2.935E-11   &         3.304E-11   &             3.721E-11   &         1.126E+00      \\
\enddata
\tablecomments{
{\it Observed resonances:} From the compilation of \citet{Tilley1998}, complemented by more recent experimental data \citep{Knee_PHD1995_o16ag,Mayer_PHD2001_o16ag,Costantini_PRC2010_o16ag,Hager_PRC2011_o16ag,Hager_PRC2012_o16ag}. The resonance strengths for $^{20}$Ne production were corrected for the $\alpha$-particle decay of higher-lying states in $^{20}$Ne.  \\
{\it Normalization:} None. \\
{\it Unobserved resonances:} None. \\
{\it High-temperature rates (in parentheses):} Matching to statistical model rate above $T$ $=$ $7.2$~GK. \\
{\it Previous rates:} \citet{NACRE,ILIADIS2010b,Costantini_PRC2010_o16ag}. \\
{\it Other:} The direct capture $S$ factor was recalculated and adjusted to \citet{Hager_PRC2011_o16ag,Hager_PRC2012_o16ag}; the results are close to previous calculations \citep{Langanke_ZPA1984_o16ag,Dufour_PRC1994_o16ag,Mohr_PRC2005_o16ag}.
}
\end{deluxetable*}

\clearpage

\begin{figure*}[hbt!]
\centering
\includegraphics[width=0.5\linewidth]{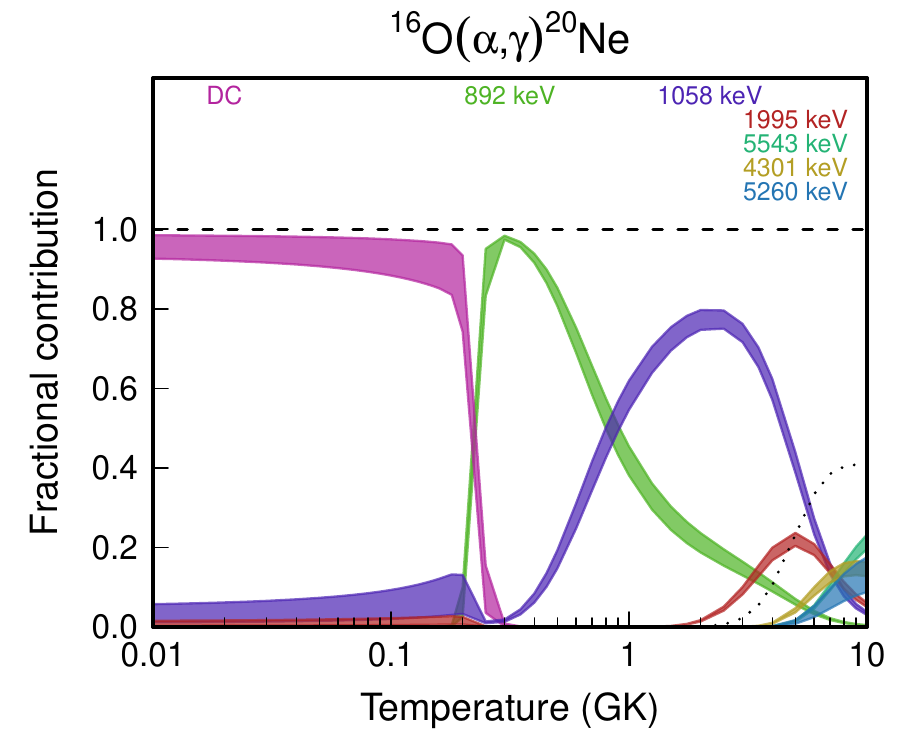}
\caption{
Fractional contributions to the total rate. ``DC'' refers to direct radiative capture. Resonance energies are given in the center-of-mass frame.  
}
\label{fig:o16ag1}
\end{figure*}
\begin{figure*}[hbt!]
\centering
\includegraphics[width=0.5\linewidth]{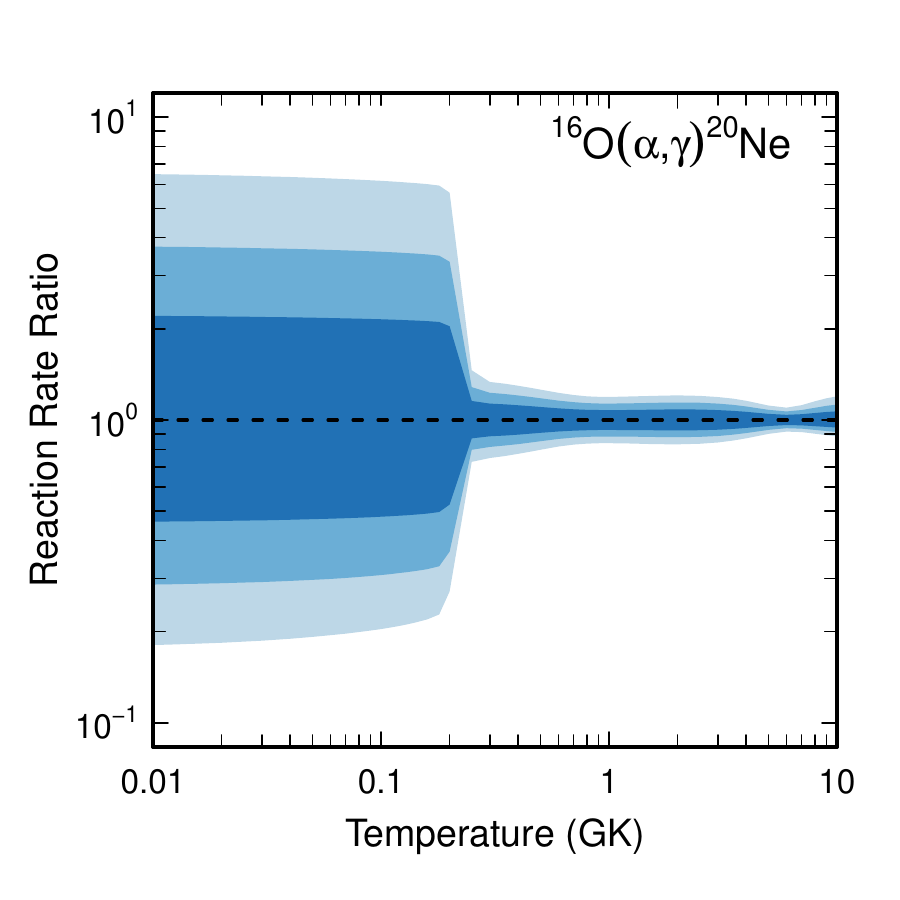}
\caption{
Reaction rate uncertainties versus temperature. The three different shades refer to coverage probabilities of 68\%, 90\%, and 98\%. 
}
\label{fig:o16ag2}
\end{figure*}

\clearpage

\startlongtable
\begin{deluxetable*}{lcccc | lcccc}
\tablecaption{Total laboratory reaction rates for $^{17}$O(p,$\gamma$)$^{18}$F.}
\label{tab:o17pg} 
\tablewidth{\columnwidth}
\tabletypesize{\footnotesize}
\tablecolumns{10}
\tablehead{
T (GK) & Low & Median & High & $f.u.$ & T (GK) & Low & Median & High & $f.u.$
}
\startdata
 0.001  &     2.033E-63   &         2.485E-63   &             3.577E-63   &         1.402E+00     &  0.140  &     4.511E-06   &         4.999E-06   &             5.575E-06   &         1.113E+00      \\ 
 0.002  &     1.138E-48   &         1.365E-48   &             1.744E-48   &         1.271E+00     &  0.150  &     9.552E-06   &         1.053E-05   &             1.168E-05   &         1.107E+00      \\ 
 0.003  &     1.623E-41   &         1.932E-41   &             2.377E-41   &         1.225E+00     &  0.160  &     1.893E-05   &         2.079E-05   &             2.295E-05   &         1.102E+00      \\ 
 0.004  &     5.277E-37   &         6.252E-37   &             7.568E-37   &         1.203E+00     &  0.180  &     6.220E-05   &         6.791E-05   &             7.451E-05   &         1.095E+00      \\ 
 0.005  &     8.534E-34   &         1.008E-33   &             1.209E-33   &         1.193E+00     &  0.200  &     1.674E-04   &         1.823E-04   &             1.996E-04   &         1.093E+00      \\ 
 0.006  &     2.390E-31   &         2.816E-31   &             3.354E-31   &         1.186E+00     &  0.250  &     1.102E-03   &         1.201E-03   &             1.317E-03   &         1.094E+00      \\ 
 0.007  &     2.149E-29   &         2.530E-29   &             3.001E-29   &         1.183E+00     &  0.300  &     4.673E-03   &         5.096E-03   &             5.590E-03   &         1.095E+00      \\ 
 0.008  &     8.789E-28   &         1.033E-27   &             1.223E-27   &         1.181E+00     &  0.350  &     1.846E-02   &         2.003E-02   &             2.180E-02   &         1.088E+00      \\
\enddata
\tablecomments{
{\it Observed resonances:} \citet{ROLFS1973257,PhysRevC.75.035810,PhysRevC.86.055801,Leva,buckner2015}. \\
{\it Normalization:} \citet{Fox,buckner2015}. \\
{\it Unobserved resonances:} \citet{MAK198079,Landre1989}. \\
{\it High-temperature rates (in parentheses):} No corrections for missing high-energy resonances are necessary up to $T$ $=$ $10$~GK. \\
{\it Previous rates:} \citet{Rapagnani2025}. \\
{\it Other:} The two interfering $1^-$ resonances at $E_r^{c.m.}$ $=$ $-1.6$ and $65$~keV were sampled with a random interference sign.
}
\end{deluxetable*}

\clearpage

\begin{figure*}[hbt!]
\centering
\includegraphics[width=0.5\linewidth]{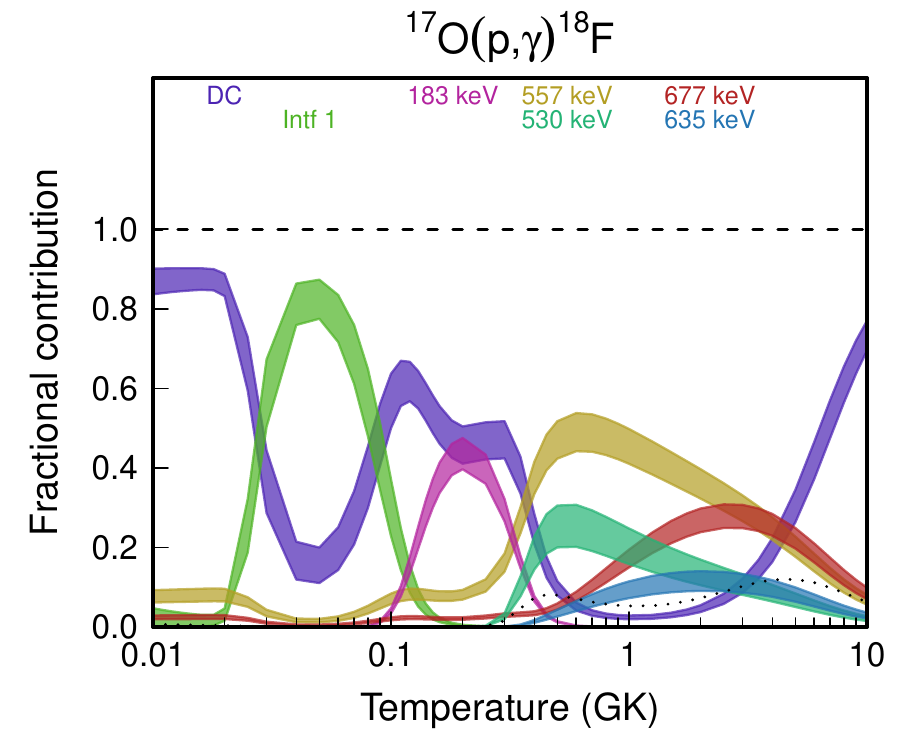}
\caption{
Fractional contributions to the total rate. ``DC'' refers to direct radiative capture. Resonance energies are given in the center-of-mass frame. ``Intf 1'' labels the combined contribution of the two interfering $1^-$ resonances at $E_r^{c.m.}$ $=$ $-1.6$ and $65$~keV.
}
\label{fig:o17pg1}
\end{figure*}
\begin{figure*}[hbt!]
\centering
\includegraphics[width=0.5\linewidth]{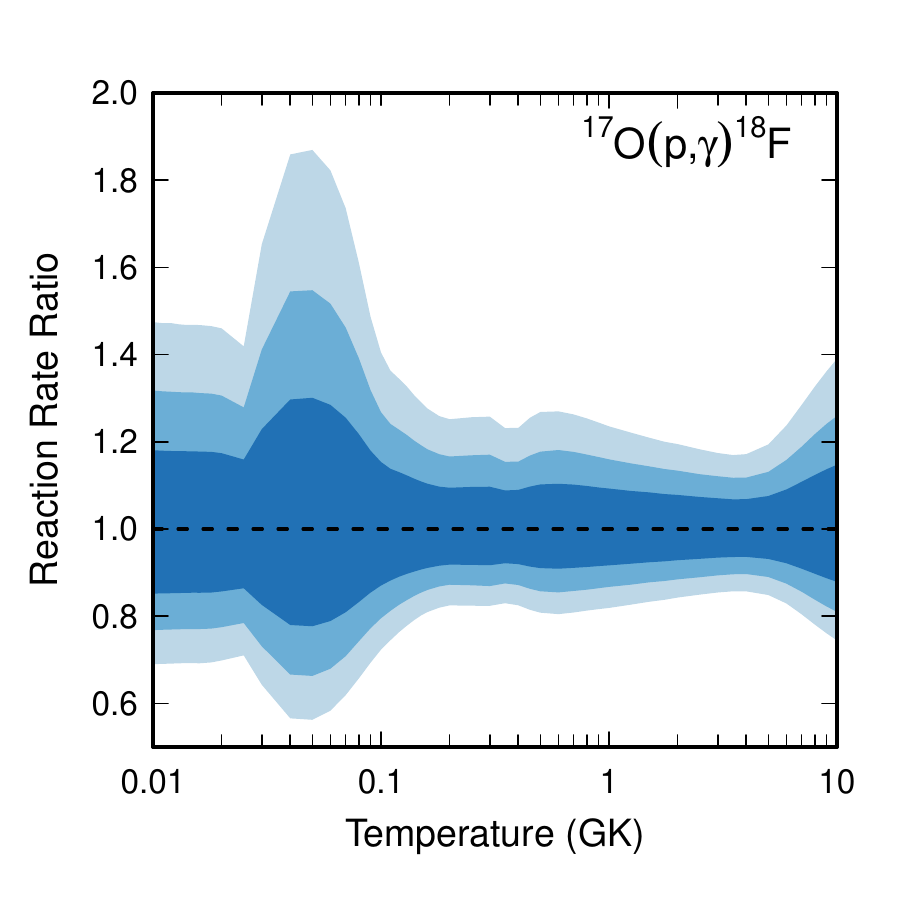}
\caption{
Reaction rate uncertainties versus temperature. The three different shades refer to coverage probabilities of 68\%, 90\%, and 98\%.}
\label{fig:o17pg2}
\end{figure*}

\clearpage

\startlongtable


\clearpage

\begin{figure*}[hbt!]
\centering
\includegraphics[width=0.5\linewidth]{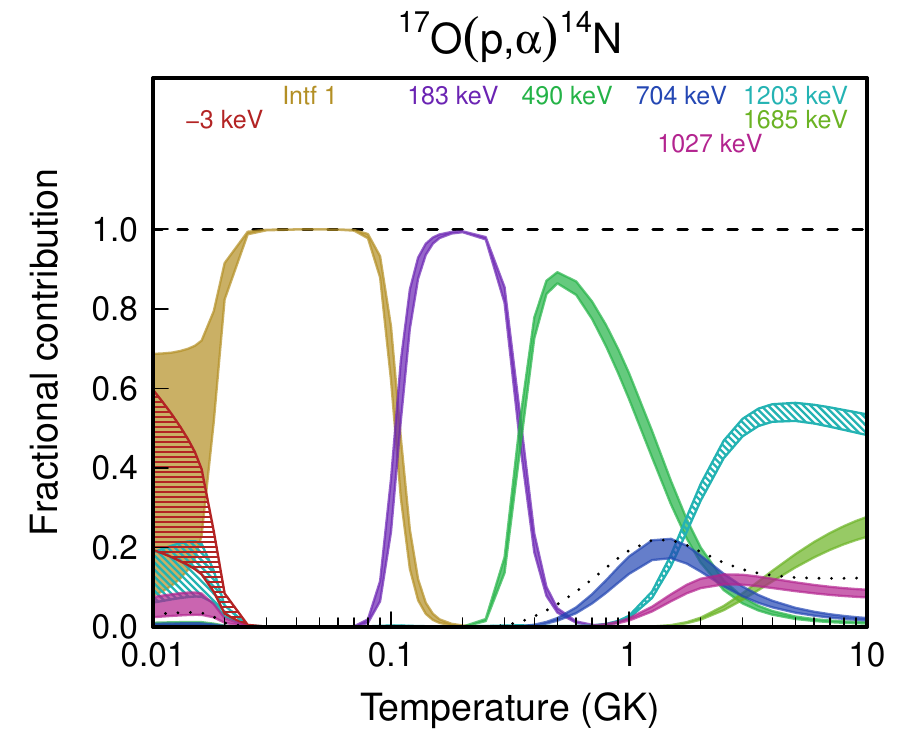}
\caption{
Fractional contributions to the total rate. Resonance energies are given in the center-of-mass frame. ``Intf 1'' labels the combined contribution of the two interfering $1^-$ resonances at $E_r^{c.m.}$ $=$ $-1.6$ and $65$~keV.
}
\label{fig:o17pa1}
\end{figure*}
\begin{figure*}[hbt!]
\centering
\includegraphics[width=0.5\linewidth]{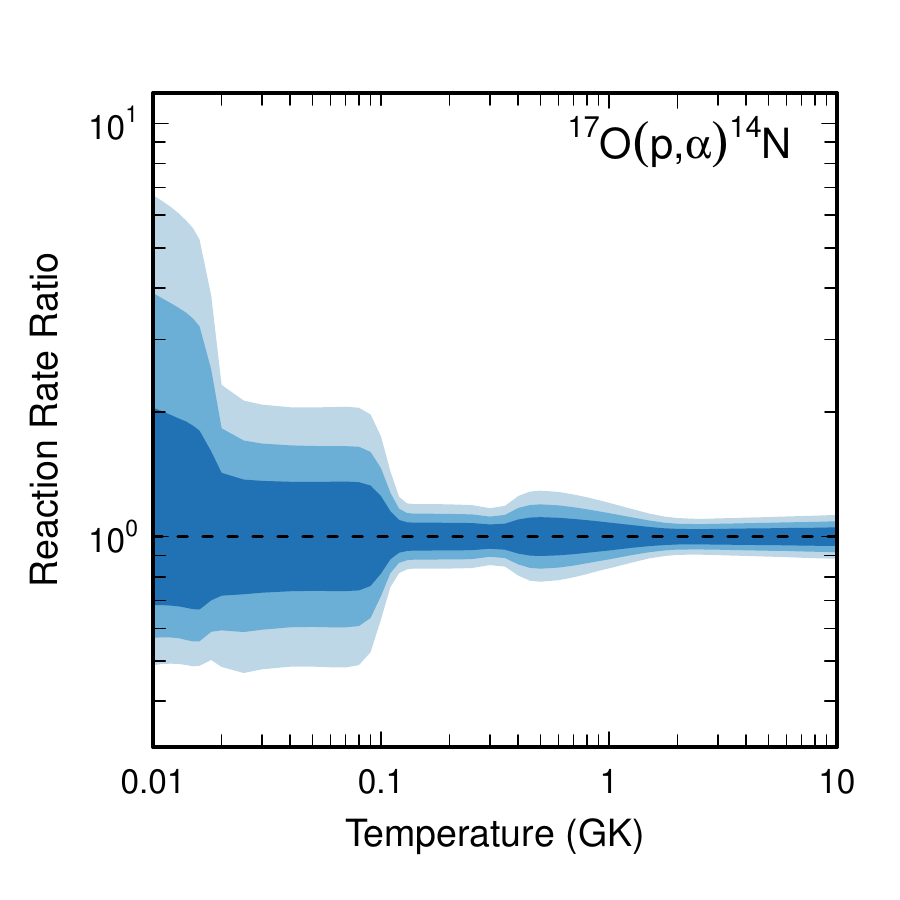}
\caption{
Reaction rate uncertainties versus temperature. The three different shades refer to coverage probabilities of 68\%, 90\%, and 98\%. 
}
\label{fig:o17pa2}
\end{figure*}

\clearpage

\startlongtable


\clearpage

\begin{figure*}[hbt!]
\centering
\includegraphics[width=0.5\linewidth]{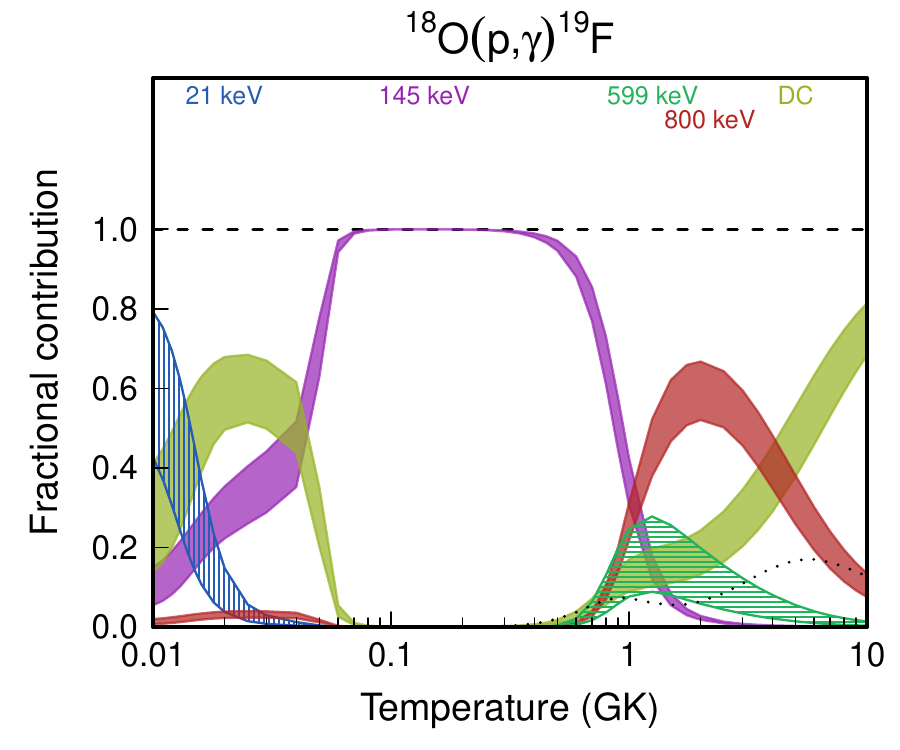}
\caption{
Fractional contributions to the total rate. ``DC'' refers to direct radiative capture. Resonance energies are given in the center-of-mass frame.  
}
\label{fig:o18pg1}
\end{figure*}
\begin{figure*}[hbt!]
\centering
\includegraphics[width=0.5\linewidth]{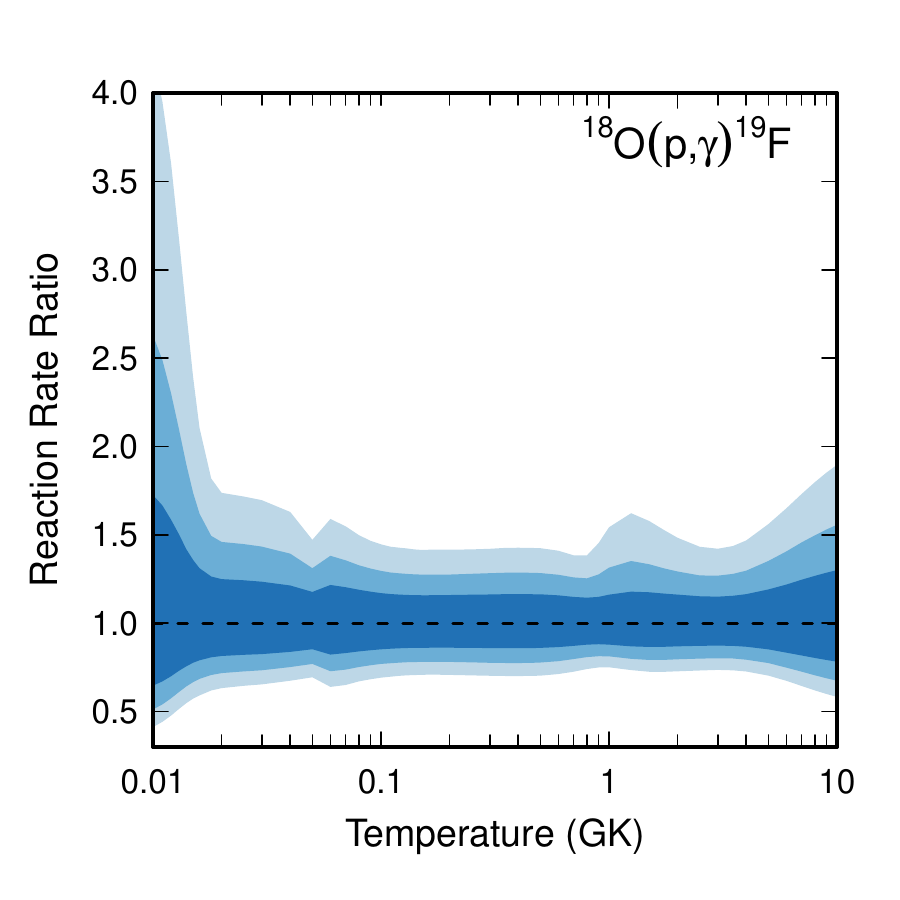}
\caption{
Reaction rate uncertainties versus temperature. The three different shades refer to coverage probabilities of 68\%, 90\%, and 98\%. 
}
\label{fig:o18pg2}
\end{figure*}

\clearpage

\startlongtable


\clearpage

\begin{figure*}[hbt!]
\centering
\includegraphics[width=0.5\linewidth]{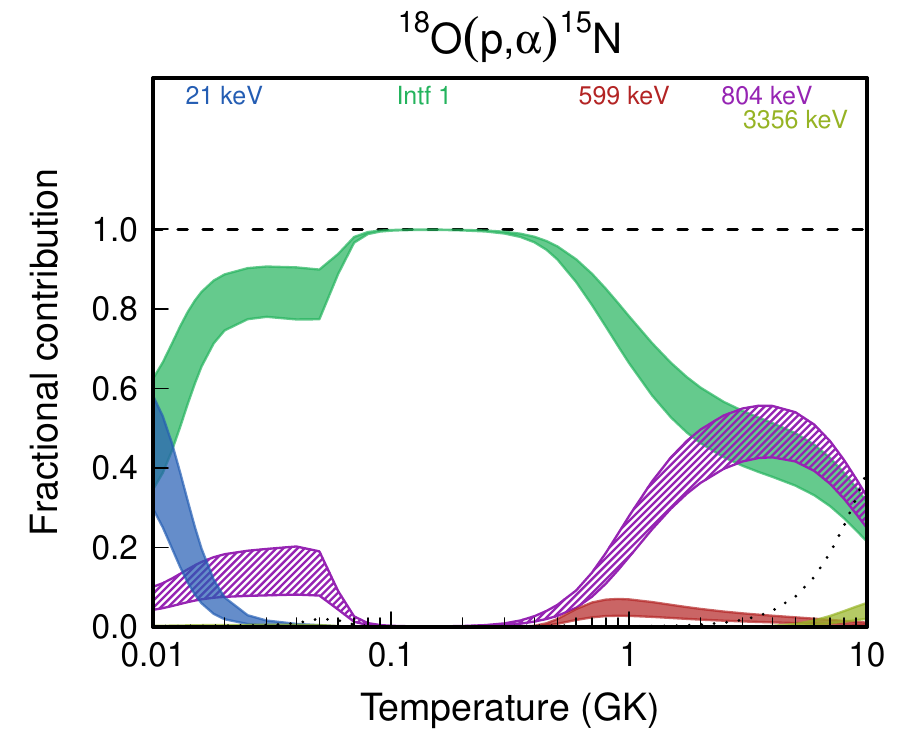}
\caption{
Fractional contributions to the total rate. Resonance energies are given in the center-of-mass frame. ``Intf 1'' labels the combined contribution of the two interfering $1/2^+$ resonances at $E_r^{c.m.}$ $=$ $145$ and $628$~keV.
}
\label{fig:o18pa1}
\end{figure*}
\begin{figure*}[hbt!]
\centering
\includegraphics[width=0.5\linewidth]{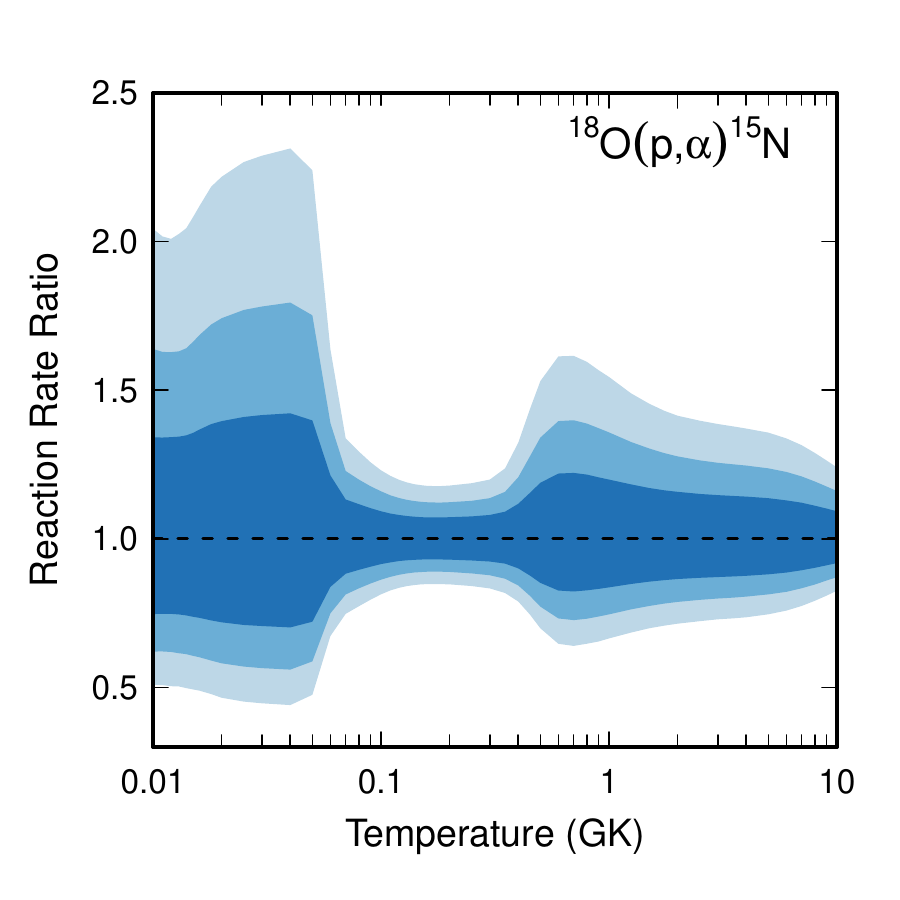}
\caption{
Reaction rate uncertainties versus temperature. The three different shades refer to coverage probabilities of 68\%, 90\%, and 98\%.}
\label{fig:o18pa2}
\end{figure*}

\clearpage

\startlongtable


\clearpage

\begin{figure*}[hbt!]
\centering
\includegraphics[width=0.5\linewidth]{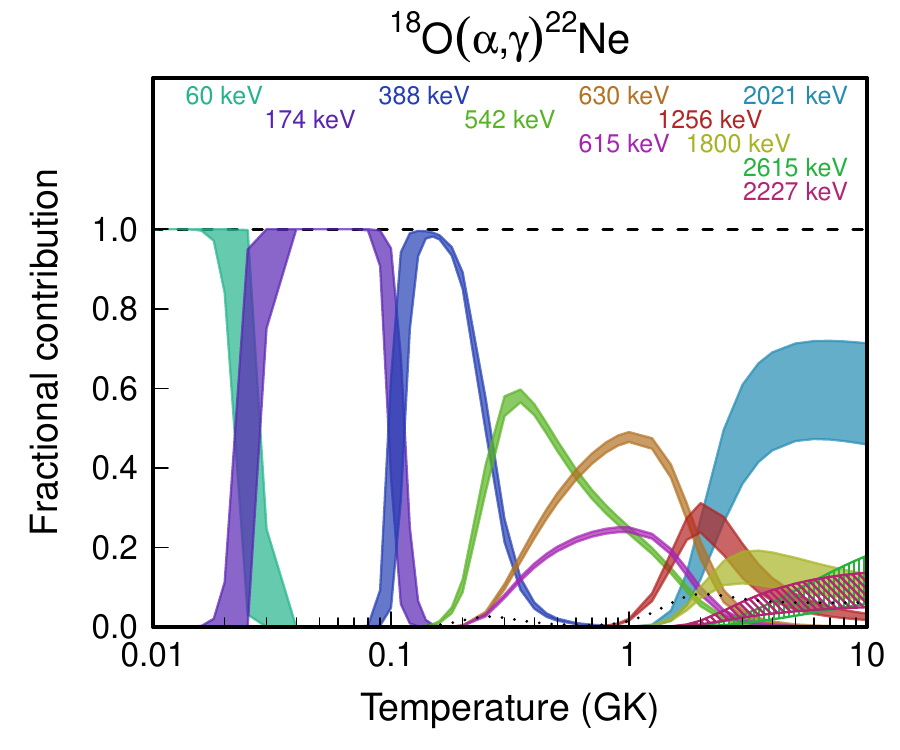}
\caption{
Fractional contributions to the total rate. Resonance energies are given in the center-of-mass frame. 
}
\label{fig:o18ag1}
\end{figure*}
\begin{figure*}[hbt!]
\centering
\includegraphics[width=0.5\linewidth]{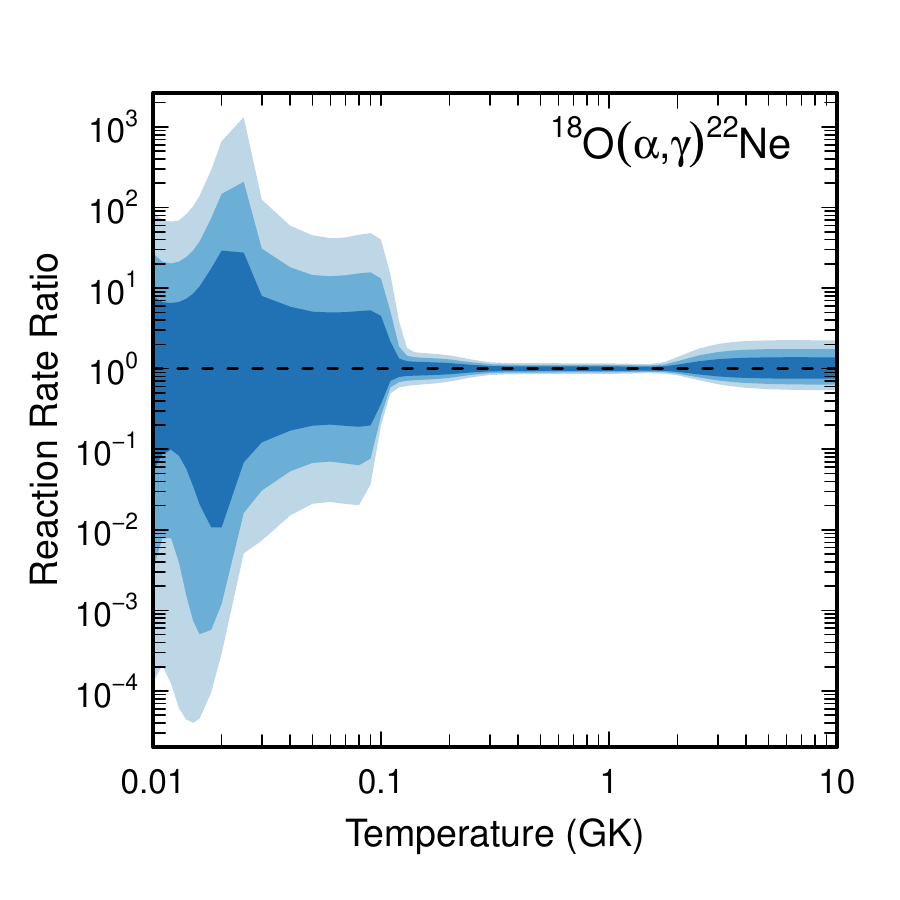}
\caption{
Reaction rate uncertainties versus temperature. The three different shades refer to coverage probabilities of 68\%, 90\%, and 98\%. 
}
\label{fig:o18ag2}
\end{figure*}

\clearpage

\startlongtable


\clearpage

\begin{figure*}[hbt!]
\centering
\includegraphics[width=0.5\linewidth]{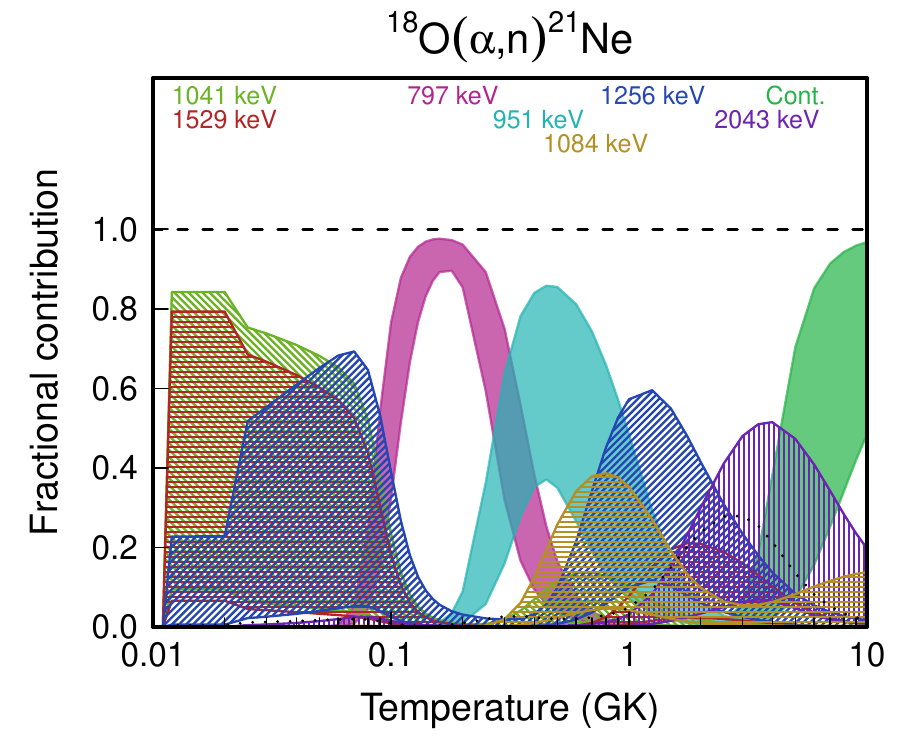}
\caption{
Fractional contributions to the total rate. ``Cont.'' refers to the continuum of unresolved resonances. Resonance energies are given in the center-of-mass frame.
}
\label{fig:o18an1}
\end{figure*}
\begin{figure*}[hbt!]
\centering
\includegraphics[width=0.5\linewidth]{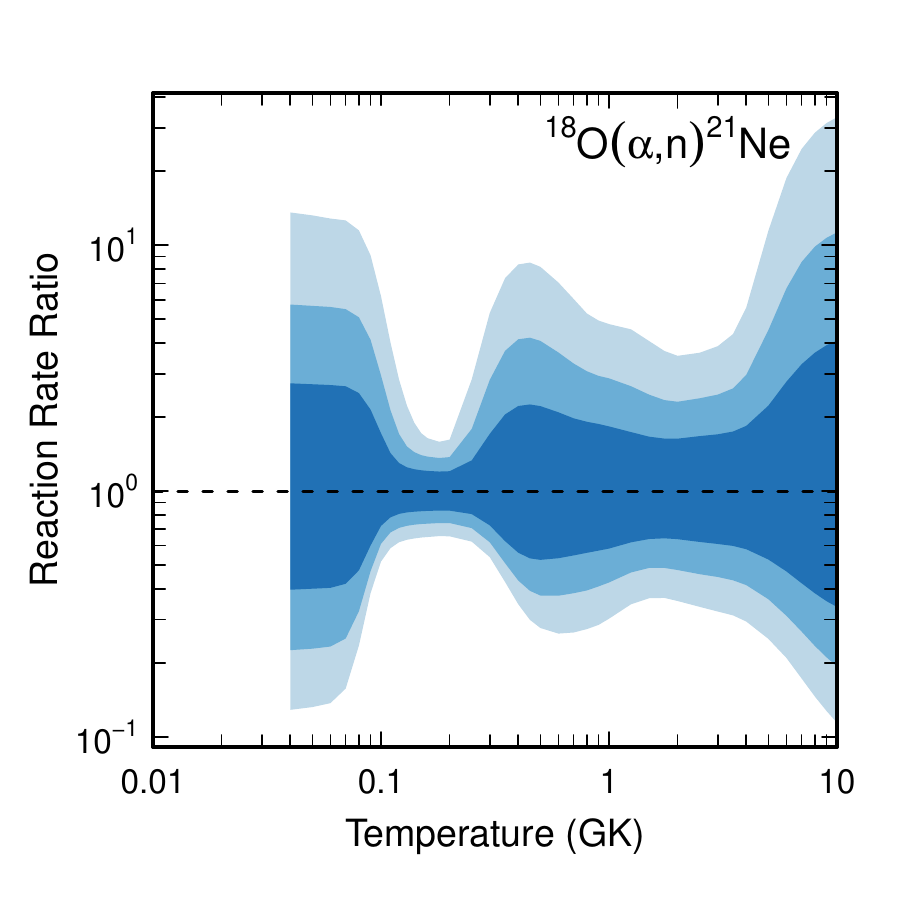}
\caption{Reaction rate uncertainties versus temperature. The three different shades refer to coverage probabilities of 68\%, 90\%, and 98\%.}
\label{fig:o18an2}
\end{figure*}

\clearpage

\startlongtable


\clearpage

\begin{figure*}[hbt!]
\centering
\includegraphics[width=0.5\linewidth]{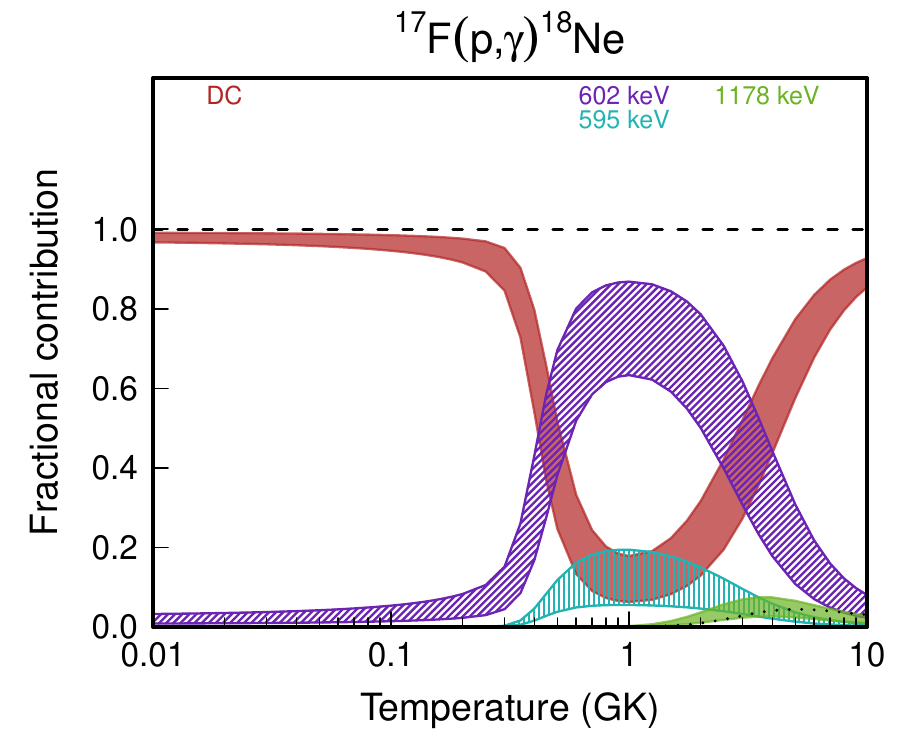}
\caption{
Fractional contributions to the total rate. ``DC'' refers to direct radiative capture. Resonance energies are given in the center-of-mass frame.  
}
\label{fig:f17pg1}
\end{figure*}
\begin{figure*}[hbt!]
\centering
\includegraphics[width=0.5\linewidth]{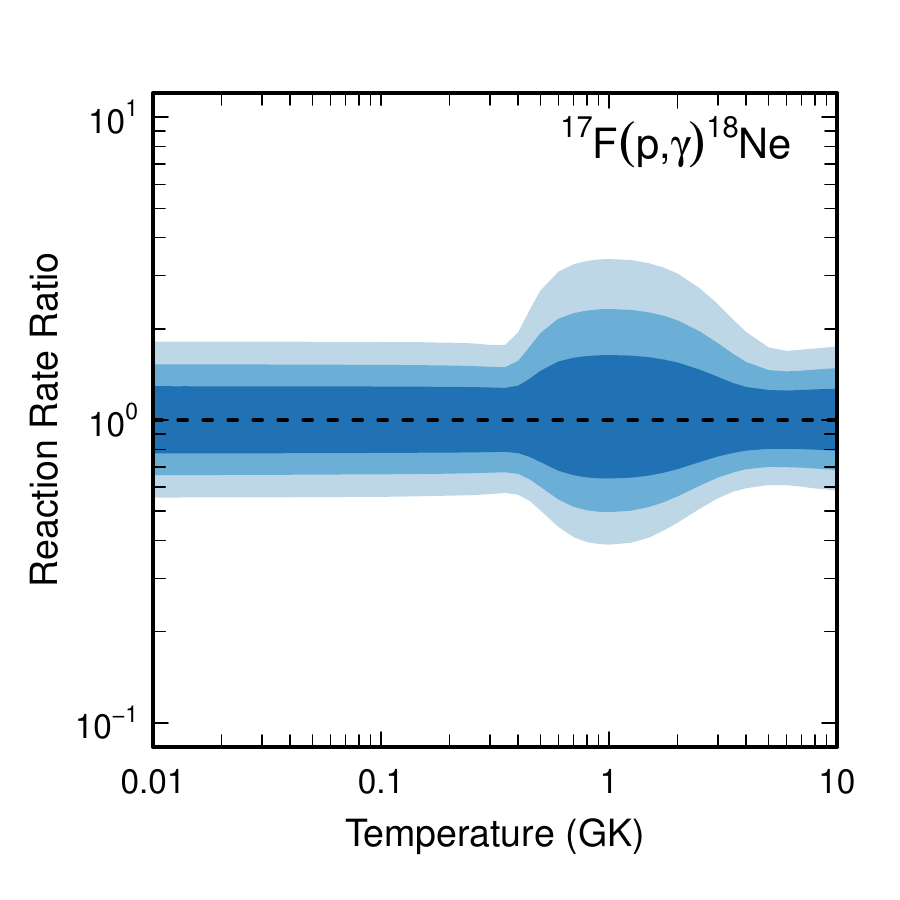}
\caption{
Reaction rate uncertainties versus temperature. The three different shades refer to coverage probabilities of 68\%, 90\%, and 98\%.}
\label{fig:f17pg2}
\end{figure*}

\clearpage

\startlongtable


\clearpage

\begin{figure*}[hbt!]
\centering
\includegraphics[width=0.5\linewidth]{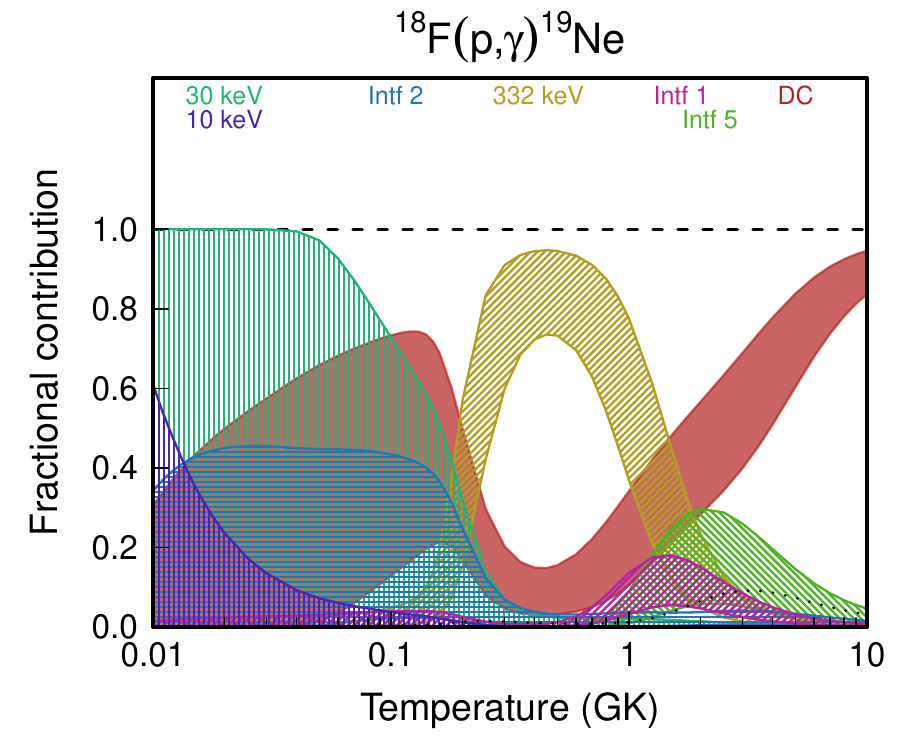}
\caption{
Fractional contributions to the total rate. ``DC'' refers to direct radiative capture. Resonance energies are given in the center-of-mass frame. Other labels refer to the interference of two resonances: $3/2^{+}$ resonances at $E_{r}^{c.m.}$ $=$ $-277$ and 663~keV (Intf 1); $1/2^{+}$ resonances at $-127$ and 1392 keV (Intf 2); and $3/2^{+}$ resonances at $833$ and 1161 keV (Intf 5).
}
\label{fig:f18pg1}
\end{figure*}
\begin{figure*}[hbt!]
\centering
\includegraphics[width=0.5\linewidth]{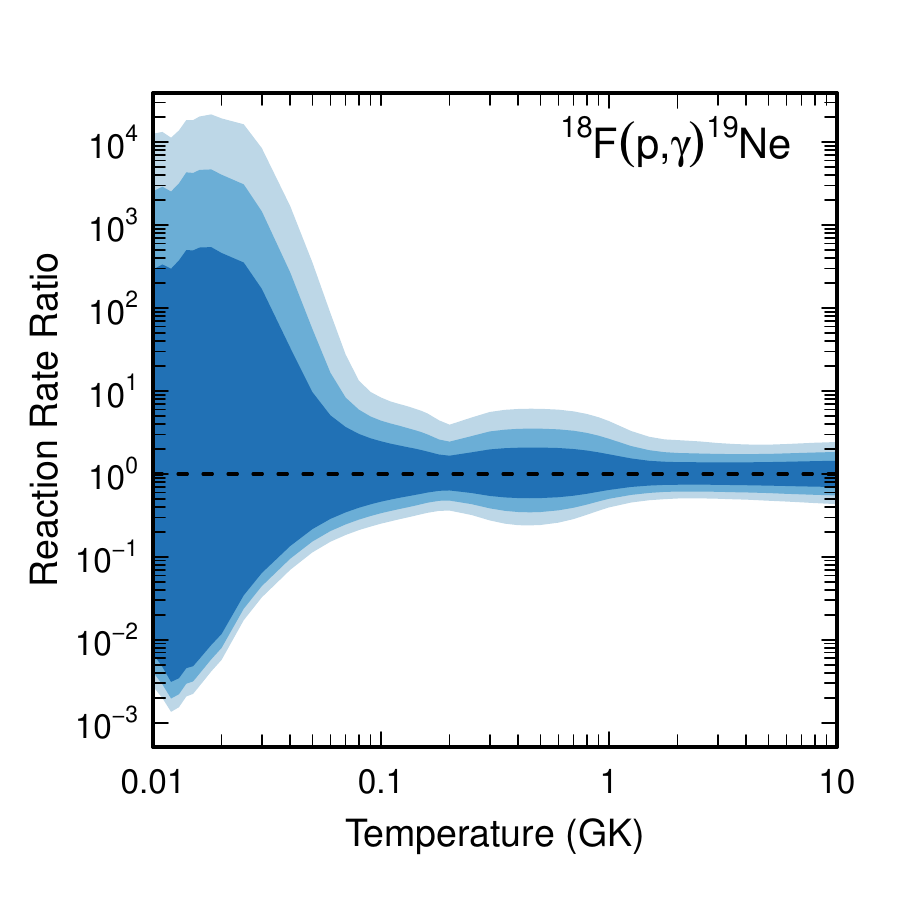}
\caption{
Reaction rate uncertainties versus temperature. The three different shades refer to coverage probabilities of 68\%, 90\%, and 98\%. 
}
\label{fig:f18pg2}
\end{figure*}

\clearpage

\startlongtable


\clearpage

\begin{figure*}[hbt!]
\centering
\includegraphics[width=0.5\linewidth]{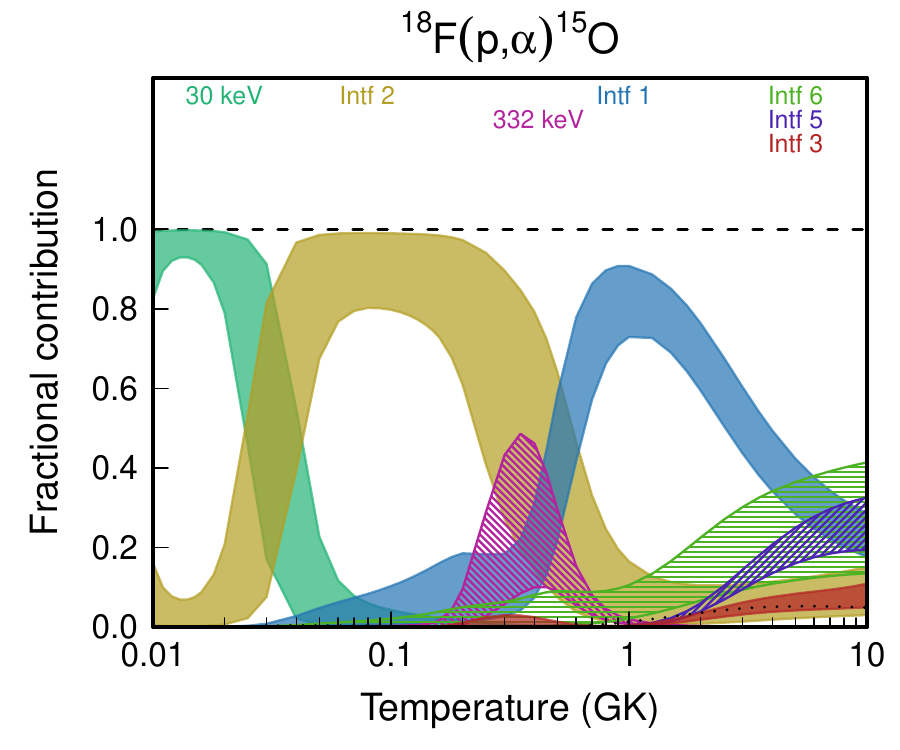}
\caption{
Fractional contributions to the total rate. Resonance energies are given in the center-of-mass frame. Other labels refer to the combined contribution of two interfering resonances: $3/2^{+}$ resonances at $E_{r}^{c.m.}$ $=$ $-277$ and 663 keV (Intf 1); $1/2^{+}$ resonances at $-127$ and 1392 keV (Intf 2); $5/2^{+}$ resonances at $291$ and 1091 keV (Intf 3); $3/2^{+}$ resonances at $1201$ and 1343 keV (Intf 5); and $1/2^{+}$ resonances at $1022$ and 1426 keV (Intf 6).
}
\label{fig:f18pa1}
\end{figure*}
\begin{figure*}[hbt!]
\centering
\includegraphics[width=0.5\linewidth]{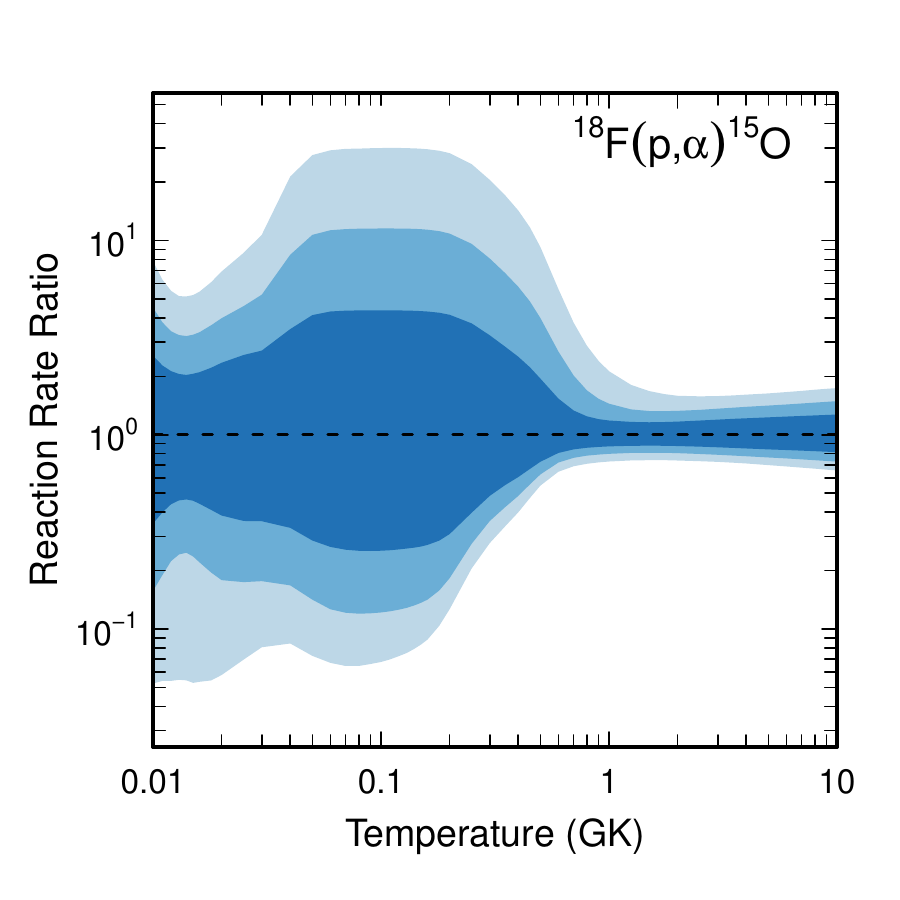}
\caption{
Reaction rate uncertainties versus temperature. The three different shades refer to coverage probabilities of 68\%, 90\%, and 98\%. 
}
\label{fig:f18pa2}
\end{figure*}

\clearpage

\startlongtable
\begin{deluxetable*}{lcccc | lcccc}
\tablecaption{Total laboratory reaction rates for $^{18}$Ne($\alpha$,p)$^{21}$Na.}
\label{tab:ne18ap} 
\tablewidth{\columnwidth}
\tabletypesize{\footnotesize}
\tablecolumns{10}
\tablehead{
T (GK) & Low & Median & High & $f.u.$ & T (GK) & Low & Median & High & $f.u.$
}
\startdata
 0.001  &  0.000E+00  &  0.000E+00  &  0.000E+00        &  1.000E+00  &  0.140  & 5.377E-22   & 1.314E-21   &  3.278E-21  &  2.510E+00   \\ 
 0.002  &  0.000E+00  &  0.000E+00  &  0.000E+00        &  1.000E+00  &  0.150  & 4.967E-21   & 1.193E-20   &  2.862E-20  &  2.421E+00   \\ 
 0.003  &  0.000E+00  &  0.000E+00  &  0.000E+00        &  1.000E+00  &  0.160  & 3.723E-20   & 8.926E-20   &  2.101E-19  &  2.382E+00   \\ 
 0.004  &  0.000E+00  &  0.000E+00  &  0.000E+00        &  1.000E+00  &  0.180  & 1.333E-18   & 3.199E-18   &  7.639E-18  &  2.409E+00   \\ 
 0.005  &  2.203E-99  &  2.614E-98  &  1.269E-97        &  1.003E+01  &  0.200  & 2.957E-17   & 7.477E-17   &  1.802E-16  &  2.493E+00   \\ 
 0.006  &  5.313E-93  &  6.898E-92  &  3.875E-91        &  1.096E+01  &  0.250  & 1.355E-14   & 3.748E-14   &  8.864E-14  &  2.600E+00   \\ 
 0.007  &  1.717E-88  &  2.519E-87  &  1.767E-86        &  1.280E+01  &  0.300  & 1.296E-12   & 3.176E-12   &  7.349E-12  &  2.406E+00   \\ 
 0.008  &  3.800E-85  &  6.486E-84  &  5.581E-83        &  1.500E+01  &  0.350  & 5.055E-11   & 1.101E-10   &  2.336E-10  &  2.176E+00   \\ 
 0.009  &  1.441E-82  &  2.852E-81  &  2.964E-80        &  1.733E+01  &  0.400  & 9.905E-10   & 2.115E-09   &  4.478E-09  &  2.153E+00   \\ 
 0.010  &  1.640E-80  &  3.660E-79  &  4.459E-78        &  1.966E+01  &  0.450  & 1.161E-08   & 2.497E-08   &  5.603E-08  &  2.237E+00   \\ 
 0.011  &  7.755E-79  &  1.913E-77  &  2.669E-76        &  2.195E+01  &  0.500  & 9.466E-08   & 2.028E-07   &  4.797E-07  &  2.287E+00   \\ 
 0.012  &  1.888E-77  &  5.096E-76  &  8.048E-75        &  2.414E+01  &  0.600  & 3.341E-06   & 6.519E-06   &  1.453E-05  &  2.124E+00   \\ 
 0.013  &  2.794E-76  &  8.176E-75  &  1.423E-73        &  2.624E+01  &  0.700  & 6.729E-05   & 1.194E-04   &  2.277E-04  &  1.871E+00   \\ 
 0.014  &  2.774E-75  &  8.738E-74  &  1.656E-72        &  2.824E+01  &  0.800  & 8.218E-04   & 1.388E-03   &  2.443E-03  &  1.748E+00   \\ 
 0.015  &  2.004E-74  &  6.793E-73  &  1.385E-71        &  3.012E+01  &  0.900  & 6.482E-03   & 1.074E-02   &  1.846E-02  &  1.709E+00   \\ 
 0.016  &  1.135E-73  &  4.056E-72  &  8.844E-71        &  3.190E+01  &  1.000  & 3.583E-02   & 5.905E-02   &  1.002E-01  &  1.691E+00   \\ 
 0.018  &  1.967E-72  &  7.797E-71  &  1.898E-69        &  3.517E+01  &  1.250  & 8.675E-01   & 1.394E+00   &  2.295E+00  &  1.639E+00   \\ 
 0.020  &  1.914E-71  &  8.220E-70  &  2.177E-68        &  3.807E+01  &  1.500  & 7.886E+00   & 1.211E+01   &  1.917E+01  &  1.569E+00   \\ 
 0.025  &  1.087E-69  &  5.412E-68  &  1.675E-66        &  4.404E+01  &  1.750  & 4.124E+01   & 6.018E+01   &  9.050E+01  &  1.494E+00   \\ 
 0.030  &  1.568E-68  &  8.359E-67  &  2.884E-65        &  3.949E+01  &  2.000  & 1.541E+02   & 2.155E+02   &  3.087E+02  &  1.426E+00   \\ 
 0.040  &  2.086E-55  &  7.170E-55  &  2.468E-54        &  3.451E+00  &  2.500  & 1.177E+03   & 1.578E+03   &  2.144E+03  &  1.354E+00   \\ 
 0.050  &  4.982E-46  &  1.621E-45  &  5.312E-45        &  3.281E+00  &  3.000  & 5.452E+03   & 7.302E+03   &  9.908E+03  &  1.352E+00   \\ 
 0.060  &  8.429E-40  &  2.659E-39  &  8.451E-39        &  3.188E+00  &  3.500  & 1.835E+04   & 2.487E+04   &  3.420E+04  &  1.370E+00   \\ 
 0.070  &  2.274E-35  &  7.049E-35  &  2.204E-34        &  3.130E+00  &  4.000  & 4.917E+04   & 6.749E+04   &  9.434E+04  &  1.392E+00   \\ 
 0.080  &  4.677E-32  &  1.427E-31  &  4.408E-31        &  3.087E+00  &  5.000  & 2.150E+05   & 3.066E+05   &  4.474E+05  &  1.451E+00   \\ 
 0.090  &  1.736E-29  &  5.227E-29  &  1.596E-28        &  3.047E+00  &  6.000  & 5.974E+05   & 9.009E+05   &  1.398E+06  &  1.540E+00   \\ 
 0.100  &  1.962E-27  &  5.821E-27  &  1.755E-26        &  2.996E+00  &  7.000  & 1.236E+06   & 2.005E+06   &  3.350E+06  &  1.659E+00   \\ 
 0.110  &  9.555E-26  &  2.738E-25  &  8.135E-25        &  2.915E+00  &  8.000  & 2.095E+06   & 3.702E+06   &  6.688E+06  &  1.800E+00   \\ 
 0.120  &  2.561E-24  &  6.905E-24  &  1.989E-23        &  2.792E+00  &  9.000  & 3.084E+06   & 5.989E+06   &  1.170E+07  &  1.959E+00   \\ 
 0.130  &  4.413E-23  &  1.119E-22  &  3.017E-22        &  2.644E+00  &  10.000 & 4.088E+06   & 8.798E+06   &  1.853E+07  &  2.140E+00   \\ 
\enddata
\tablecomments{
{\it Observed resonances:} \cite{Mohr2014}. \\
{\it Normalization:} None. \\
{\it Unobserved resonances:} \citet{Mohr2014}. \\
{\it High-temperature rates (in parentheses):} No corrections for missing high-energy resonances are necessary up to $T$ $=$ $10$~GK.  \\
{\it Previous rates:} \citet{Mohr2014,Zhang2014}. \\
{\it Other:} The rate for $^{18}$Ne($\alpha$,p)$^{21}$Na was calculated from the (total) experimental $^{18}$Ne $+$ $\alpha$ rate (i.e., including the $p$ and $2p$ exit channels) using the calculated branching ratio for the ($\alpha$,p) channel from TALYS, which deviates from unity by less than 10\% up to $T$ $=$ $2$~GK.}
\end{deluxetable*}

\clearpage

\begin{figure*}[hbt!]
\centering
\includegraphics[width=0.5\linewidth]{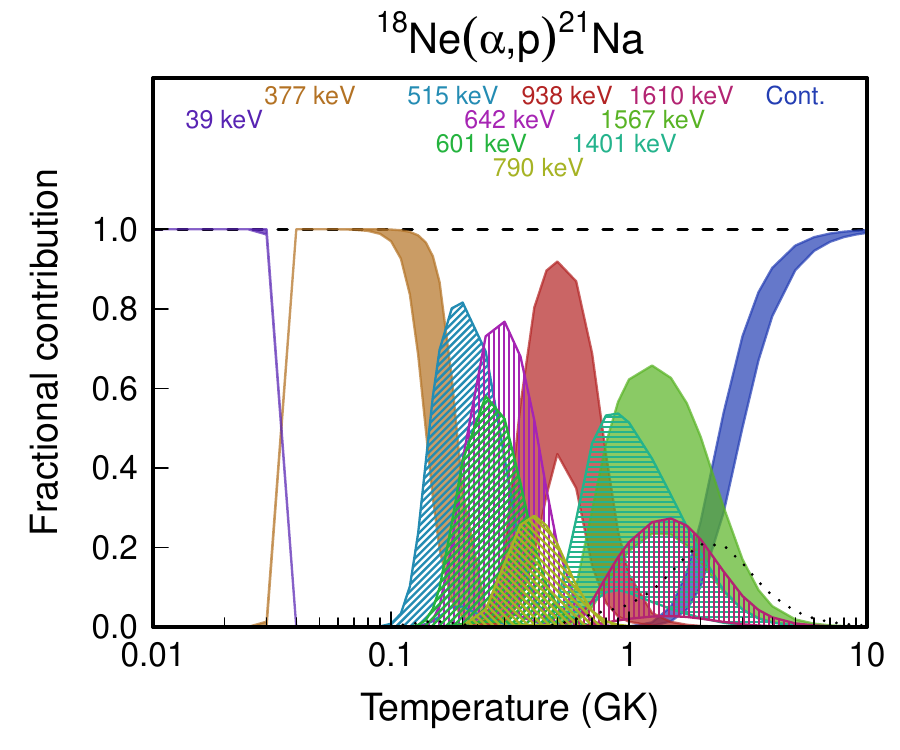}
\caption{
Fractional contributions to the total rate. ``Cont.'' refers to the continuum of higher-lying, unresolved resonances. Resonance energies are given in the center-of-mass frame. 
}
\label{fig:ne18ap1}
\end{figure*}
\begin{figure*}[hbt!]
\centering
\includegraphics[width=0.5\linewidth]{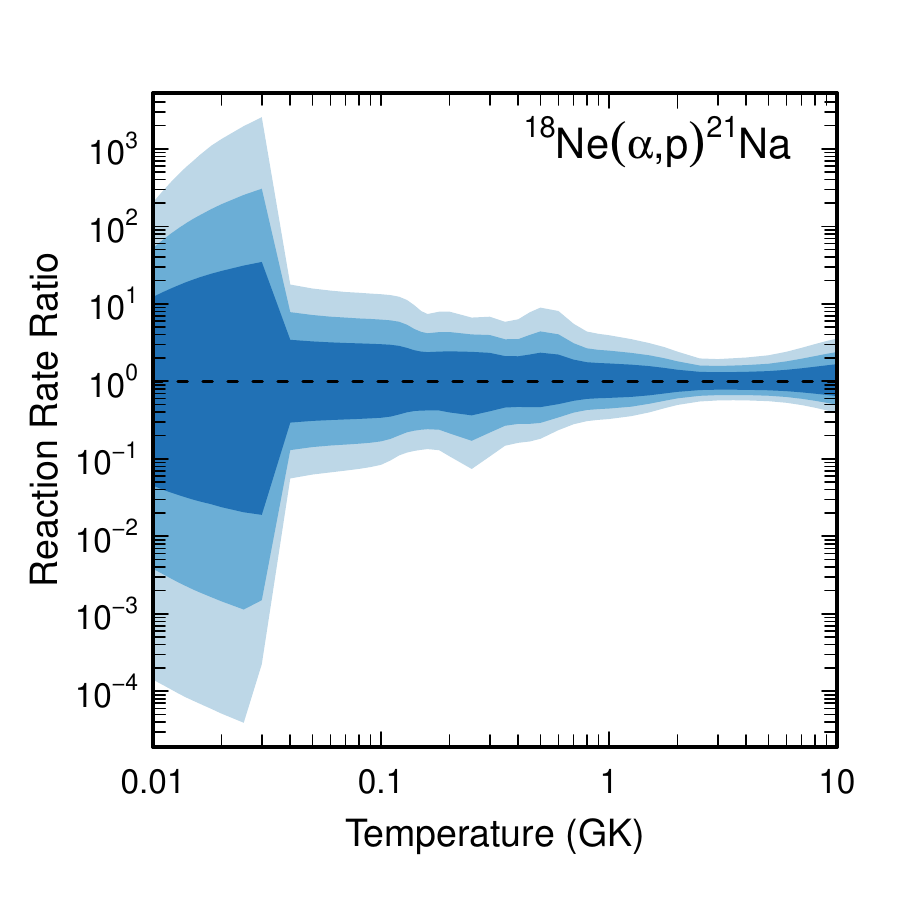}
\caption{
Reaction rate uncertainties versus temperature. The three different shades refer to coverage probabilities of 68\%, 90\%, and 98\%.}
\label{fig:ne18ap2}
\end{figure*}

\clearpage

\startlongtable


\clearpage

\begin{figure*}[hbt!]
\centering
\includegraphics[width=0.5\linewidth]{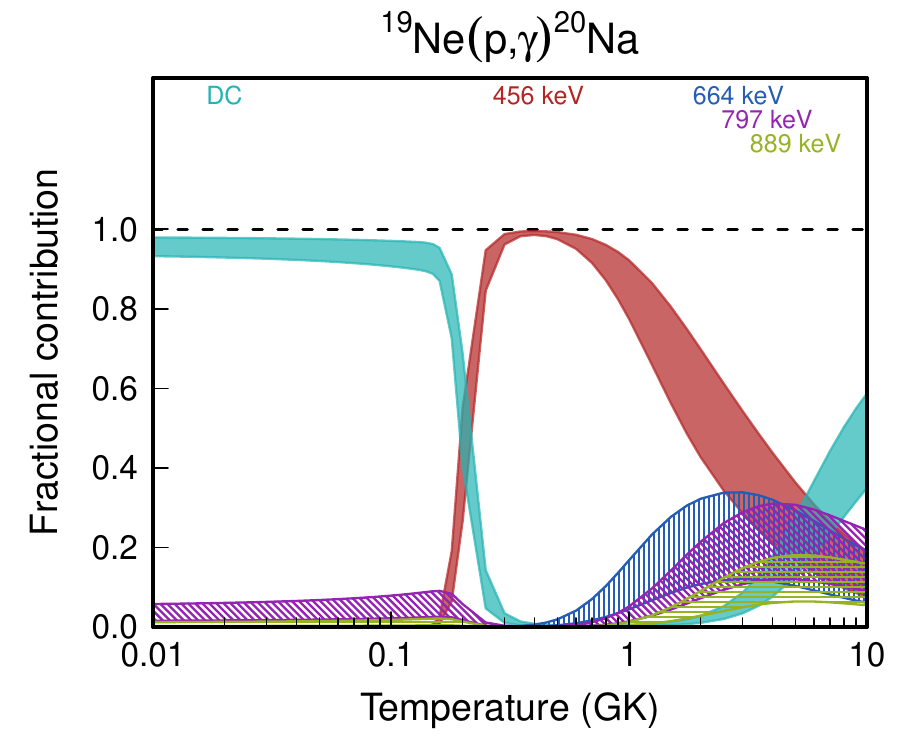}
\caption{
Fractional contributions to the total rate. ``DC'' refers to direct radiative capture. Resonance energies are given in the center-of-mass frame. 
}
\label{fig:ne19pg1}
\end{figure*}
\begin{figure*}[hbt!]
\centering
\includegraphics[width=0.5\linewidth]{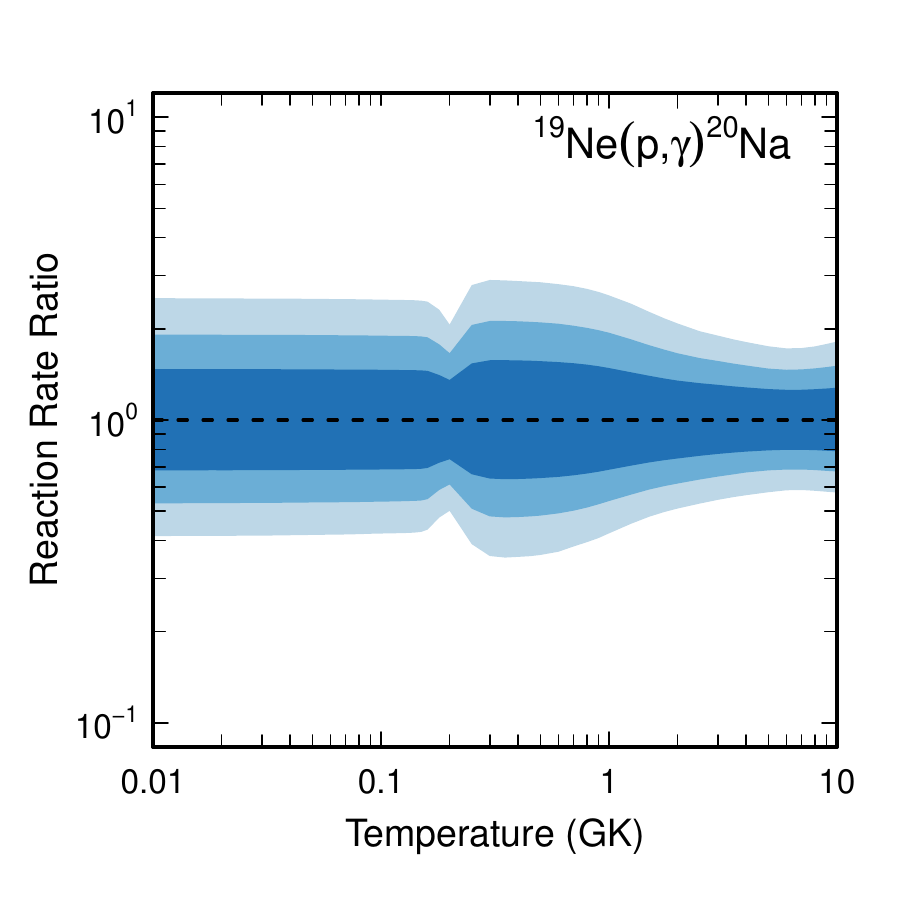}
\caption{
Reaction rate uncertainties versus temperature. The three different shades refer to coverage probabilities of 68\%, 90\%, and 98\%. 
}
\label{fig:ne19pg2}
\end{figure*}

\clearpage

\startlongtable


\clearpage

\begin{figure*}[hbt!]
\centering
\includegraphics[width=0.5\linewidth]{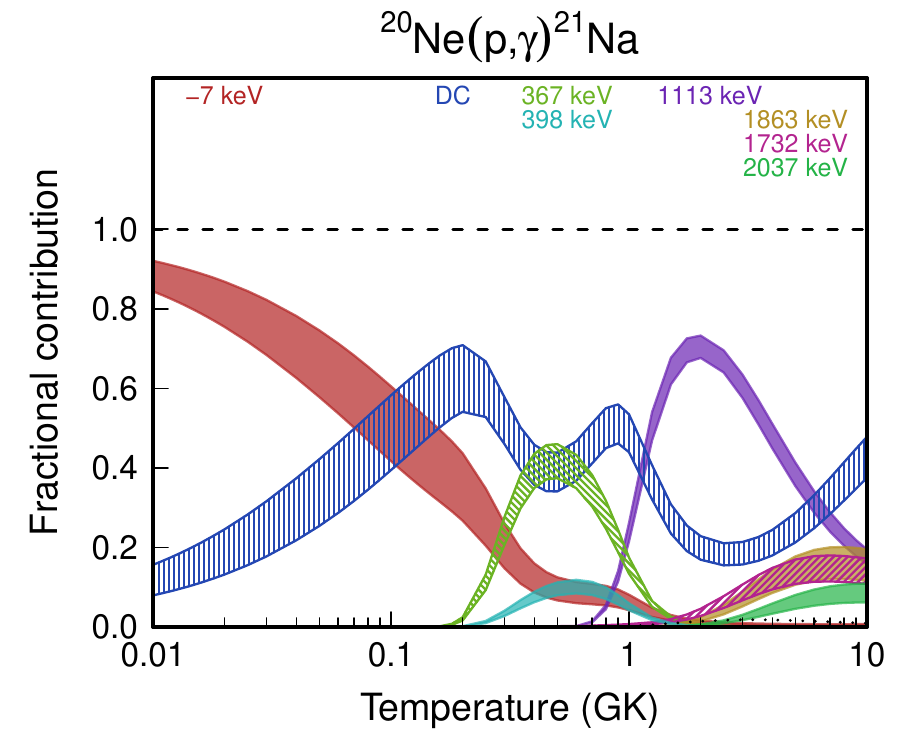}
\caption{
Fractional contributions to the total rate. ``DC'' refers to direct radiative capture. Resonance energies are given in the center-of-mass frame. 
}
\label{fig:ne20pg1}
\end{figure*}
\begin{figure*}[hbt!]
\centering
\includegraphics[width=0.5\linewidth]{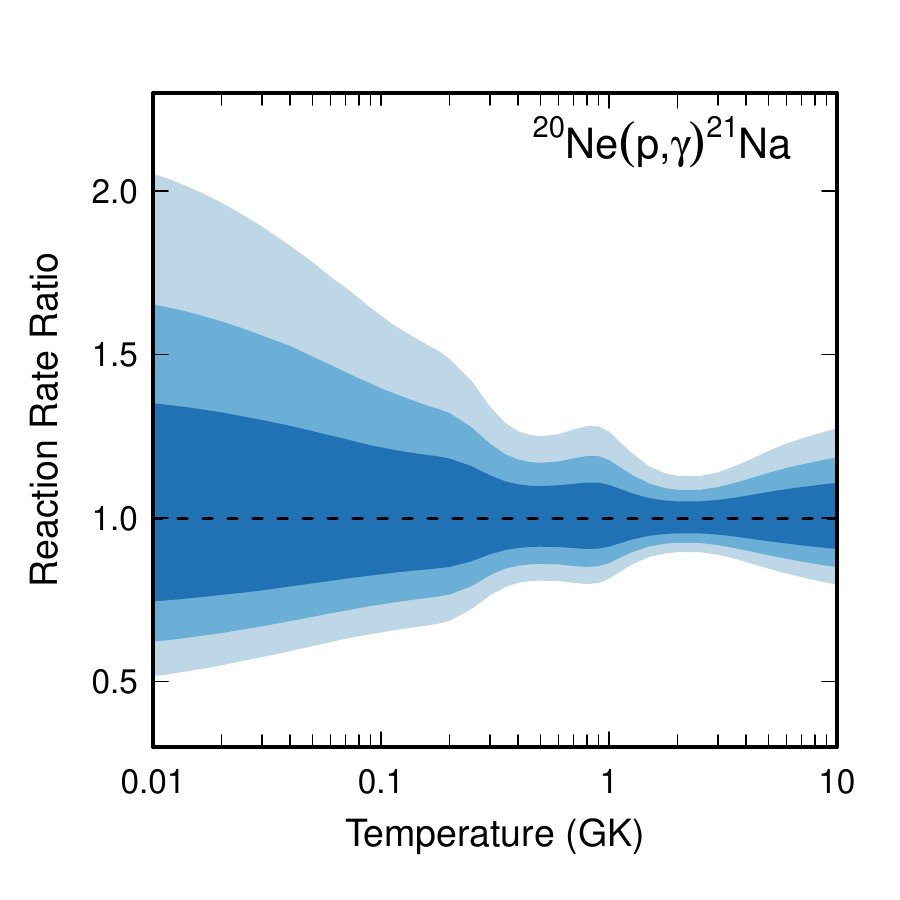}
\caption{
Reaction rate uncertainties versus temperature. The three different shades refer to coverage probabilities of 68\%, 90\%, and 98\%. 
}
\label{fig:ne20pg2}
\end{figure*}

\clearpage

\startlongtable


\clearpage

\begin{figure*}[hbt!]
\centering
\includegraphics[width=0.5\linewidth]{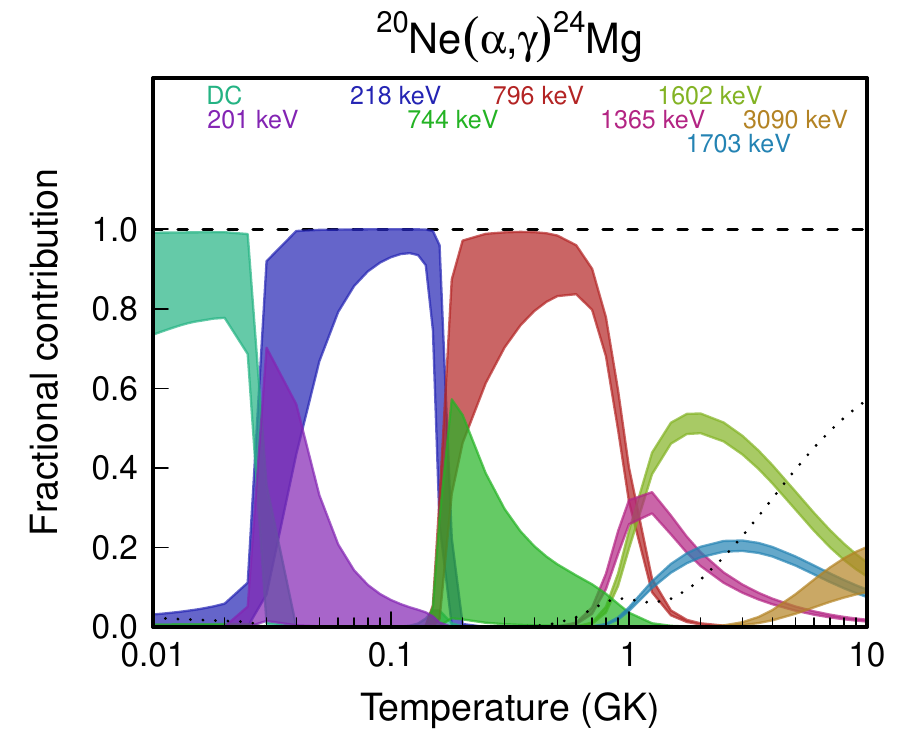}
\caption{
Fractional contributions to the total rate. ``DC'' refers to direct radiative capture. Resonance energies are given in the center-of-mass frame.  
}
\label{fig:ne20ag1}
\end{figure*}
\begin{figure*}[hbt!]
\centering
\includegraphics[width=0.5\linewidth]{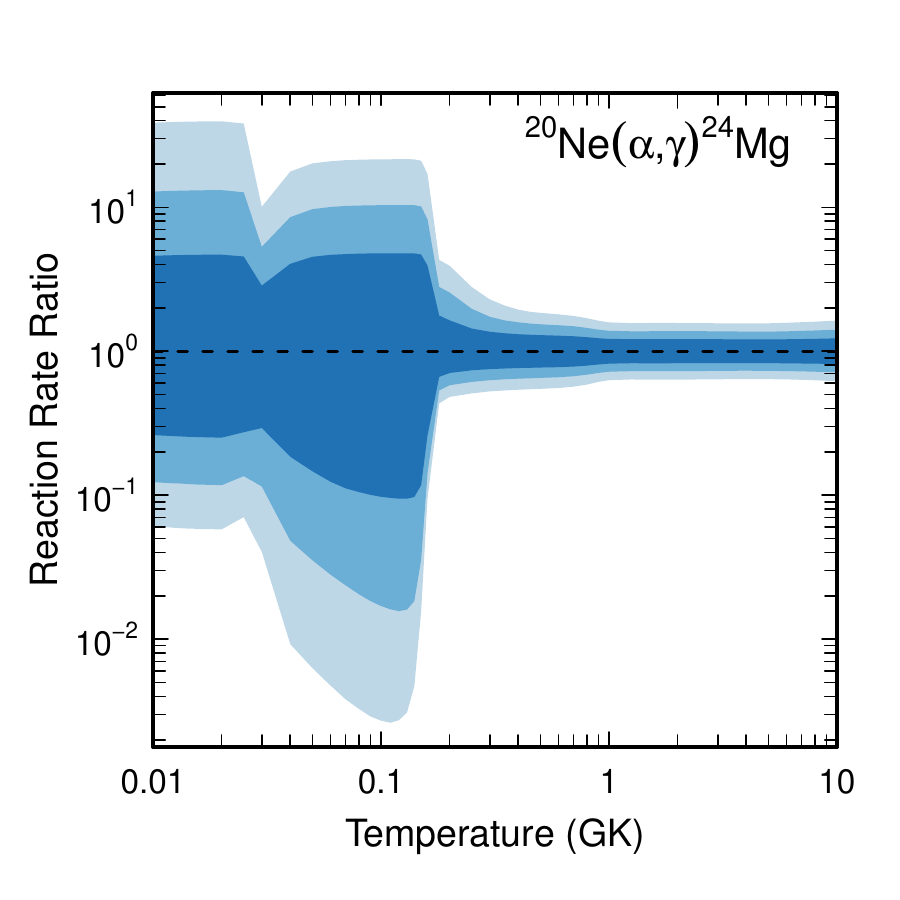}
\caption{
Reaction rate uncertainties versus temperature. The three different shades refer to coverage probabilities of 68\%, 90\%, and 98\%.}
\label{fig:ne20ag2}
\end{figure*}

\clearpage

\startlongtable


\clearpage

\begin{figure*}[hbt!]
\centering
\includegraphics[width=0.5\linewidth]{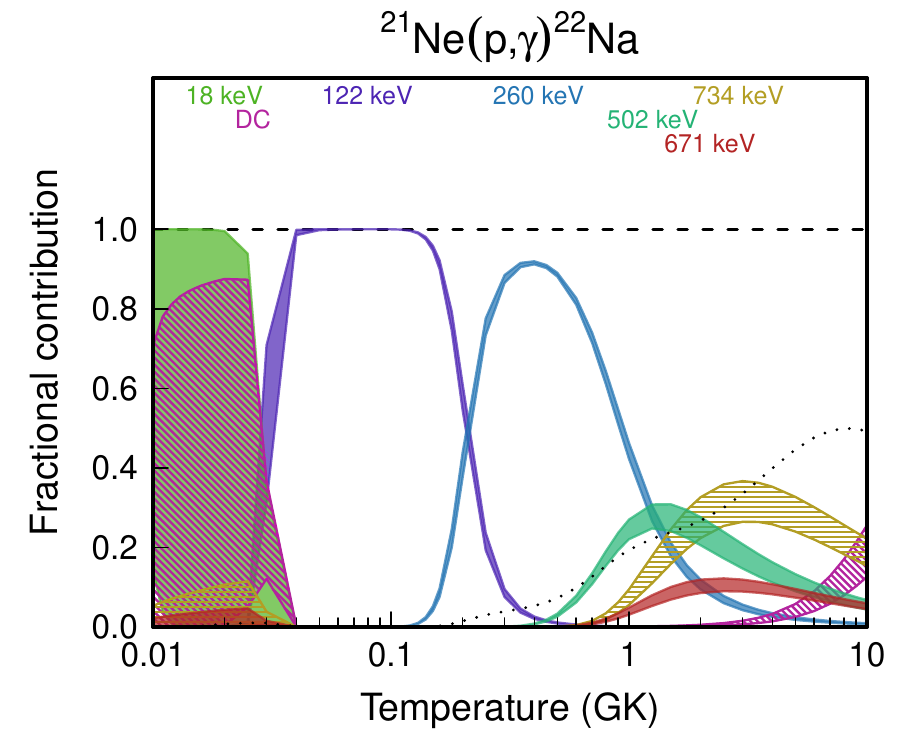}
\caption{
Fractional contributions to the total rate. ``DC'' refers to direct radiative capture. Resonance energies are given in the center-of-mass frame.  
}
\label{fig:ne21pg1}
\end{figure*}
\begin{figure*}[hbt!]
\centering
\includegraphics[width=0.5\linewidth]{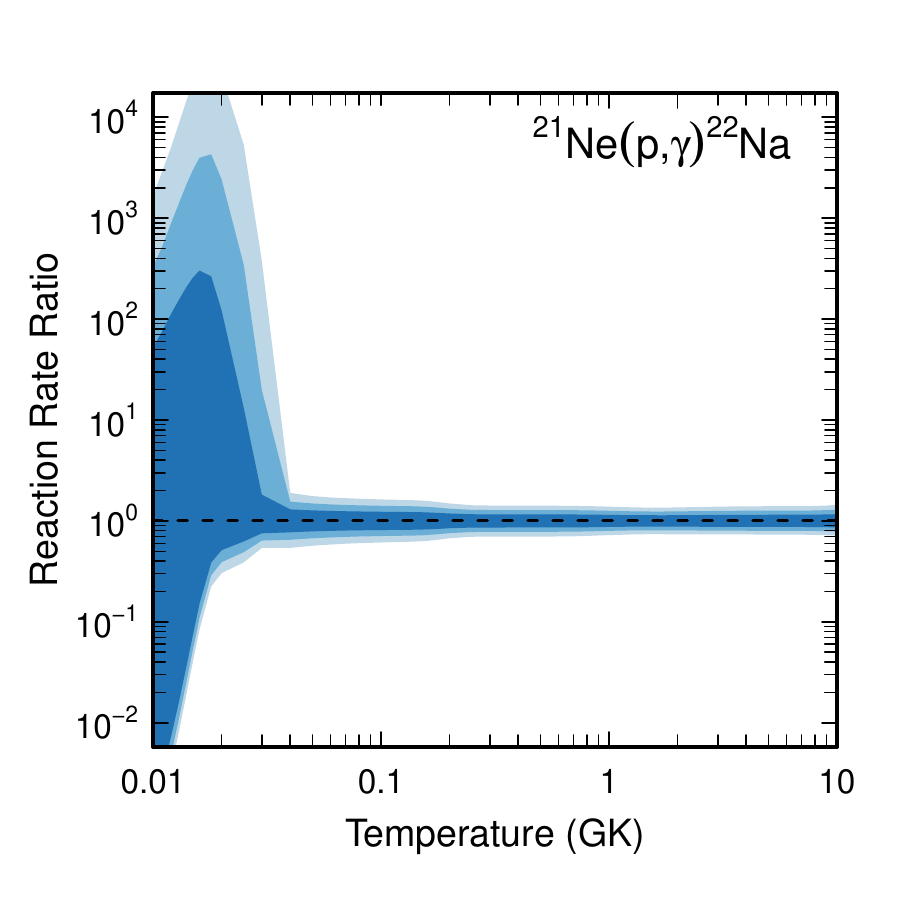}
\caption{
Reaction rate uncertainties versus temperature. The three different shades refer to coverage probabilities of 68\%, 90\%, and 98\%. 
}
\label{fig:ne21pg2}
\end{figure*}

\clearpage

\startlongtable


\clearpage

\begin{figure*}[hbt!]
\centering
\includegraphics[width=0.5\linewidth]{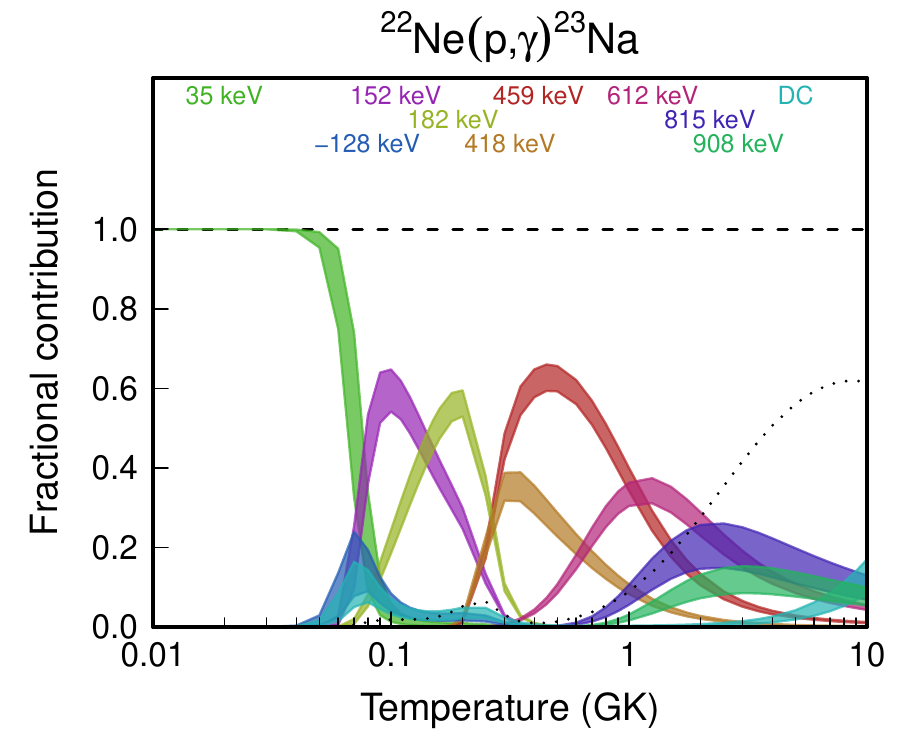}
\caption{
Fractional contributions to the total rate. ``DC'' refers to direct radiative capture. Resonance energies are given in the center-of-mass frame.  
}
\label{fig:ne22pg1}
\end{figure*}
\begin{figure*}[hbt!]
\centering
\includegraphics[width=0.5\linewidth]{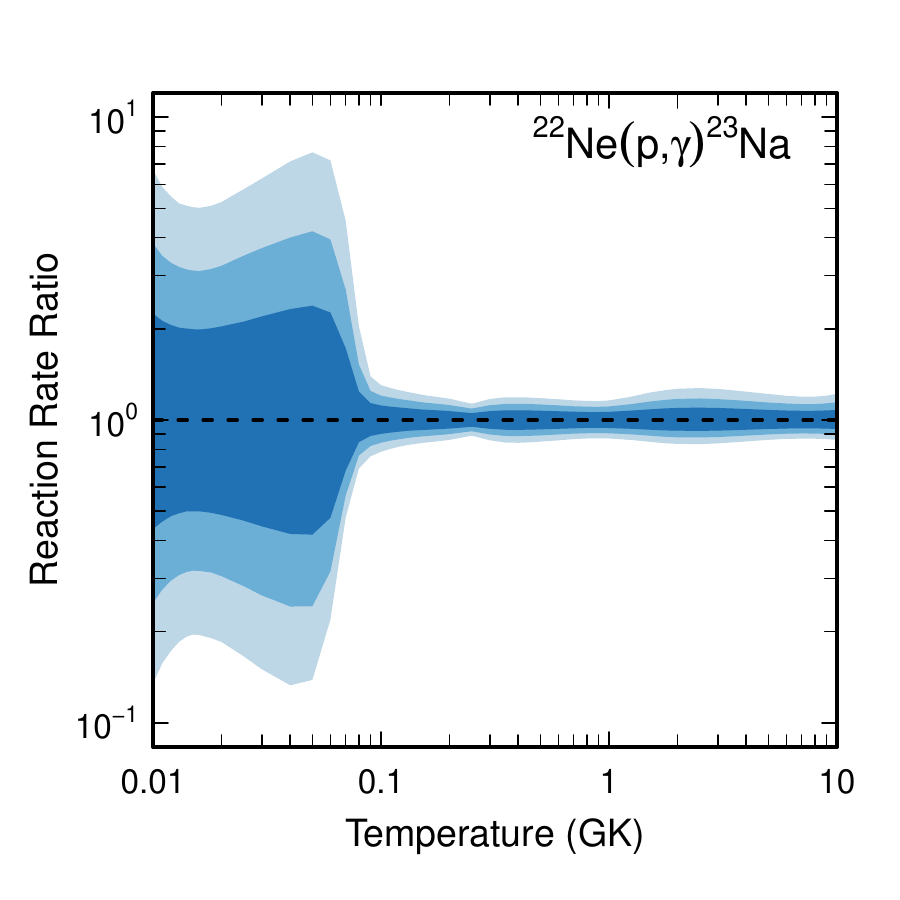}
\caption{
Reaction rate uncertainties versus temperature. The three different shades refer to coverage probabilities of 68\%, 90\%, and 98\%. 
}
\label{fig:ne22pg2}
\end{figure*}

\clearpage

\startlongtable


\clearpage

\begin{figure*}[hbt!]
\centering
\includegraphics[width=0.5\linewidth]{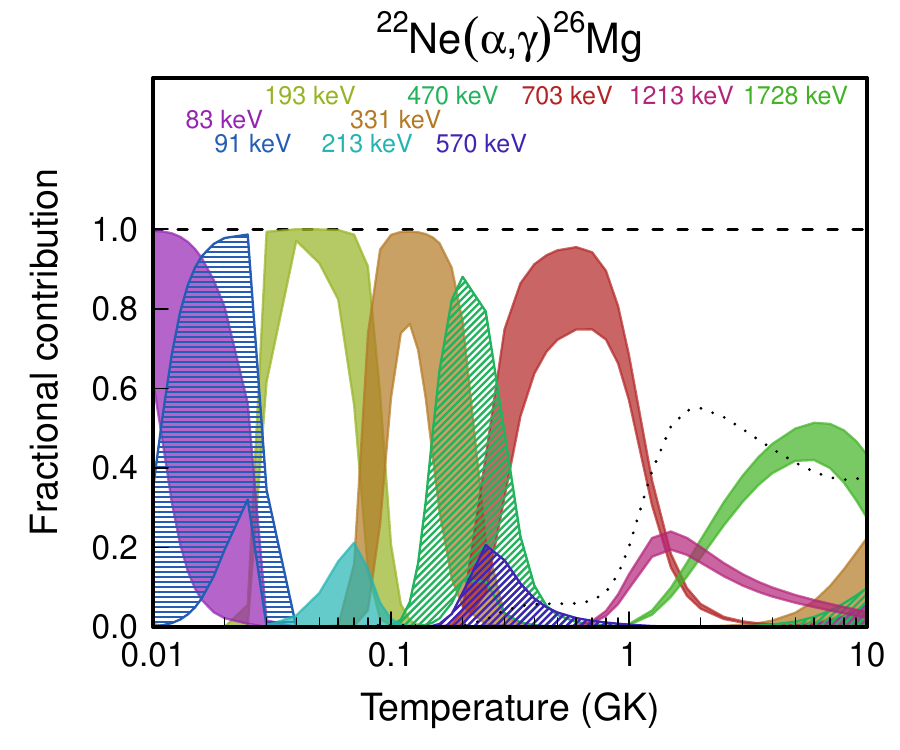}
\caption{
Fractional contributions to the total rate. Resonance energies are given in the center-of-mass frame.  
}
\label{fig:ne22ag1}
\end{figure*}
\begin{figure*}[hbt!]
\centering
\includegraphics[width=0.5\linewidth]{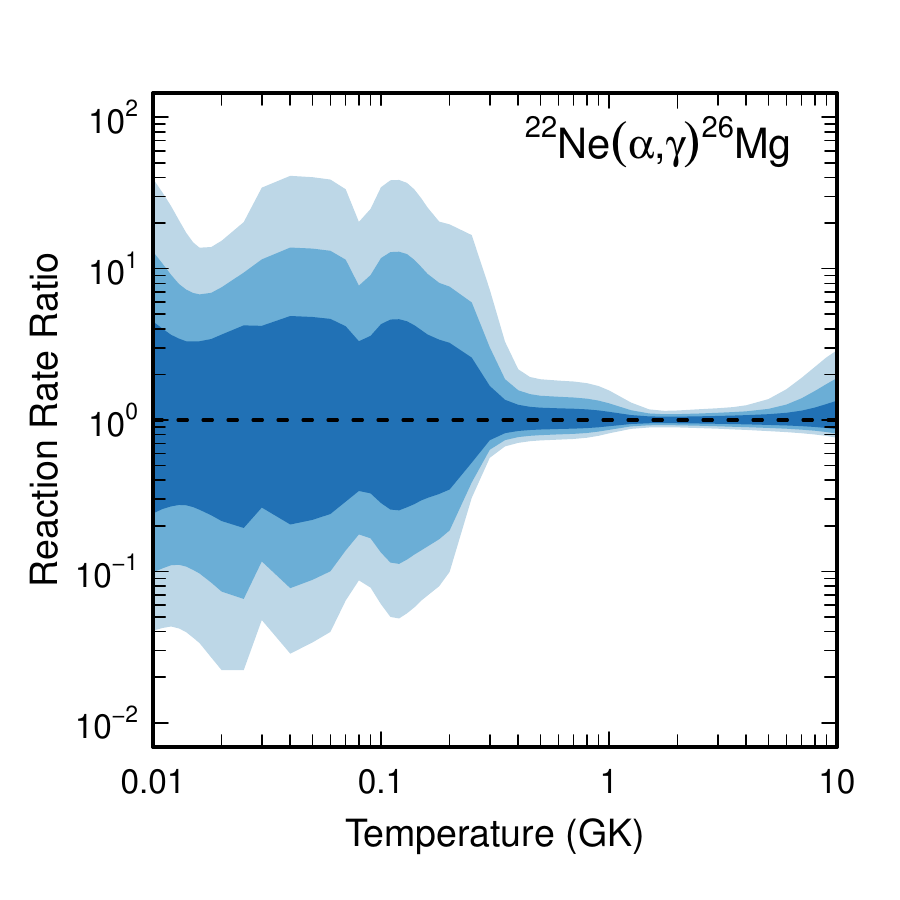}
\caption{
Reaction rate uncertainties versus temperature. The three different shades refer to coverage probabilities of 68\%, 90\%, and 98\%. 
}
\label{fig:ne22ag2}
\end{figure*}

\clearpage

\startlongtable


\clearpage

\begin{figure*}[hbt!]
\centering
\includegraphics[width=0.5\linewidth]{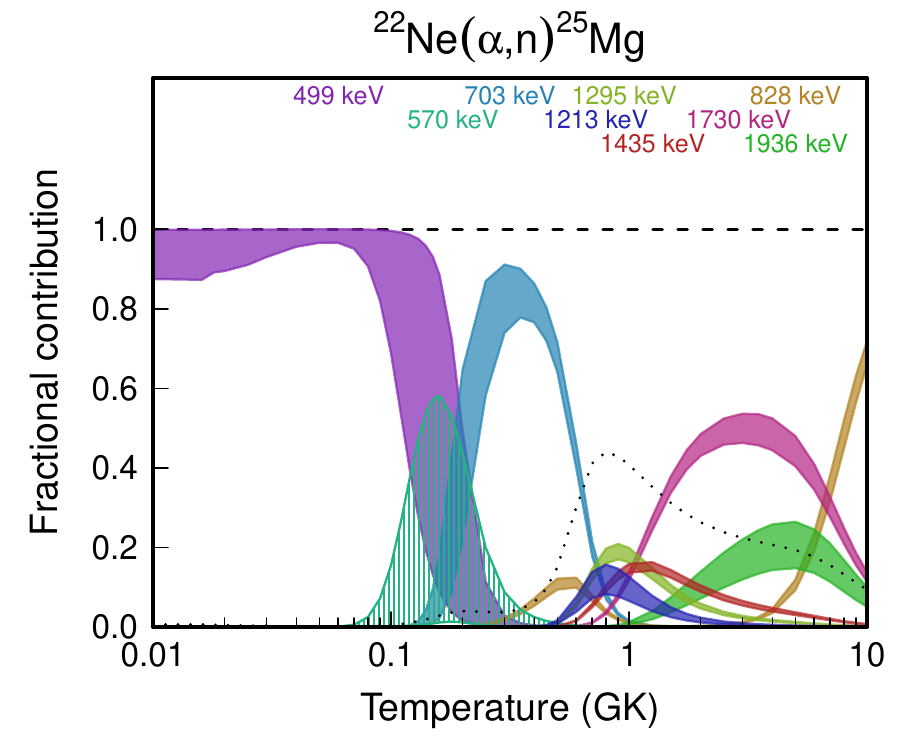}
\caption{
Fractional contributions to the total rate. Resonance energies are given in the center-of-mass frame.  
}
\label{fig:ne22an1}
\end{figure*}
\begin{figure*}[hbt!]
\centering
\includegraphics[width=0.5\linewidth]{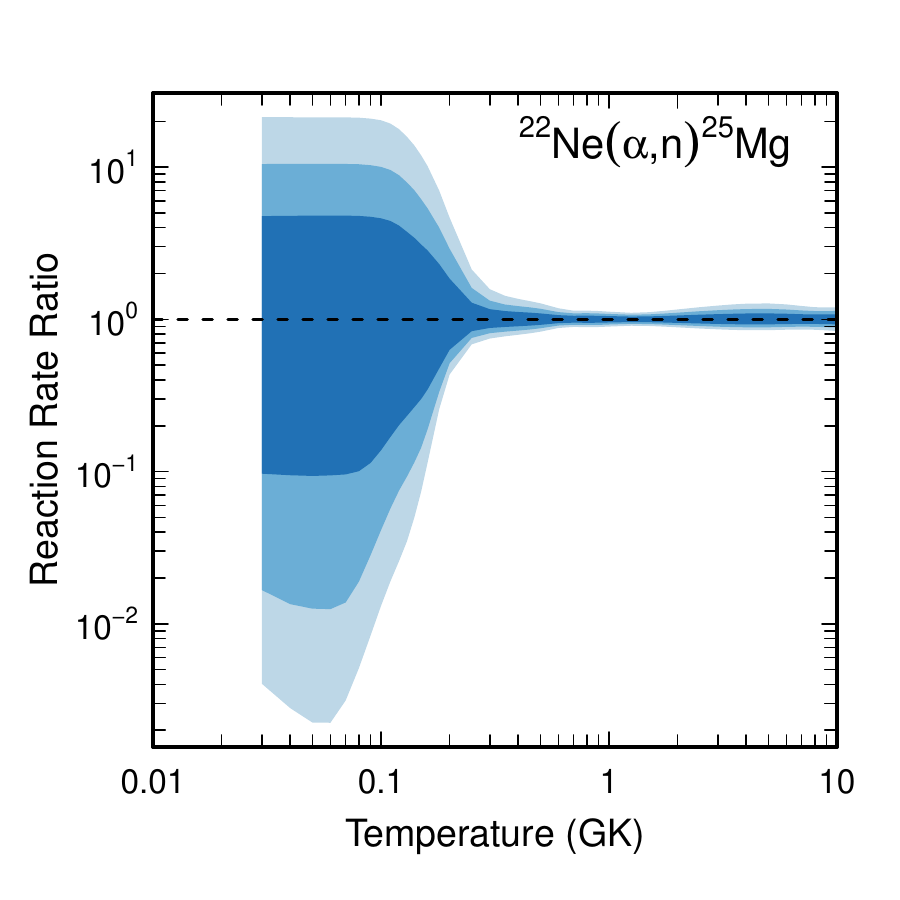}
\caption{
Reaction rate uncertainties versus temperature. The three different shades refer to coverage probabilities of 68\%, 90\%, and 98\%. 
}
\label{fig:ne22an2}
\end{figure*}

\clearpage

\startlongtable


\clearpage

\begin{figure*}[hbt!]
\centering
\includegraphics[width=0.5\linewidth]{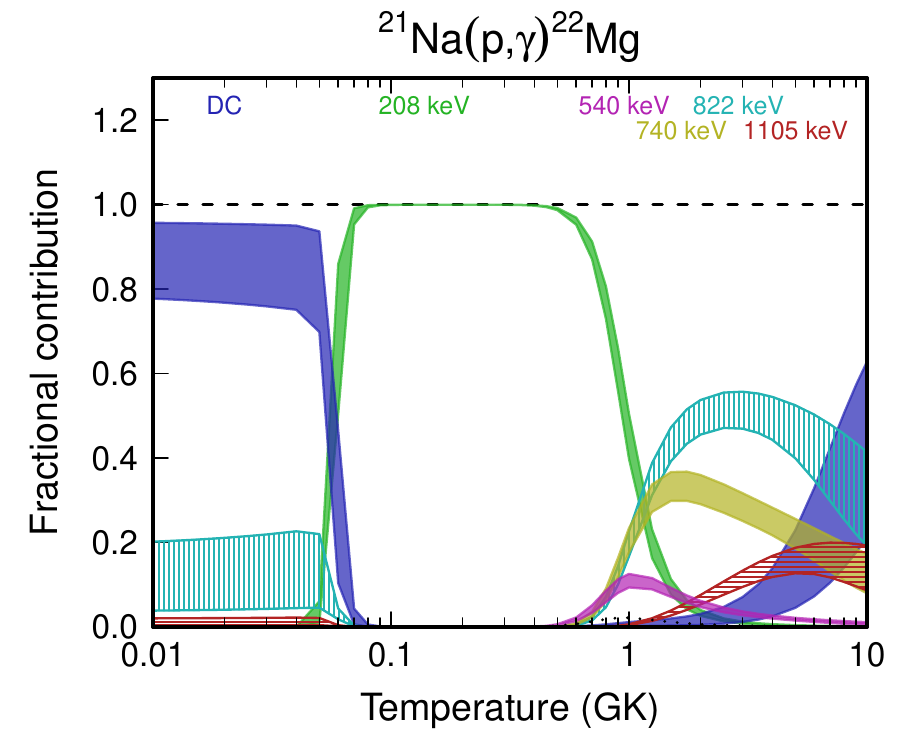}
\caption{
Fractional contributions to the total rate. ``DC'' refers to direct radiative capture. Resonance energies are given in the center-of-mass frame. 
}
\label{fig:na21pg1}
\end{figure*}
\begin{figure*}[hbt!]
\centering
\includegraphics[width=0.5\linewidth]{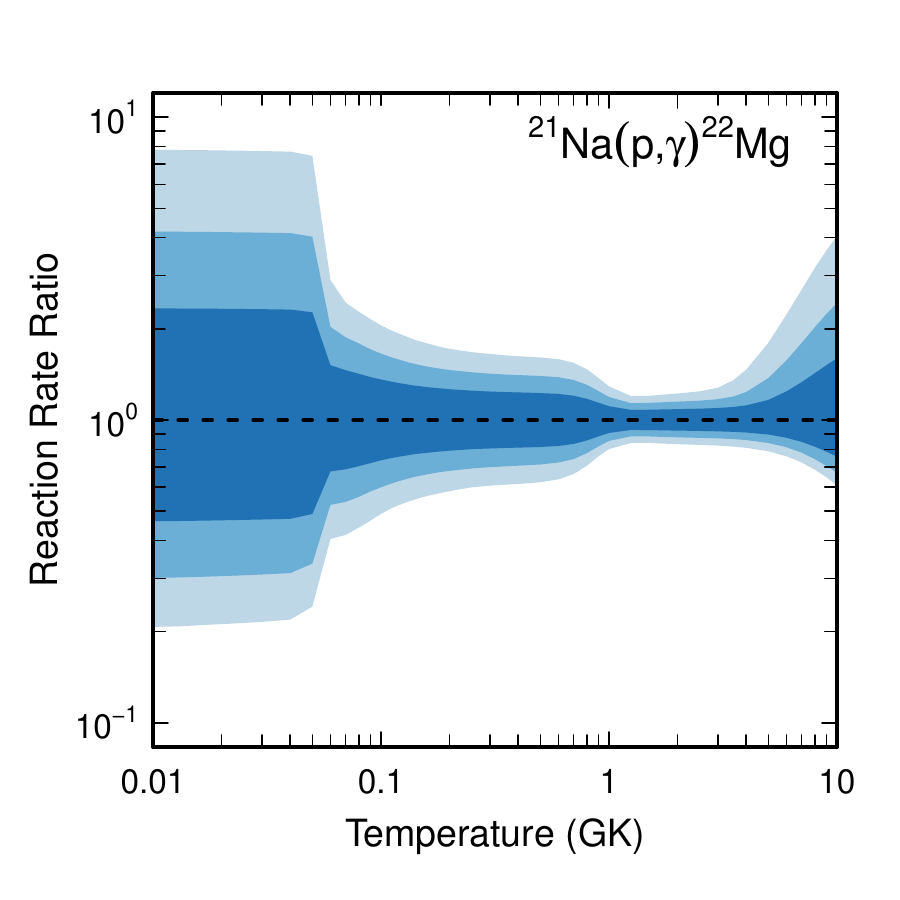}
\caption{
Reaction rate uncertainties versus temperature. The three different shades refer to coverage probabilities of 68\%, 90\%, and 98\%. 
}
\label{fig:na21pg2}
\end{figure*}

\clearpage

\startlongtable


\clearpage

\begin{figure*}[hbt!]
\centering
\includegraphics[width=0.5\linewidth]{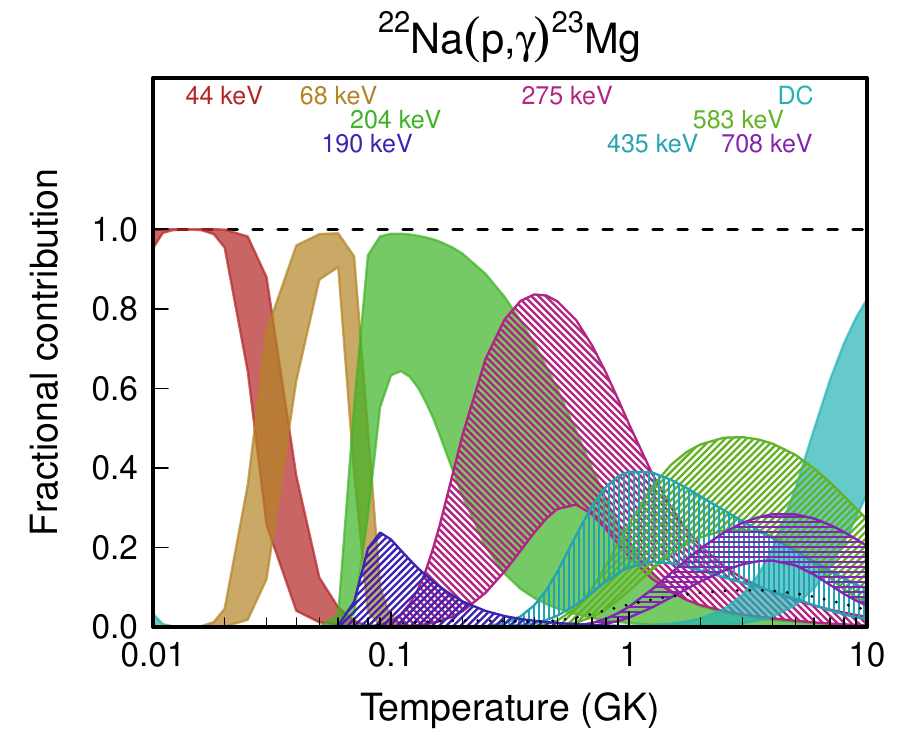}
\caption{
Fractional contributions to the total rate. ``DC'' refers to direct radiative capture. Resonance energies are given in the center-of-mass frame. 
}
\label{fig:na22pg1}
\end{figure*}
\begin{figure*}[hbt!]
\centering
\includegraphics[width=0.5\linewidth]{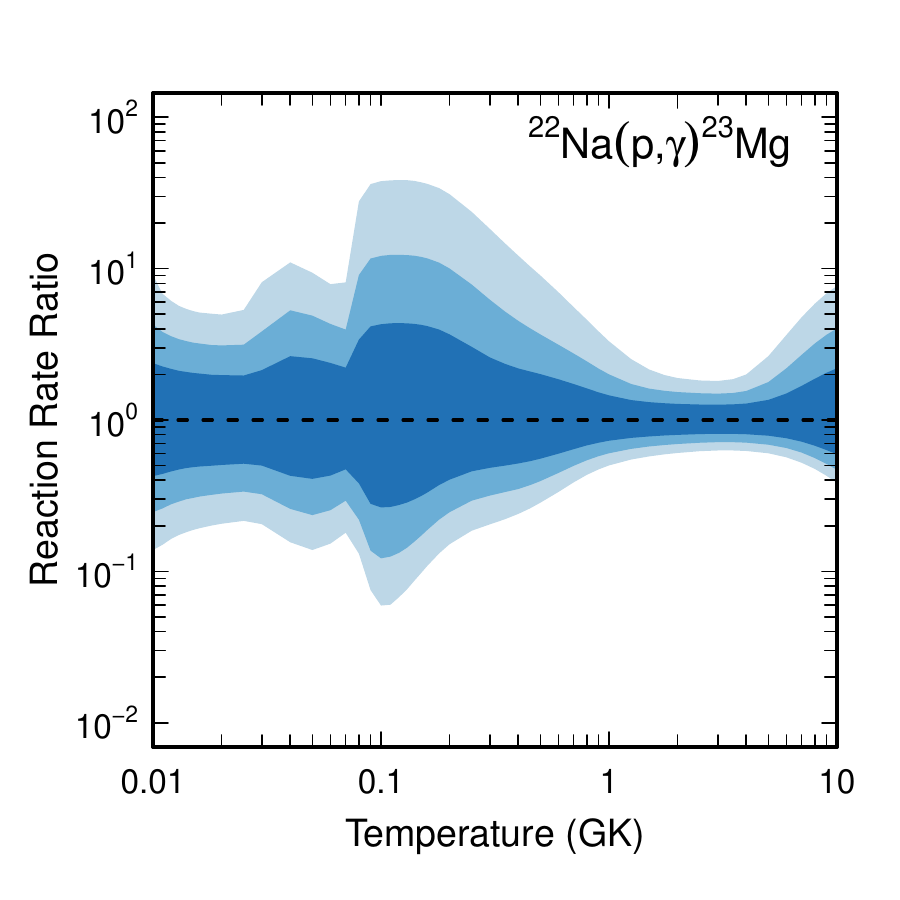}
\caption{
Reaction rate uncertainties versus temperature. The three different shades refer to coverage probabilities of 68\%, 90\%, and 98\%. 
}
\label{fig:na22pg2}
\end{figure*}

\clearpage

\startlongtable


\clearpage

\begin{figure*}[hbt!]
\centering
\includegraphics[width=0.5\linewidth]{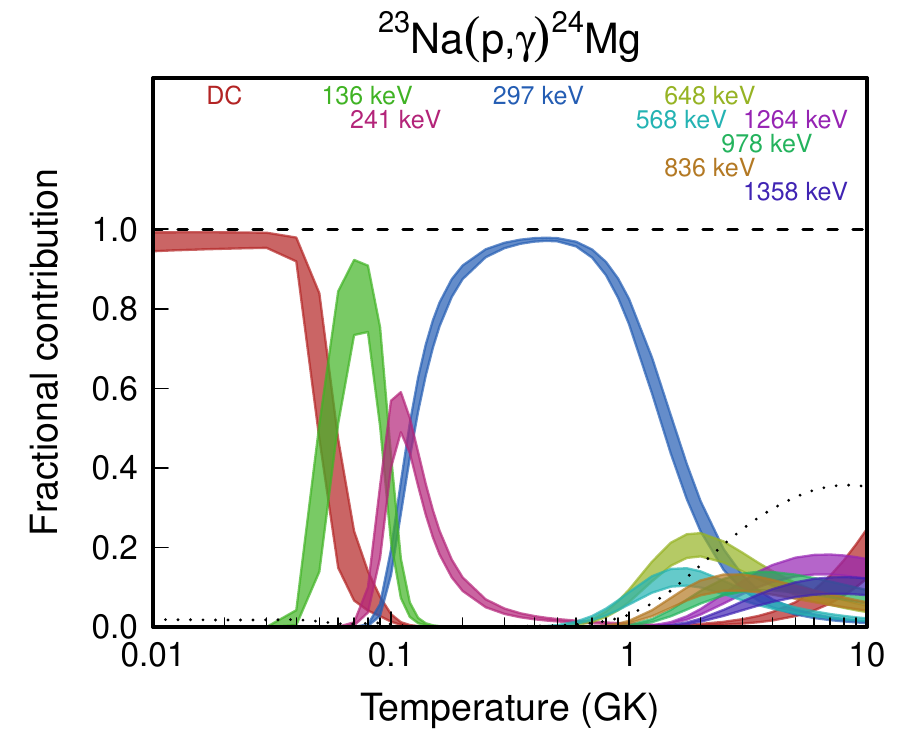}
\caption{
Fractional contributions to the total rate. ``DC'' refers to direct radiative capture. Resonance energies are given in the center-of-mass frame.  
}
\label{fig:na23pg1}
\end{figure*}
\begin{figure*}[hbt!]
\centering
\includegraphics[width=0.5\linewidth]{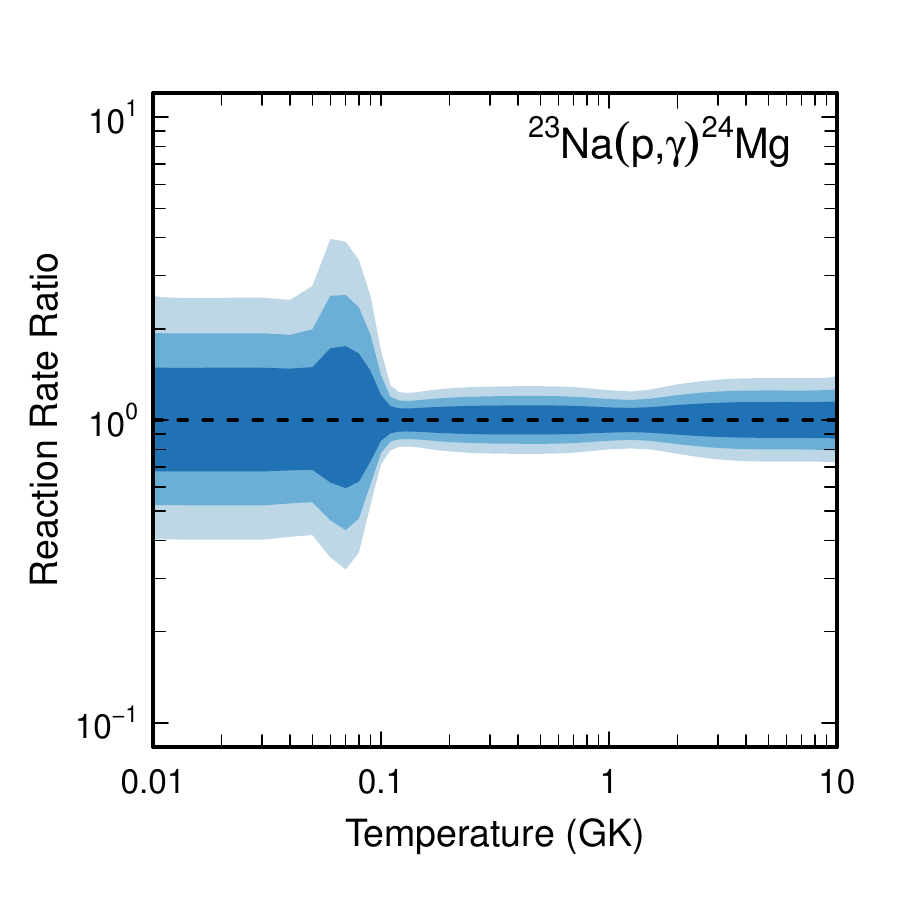}
\caption{
Reaction rate uncertainties versus temperature. The three different shades refer to coverage probabilities of 68\%, 90\%, and 98\%. 
}
\label{fig:na23pg2}
\end{figure*}

\clearpage

\startlongtable


\clearpage

\begin{figure*}[hbt!]
\centering
\includegraphics[width=0.5\linewidth]{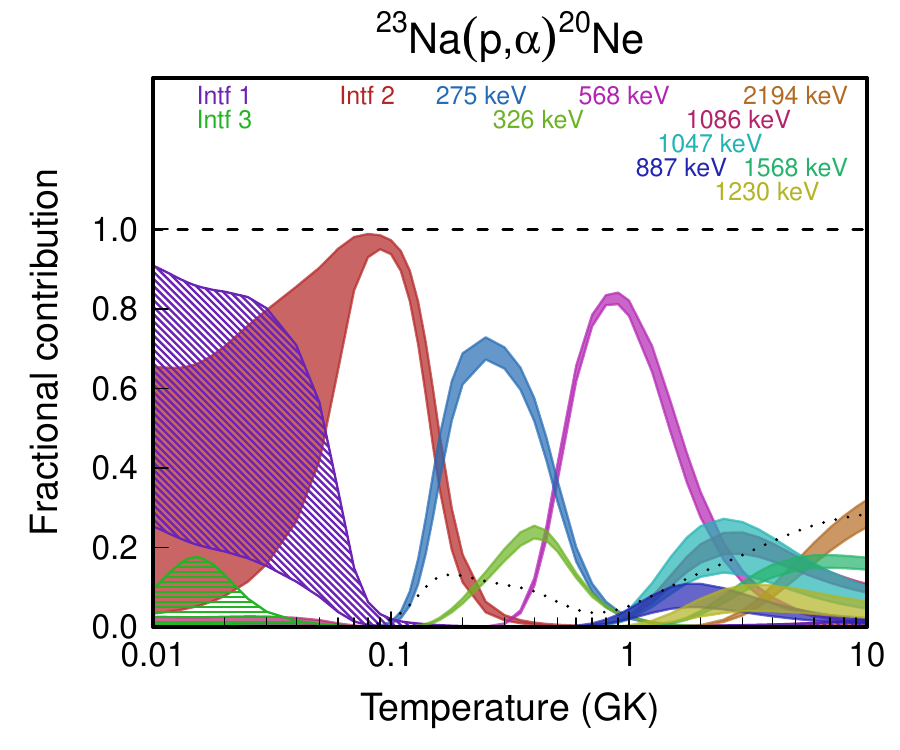}
\caption{
Fractional contributions to the total rate. Resonance energies are given in the center-of-mass frame. ``Inf1'' refers to the combined contribution of the two interfering $1^-$ resonances at $-303$ and $170.2$ keV. ``Inf2" stands for for the two interfering $2^+$ resonances at $-170.4$ and $-239$ keV. ``Inf3" denotes the two interfering $0^+$ resonances at $-235$ and $38$ keV. 
}
\label{fig:na23pa1}
\end{figure*}
\begin{figure*}[hbt!]
\centering
\includegraphics[width=0.5\linewidth]{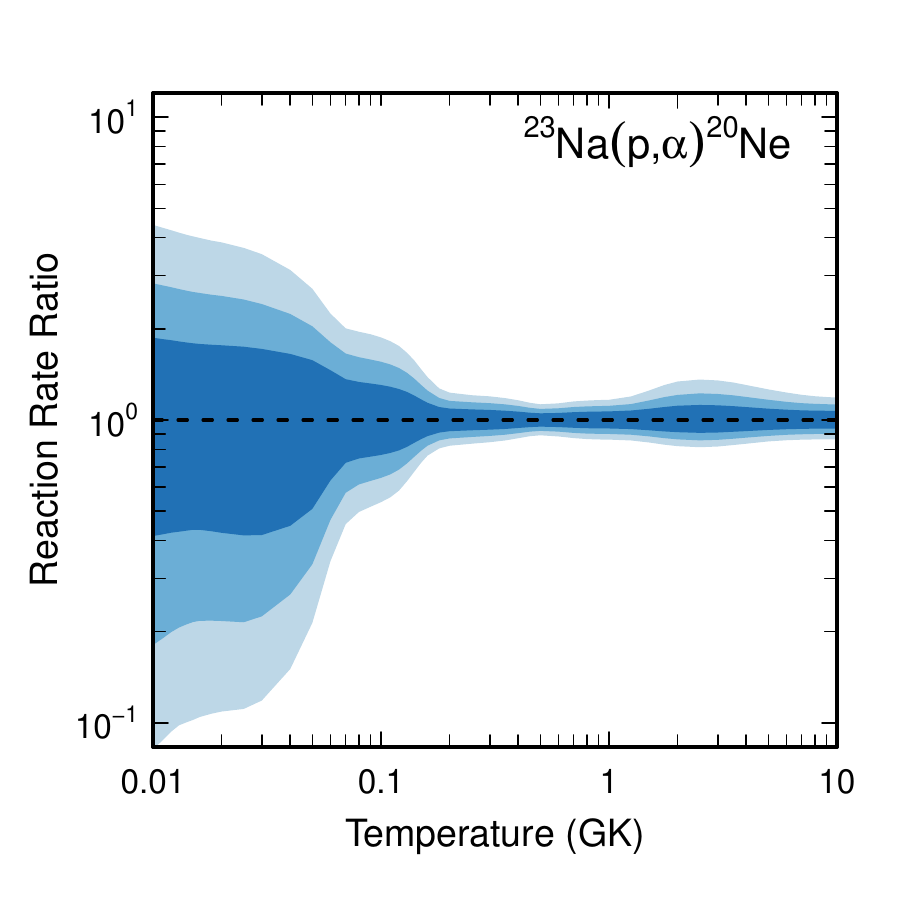}
\caption{
Reaction rate uncertainties versus temperature. The three different shades refer to coverage probabilities of 68\%, 90\%, and 98\%. 
}
\label{fig:na23pa2}
\end{figure*}

\clearpage

\startlongtable


\clearpage

\begin{figure*}[hbt!]
\centering
\includegraphics[width=0.5\linewidth]{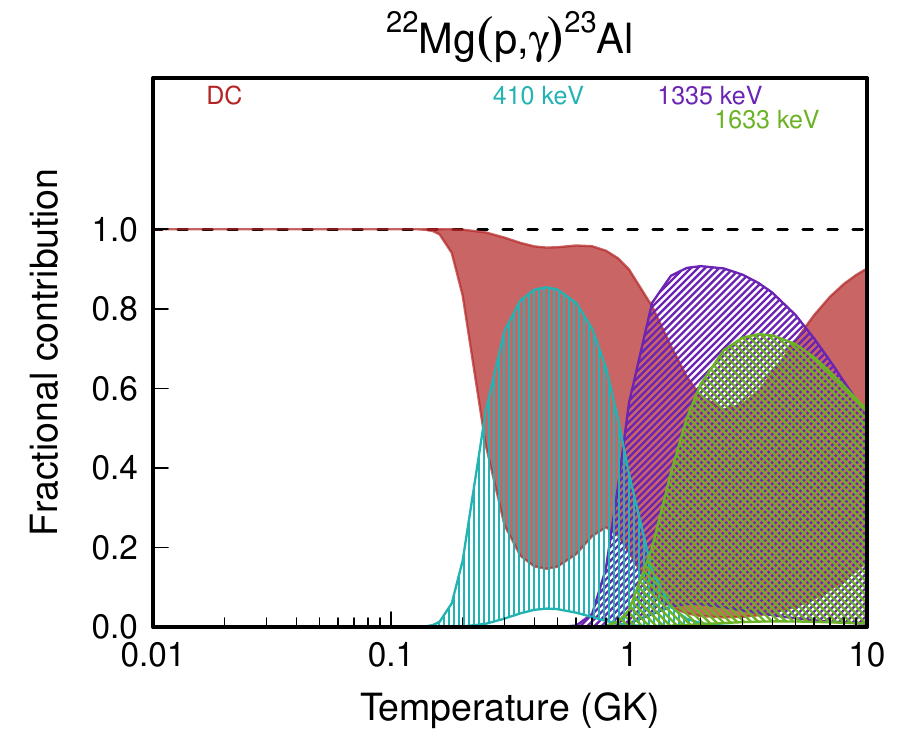}
\caption{
Fractional contributions to the total rate. ``DC'' refers to direct radiative capture. Resonance energies are given in the center-of-mass frame. 
}
\label{fig:mg22pg1}
\end{figure*}
\begin{figure*}[hbt!]
\centering
\includegraphics[width=0.5\linewidth]{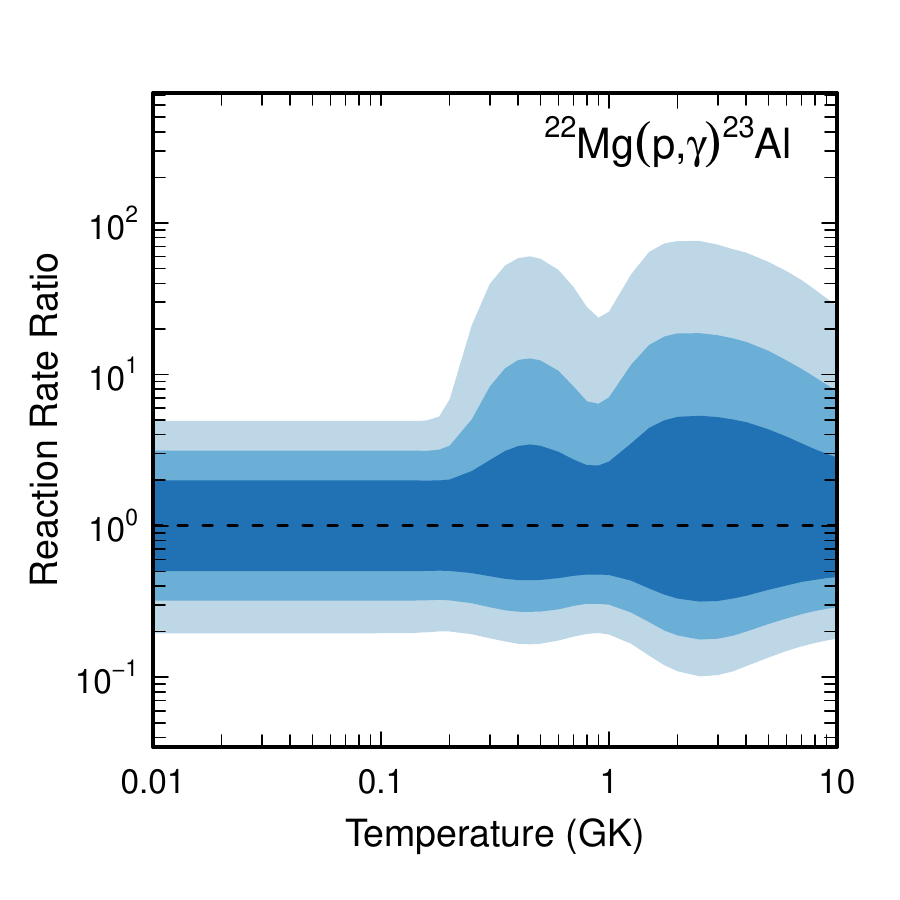}
\caption{
Reaction rate uncertainties versus temperature. The three different shades refer to coverage probabilities of 68\%, 90\%, and 98\%. 
}
\label{fig:mg22pg2}
\end{figure*}

\clearpage

\startlongtable


\clearpage

\begin{figure*}[hbt!]
\centering
\includegraphics[width=0.5\linewidth]{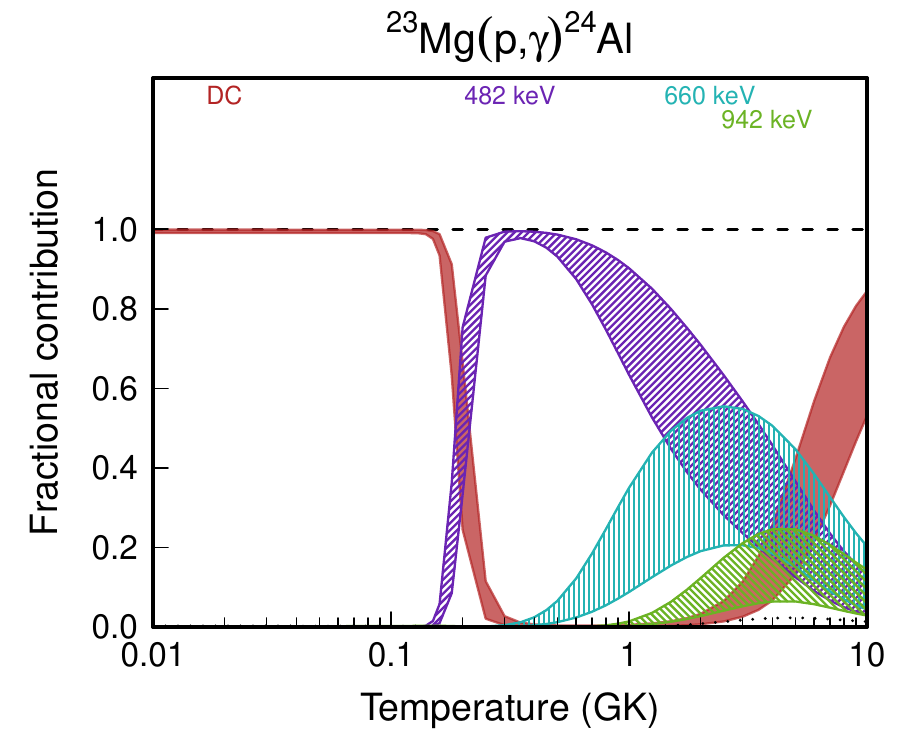}
\caption{
Fractional contributions to the total rate. ``DC'' refers to direct radiative capture. Resonance energies are given in the center-of-mass frame.  
}
\label{fig:mg23pg1}
\end{figure*}
\begin{figure*}[hbt!]
\centering
\includegraphics[width=0.5\linewidth]{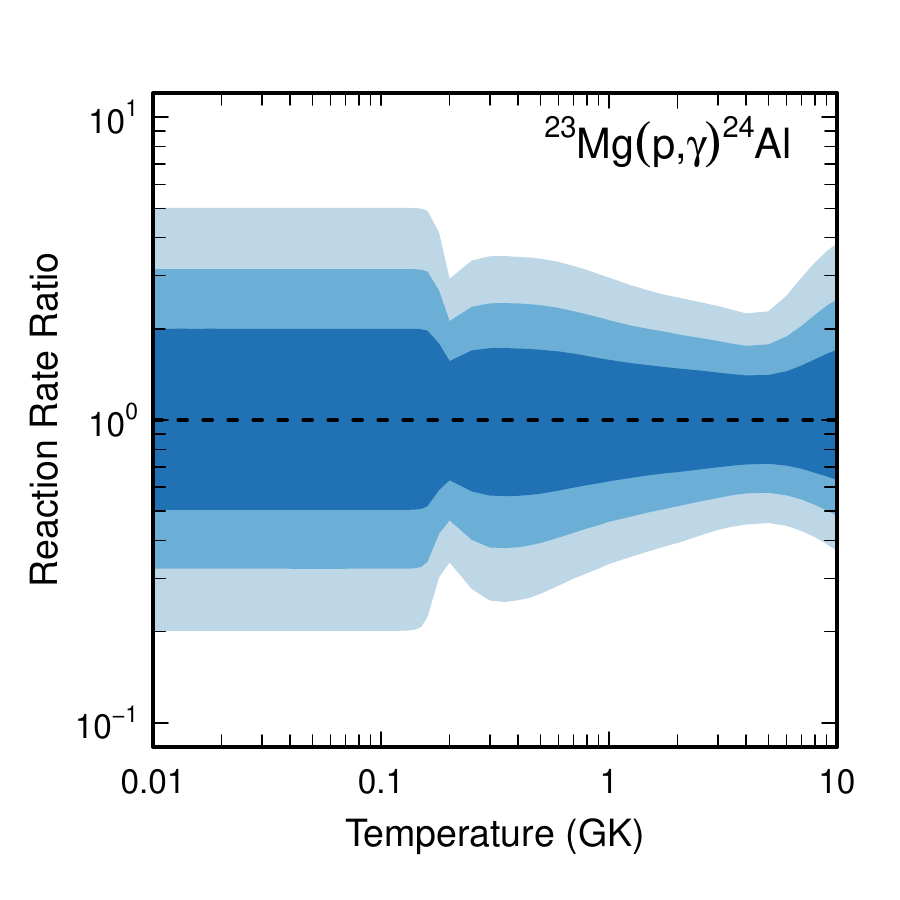}
\caption{
Reaction rate uncertainties versus temperature. The three different shades refer to coverage probabilities of 68\%, 90\%, and 98\%. 
}
\label{fig:mg23pg2}
\end{figure*}

\clearpage

\startlongtable


\clearpage

\begin{figure*}[hbt!]
\centering
\includegraphics[width=0.5\linewidth]{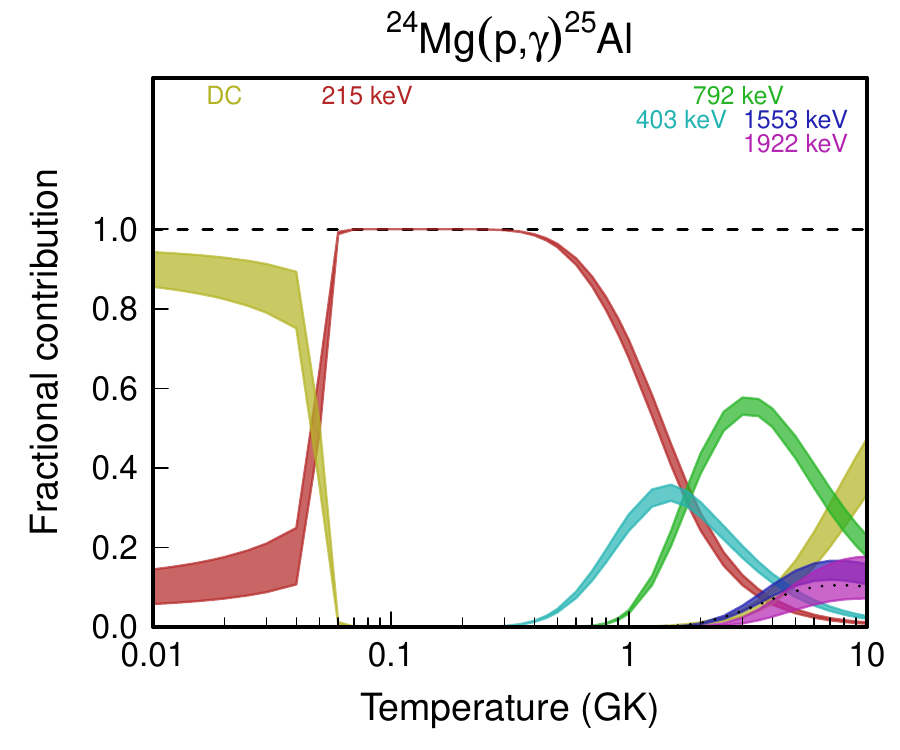}
\caption{
Fractional contributions to the total rate. ``DC'' refers to direct radiative capture. Resonance energies are given in the center-of-mass frame.  
}
\label{fig:mg24pg1}
\end{figure*}
\begin{figure*}[hbt!]
\centering
\includegraphics[width=0.5\linewidth]{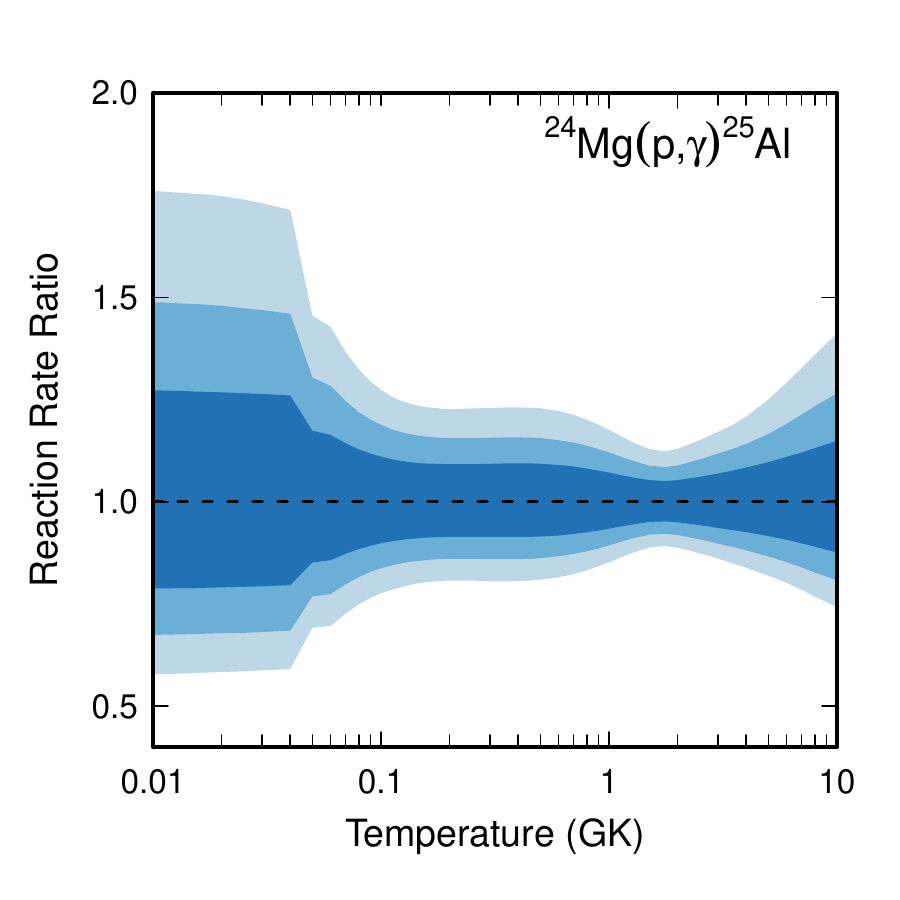}
\caption{
Reaction rate uncertainties versus temperature. The three different shades refer to coverage probabilities of 68\%, 90\%, and 98\%. 
}
\label{fig:mg24pg2}
\end{figure*}

\clearpage

\startlongtable


\clearpage

\begin{figure*}[hbt!]
\centering
\includegraphics[width=0.5\linewidth]{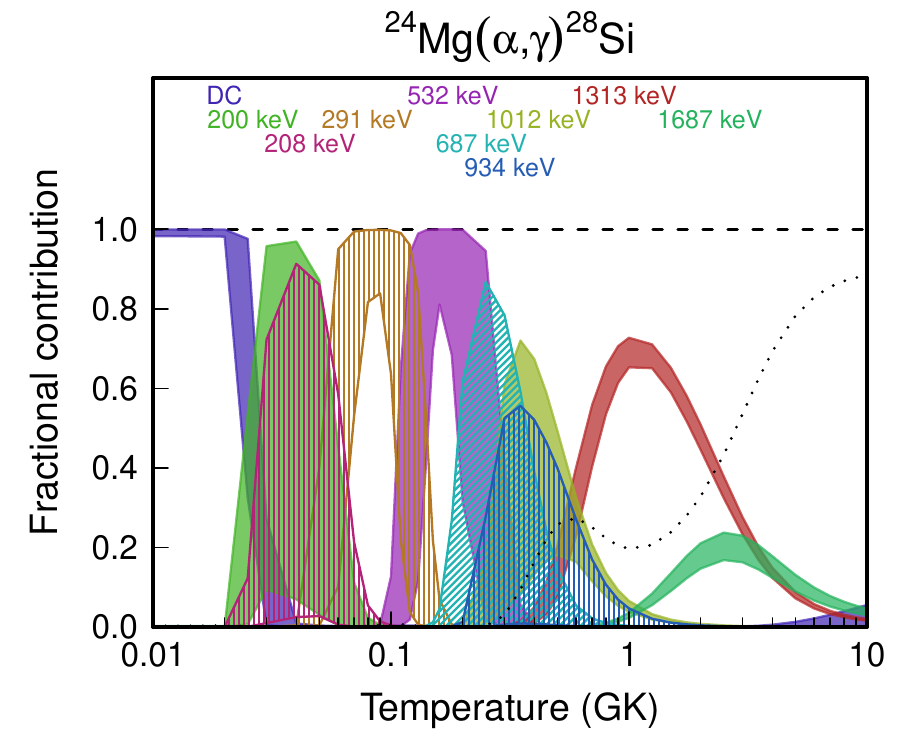}
\caption{
Fractional contributions to the total rate. ``DC'' refers to direct radiative capture. Resonance energies are given in the center-of-mass frame.  
}
\label{fig:mg24ag1}
\end{figure*}
\begin{figure*}[hbt!]
\centering
\includegraphics[width=0.5\linewidth]{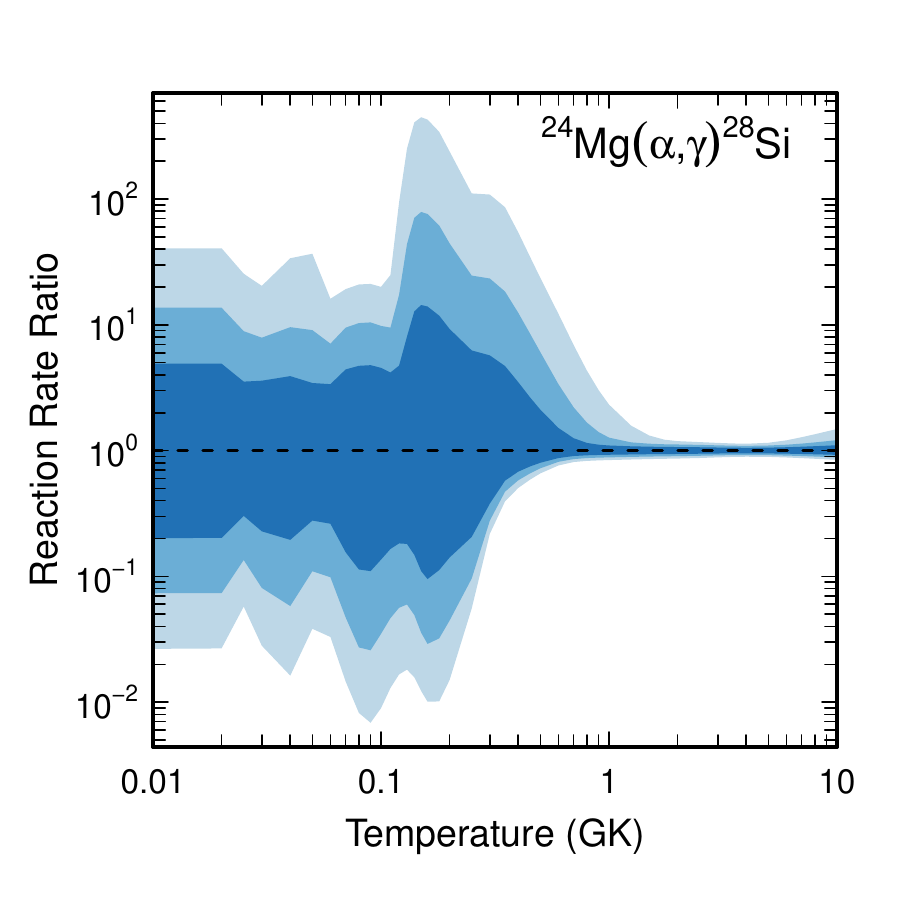}
\caption{
Reaction rate uncertainties versus temperature. The three different shades refer to coverage probabilities of 68\%, 90\%, and 98\%. 
}
\label{fig:mg24ag2}
\end{figure*}

\clearpage

\startlongtable


\clearpage

\begin{figure*}[hbt!]
\centering
\includegraphics[width=0.5\linewidth]{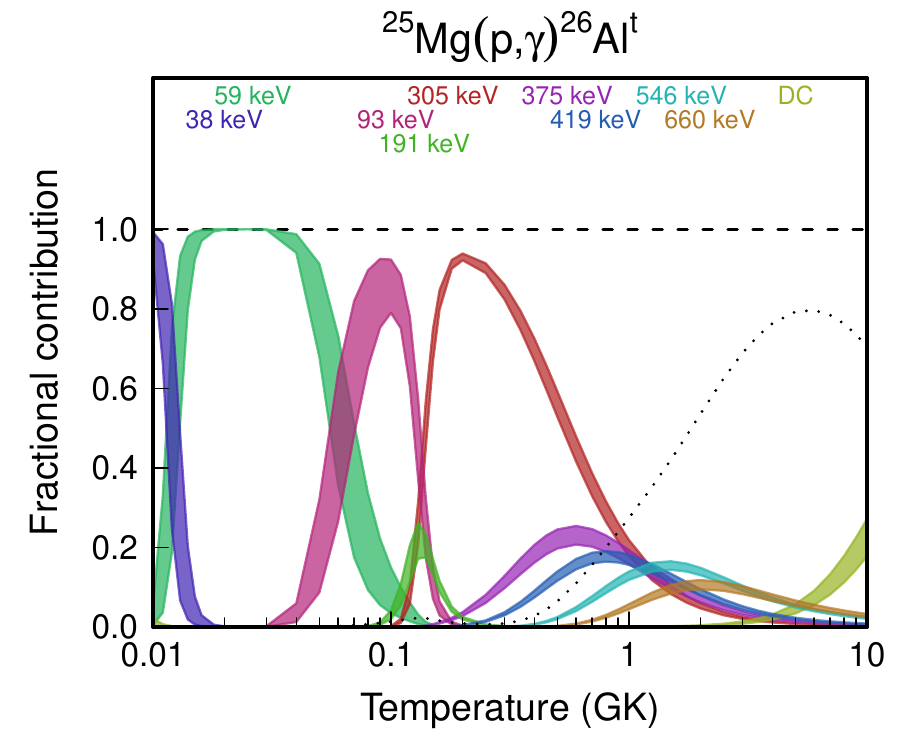}
\caption{
Fractional contributions to the total rate. ``DC'' refers to direct radiative capture. Resonance energies are given in the center-of-mass frame.  
}
\label{fig:mg25pgt1}
\end{figure*}
\begin{figure*}[hbt!]
\centering
\includegraphics[width=0.5\linewidth]{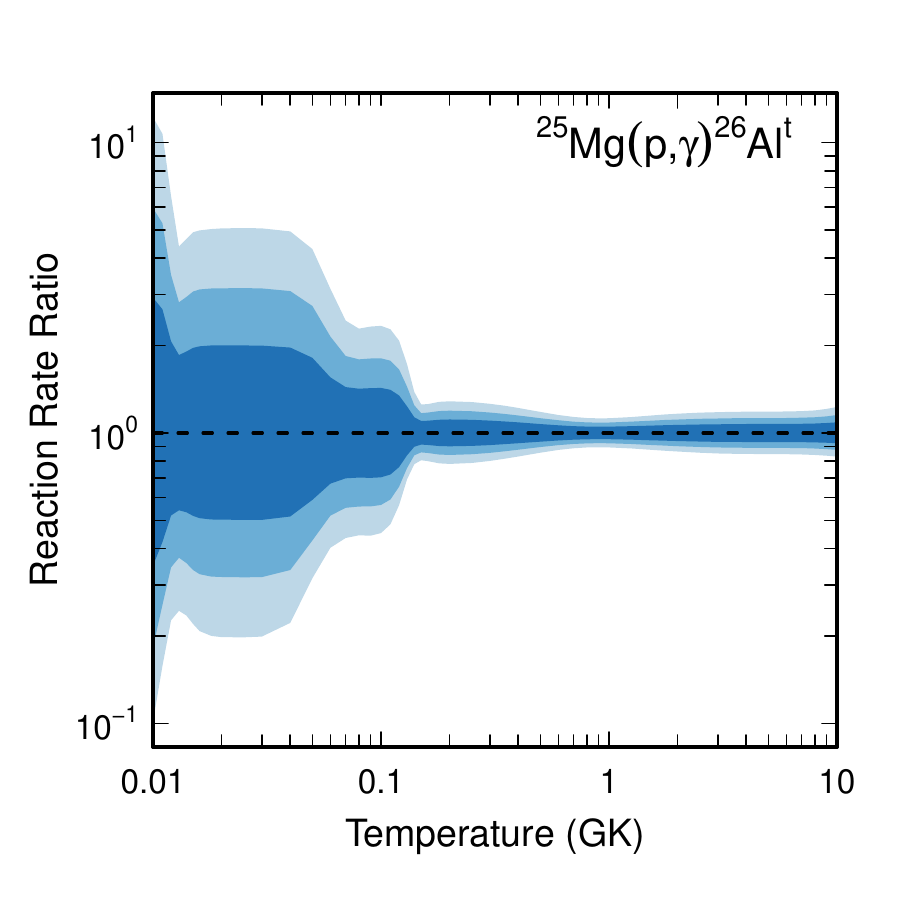}
\caption{
Reaction rate uncertainties versus temperature. The three different shades refer to coverage probabilities of 68\%, 90\%, and 98\%. 
}
\label{fig:mg25pgt2}
\end{figure*}

\clearpage

\startlongtable


\clearpage

\begin{figure*}[hbt!]
\centering
\includegraphics[width=0.5\linewidth]{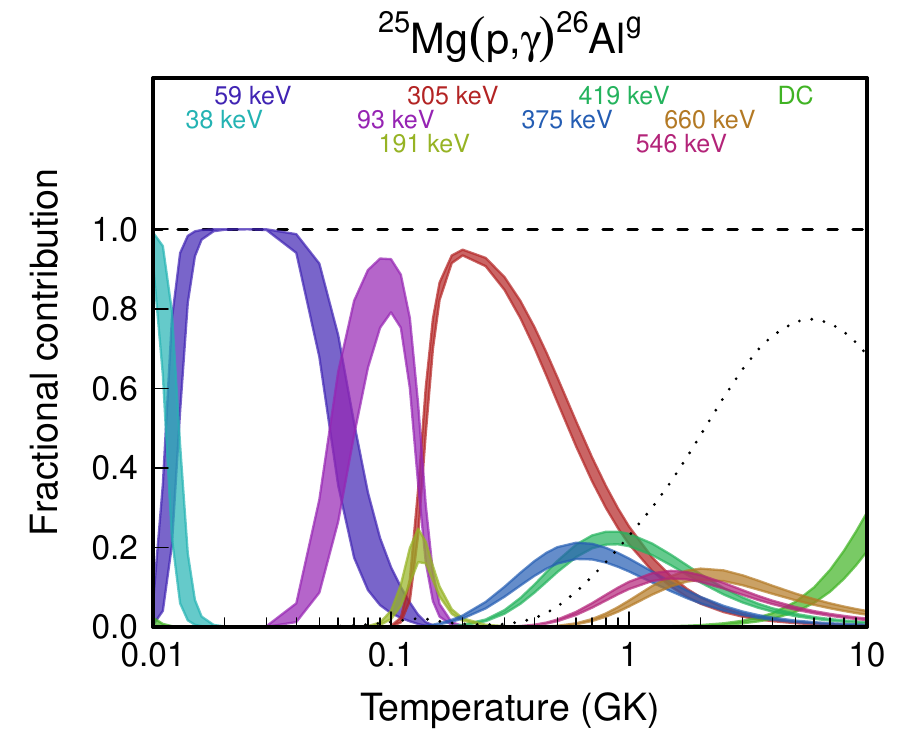}
\caption{
Fractional contributions to the total rate. ``DC'' refers to direct radiative capture. Resonance energies are given in the center-of-mass frame.  
}
\label{fig:mg25pgg1}
\end{figure*}
\begin{figure*}[hbt!]
\centering
\includegraphics[width=0.5\linewidth]{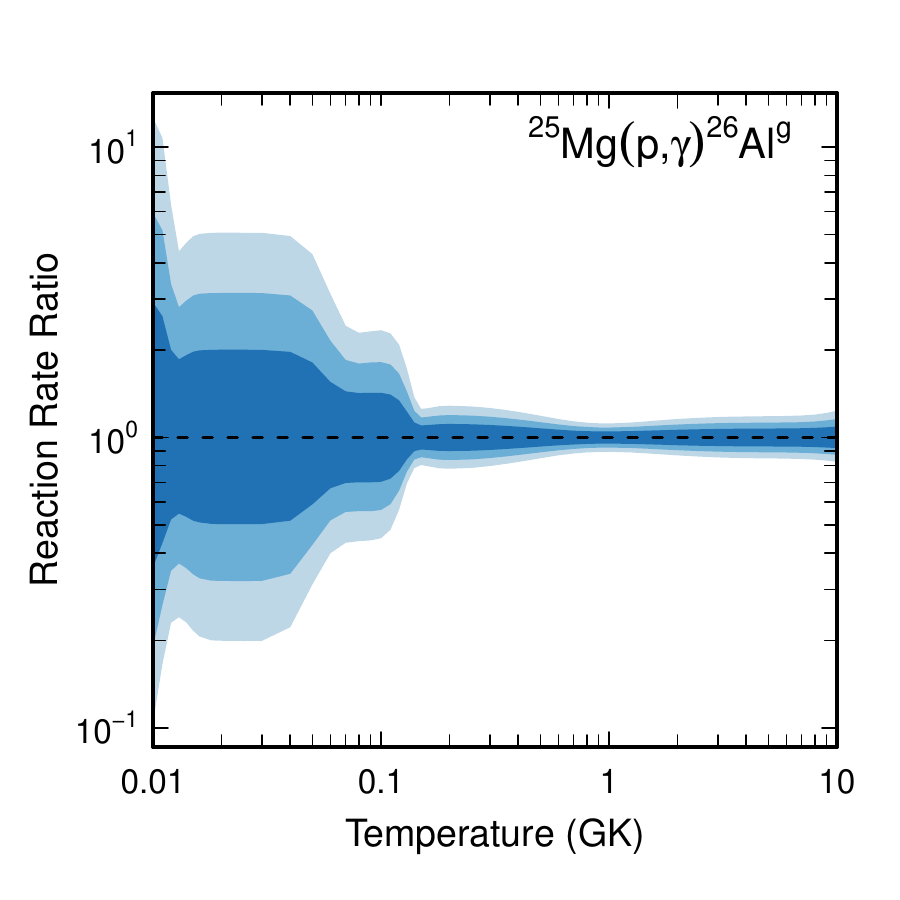}
\caption{
Reaction rate uncertainties versus temperature. The three different shades refer to coverage probabilities of 68\%, 90\%, and 98\%. 
}
\label{fig:mg25pgg2}
\end{figure*}

\clearpage

\startlongtable


\clearpage

\begin{figure*}[hbt!]
\centering
\includegraphics[width=0.5\linewidth]{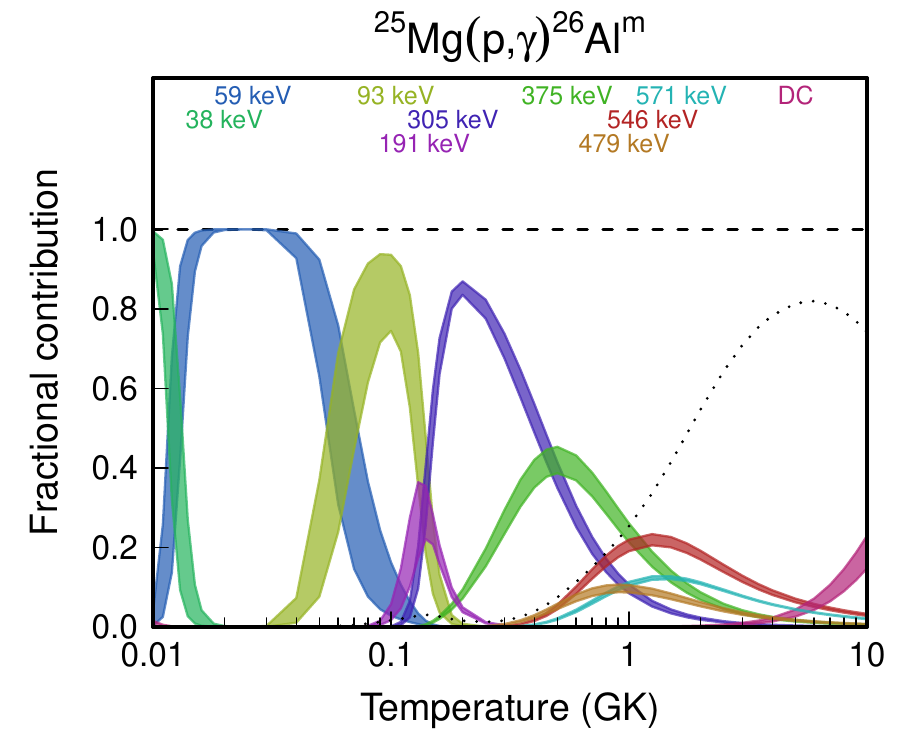}
\caption{
Fractional contributions to the total rate. ``DC'' refers to direct radiative capture. Resonance energies are given in the center-of-mass frame.  
}
\label{fig:mg25pgm1}
\end{figure*}
\begin{figure*}[hbt!]
\centering
\includegraphics[width=0.5\linewidth]{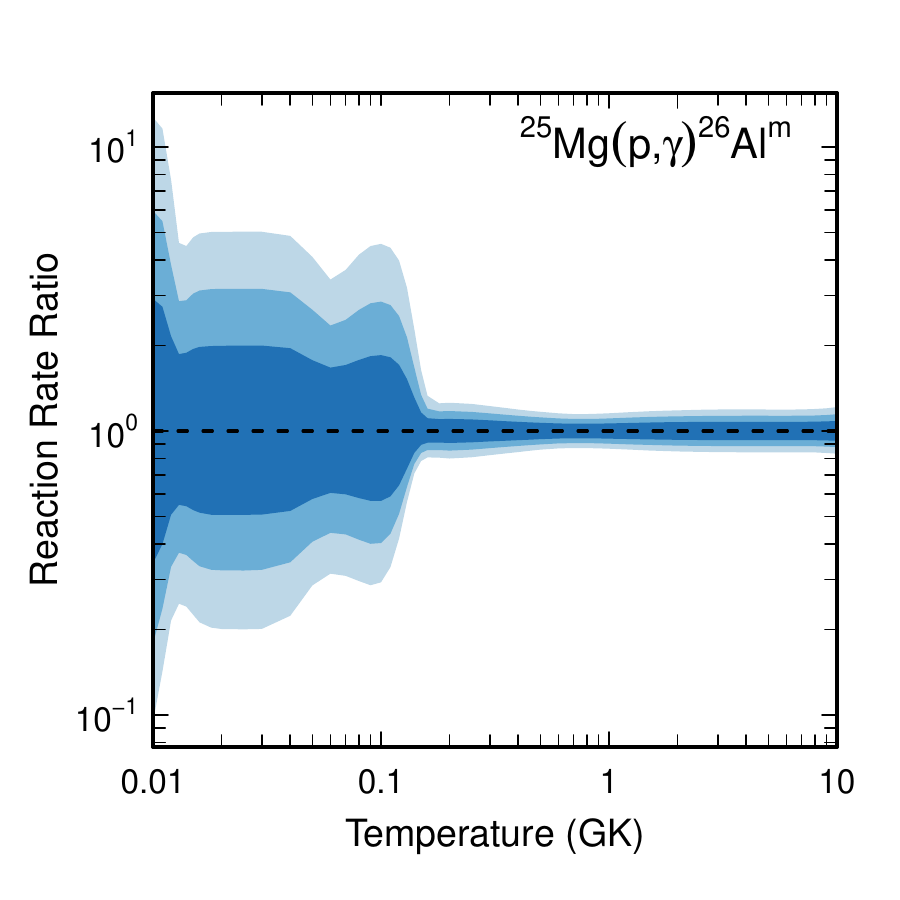}
\caption{
Reaction rate uncertainties versus temperature. The three different shades refer to coverage probabilities of 68\%, 90\%, and 98\%. 
}
\label{fig:mg25pgm2}
\end{figure*}

\clearpage

\startlongtable


\clearpage

\begin{figure*}[hbt!]
\centering
\includegraphics[width=0.5\linewidth]{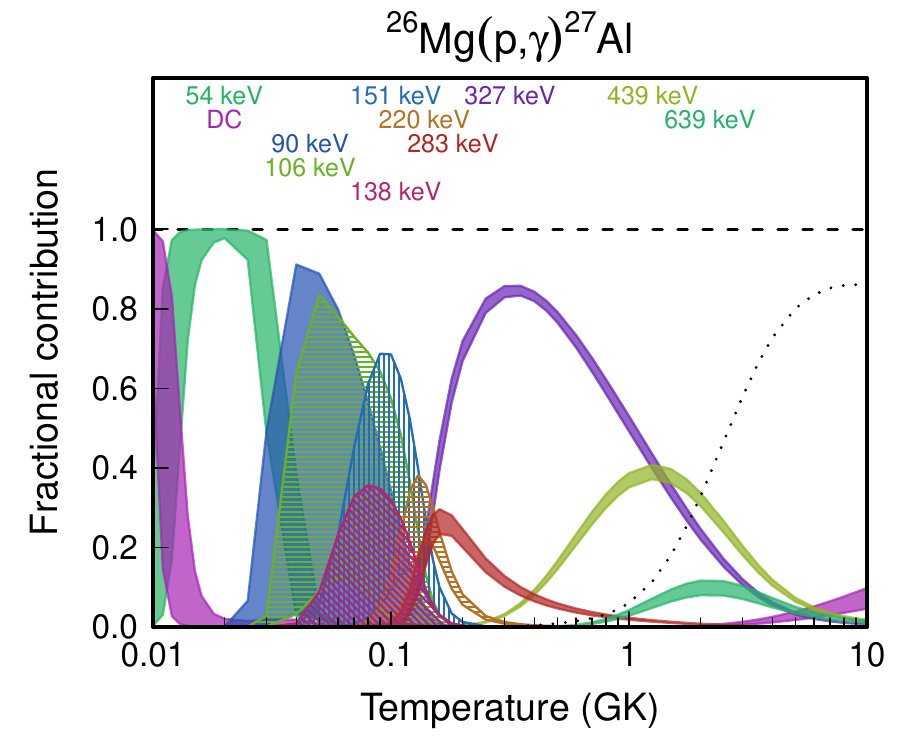}
\caption{
Fractional contributions to the total rate. ``DC'' refers to direct radiative capture. Resonance energies are given in the center-of-mass frame.  
}
\label{fig:mg26pg1}
\end{figure*}
\begin{figure*}[hbt!]
\centering
\includegraphics[width=0.5\linewidth]{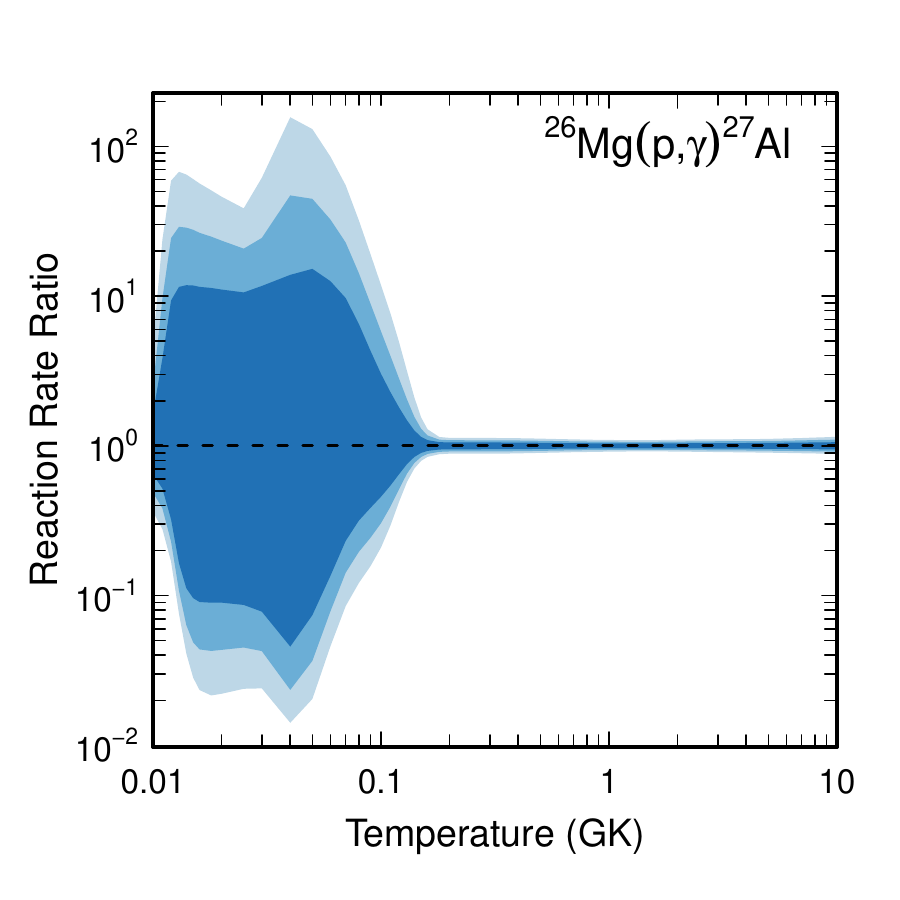}
\caption{
Reaction rate uncertainties versus temperature. The three different shades refer to coverage probabilities of 68\%, 90\%, and 98\%. 
}
\label{fig:mg26pg2}
\end{figure*}

\clearpage

\startlongtable


\clearpage

\begin{figure*}[hbt!]
\centering
\includegraphics[width=0.5\linewidth]{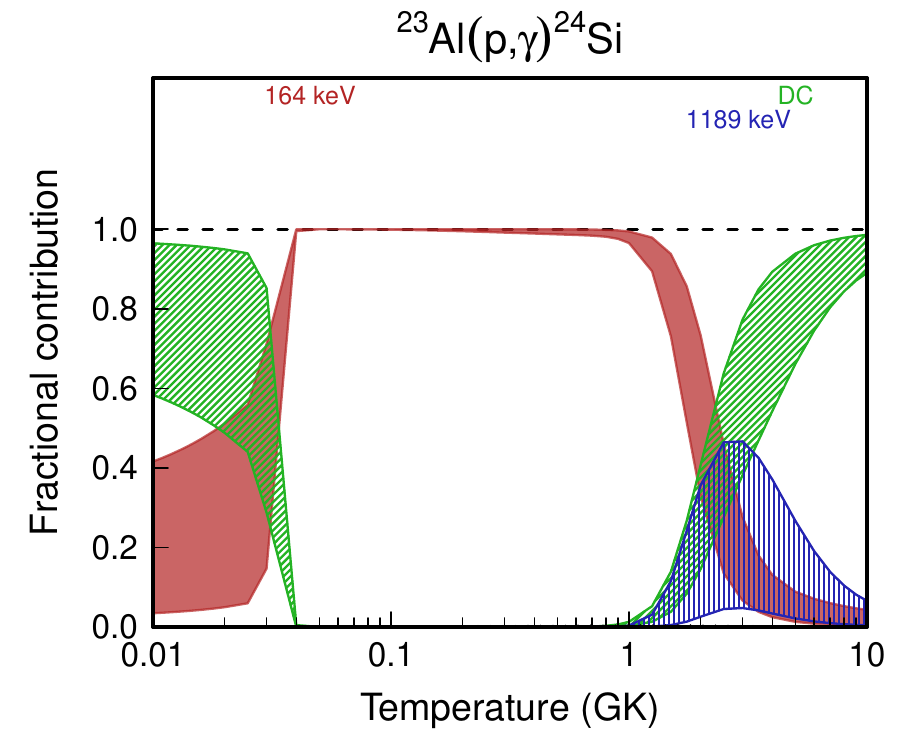}
\caption{
Fractional contributions to the total rate. ``DC'' refers to direct radiative capture. Resonance energies are given in the center-of-mass frame.  
}
\label{fig:al23pg1}
\end{figure*}
\begin{figure*}[hbt!]
\centering
\includegraphics[width=0.5\linewidth]{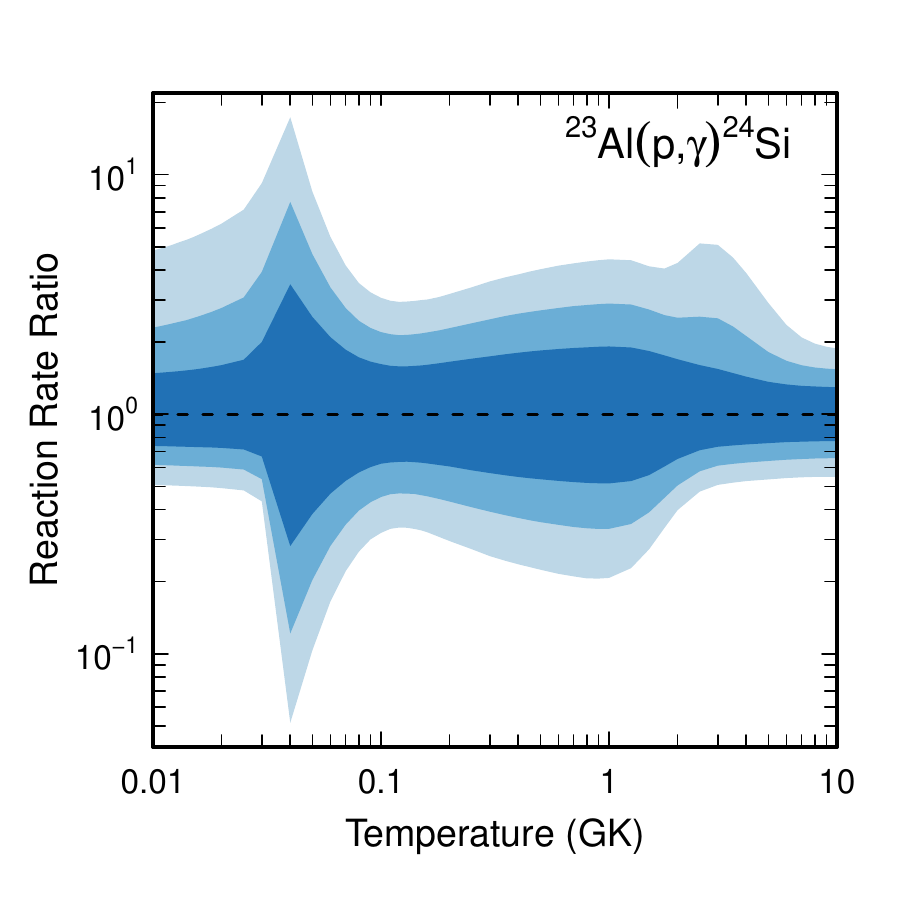}
\caption{
Reaction rate uncertainties versus temperature. The three different shades refer to coverage probabilities of 68\%, 90\%, and 98\%. 
}
\label{fig:al23pg2}
\end{figure*}

\clearpage

\startlongtable


\clearpage

\begin{figure*}[hbt!]
\centering
\includegraphics[width=0.5\linewidth]{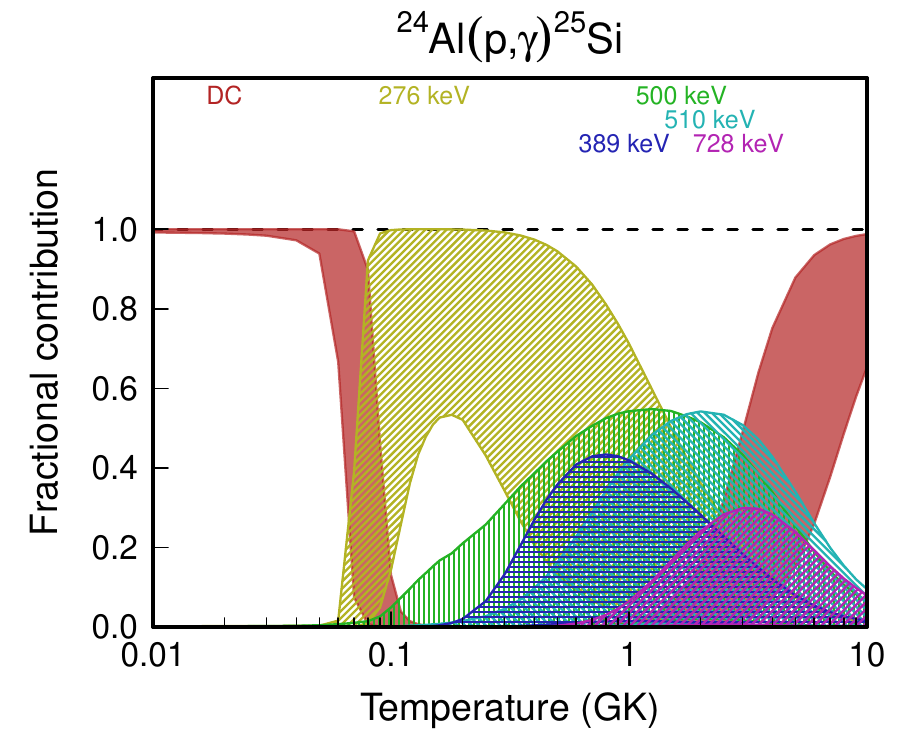}
\caption{
Fractional contributions to the total rate. ``DC'' refers to direct radiative capture. Resonance energies are given in the center-of-mass frame.  
}
\label{fig:al24pg1}
\end{figure*}
\begin{figure*}[hbt!]
\centering
\includegraphics[width=0.5\linewidth]{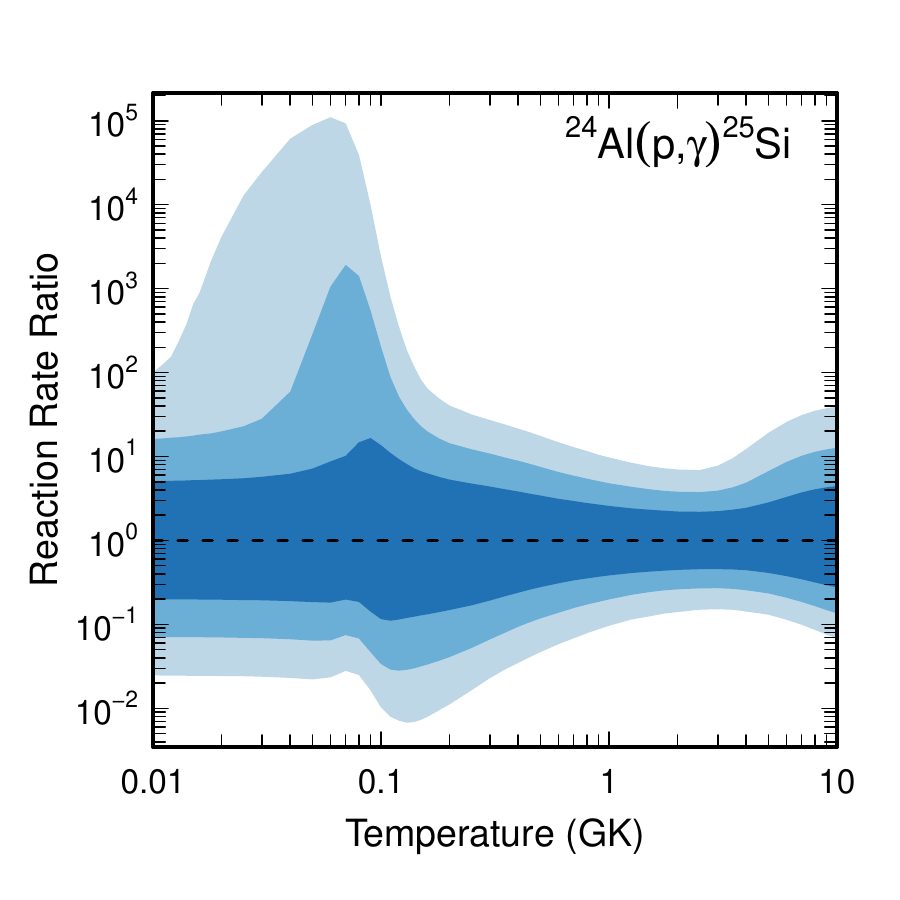}
\caption{
Reaction rate uncertainties versus temperature. The three different shades refer to coverage probabilities of 68\%, 90\%, and 98\%. 
}
\label{fig:al24pg2}
\end{figure*}

\clearpage

\startlongtable


\clearpage

\begin{figure*}[hbt!]
\centering
\includegraphics[width=0.5\linewidth]{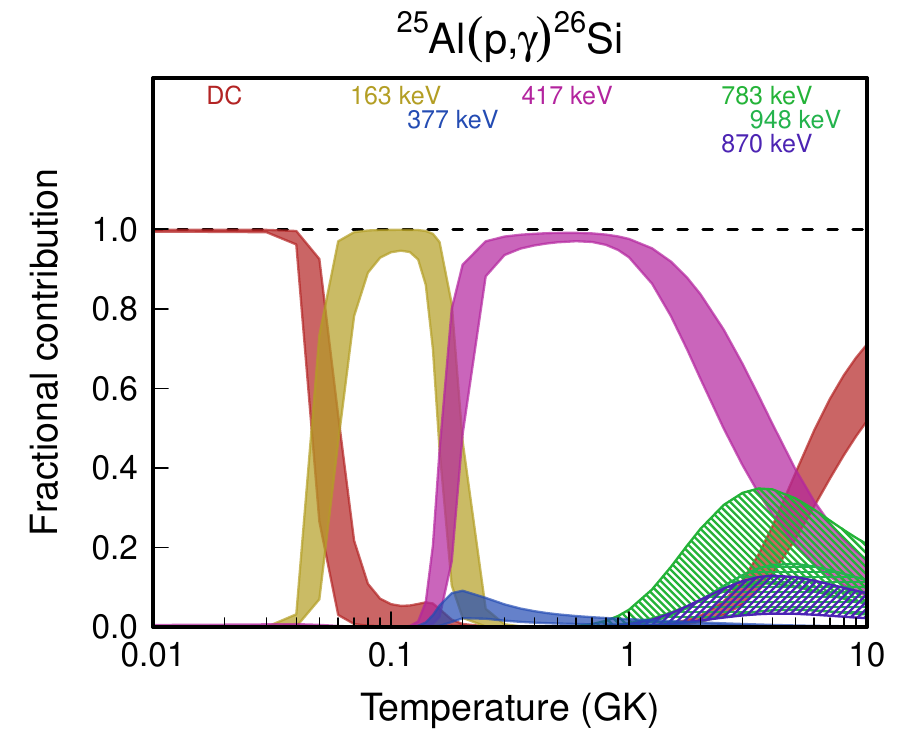}
\caption{
Fractional contributions to the total rate. ``DC'' refers to direct radiative capture. Resonance energies are given in the center-of-mass frame. 
}
\label{fig:al25pg1}
\end{figure*}
\begin{figure*}[hbt!]
\centering
\includegraphics[width=0.5\linewidth]{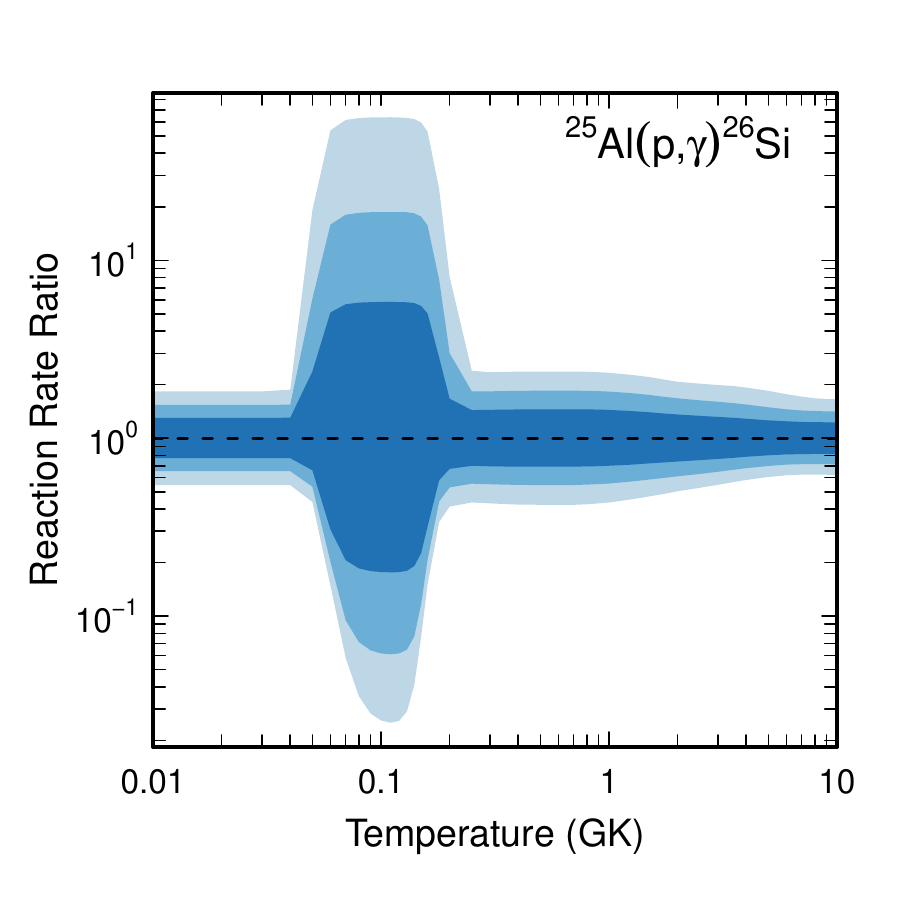}
\caption{
Reaction rate uncertainties versus temperature. The three different shades refer to coverage probabilities of 68\%, 90\%, and 98\%. 
}
\label{fig:al25pg2}
\end{figure*}

\clearpage

\startlongtable


\clearpage

\begin{figure*}[hbt!]
\centering
\includegraphics[width=0.5\linewidth]{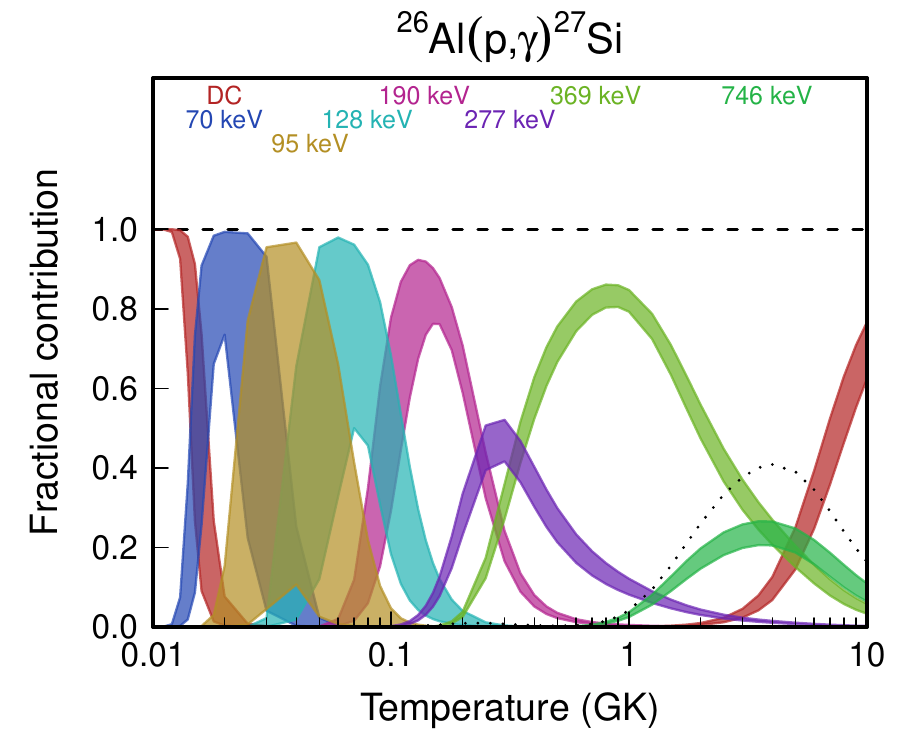}
\caption{
Fractional contributions to the total rate. ``DC'' refers to direct radiative capture. Resonance energies are given in the center-of-mass frame.  
}
\label{fig:al26pg1}
\end{figure*}
\begin{figure*}[hbt!]
\centering
\includegraphics[width=0.5\linewidth]{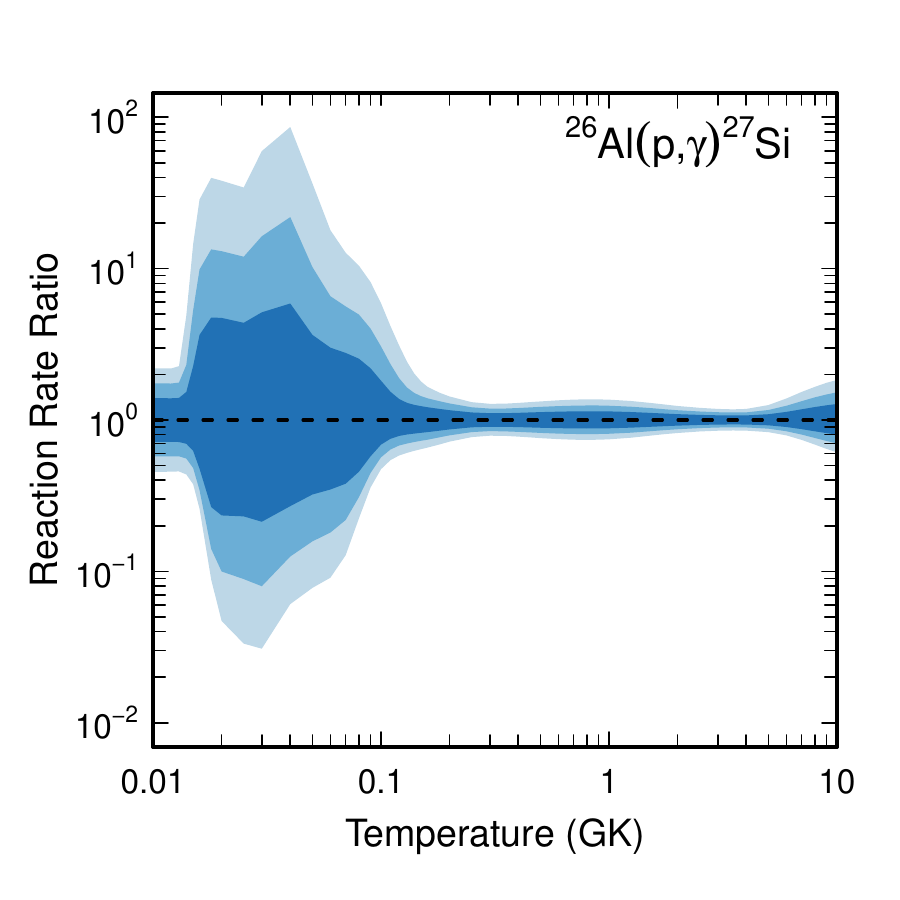}
\caption{
Reaction rate uncertainties versus temperature. The three different shades refer to coverage probabilities of 68\%, 90\%, and 98\%. 
}
\label{fig:al26pg2}
\end{figure*}

\clearpage

\startlongtable


\clearpage

\begin{figure*}[hbt!]
\centering
\includegraphics[width=0.5\linewidth]{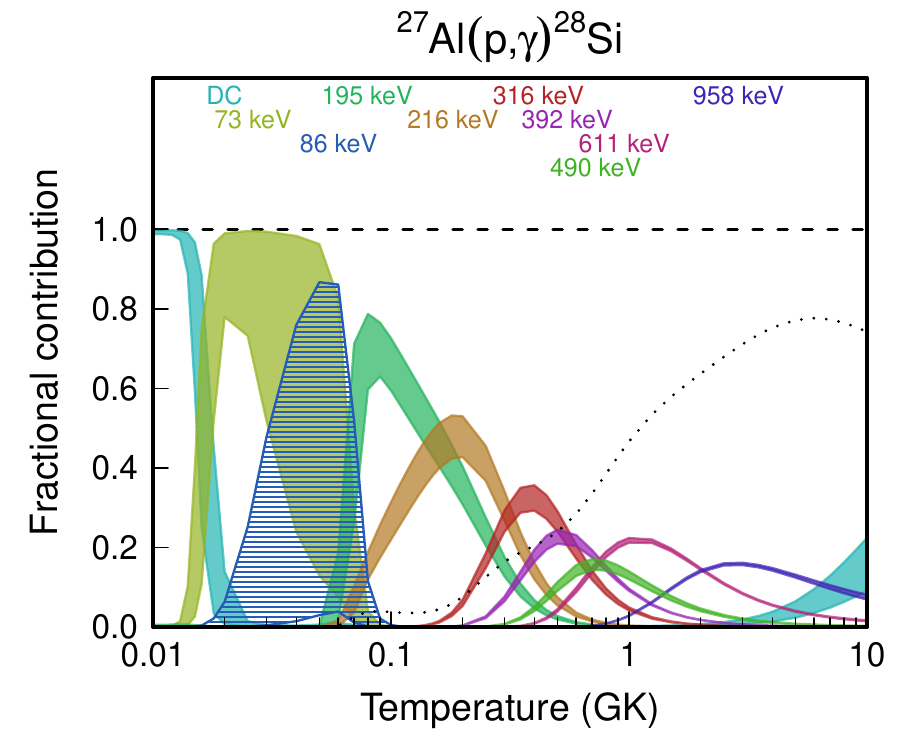}
\caption{
Fractional contributions to the total rate. ``DC'' refers to direct radiative capture. Resonance energies are given in the center-of-mass frame.  
}
\label{fig:al27pg1}
\end{figure*}
\begin{figure*}[hbt!]
\centering
\includegraphics[width=0.5\linewidth]{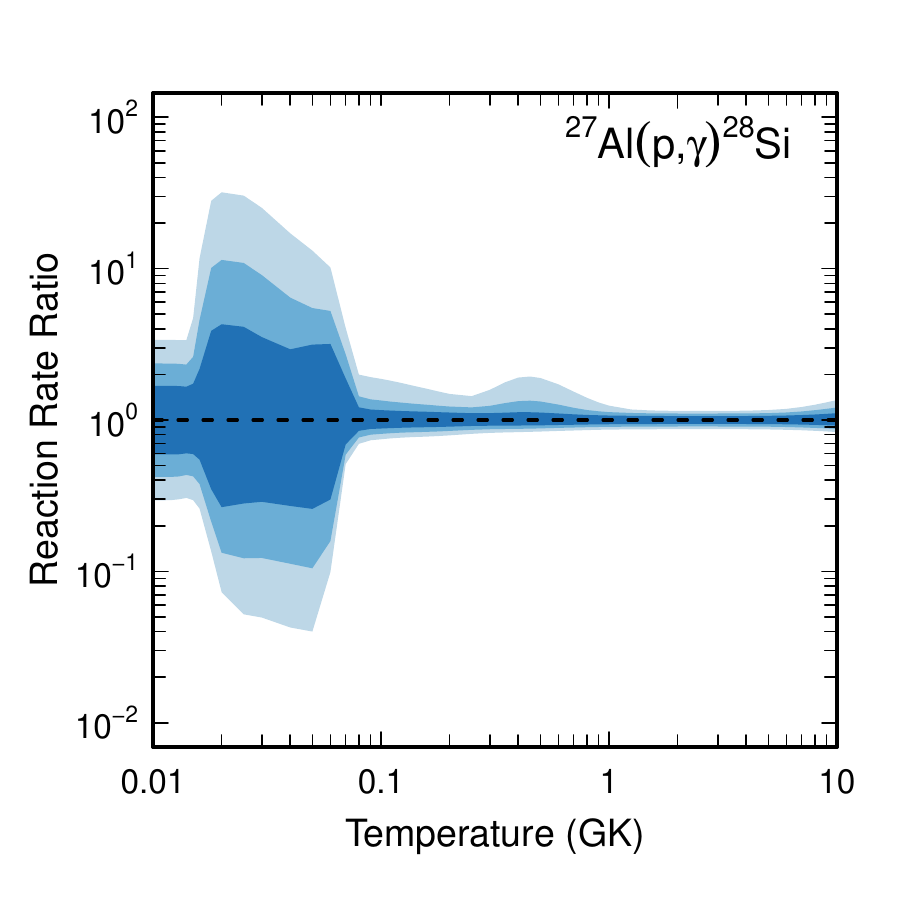}
\caption{
Reaction rate uncertainties versus temperature. The three different shades refer to coverage probabilities of 68\%, 90\%, and 98\%.}
\label{fig:al27pg2}
\end{figure*}

\clearpage

\startlongtable


\clearpage

\begin{figure*}[hbt!]
\centering
\includegraphics[width=0.5\linewidth]{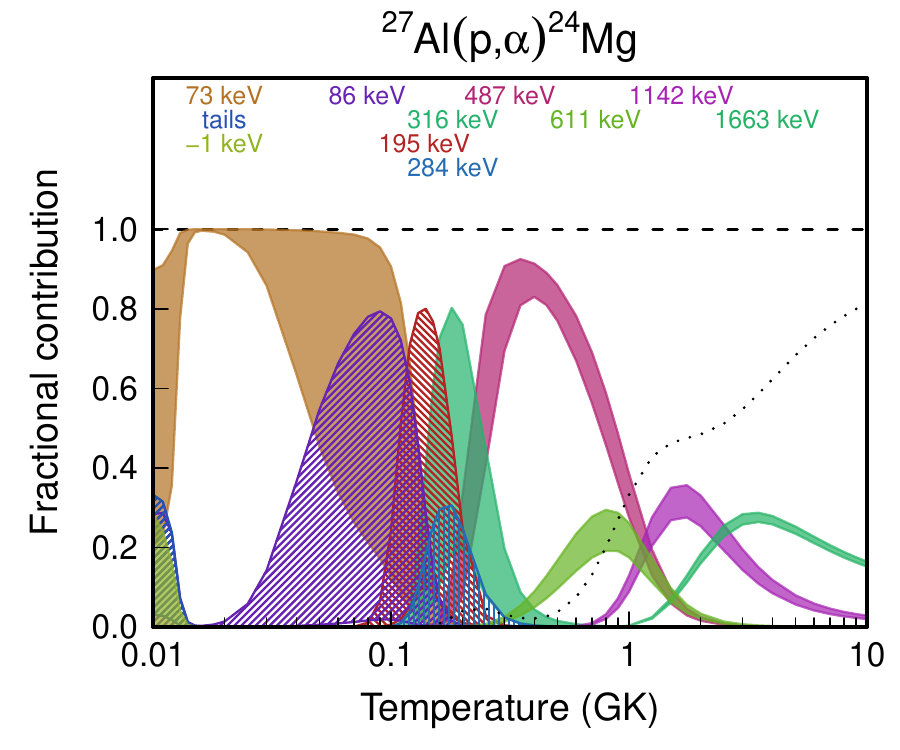}
\caption{
Fractional contributions to the total rate. ``tails'' refers to low-energy tails of broad resonances. Resonance energies are given in the center-of-mass frame.  
}
\label{fig:al27pa1}
\end{figure*}
\begin{figure*}[hbt!]
\centering
\includegraphics[width=0.5\linewidth]{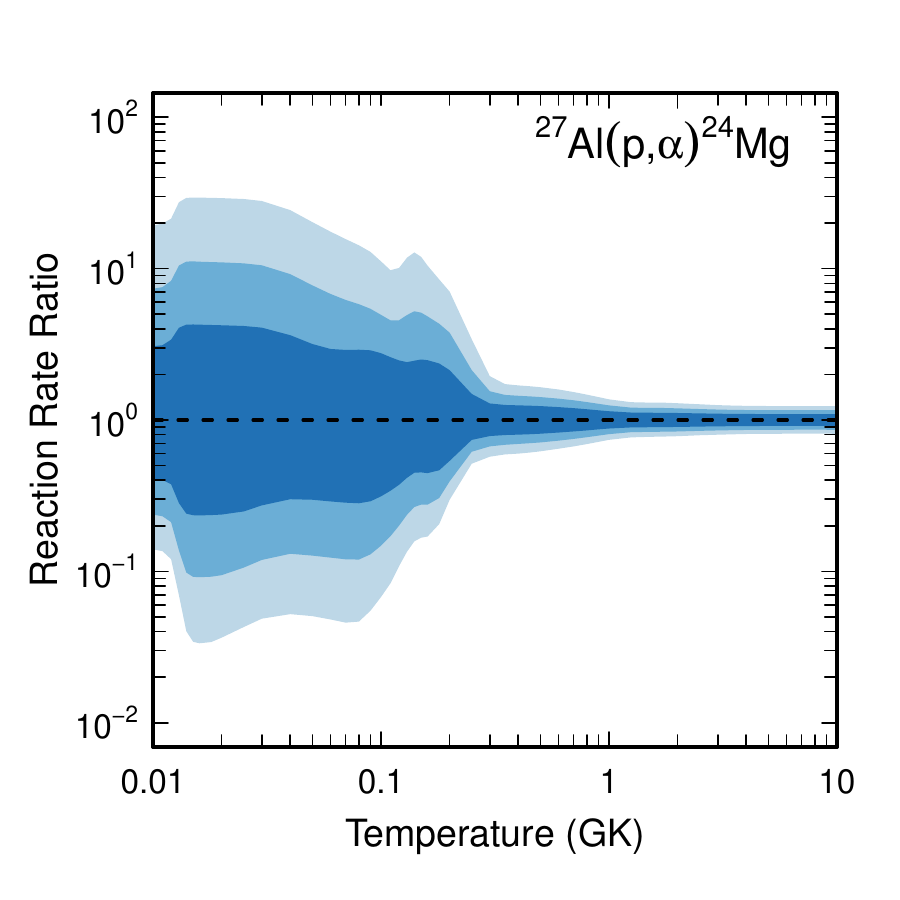}
\caption{
Reaction rate uncertainties versus temperature. The three different shades refer to coverage probabilities of 68\%, 90\%, and 98\%. 
}
\label{fig:al27pa2}
\end{figure*}

\clearpage

\startlongtable


\clearpage

\begin{figure*}[hbt!]
\centering
\includegraphics[width=0.5\linewidth]{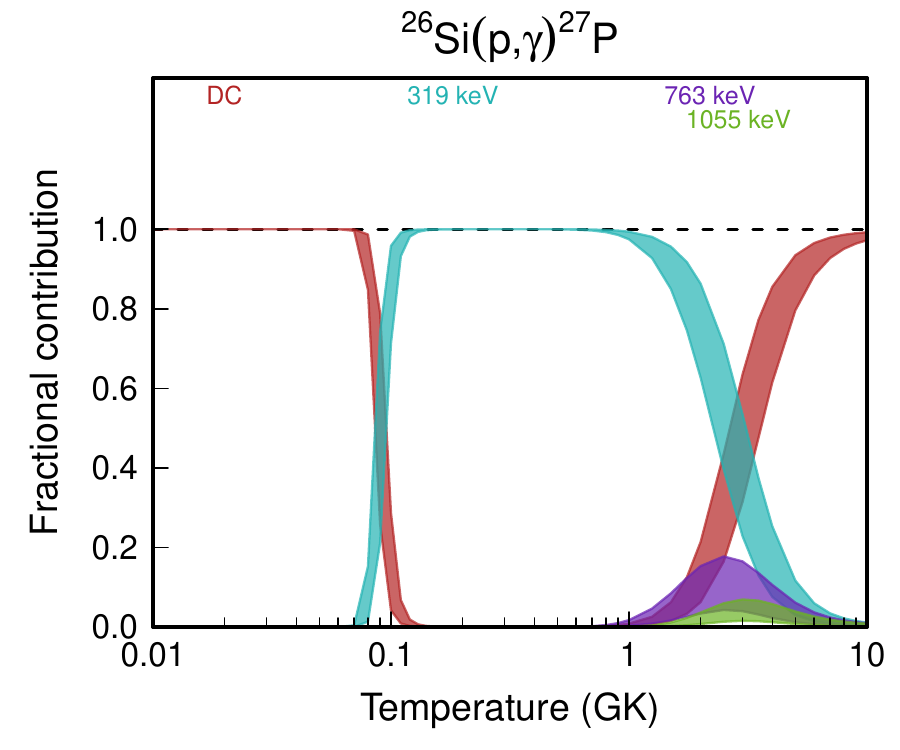}
\caption{
Fractional contributions to the total rate. ``DC'' refers to direct radiative capture. Resonance energies are given in the center-of-mass frame.  
}
\label{fig:si26pg1}
\end{figure*}
\begin{figure*}[hbt!]
\centering
\includegraphics[width=0.5\linewidth]{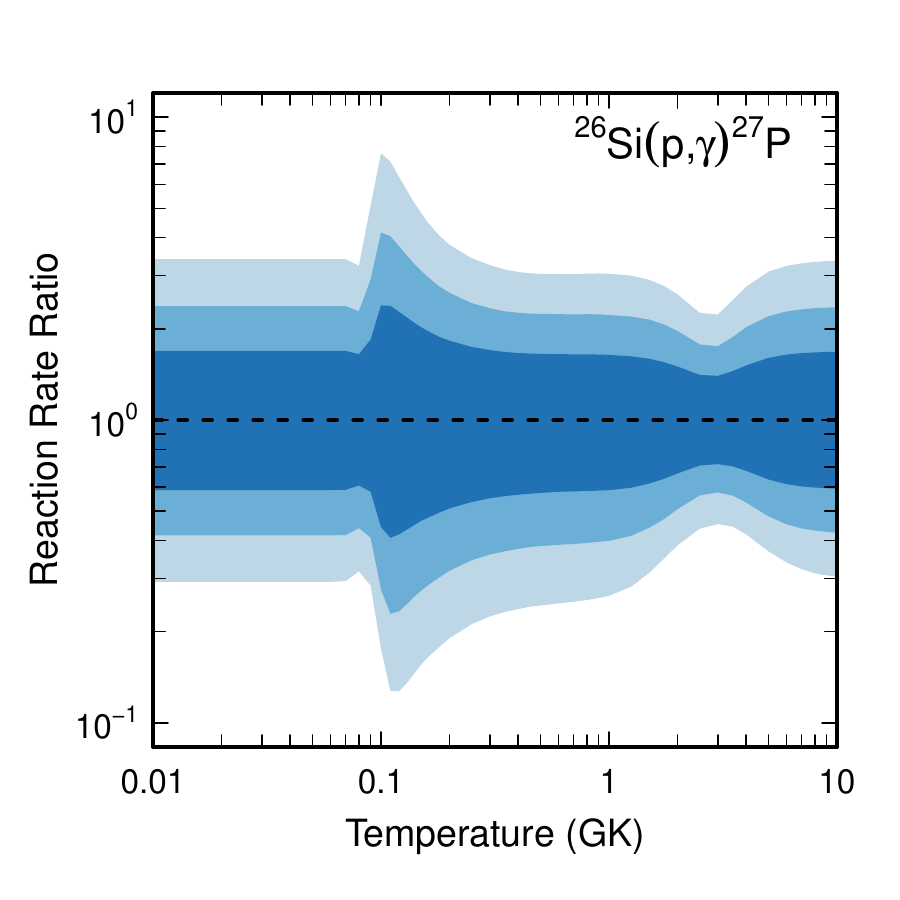}
\caption{
Reaction rate uncertainties versus temperature. The three different shades refer to coverage probabilities of 68\%, 90\%, and 98\%. 
}
\label{fig:si26pg2}
\end{figure*}

\clearpage

\startlongtable


\clearpage

\begin{figure*}[hbt!]
\centering
\includegraphics[width=0.5\linewidth]{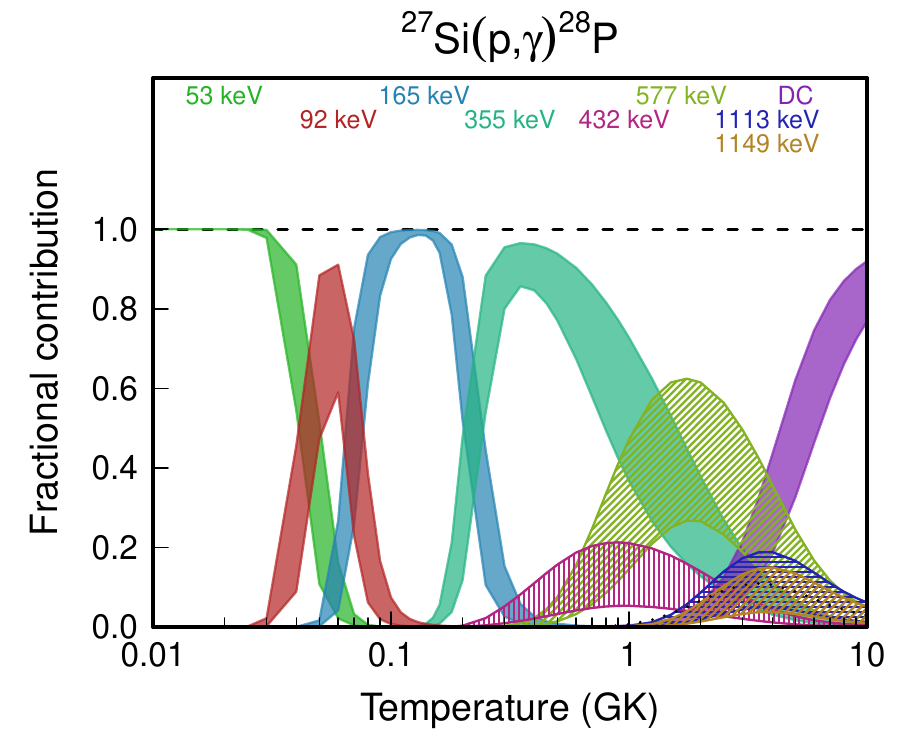}
\caption{
Fractional contributions to the total rate. ``DC'' refers to direct radiative capture. Resonance energies are given in the center-of-mass frame.  
}
\label{fig:si27pg1}
\end{figure*}
\begin{figure*}[hbt!]
\centering
\includegraphics[width=0.5\linewidth]{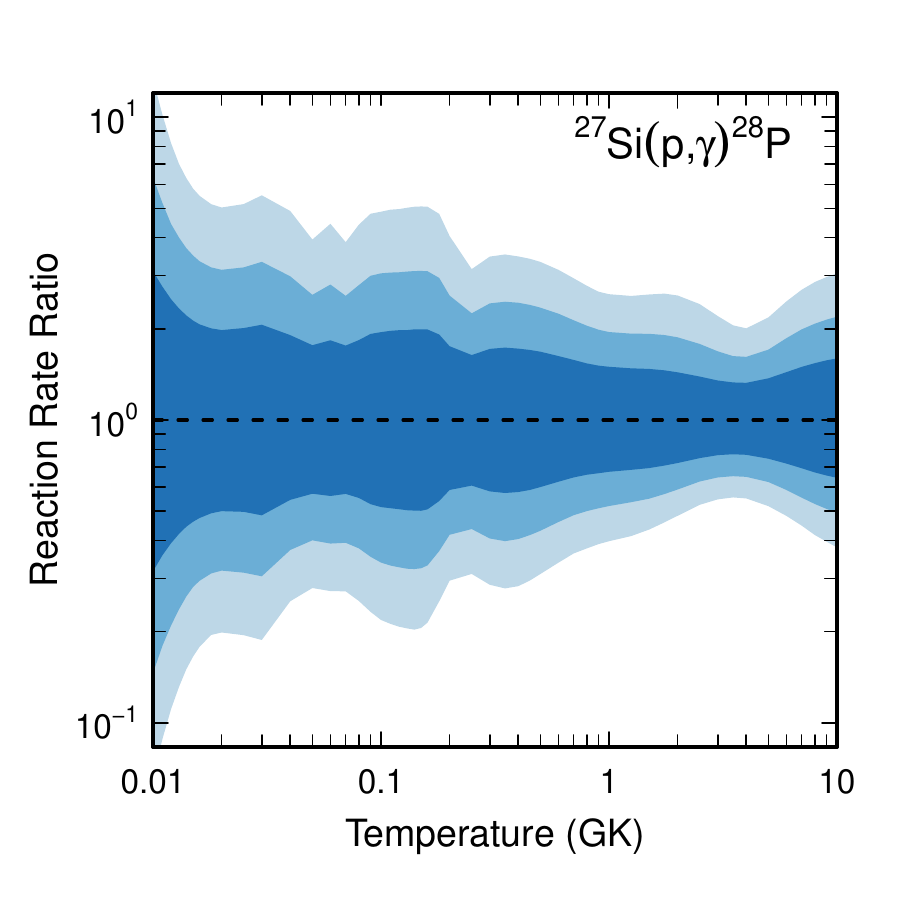}
\caption{
Reaction rate uncertainties versus temperature. The three different shades refer to coverage probabilities of 68\%, 90\%, and 98\%. 
}
\label{fig:si27pg2}
\end{figure*}

\clearpage

\startlongtable


\clearpage

\begin{figure*}[hbt!]
\centering
\includegraphics[width=0.5\linewidth]{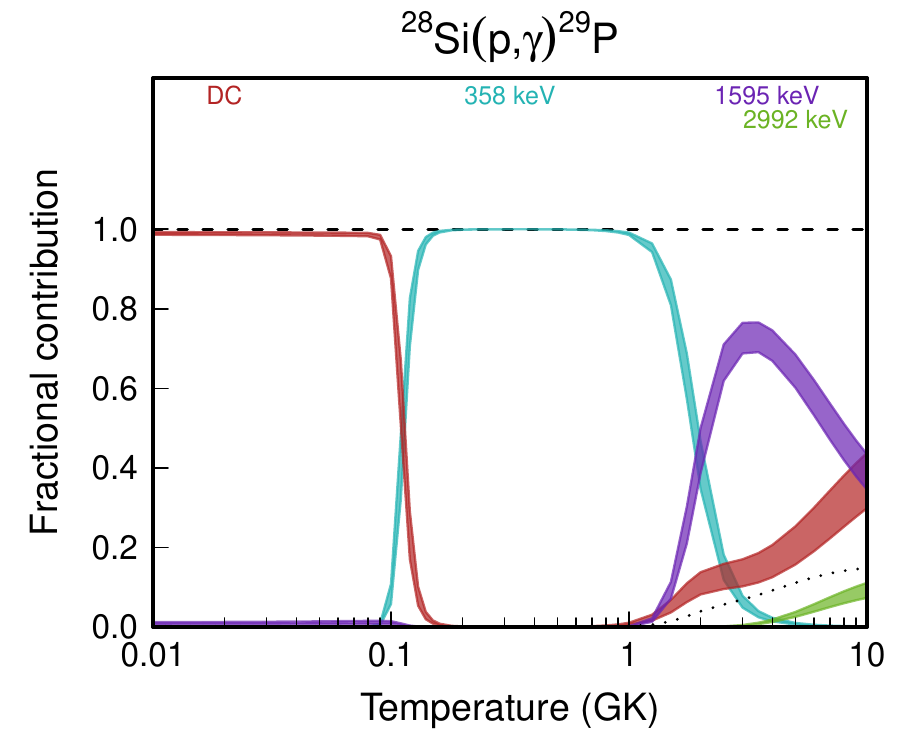}
\caption{
Fractional contributions to the total rate. ``DC'' refers to direct radiative capture. Resonance energies are given in the center-of-mass frame.  
}
\label{fig:si28pg1}
\end{figure*}
\begin{figure*}[hbt!]
\centering
\includegraphics[width=0.5\linewidth]{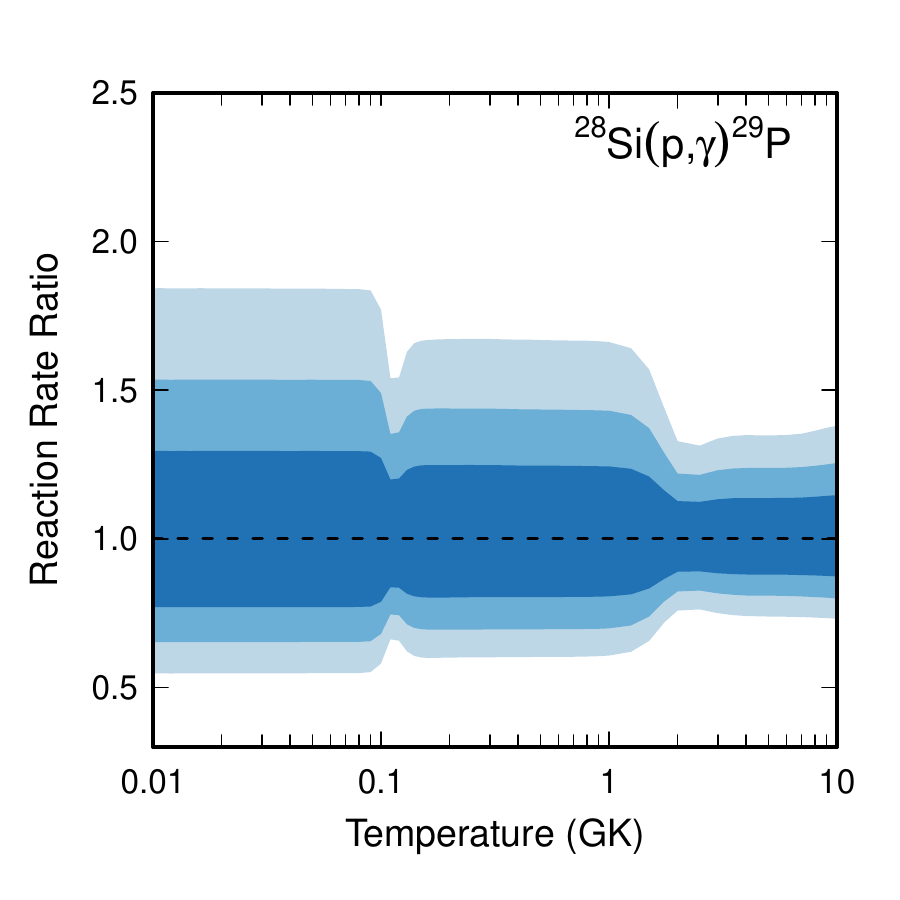}
\caption{
Reaction rate uncertainties versus temperature. The three different shades refer to coverage probabilities of 68\%, 90\%, and 98\%. 
}
\label{fig:si28pg2}
\end{figure*}

\clearpage

\startlongtable


\clearpage

\begin{figure*}[hbt!]
\centering
\includegraphics[width=0.5\linewidth]{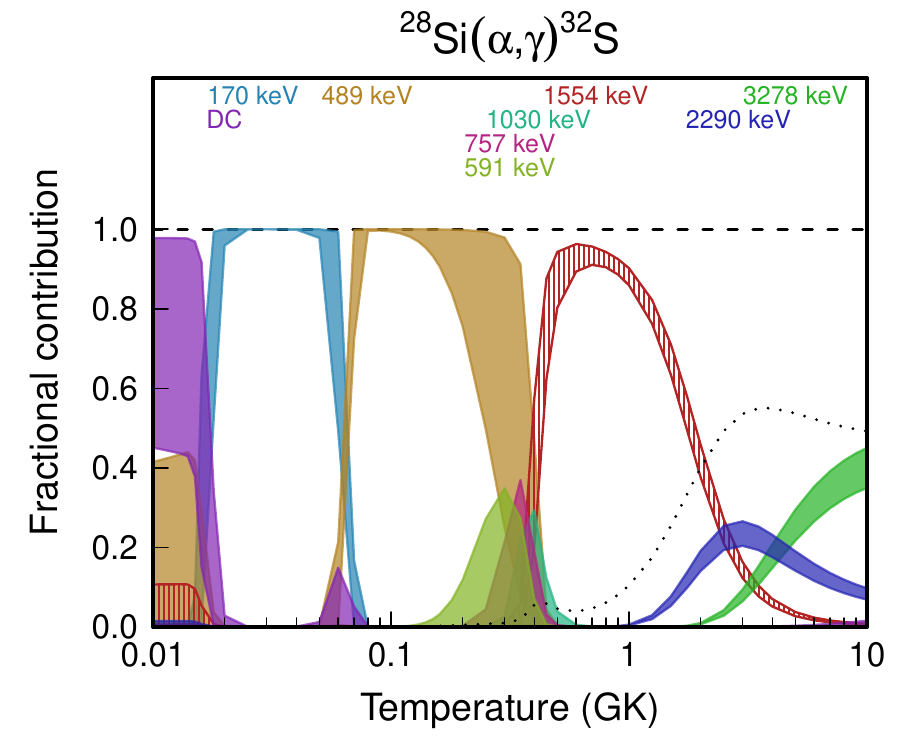}
\caption{
Fractional contributions to the total rate. ``DC'' refers to direct radiative capture. Resonance energies are given in the center-of-mass frame.  
}
\label{fig:si28ag1}
\end{figure*}
\begin{figure*}[hbt!]
\centering
\includegraphics[width=0.5\linewidth]{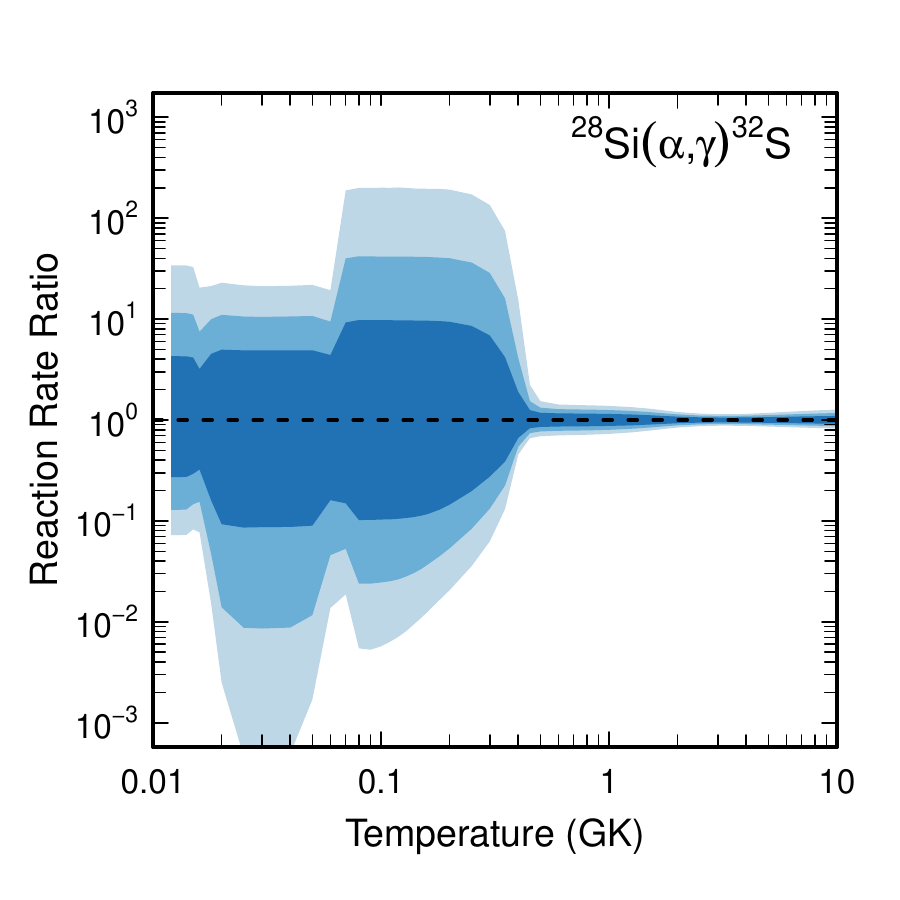}
\caption{
Reaction rate uncertainties versus temperature. The three different shades refer to coverage probabilities of 68\%, 90\%, and 98\%. 
}
\label{fig:si28ag2}
\end{figure*}

\clearpage

\startlongtable


\clearpage

\begin{figure*}[hbt!]
\centering
\includegraphics[width=0.5\linewidth]{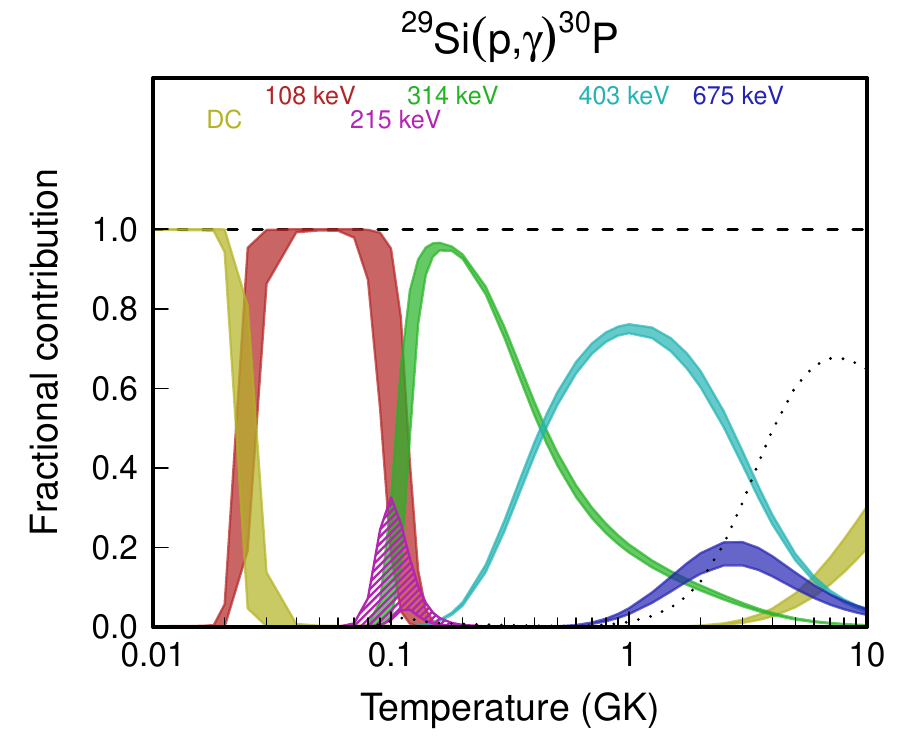}
\caption{
Fractional contributions to the total rate. ``DC'' refers to direct radiative capture. Resonance energies are given in the center-of-mass frame.  
}
\label{fig:si29pg1}
\end{figure*}
\begin{figure*}[hbt!]
\centering
\includegraphics[width=0.5\linewidth]{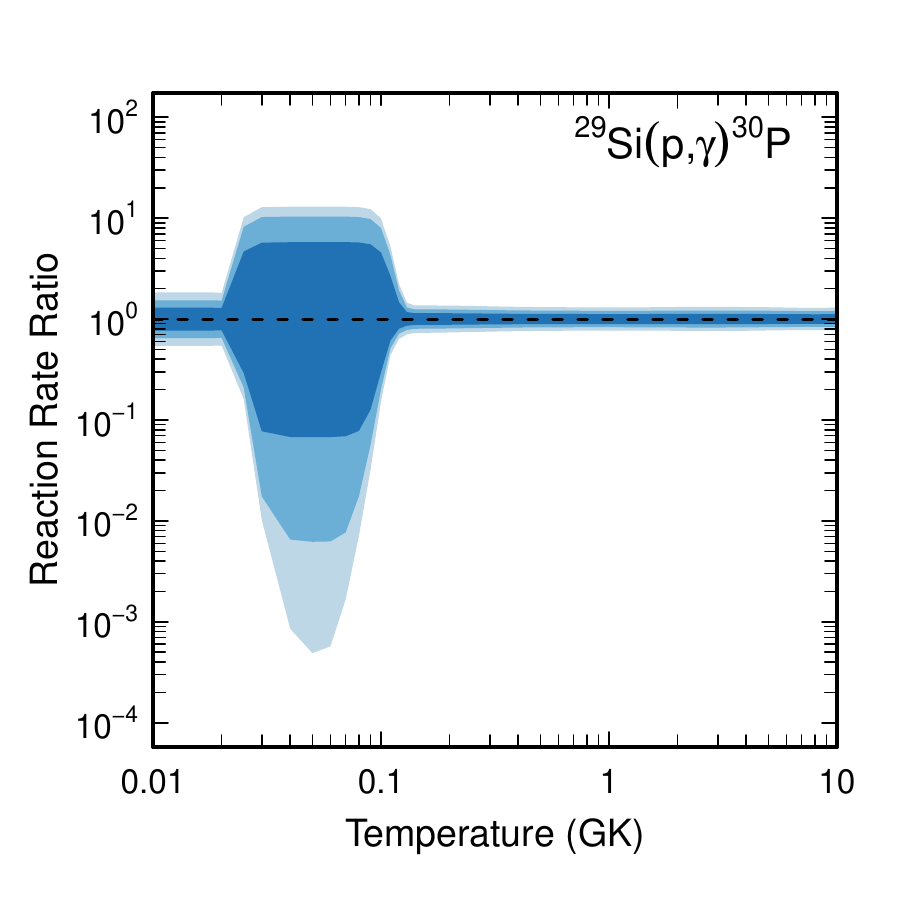}
\caption{
Reaction rate uncertainties versus temperature. The three different shades refer to coverage probabilities of 68\%, 90\%, and 98\%. 
}
\label{fig:si29pg2}
\end{figure*}

\clearpage

\startlongtable


\clearpage

\begin{figure*}[hbt!]
\centering
\includegraphics[width=0.5\linewidth]{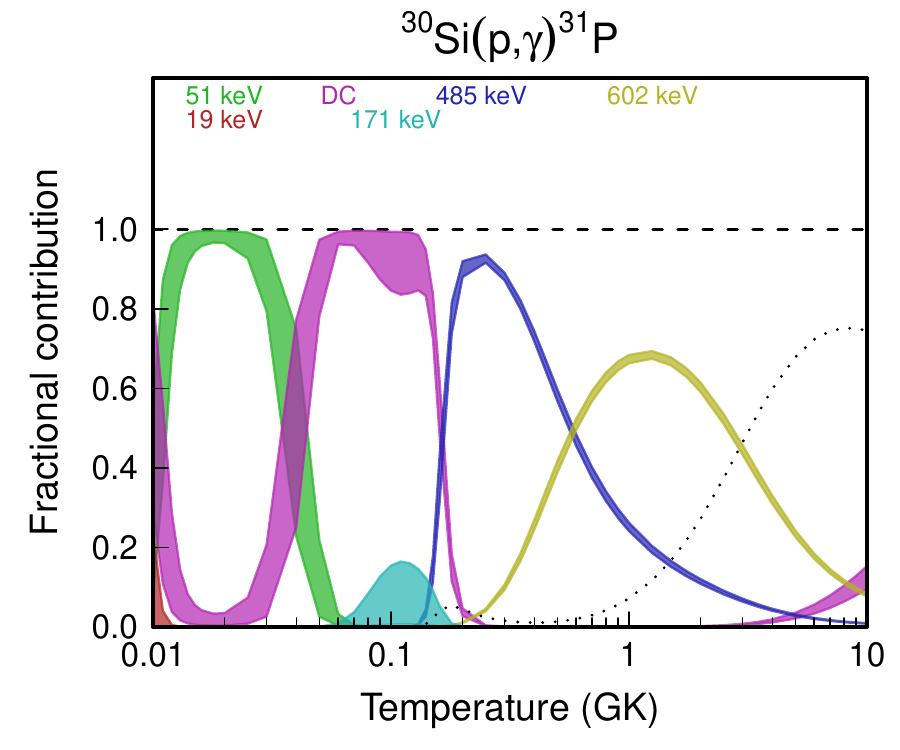}
\caption{
Fractional contributions to the total rate. ``DC'' refers to direct radiative capture. Resonance energies are given in the center-of-mass frame.  
}
\label{fig:si30pg1}
\end{figure*}
\begin{figure*}[hbt!]
\centering
\includegraphics[width=0.5\linewidth]{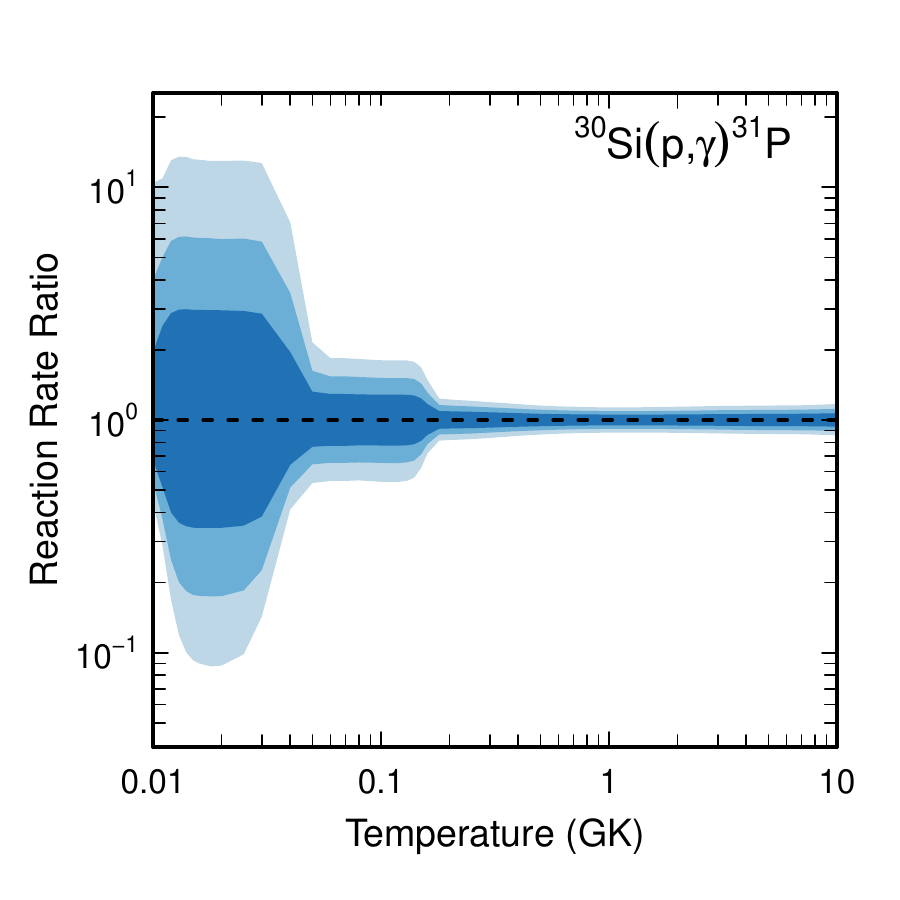}
\caption{
Reaction rate uncertainties versus temperature. The three different shades refer to coverage probabilities of 68\%, 90\%, and 98\%. 
}
\label{fig:si30pg2}
\end{figure*}

\clearpage

\startlongtable


\clearpage

\begin{figure*}[hbt!]
\centering
\includegraphics[width=0.5\linewidth]{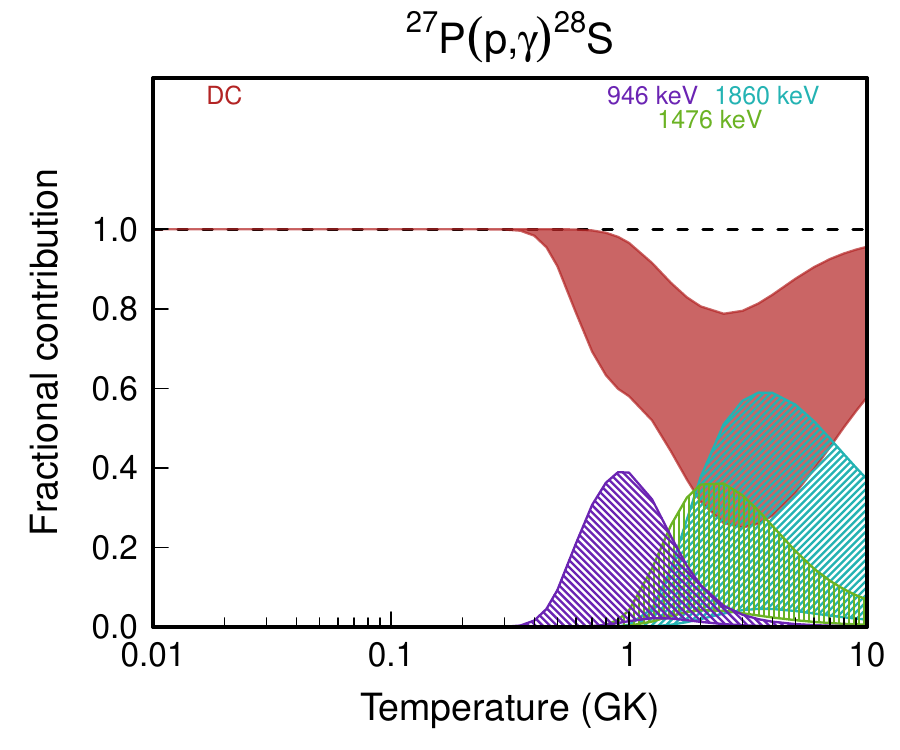}
\caption{
Fractional contributions to the total rate. ``DC'' refers to direct radiative capture. Resonance energies are given in the center-of-mass frame.  
}
\label{fig:p27pg1}
\end{figure*}
\begin{figure*}[hbt!]
\centering
\includegraphics[width=0.5\linewidth]{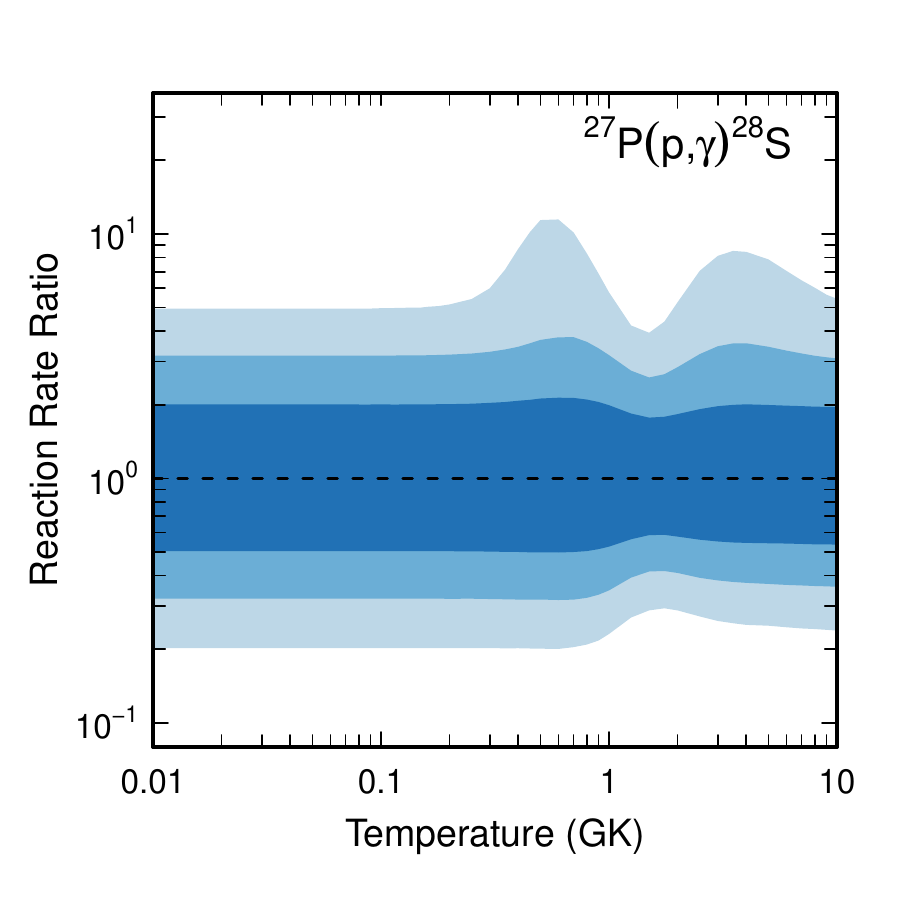}
\caption{
Reaction rate uncertainties versus temperature. The three different shades refer to coverage probabilities of 68\%, 90\%, and 98\%. 
}
\label{fig:p27pg2}
\end{figure*}

\clearpage

\startlongtable


\clearpage

\begin{figure*}[hbt!]
\centering
\includegraphics[width=0.5\linewidth]{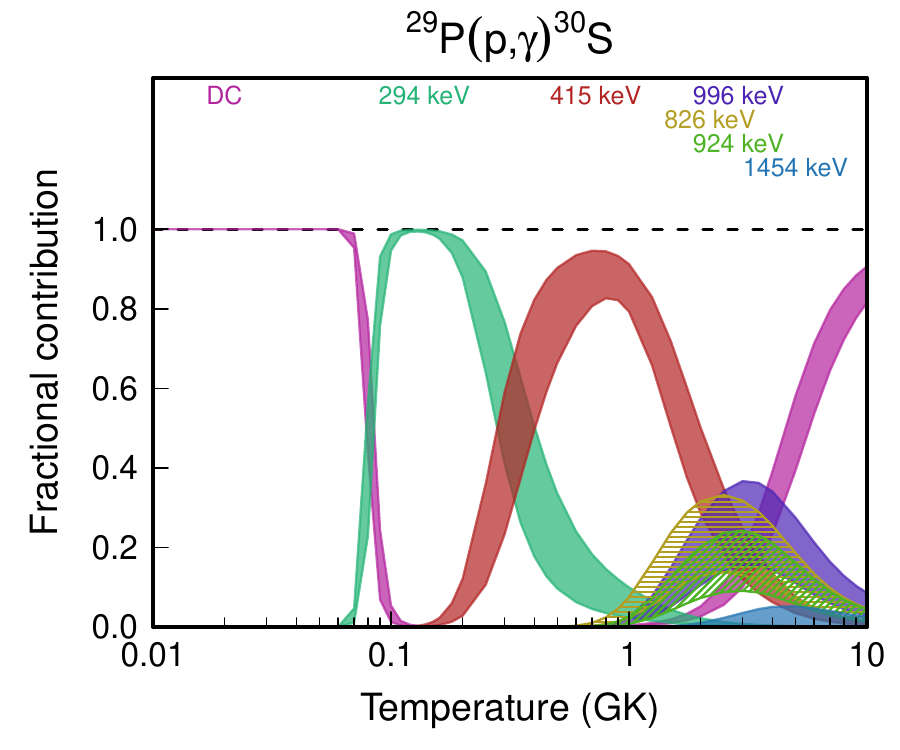}
\caption{
Fractional contributions to the total rate. ``DC'' refers to direct radiative capture. Resonance energies are given in the center-of-mass frame.  
}
\label{fig:p29pg1}
\end{figure*}
\begin{figure*}[hbt!]
\centering
\includegraphics[width=0.5\linewidth]{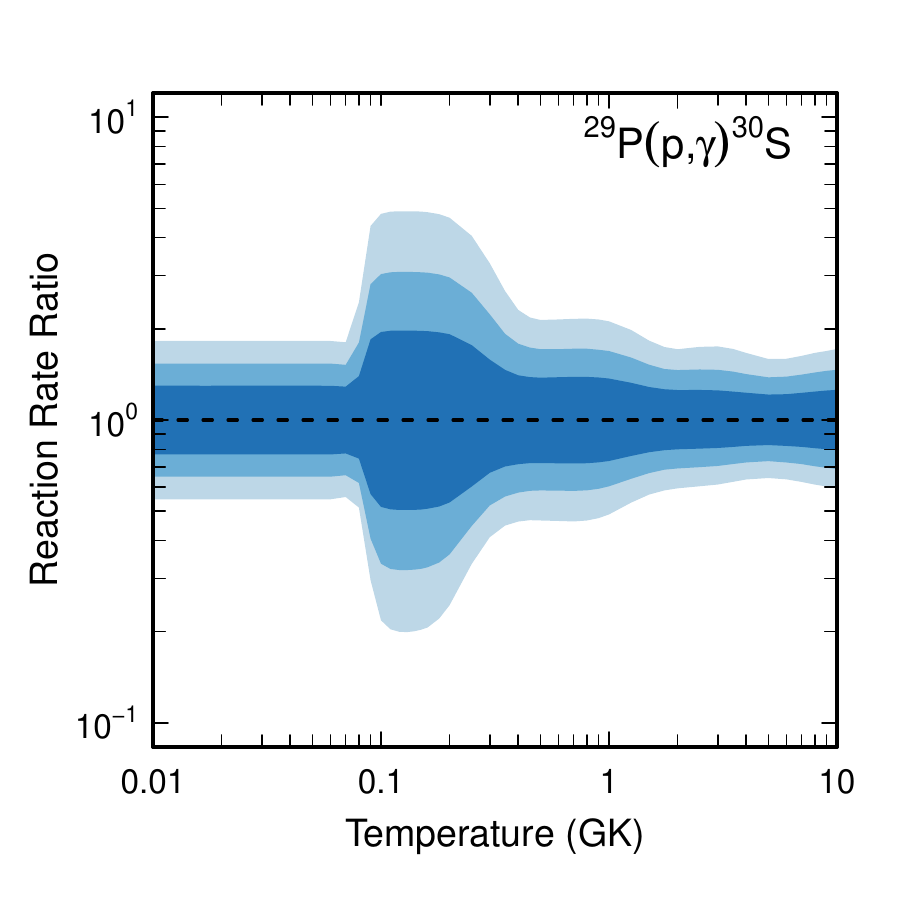}
\caption{
Reaction rate uncertainties versus temperature. The three different shades refer to coverage probabilities of 68\%, 90\%, and 98\%. 
}
\label{fig:p29pg2}
\end{figure*}

\clearpage

\startlongtable


\clearpage

\begin{figure*}[hbt!]
\centering
\includegraphics[width=0.5\linewidth]{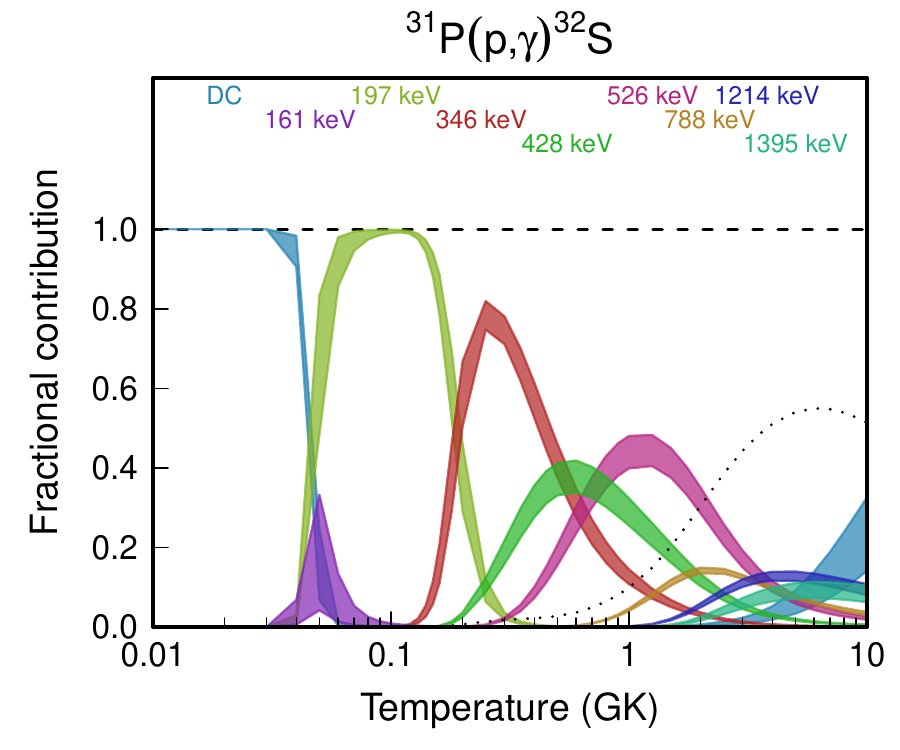}
\caption{
Fractional contributions to the total rate. ``DC'' refers to direct radiative capture. Resonance energies are given in the center-of-mass frame.  
}
\label{fig:p31pg1}
\end{figure*}
\begin{figure*}[hbt!]
\centering
\includegraphics[width=0.5\linewidth]{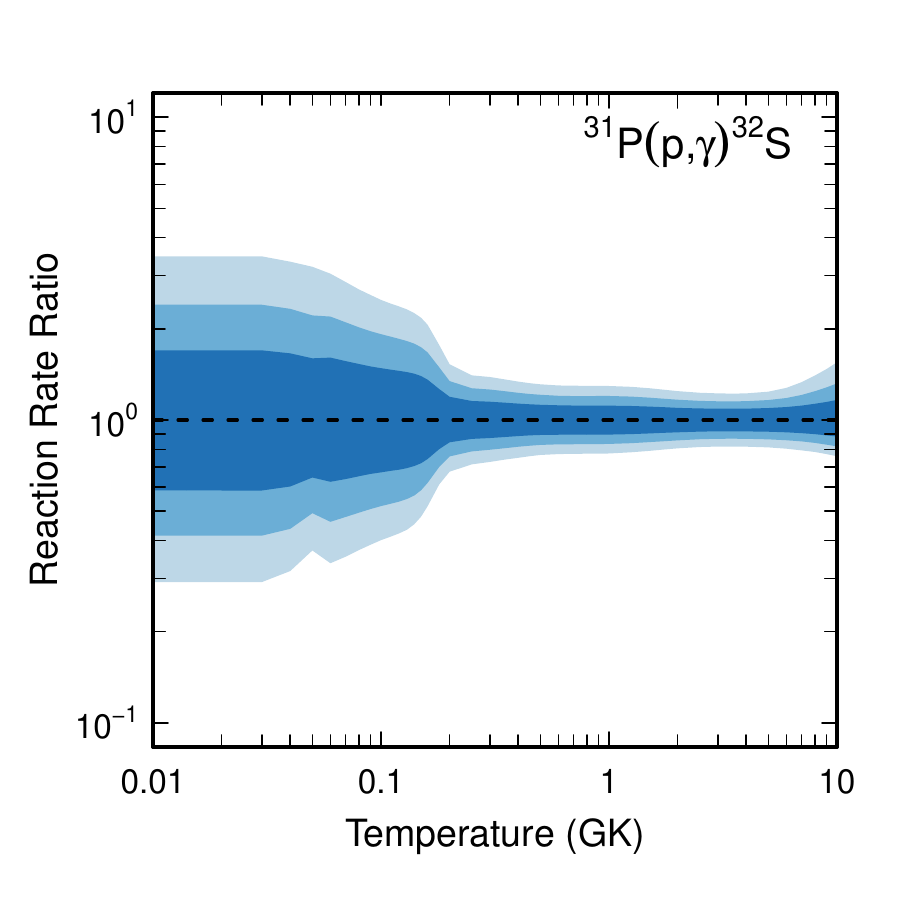}
\caption{
Reaction rate uncertainties versus temperature. The three different shades refer to coverage probabilities of 68\%, 90\%, and 98\%. 
}
\label{fig:p31pg2}
\end{figure*}

\clearpage

\startlongtable


\clearpage

\begin{figure*}[hbt!]
\centering
\includegraphics[width=0.5\linewidth]{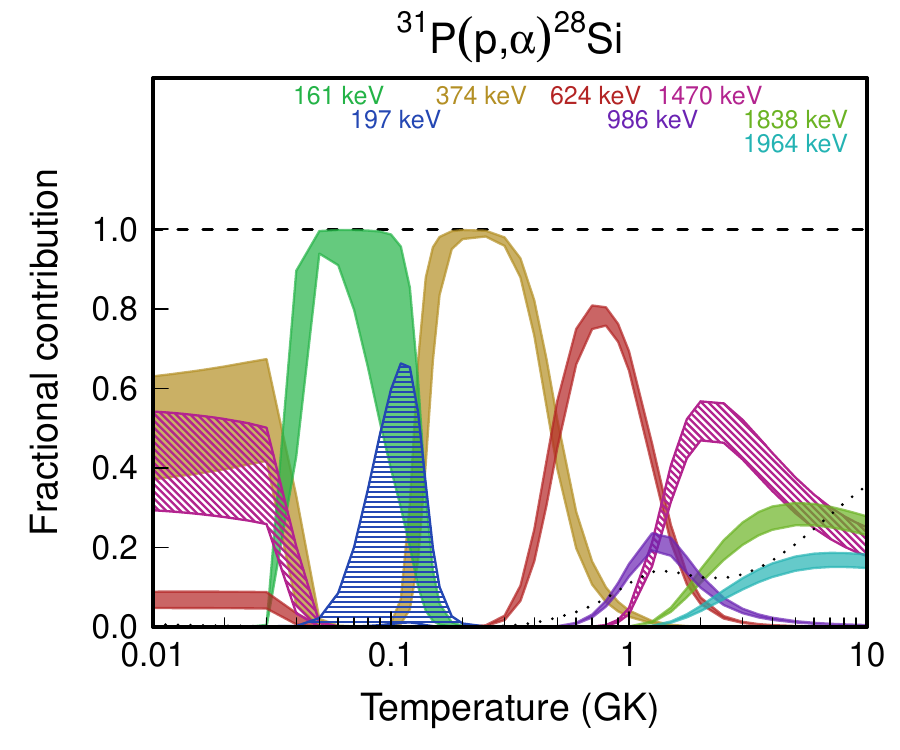}
\caption{
Fractional contributions to the total rate. Resonance energies are given in the center-of-mass frame.  
}
\label{fig:p31pa}
\end{figure*}
\begin{figure*}[hbt!]
\centering
\includegraphics[width=0.5\linewidth]{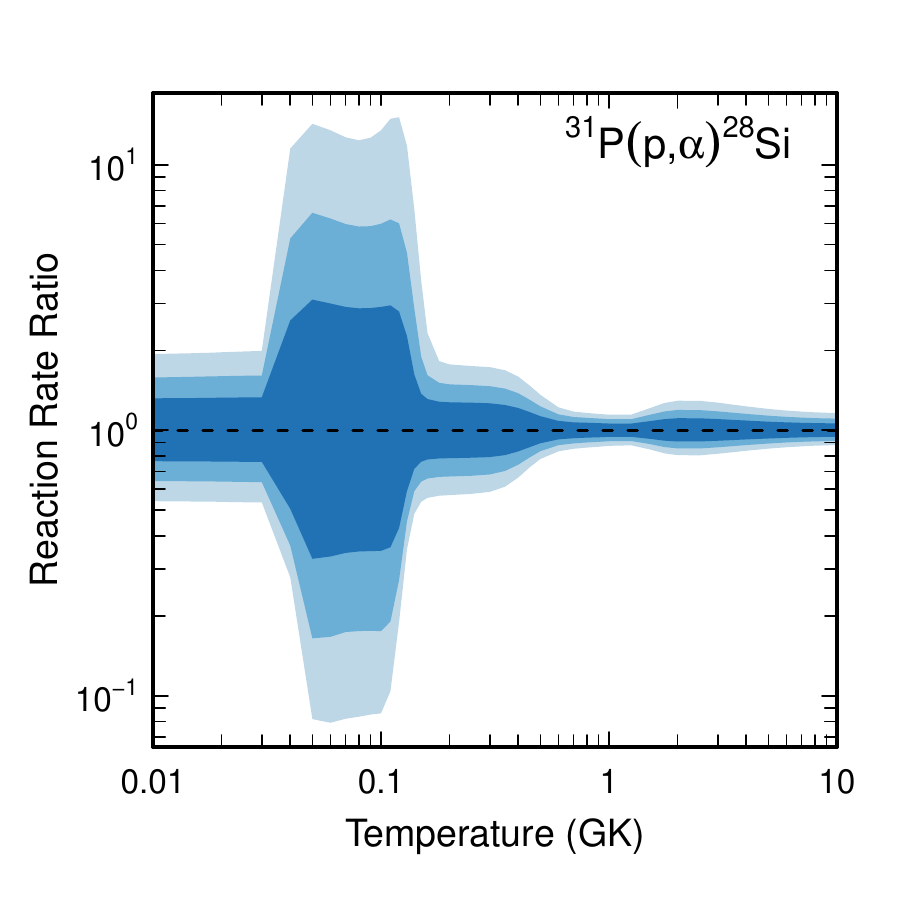}
\caption{
Reaction rate uncertainties versus temperature. The three different shades refer to coverage probabilities of 68\%, 90\%, and 98\%. 
}
\label{fig:p31pa2}
\end{figure*}

\clearpage

\startlongtable


\clearpage

\begin{figure*}[hbt!]
\centering
\includegraphics[width=0.5\linewidth]{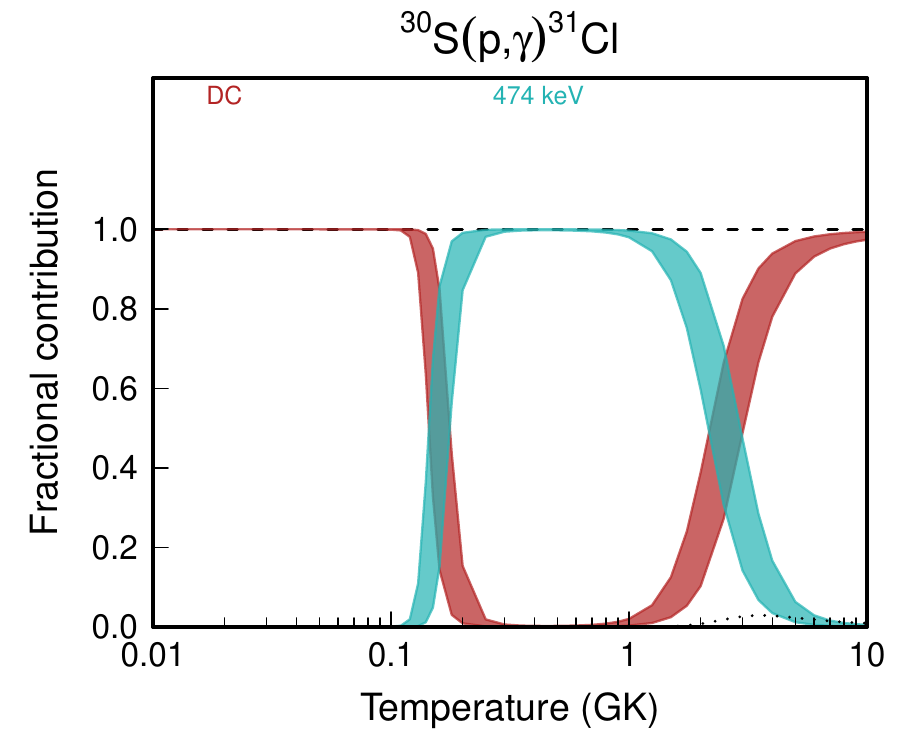}
\caption{
Fractional contributions to the total rate. ``DC'' refers to direct radiative capture. Resonance energies are given in the center-of-mass frame.  
}
\label{fig:s30pg1}
\end{figure*}
\begin{figure*}[hbt!]
\centering
\includegraphics[width=0.5\linewidth]{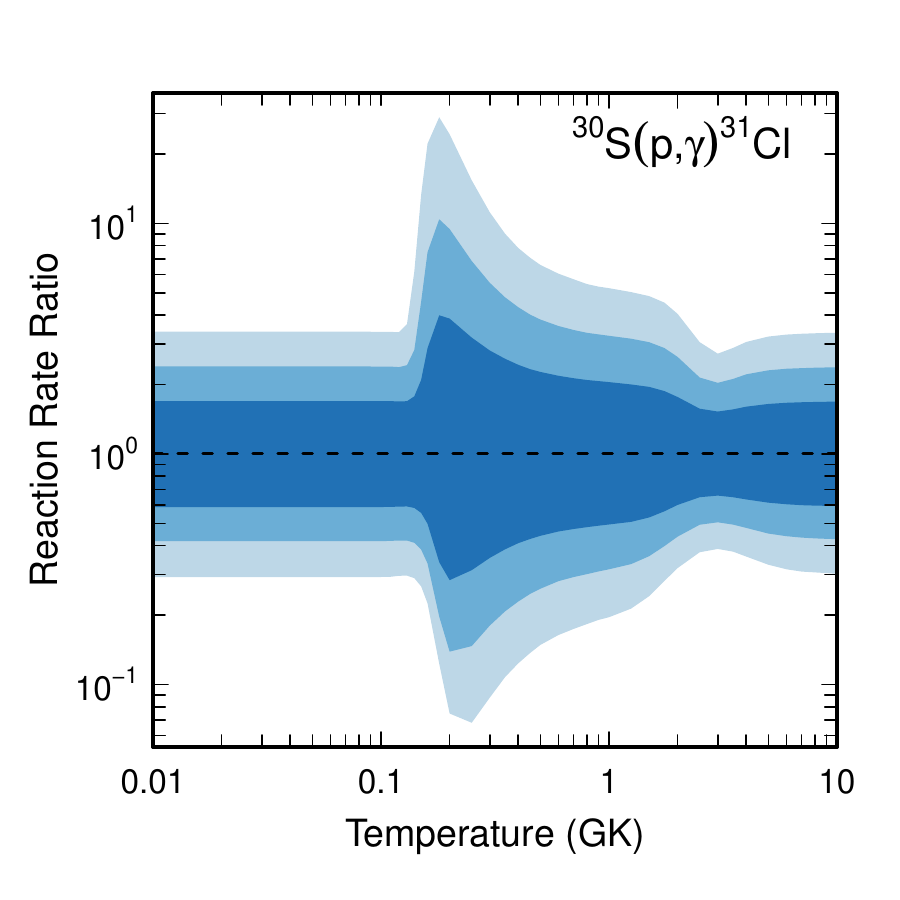}
\caption{
Reaction rate uncertainties versus temperature. The three different shades refer to coverage probabilities of 68\%, 90\%, and 98\%. 
}
\label{fig:s30pg2}
\end{figure*}

\clearpage

\startlongtable


\clearpage

\begin{figure*}[hbt!]
\centering
\includegraphics[width=0.5\linewidth]{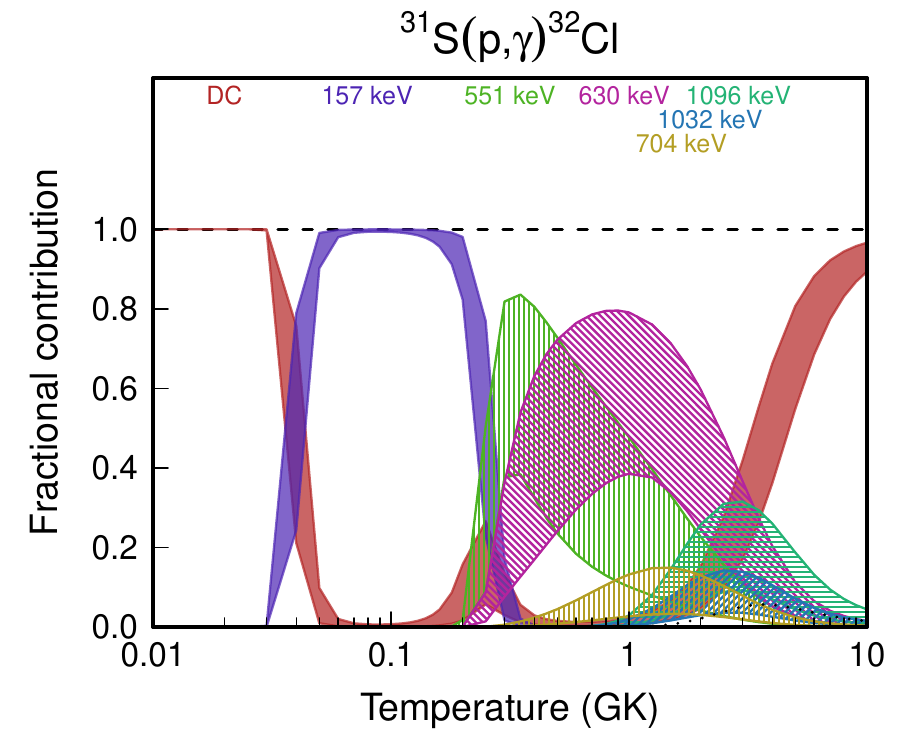}
\caption{
Fractional contributions to the total rate. ``DC'' refers to direct radiative capture. Resonance energies are given in the center-of-mass frame.  
}
\label{fig:s31pg1}
\end{figure*}
\begin{figure*}[hbt!]
\centering
\includegraphics[width=0.5\linewidth]{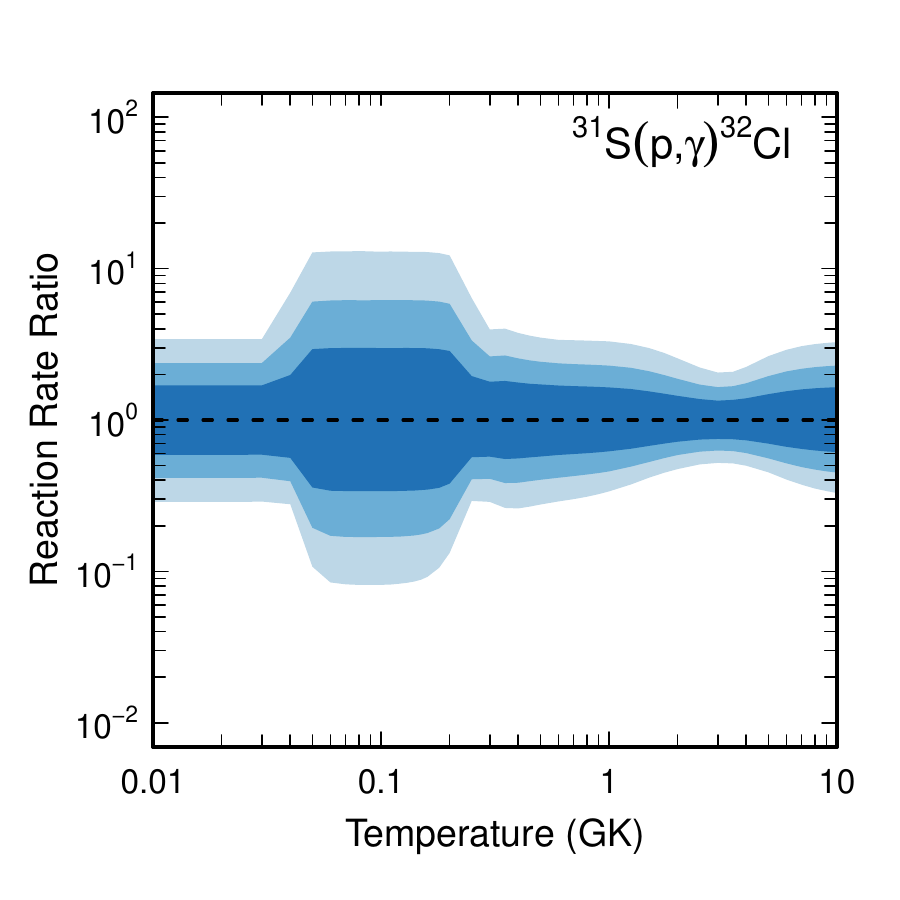}
\caption{
Reaction rate uncertainties versus temperature. The three different shades refer to coverage probabilities of 68\%, 90\%, and 98\%. 
}
\label{fig:s31pg2}
\end{figure*}

\clearpage

\startlongtable


\clearpage

\begin{figure*}[hbt!]
\centering
\includegraphics[width=0.5\linewidth]{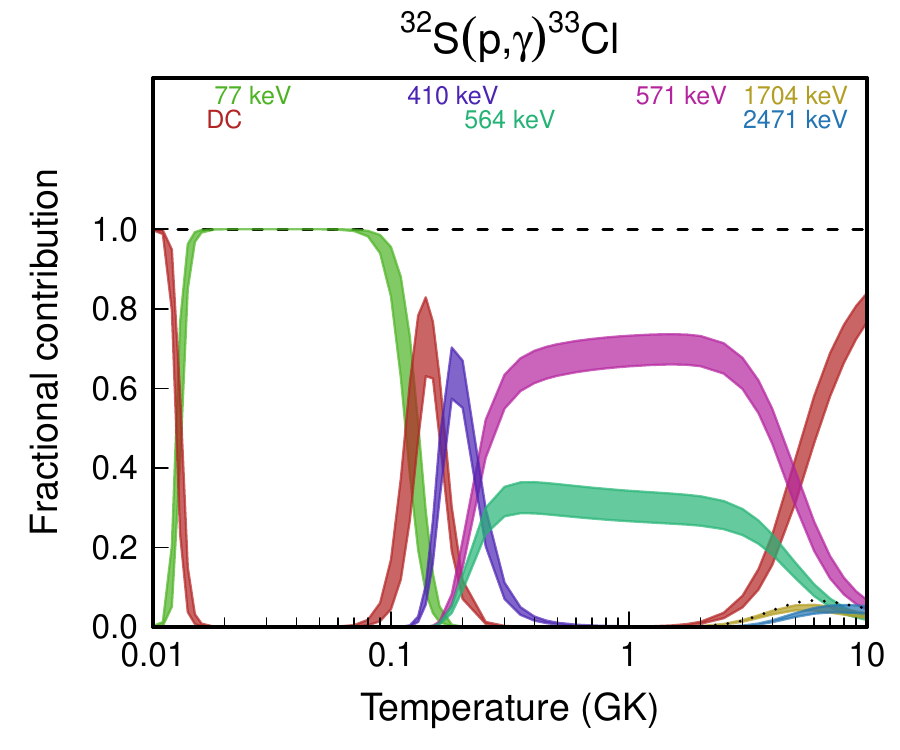}
\caption{
Fractional contributions to the total rate. ``DC'' refers to direct radiative capture. Resonance energies are given in the center-of-mass frame.  
}
\label{fig:s32pg1}
\end{figure*}
\begin{figure*}[hbt!]
\centering
\includegraphics[width=0.5\linewidth]{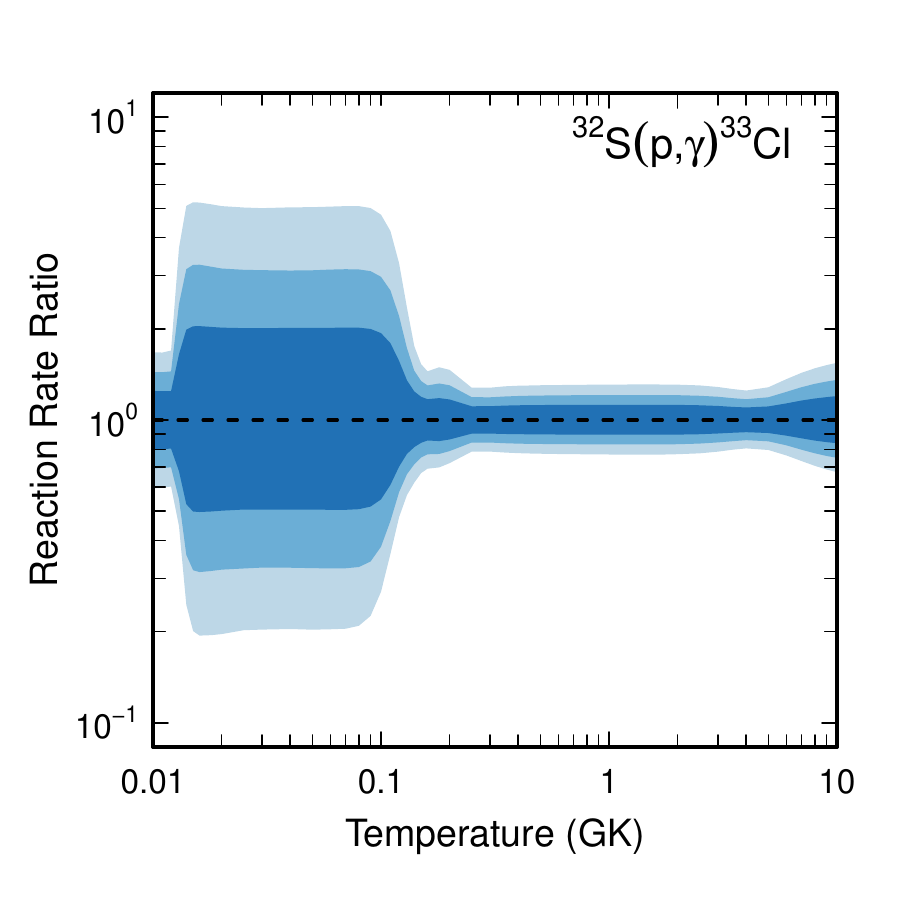}
\caption{
Reaction rate uncertainties versus temperature. The three different shades refer to coverage probabilities of 68\%, 90\%, and 98\%. 
}
\label{fig:s32pg2}
\end{figure*}

\clearpage

\startlongtable


\clearpage

\begin{figure*}[hbt!]
\centering
\includegraphics[width=0.5\linewidth]{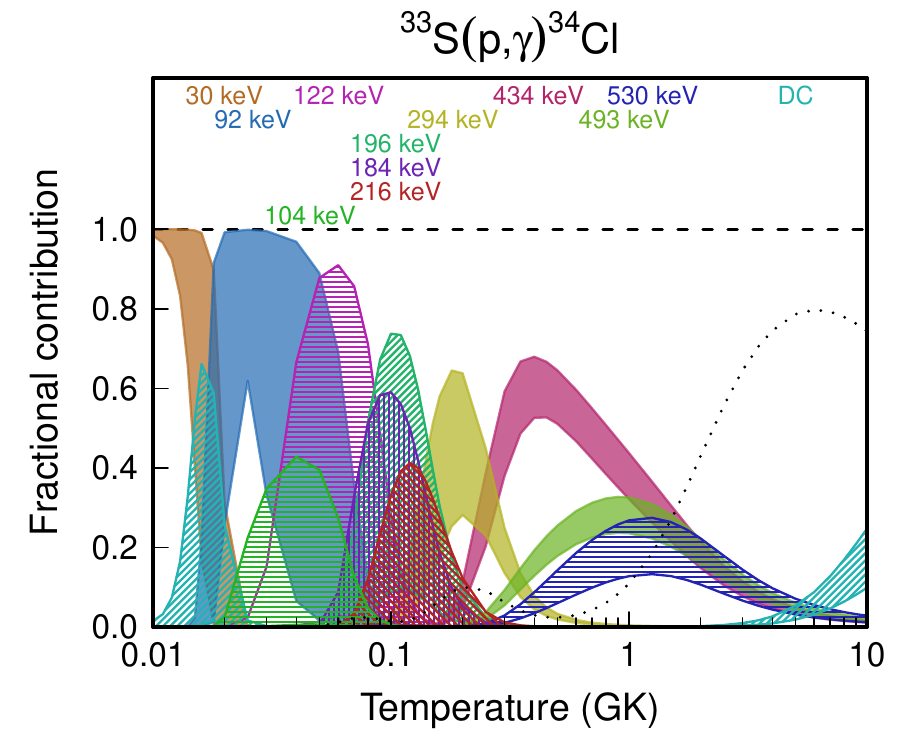}
\caption{
Fractional contributions to the total rate. ``DC'' refers to direct radiative capture. Resonance energies are given in the center-of-mass frame.  
}
\label{fig:s33pg1}
\end{figure*}
\begin{figure*}[hbt!]
\centering
\includegraphics[width=0.5\linewidth]{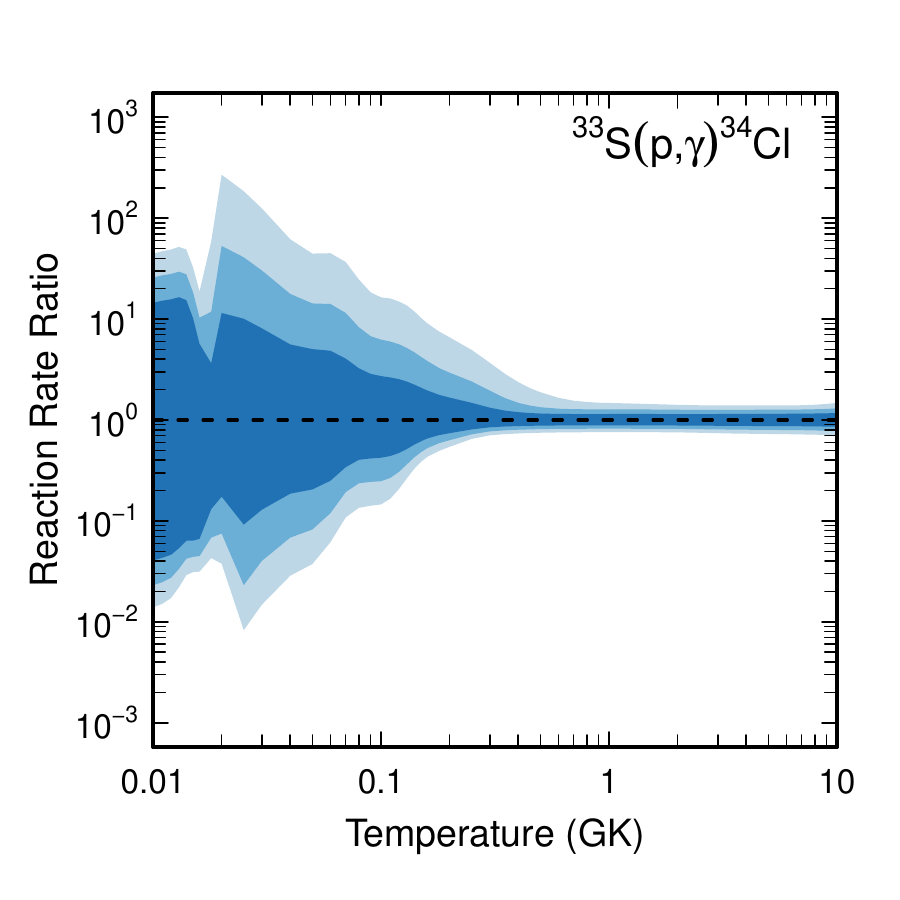}
\caption{
Reaction rate uncertainties versus temperature. The three different shades refer to coverage probabilities of 68\%, 90\%, and 98\%. 
}
\label{fig:s33pg2}
\end{figure*}

\clearpage

\startlongtable


\clearpage

\begin{figure*}[hbt!]
\centering
\includegraphics[width=0.5\linewidth]{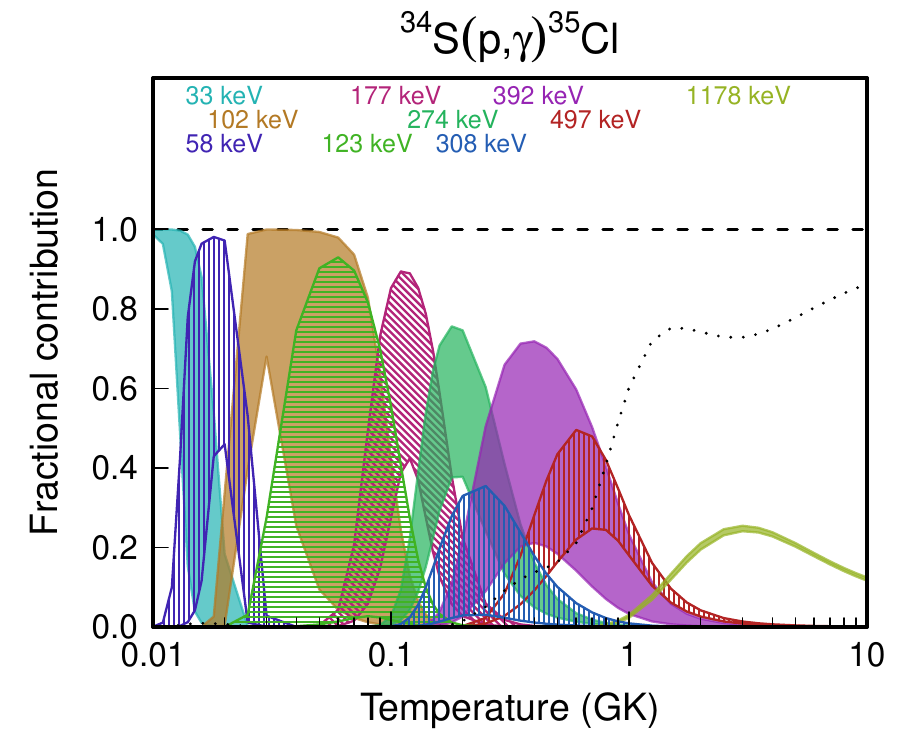}
\caption{
Fractional contributions to the total rate. Resonance energies are given in the center-of-mass frame.  
}
\label{fig:s34pg1}
\end{figure*}
\begin{figure*}[hbt!]
\centering
\includegraphics[width=0.5\linewidth]{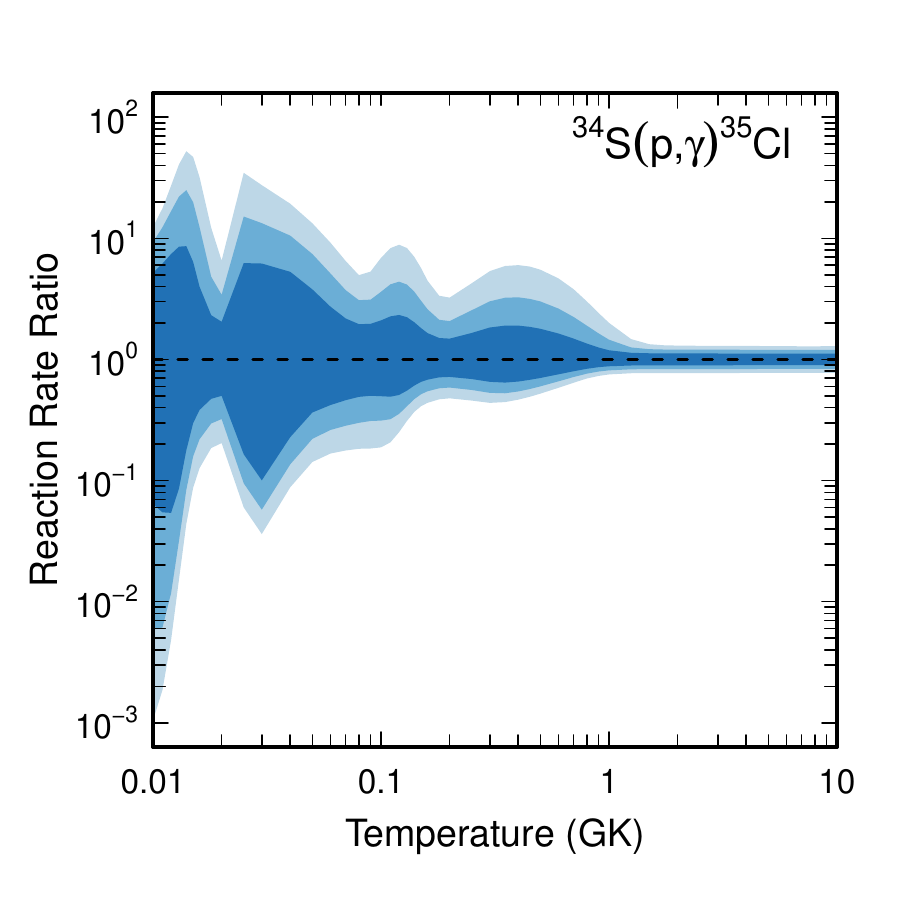}
\caption{
Reaction rate uncertainties versus temperature. The three different shades refer to coverage probabilities of 68\%, 90\%, and 98\%. 
}
\label{fig:s34pg2}
\end{figure*}

\clearpage

\startlongtable


\clearpage

\begin{figure*}[hbt!]
\centering
\includegraphics[width=0.5\linewidth]{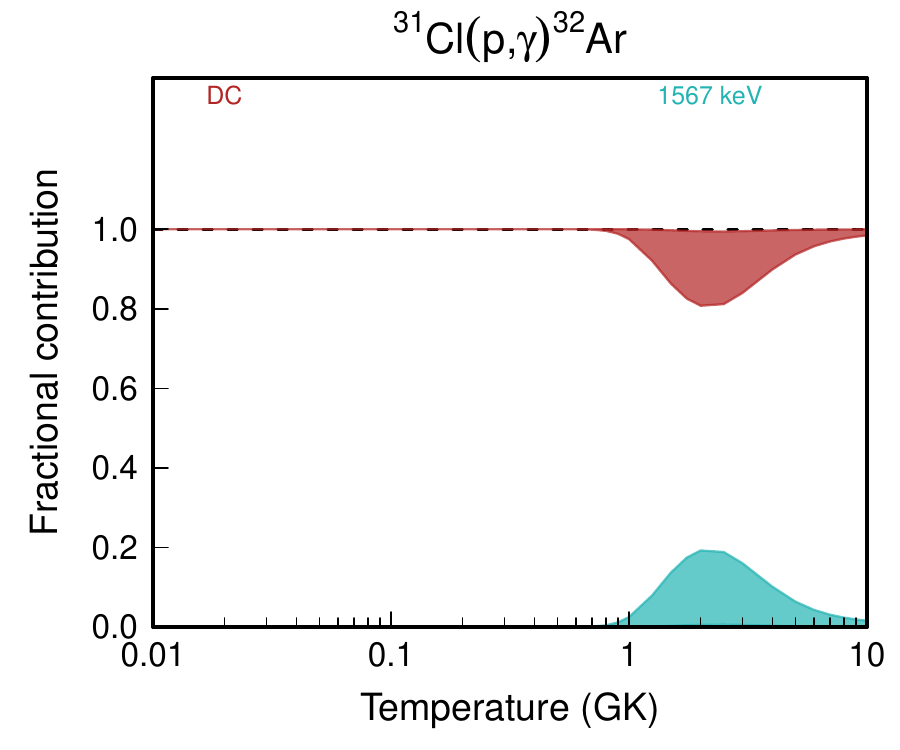}
\caption{
Fractional contributions to the total rate. ``DC'' refers to direct radiative capture. Resonance energies are given in the center-of-mass frame. 
}
\label{fig:cl31pg1}
\end{figure*}
\begin{figure*}[hbt!]
\centering
\includegraphics[width=0.5\linewidth]{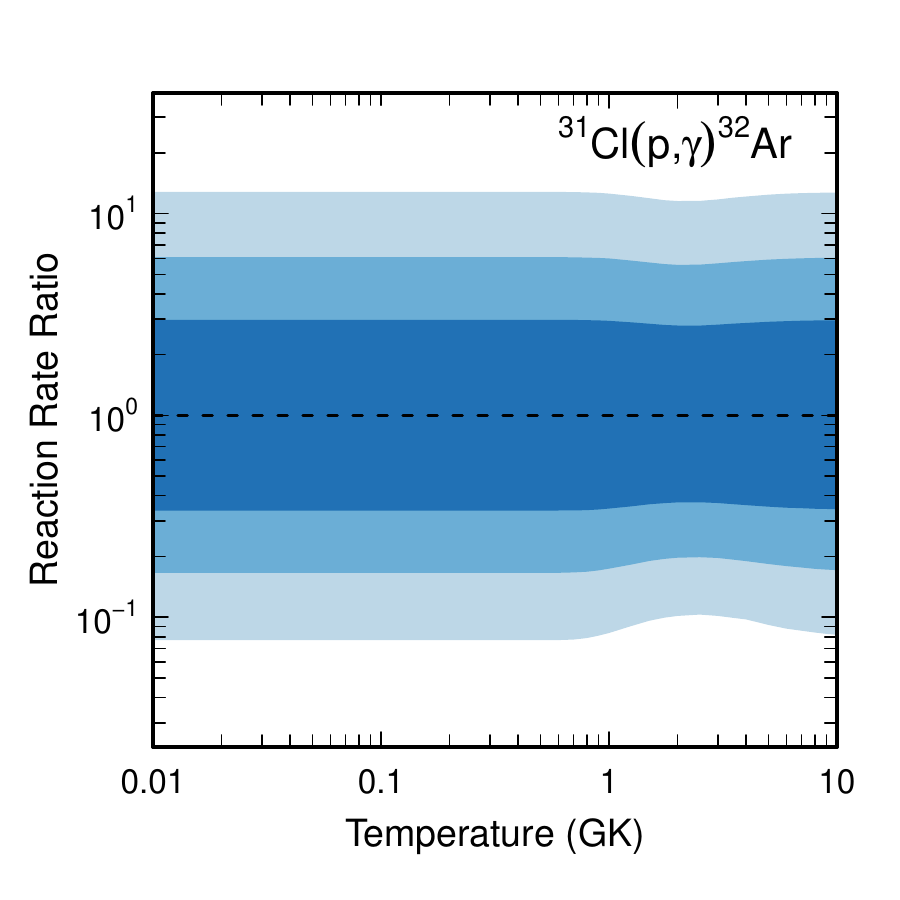}
\caption{
Reaction rate uncertainties versus temperature. The three different shades refer to coverage probabilities of 68\%, 90\%, and 98\%.}
\label{fig:cl31pg2}
\end{figure*}

\clearpage

\startlongtable


\clearpage

\begin{figure*}[hbt!]
\centering
\includegraphics[width=0.5\linewidth]{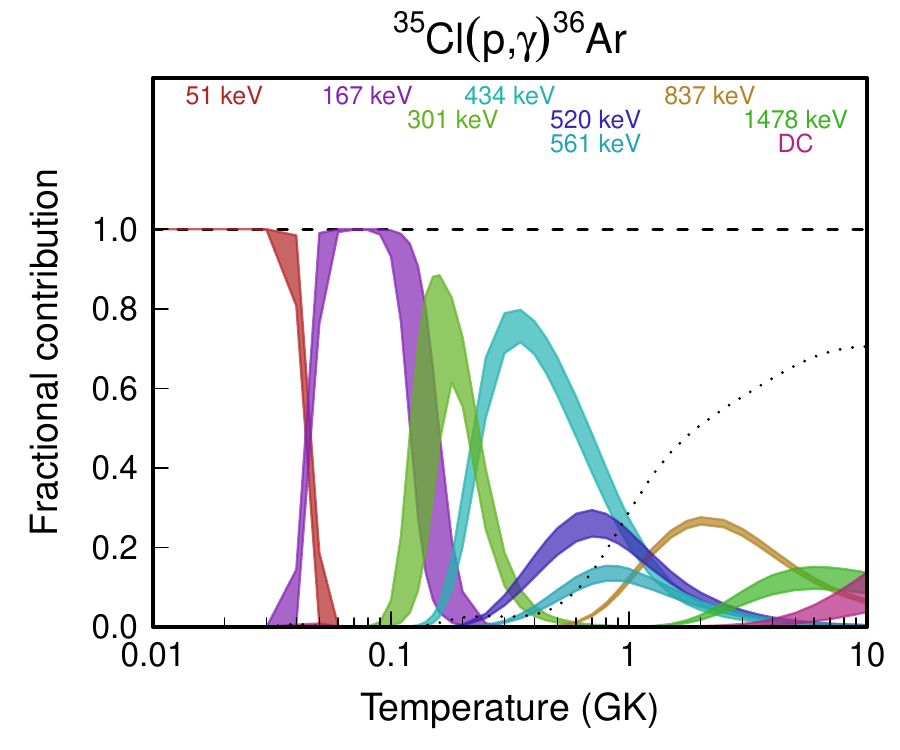}
\caption{
Fractional contributions to the total rate. ``DC'' refers to direct radiative capture. Resonance energies are given in the center-of-mass frame.  
}
\label{fig:cl35pg1}
\end{figure*}
\begin{figure*}[hbt!]
\centering
\includegraphics[width=0.5\linewidth]{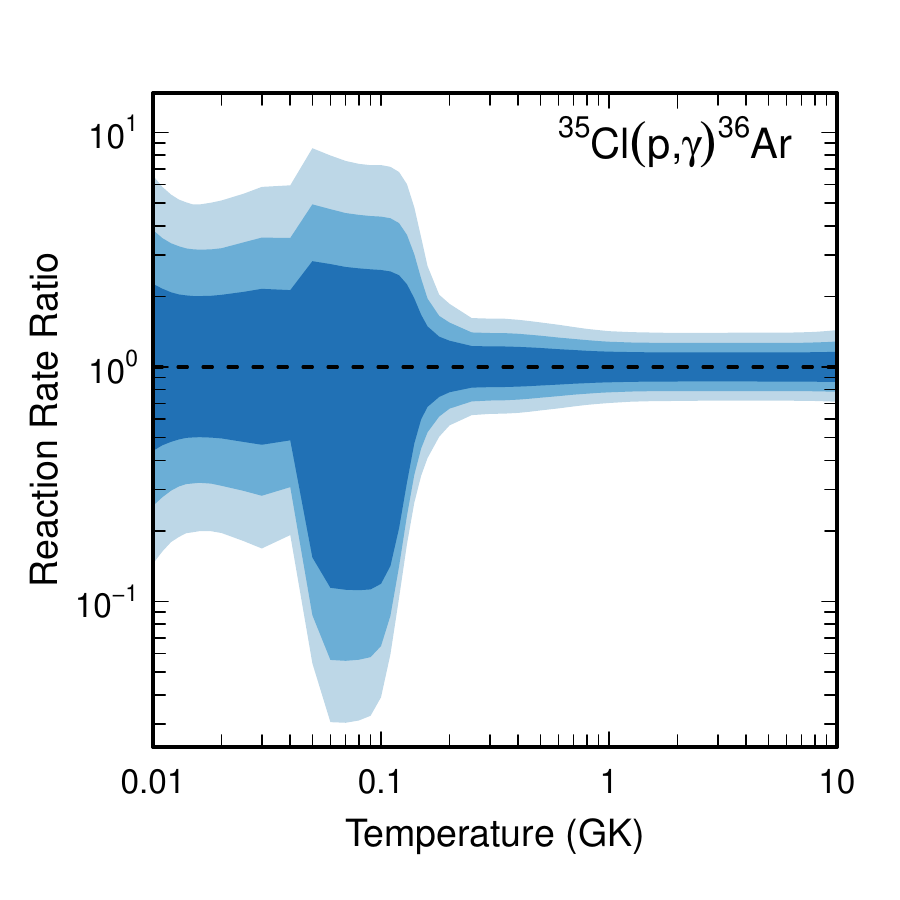}
\caption{
Reaction rate uncertainties versus temperature. The three different shades refer to coverage probabilities of 68\%, 90\%, and 98\%. 
}
\label{fig:cl35pg2}
\end{figure*}

\clearpage

\startlongtable


\clearpage

\begin{figure*}[hbt!]
\centering
\includegraphics[width=0.5\linewidth]{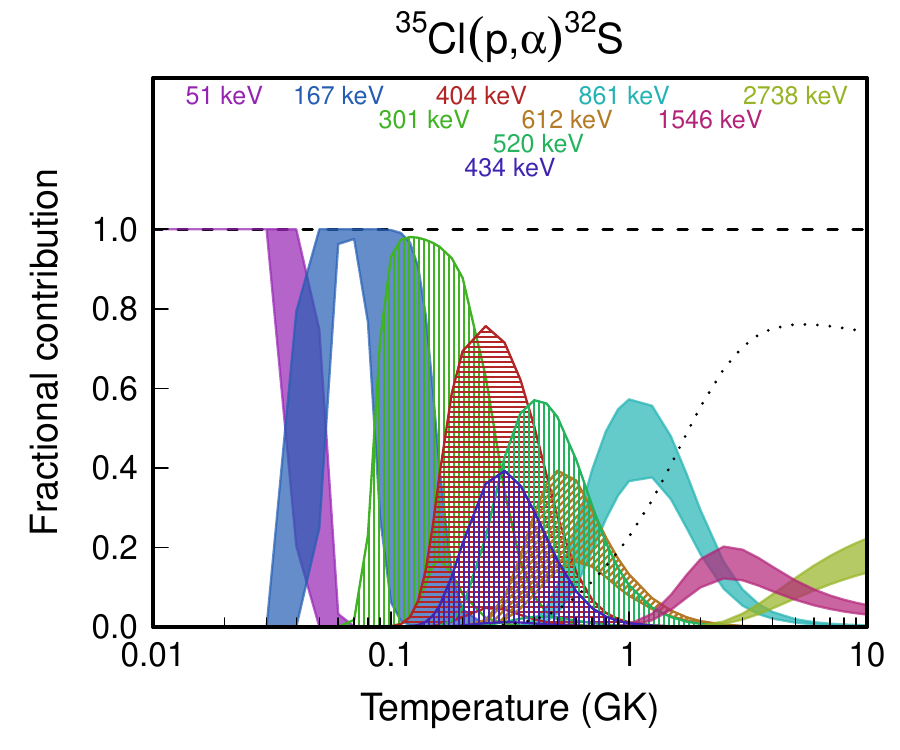}
\caption{
Fractional contributions to the total rate. Resonance energies are given in the center-of-mass frame.  
}
\label{fig:cl35pa1}
\end{figure*}
\begin{figure*}[hbt!]
\centering
\includegraphics[width=0.5\linewidth]{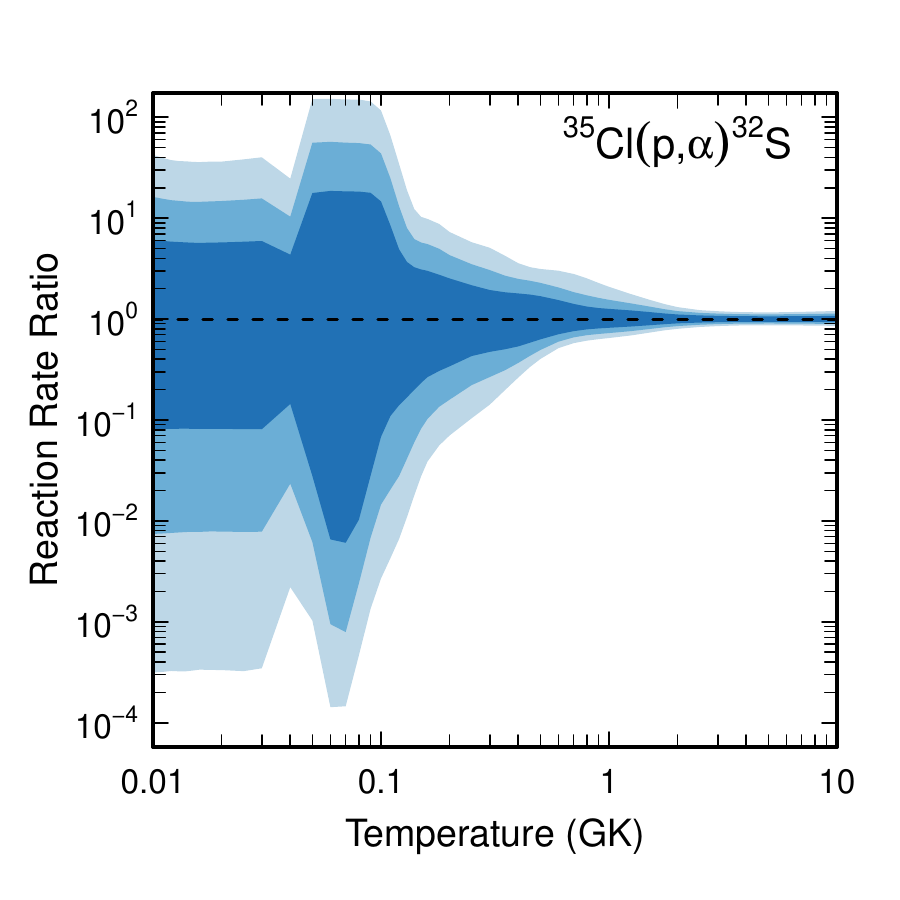}
\caption{
Reaction rate uncertainties versus temperature. The three different shades refer to coverage probabilities of 68\%, 90\%, and 98\%. 
}
\label{fig:cl35pa2}
\end{figure*}

\clearpage

\startlongtable


\clearpage

\begin{figure*}[hbt!]
\centering
\includegraphics[width=0.5\linewidth]{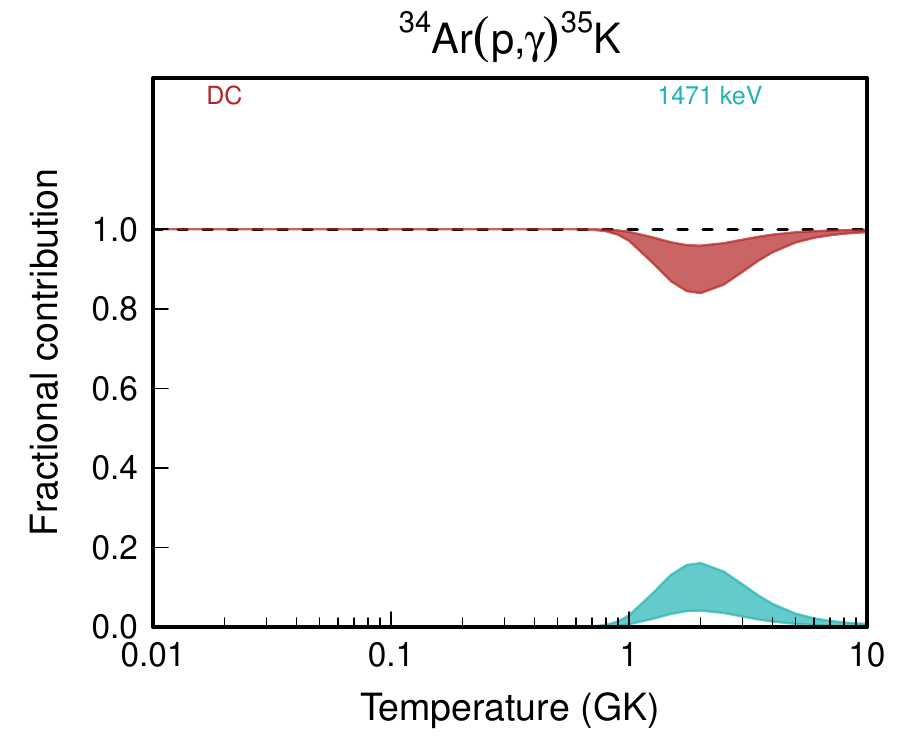}
\caption{
Fractional contributions to the total rate. ``DC'' refers to direct radiative capture. Resonance energies are given in the center-of-mass frame. 
}
\label{fig:ar34pg1}
\end{figure*}
\begin{figure*}[hbt!]
\centering
\includegraphics[width=0.5\linewidth]{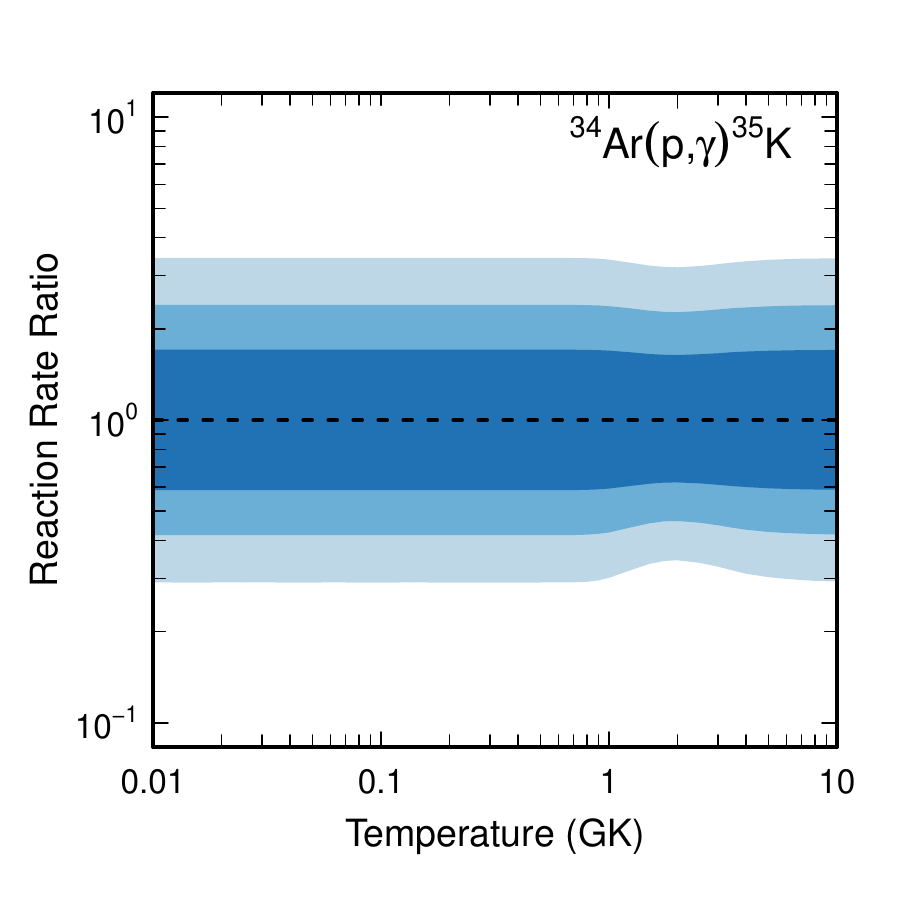}
\caption{
Reaction rate uncertainties versus temperature. The three different shades refer to coverage probabilities of 68\%, 90\%, and 98\%.}
\label{fig:ar34pg}
\end{figure*}

\clearpage

\startlongtable


\clearpage

\begin{figure*}[hbt!]
\centering
\includegraphics[width=0.5\linewidth]{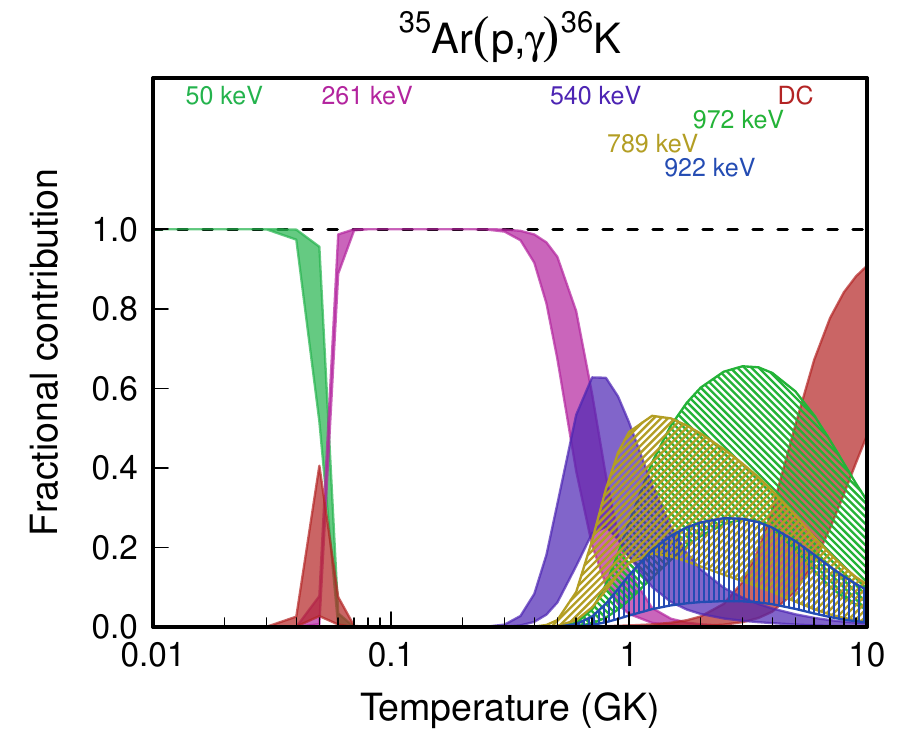}
\caption{
Fractional contributions to the total rate. ``DC'' refers to direct radiative capture. Resonance energies are given in the center-of-mass frame. 
}
\label{fig:ar35pg1}
\end{figure*}
\begin{figure*}[hbt!]
\centering
\includegraphics[width=0.5\linewidth]{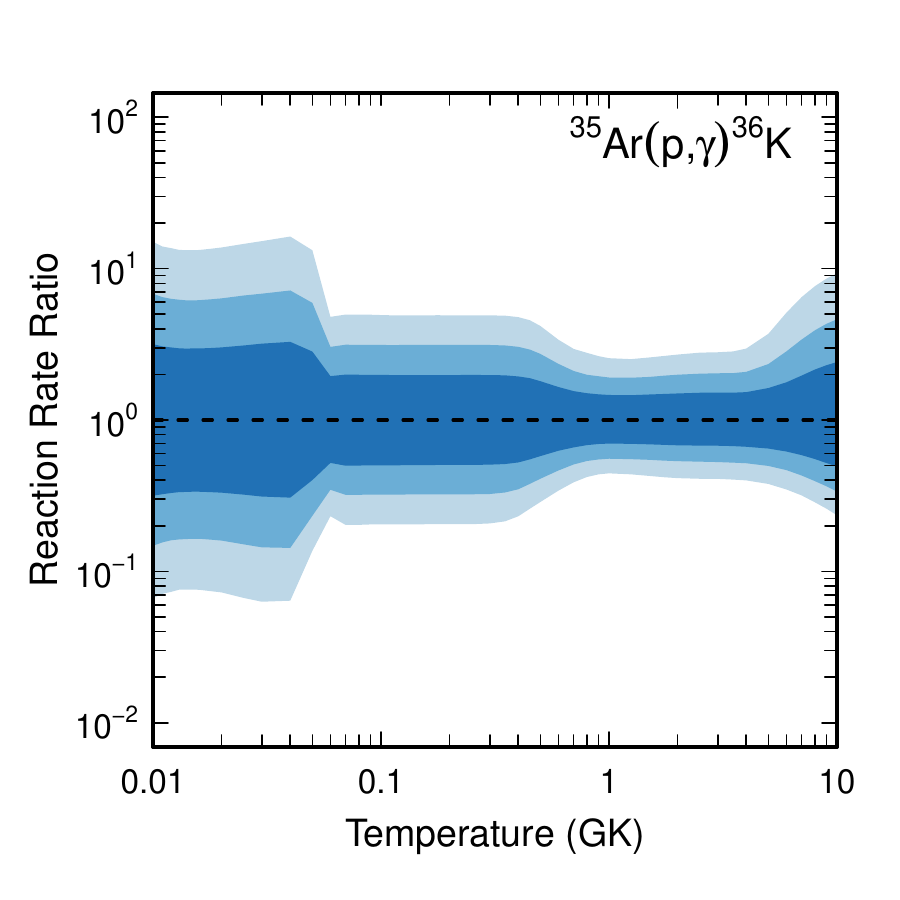}
\caption{
Reaction rate uncertainties versus temperature. The three different shades refer to coverage probabilities of 68\%, 90\%, and 98\%. 
}
\label{fig:ar35pg2}
\end{figure*}

\clearpage

\startlongtable


\clearpage

\begin{figure*}[hbt!]
\centering
\includegraphics[width=0.5\linewidth]{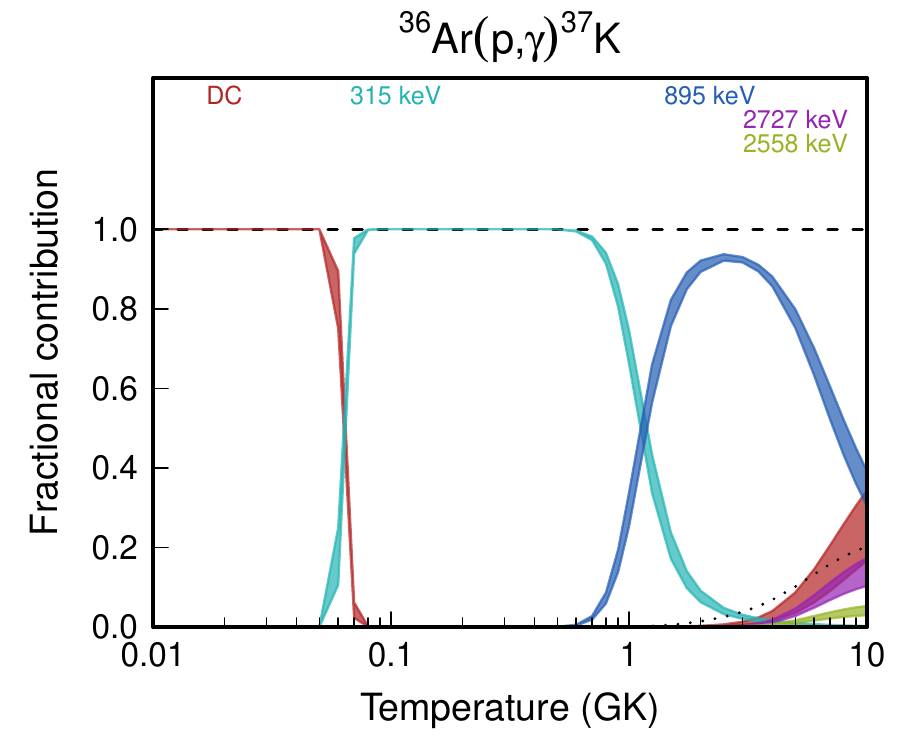}
\caption{
Fractional contributions to the total rate. ``DC'' refers to direct radiative capture. Resonance energies are given in the center-of-mass frame.  
}
\label{fig:ar36pg1}
\end{figure*}
\begin{figure*}[hbt!]
\centering
\includegraphics[width=0.5\linewidth]{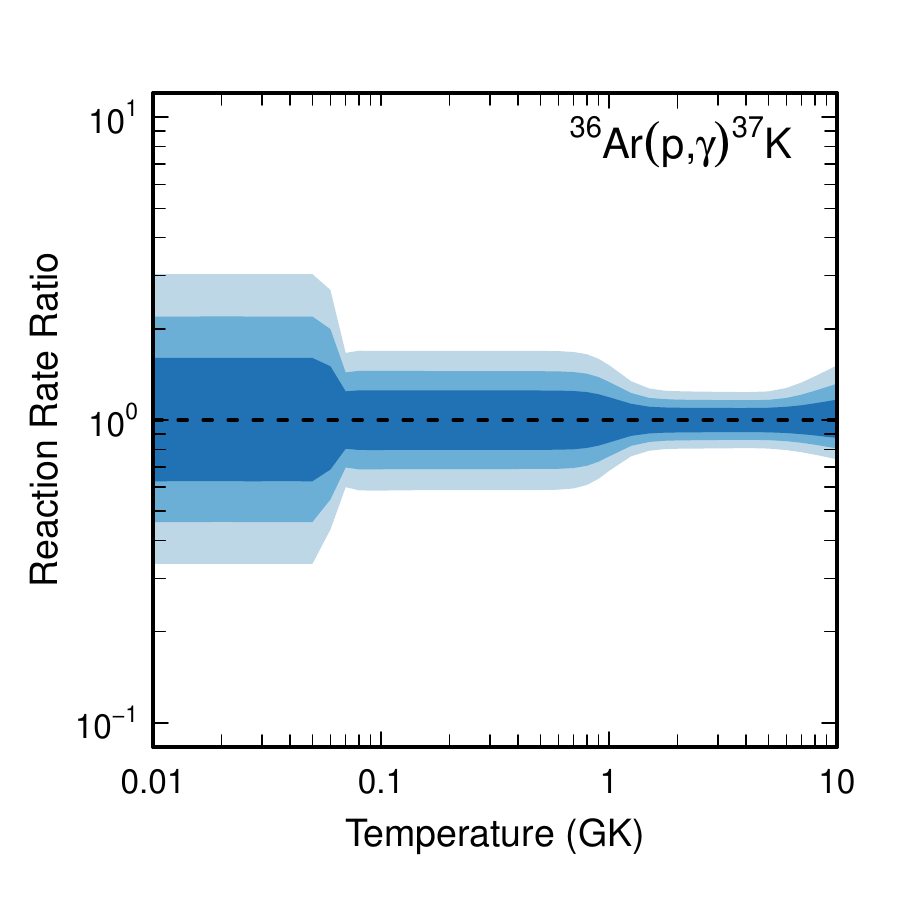}
\caption{
Reaction rate uncertainties versus temperature. The three different shades refer to coverage probabilities of 68\%, 90\%, and 98\%. 
}
\label{fig:ar36pg2}
\end{figure*}

\clearpage

\startlongtable


\clearpage

\begin{figure*}[hbt!]
\centering
\includegraphics[width=0.5\linewidth]{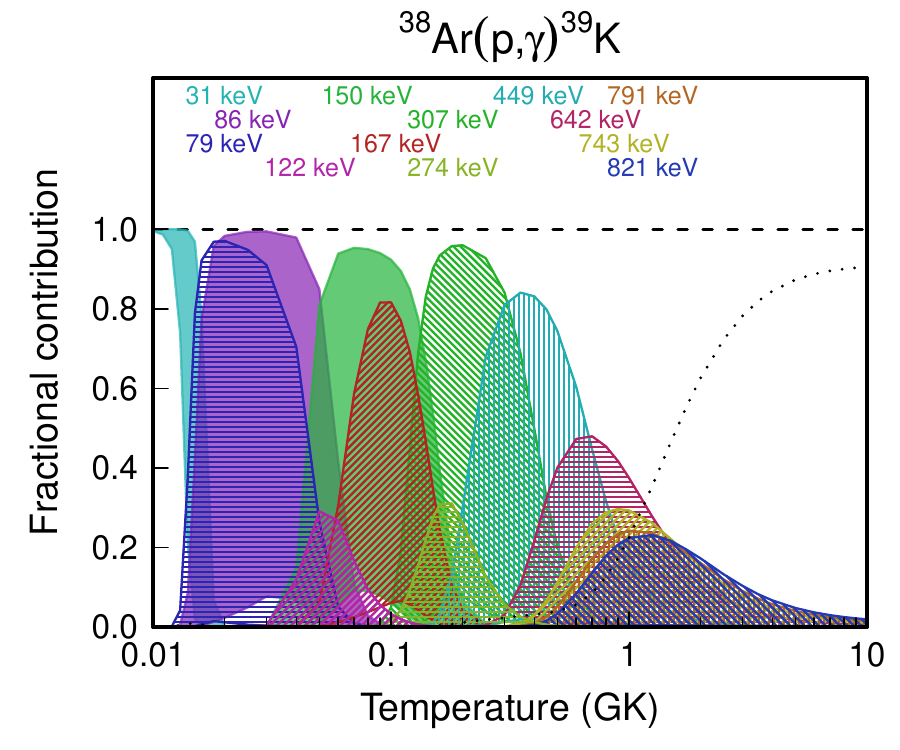}
\caption{
Fractional contributions to the total rate. Resonance energies are given in the center-of-mass frame.  
}
\label{fig:ar38pg1}
\end{figure*}
\begin{figure*}[hbt!]
\centering
\includegraphics[width=0.5\linewidth]{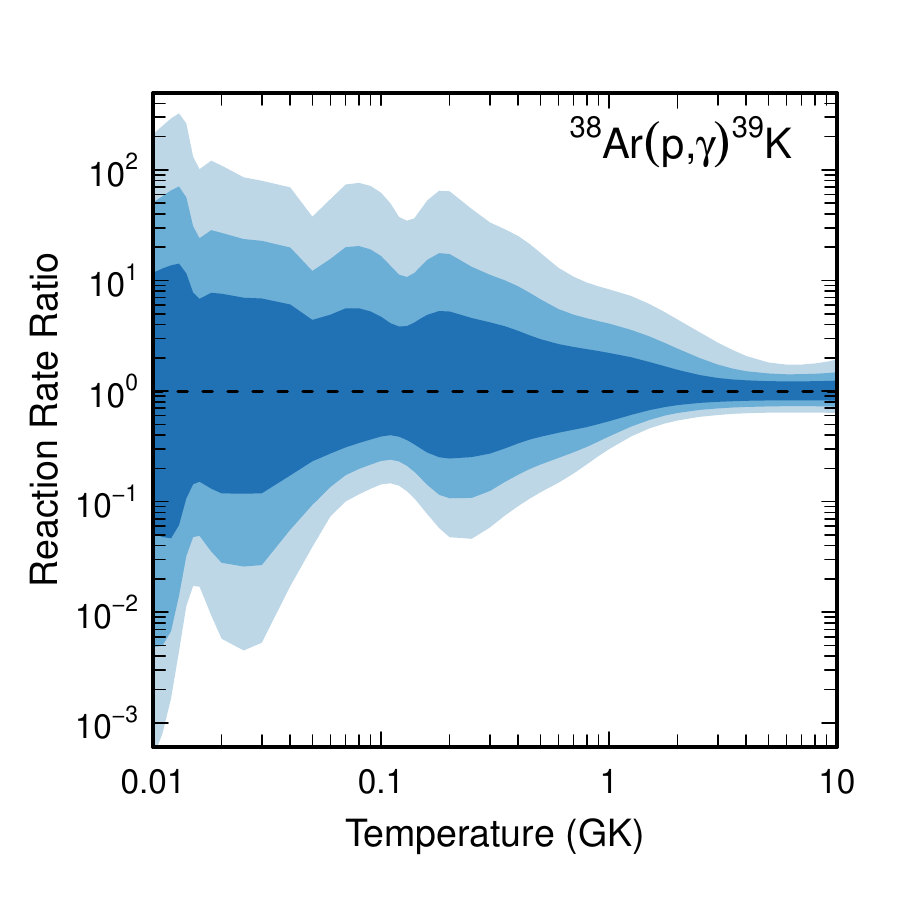}
\caption{
Reaction rate uncertainties versus temperature. The three different shades refer to coverage probabilities of 68\%, 90\%, and 98\%. 
}
\label{fig:ar38pg2}
\end{figure*}

\clearpage

\startlongtable


\clearpage

\begin{figure*}[hbt!]
\centering
\includegraphics[width=0.5\linewidth]{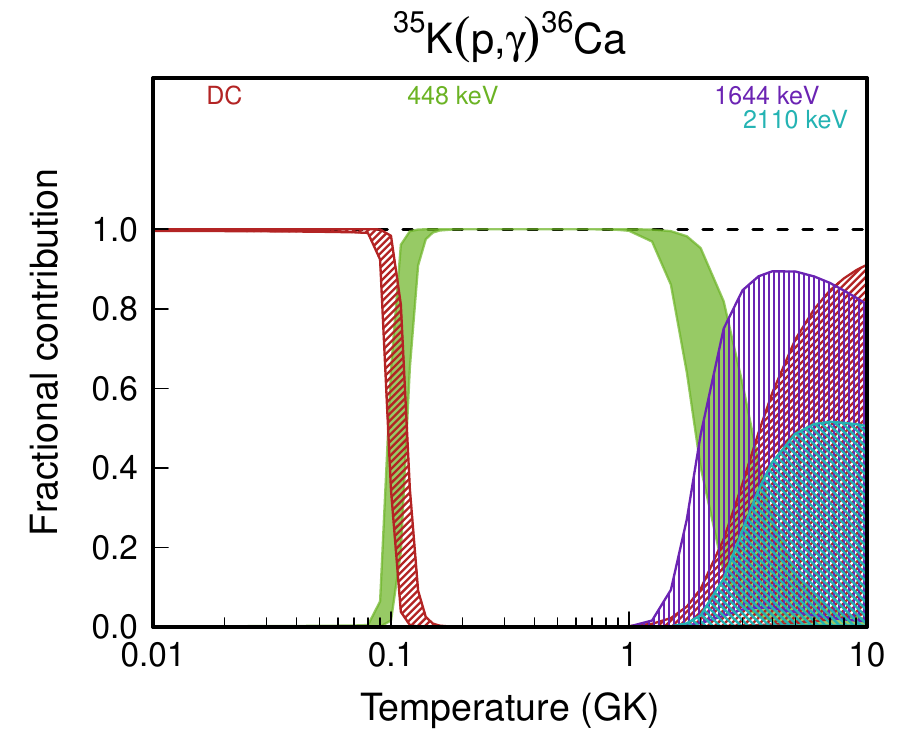}
\caption{
Fractional contributions to the total rate. ``DC'' refers to direct radiative capture. Resonance energies are given in the center-of-mass frame.  
}
\label{fig:k35pg1}
\end{figure*}
\begin{figure*}[hbt!]
\centering
\includegraphics[width=0.5\linewidth]{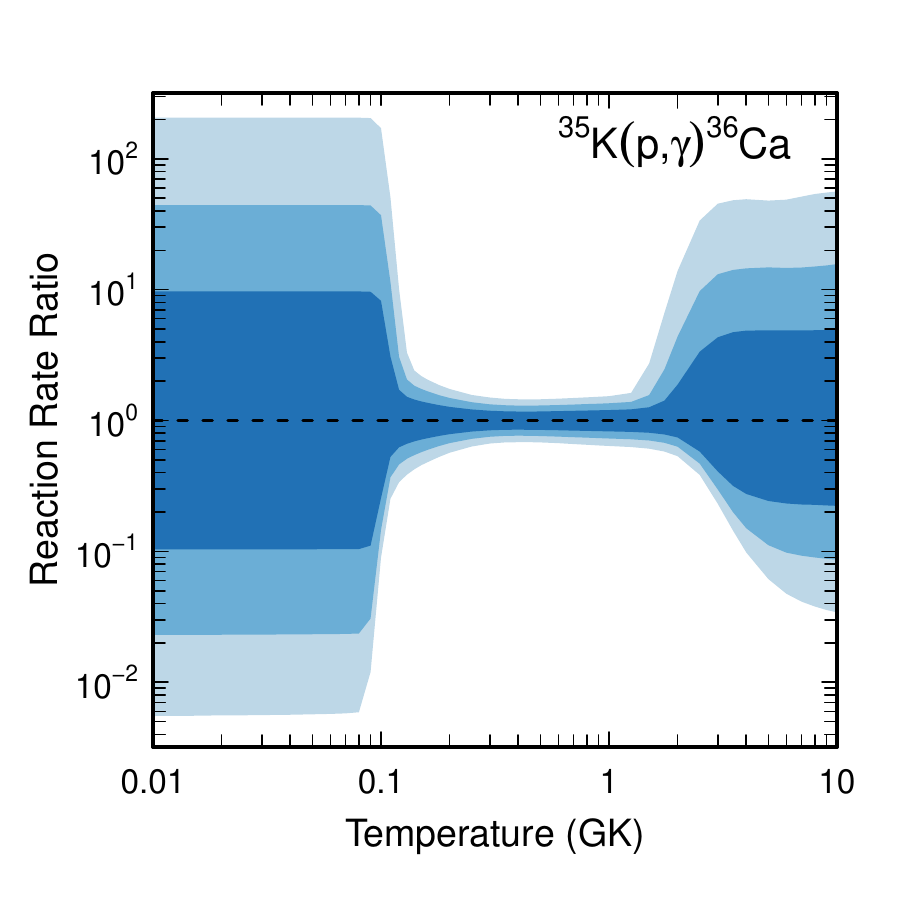}
\caption{
Reaction rate uncertainties versus temperature. The three different shades refer to coverage probabilities of 68\%, 90\%, and 98\%. 
}
\label{fig:k35pg2}
\end{figure*}

\clearpage

\startlongtable


\clearpage

\begin{figure*}[hbt!]
\centering
\includegraphics[width=0.5\linewidth]{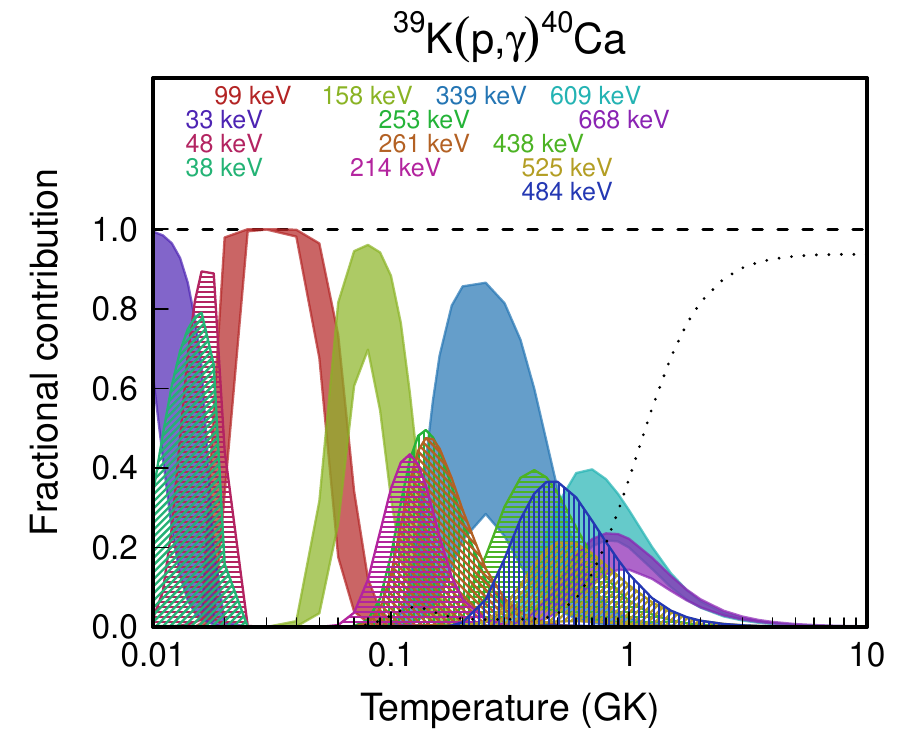}
\caption{
Fractional contributions to the total rate. Resonance energies are given in the center-of-mass frame.  
}
\label{fig:k39pg1}
\end{figure*}
\begin{figure*}[hbt!]
\centering
\includegraphics[width=0.5\linewidth]{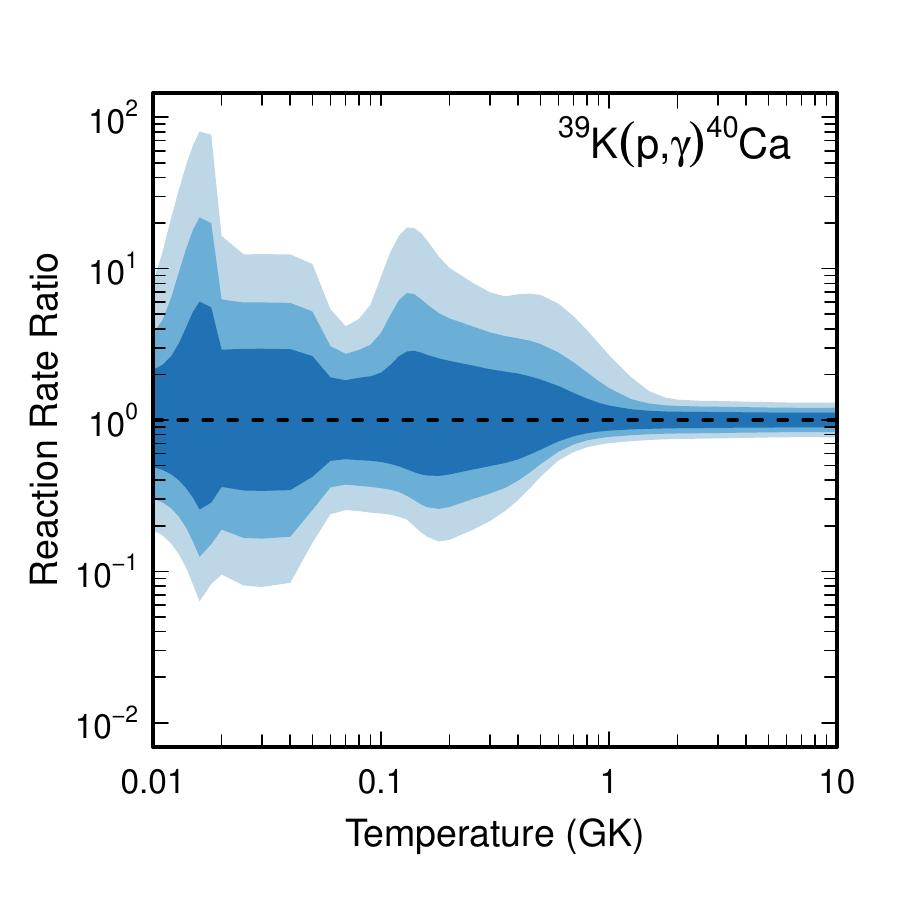}
\caption{
Reaction rate uncertainties versus temperature. The three different shades refer to coverage probabilities of 68\%, 90\%, and 98\%. 
}
\label{fig:k39pg2}
\end{figure*}

\clearpage
\startlongtable


\clearpage

\begin{figure*}[hbt!]
\centering
\includegraphics[width=0.5\linewidth]{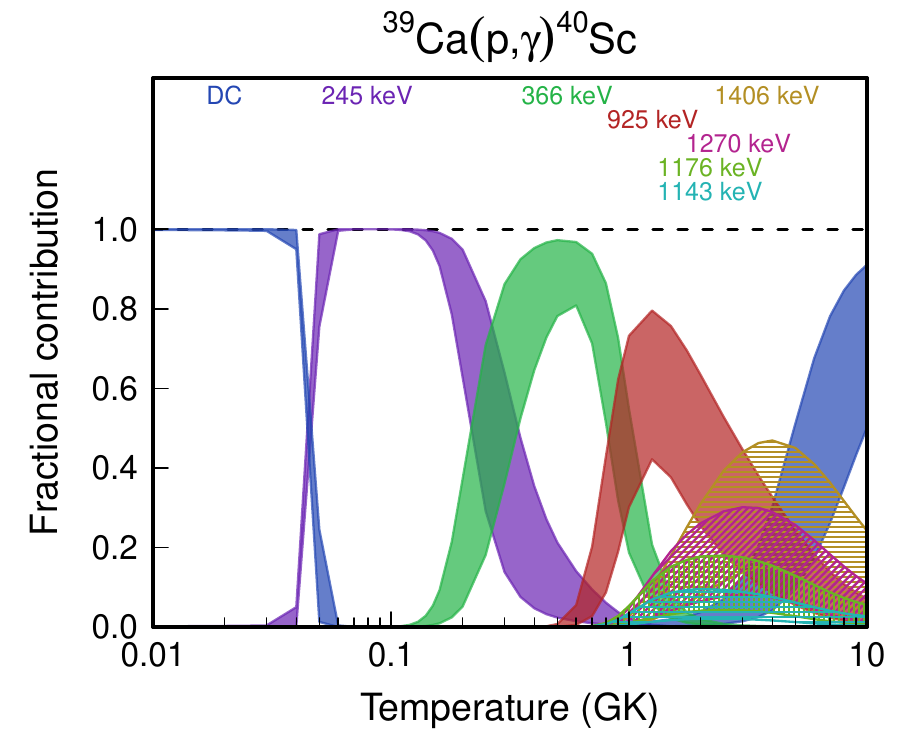}
\caption{
Fractional contributions to the total rate. ``DC'' refers to direct radiative capture. Resonance energies are given in the center-of-mass frame. 
}
\label{fig:ca39pg1}
\end{figure*}
\begin{figure*}[hbt!]
\centering
\includegraphics[width=0.5\linewidth]{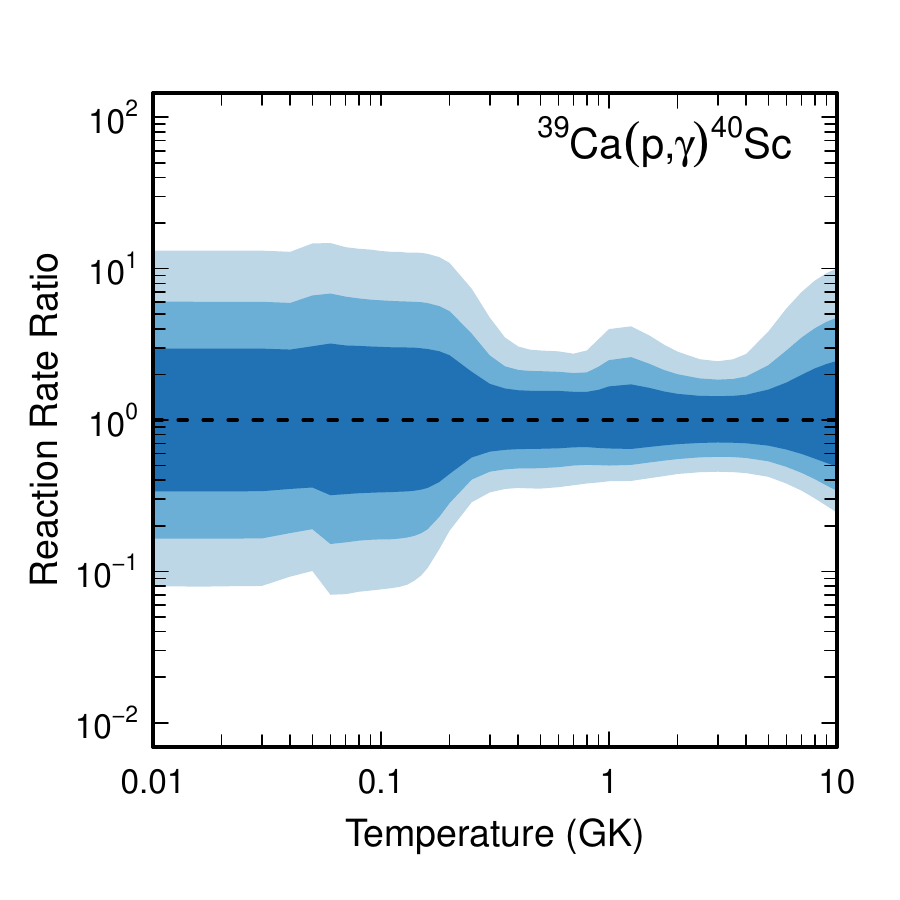}
\caption{
Reaction rate uncertainties versus temperature. The three different shades refer to coverage probabilities of 68\%, 90\%, and 98\%. 
}
\label{fig:ca39pg2}
\end{figure*}

\clearpage

\startlongtable


\clearpage

\begin{figure*}[hbt!]
\centering
\includegraphics[width=0.5\linewidth]{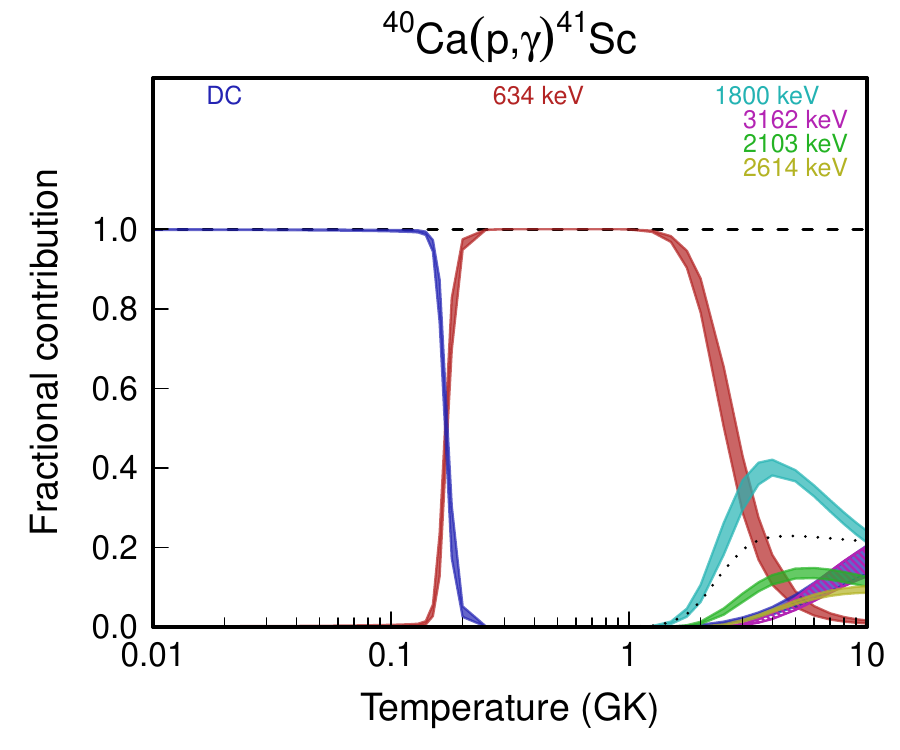}
\caption{
Fractional contributions to the total rate. ``DC'' refers to direct radiative capture. Resonance energies are given in the center-of-mass frame.  
}
\label{fig:ca40pg1}
\end{figure*}
\begin{figure*}[hbt!]
\centering
\includegraphics[width=0.5\linewidth]{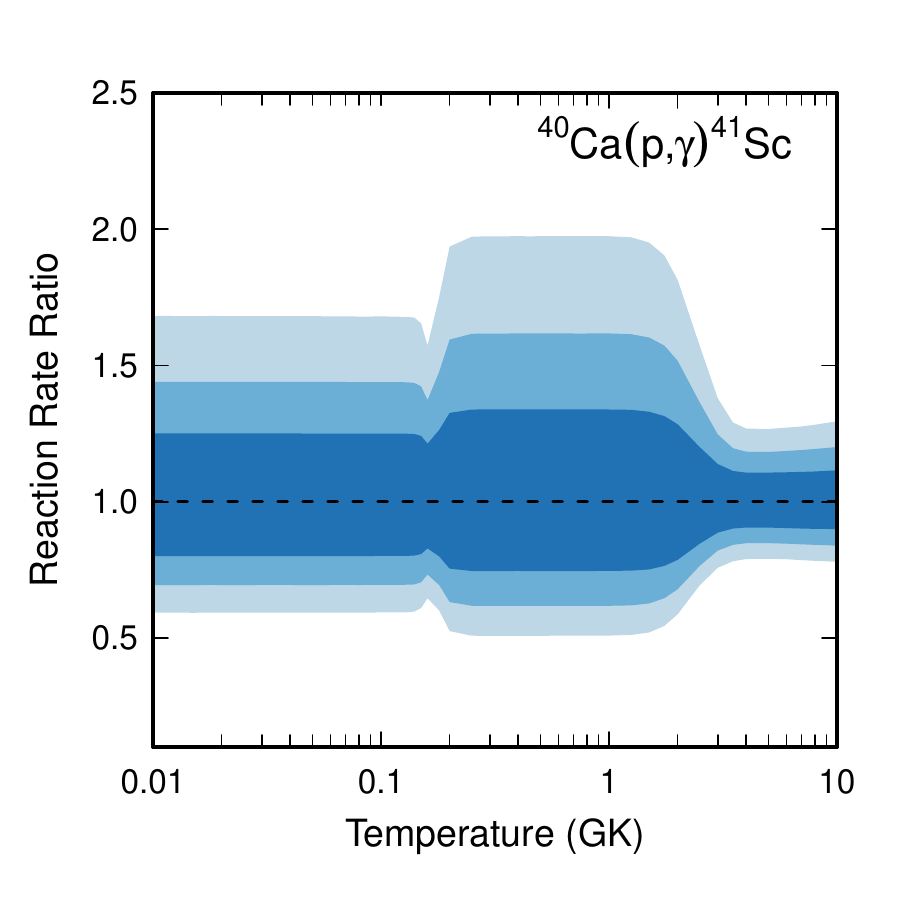}
\caption{
Reaction rate uncertainties versus temperature. The three different shades refer to coverage probabilities of 68\%, 90\%, and 98\%. 
}
\label{fig:ca40pg2}
\end{figure*}

\clearpage
\section{Comparison to previous Monte-Carlo-based reaction rates}\label{sec:compTab}
The figures below provide a comparison of our new rates (blue) with the Monte-Carlo–based rates from 2010 (gray; \citealp{ILIADIS2010b}; ETR10). Three reactions were not included in the 2010 evaluation: for $^{18}$Ne($\alpha$,p)$^{21}$Mg we compare to \citet{Mohr2014}, for $^{38}$Ar(p,$\gamma$)$^{39}$K we compare to \citet{Sallaska}, and for $^{39}$K(p,$\gamma$)$^{40}$Ca we compare to \citet{Longland18}. In each panel, the two bands represent 68\% coverage probabilities\footnote{
Note that above the matching temperature, the rate uncertainties shown in the figures below agree with the numerical values listed in the Tables of Appendix~\ref{app:reacRatesTab}, but not with the uncertainty figures that accompany each table. The latter figures display the unmatched high-temperature rates, whereas the tables list the matched values.
}, and all rates are normalized to our newly recommended values. The solid black line depicts the ratio of previous to present rates.

A visual inspection of the panels reveals not only how much the recommended rates (solid black lines) differ from previous estimates, but also how the associated uncertainties have changed. Most differences arise from newly available nuclear-physics input (e.g., updated excitation and resonance energies, resonance strengths, spectroscopic factors, and revised spin–parity assignments). Additional changes reflect updated procedural choices relative to the 2010 evaluation (ETR10), as discussed in Section \ref{sec:proc}.

Beyond these general reasons, two systematic features stand out. First, for several capture reactions (e.g., $^{14}$N($\alpha$,$\gamma$)$^{18}$F, $^{16}$O($\alpha$,$\gamma$)$^{20}$Ne, and $^{23}$Mg(p,$\gamma$)$^{24}$Al) the uncertainties of the new rates exceed the previous ones at the lowest temperatures. This mainly results from our adoption of larger uncertainties for the bound-state spectroscopic factors, which directly scale the direct-capture cross section (Section \ref{sec:dc}).

Second, when high-temperature matching is required, the new rates generally exhibit larger uncertainties than those reported in 2010. This difference stems from a fundamental change in methodology. In ETR10, the Monte-Carlo median, low, and high rates were extended beyond the matching temperature by scaling the TALYS curve. The entire TALYS rate curve up to 10 GK was multiplied by a single scale factor determined at the matching temperature: one factor for the median rate, another for the low rate, and another for the high rate. Each factor was the ratio between the corresponding Monte-Carlo value and the TALYS value at the matching point. This produced extrapolated curves that followed the TALYS temperature dependence but were shifted to agree with the Monte-Carlo results at the matching temperature. As a result, the rate uncertainty remained relatively small and essentially fixed (i.e., equal to the Monte-Carlo uncertainty at the matching temperature) throughout the extrapolated region.

In the present evaluation, we adopted a different approach (Section \ref{sec:tmatch}). We assumed the unscaled TALYS value at 10 GK to be correct within a factor-of-10 uncertainty and constructed smooth connections between this 10 GK anchor point and the Monte-Carlo low, median, and high rates at the matching temperature. Although we cannot conclusively state that one extrapolation scheme is superior, we regard the present method as more conservative and therefore preferable for representing realistic high-temperature uncertainties.

\begin{figure*}[hbt!]
\centering
\includegraphics[width=1\linewidth]{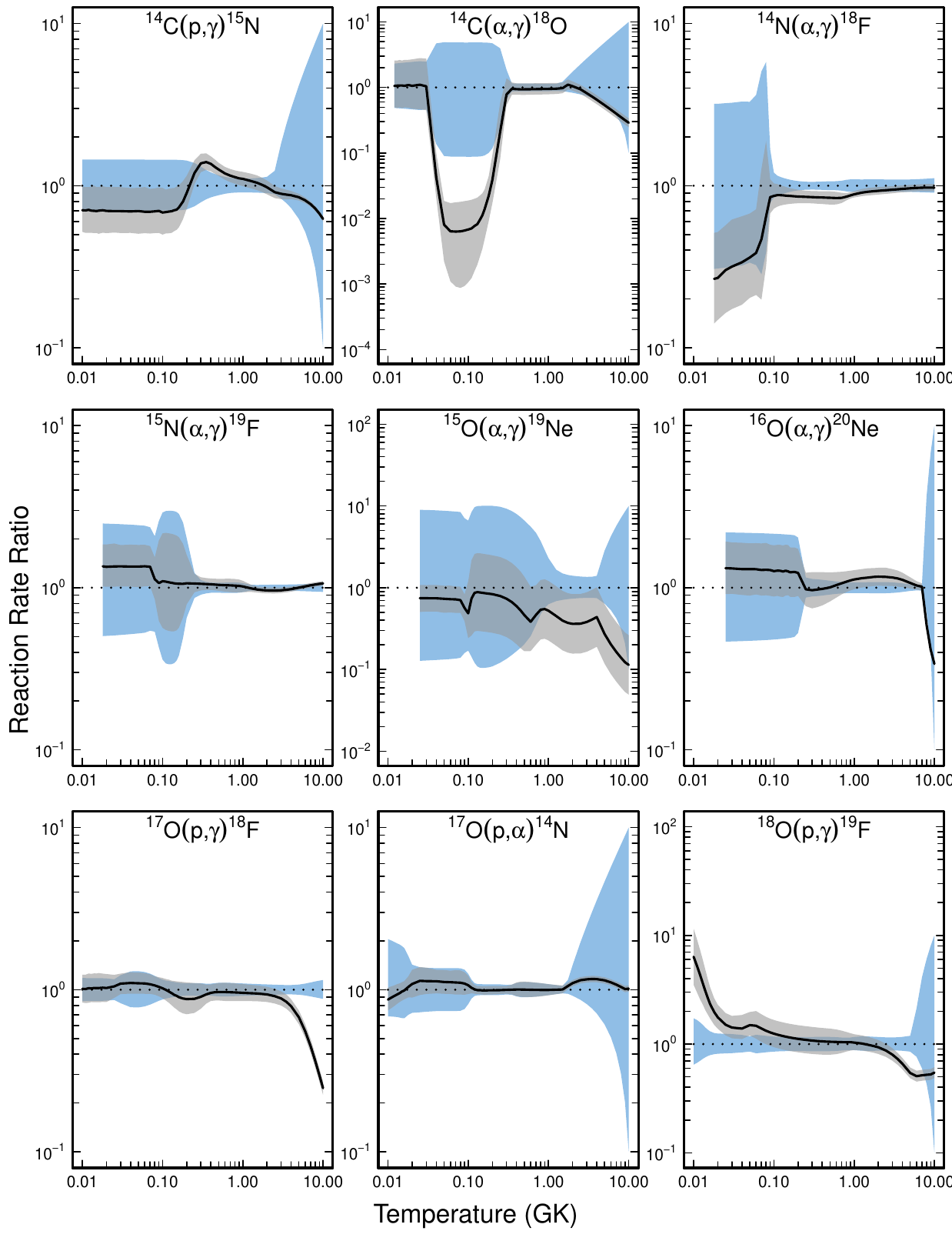}
\caption{
Comparison of our new rates (blue) with previous Monte-Carlo rates (gray; \citealp{ILIADIS2010b}). In each panel, the two bands show 68\% coverage, and rates are normalized to our new recommended values; the solid black line depicts the ratio of previous to present rates.}
\label{fig:comp1}
\end{figure*}
\begin{figure*}[hbt!]
\centering
\includegraphics[width=1\linewidth]{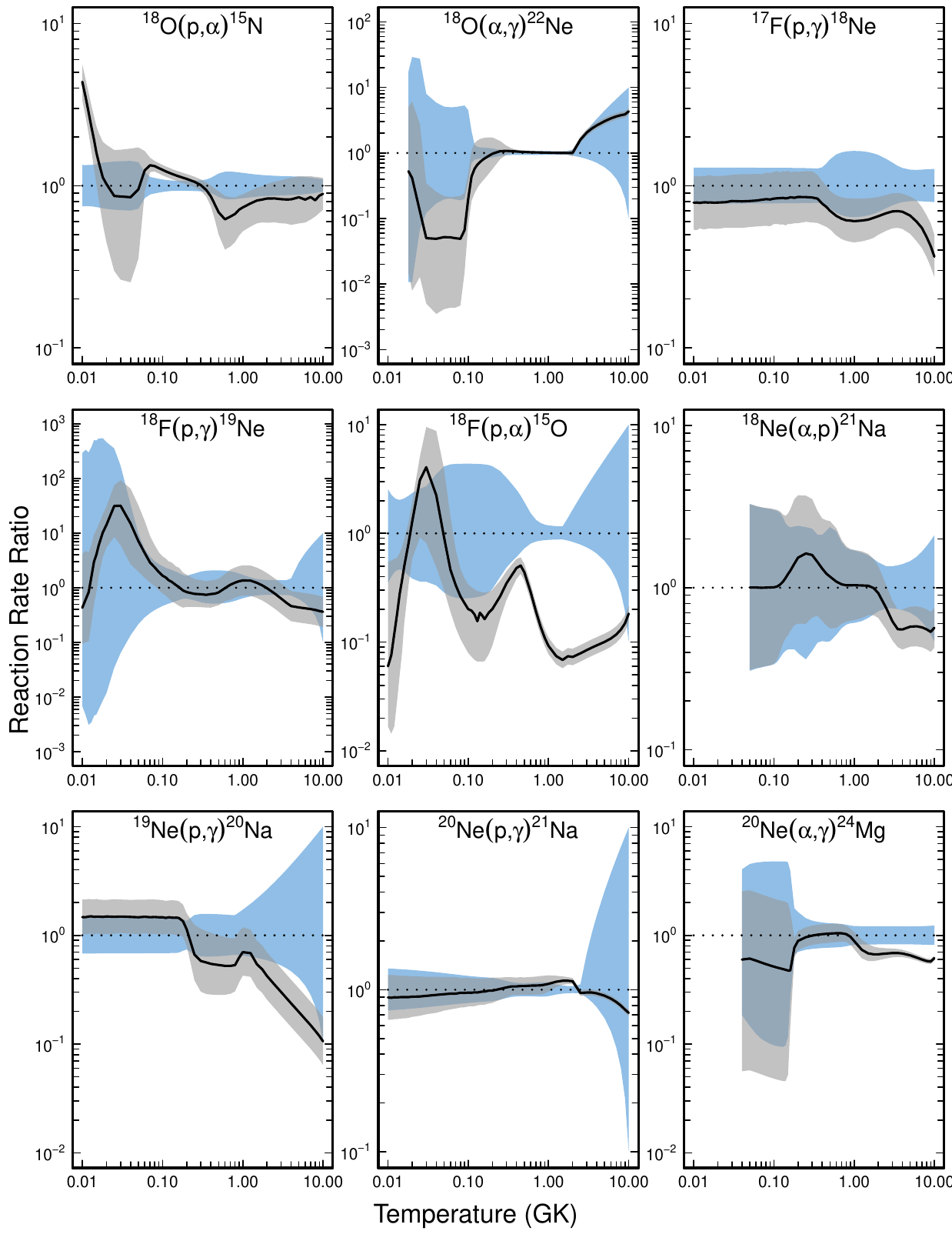}
\caption{
Comparison of our new rates (blue) with previous Monte-Carlo rates (gray; \citealp{ILIADIS2010b,Mohr2014}). In each panel, the two bands show 68\% coverage, and rates are normalized to our new recommended values; the solid black line depicts the ratio of previous to present rates.}
\label{fig:comp1}
\end{figure*}
\begin{figure*}[hbt!]
\centering
\includegraphics[width=1\linewidth]{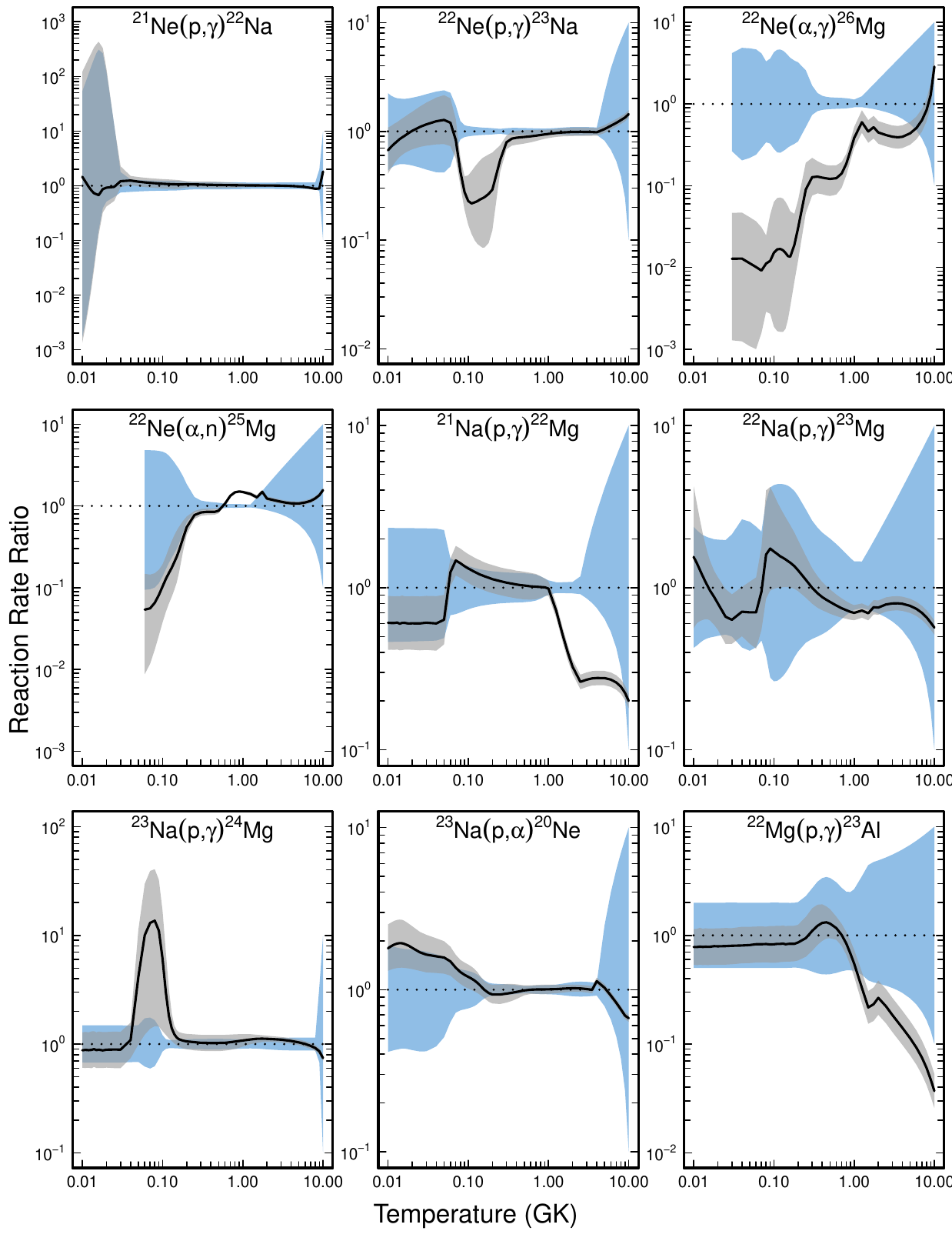}
\caption{
Comparison of our new rates (blue) with previous Monte-Carlo rates (gray; \citealp{ILIADIS2010b}). In each panel, the two bands show 68\% coverage, and rates are normalized to our new recommended values; the solid black line depicts the ratio of previous to present rates.}
\label{fig:comp1}
\end{figure*}
\begin{figure*}[hbt!]
\centering
\includegraphics[width=1\linewidth]{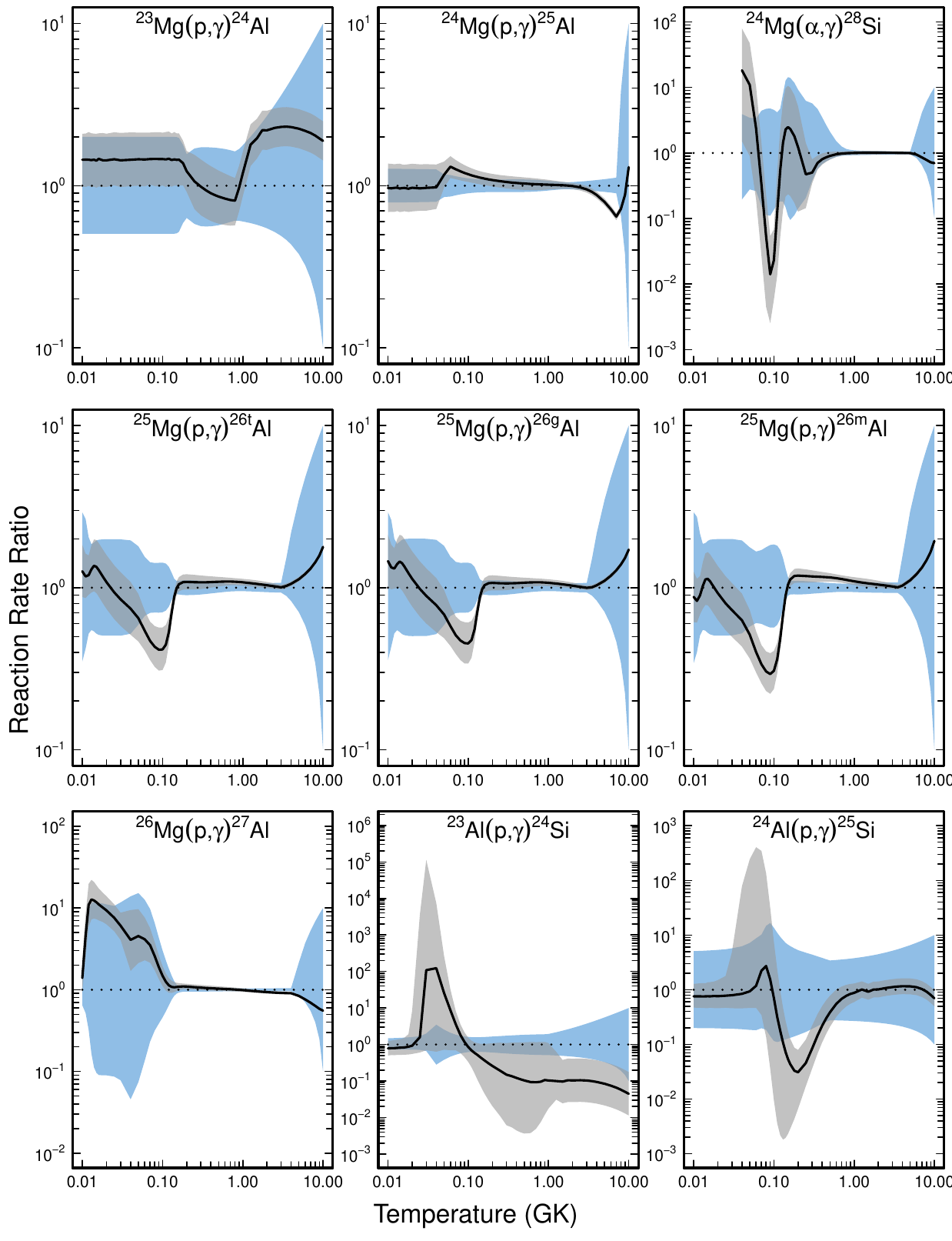}
\caption{
Comparison of our new rates (blue) with previous Monte-Carlo rates (gray; \citealp{ILIADIS2010b}). In each panel, the two bands show 68\% coverage, and rates are normalized to our new recommended values; the solid black line depicts the ratio of previous to present rates.}
\label{fig:comp1}
\end{figure*}
\begin{figure*}[hbt!]
\centering
\includegraphics[width=1\linewidth]{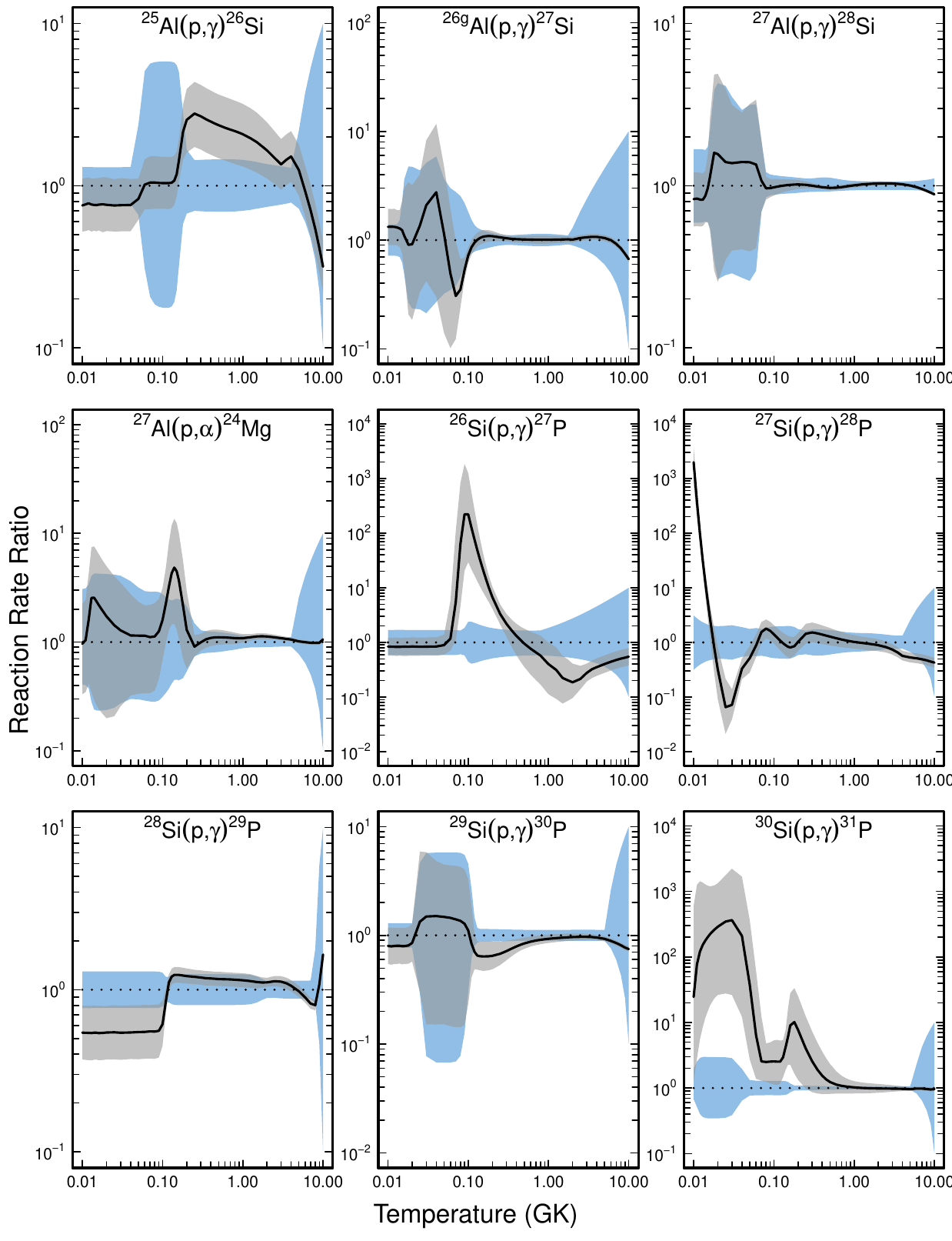}
\caption{
Comparison of our new rates (blue) with previous Monte-Carlo rates (gray; \citealp{ILIADIS2010b}). In each panel, the two bands show 68\% coverage, and rates are normalized to our new recommended values; the solid black line depicts the ratio of previous to present rates.}
\label{fig:comp1}
\end{figure*}
\begin{figure*}[hbt!]
\centering
\includegraphics[width=1\linewidth]{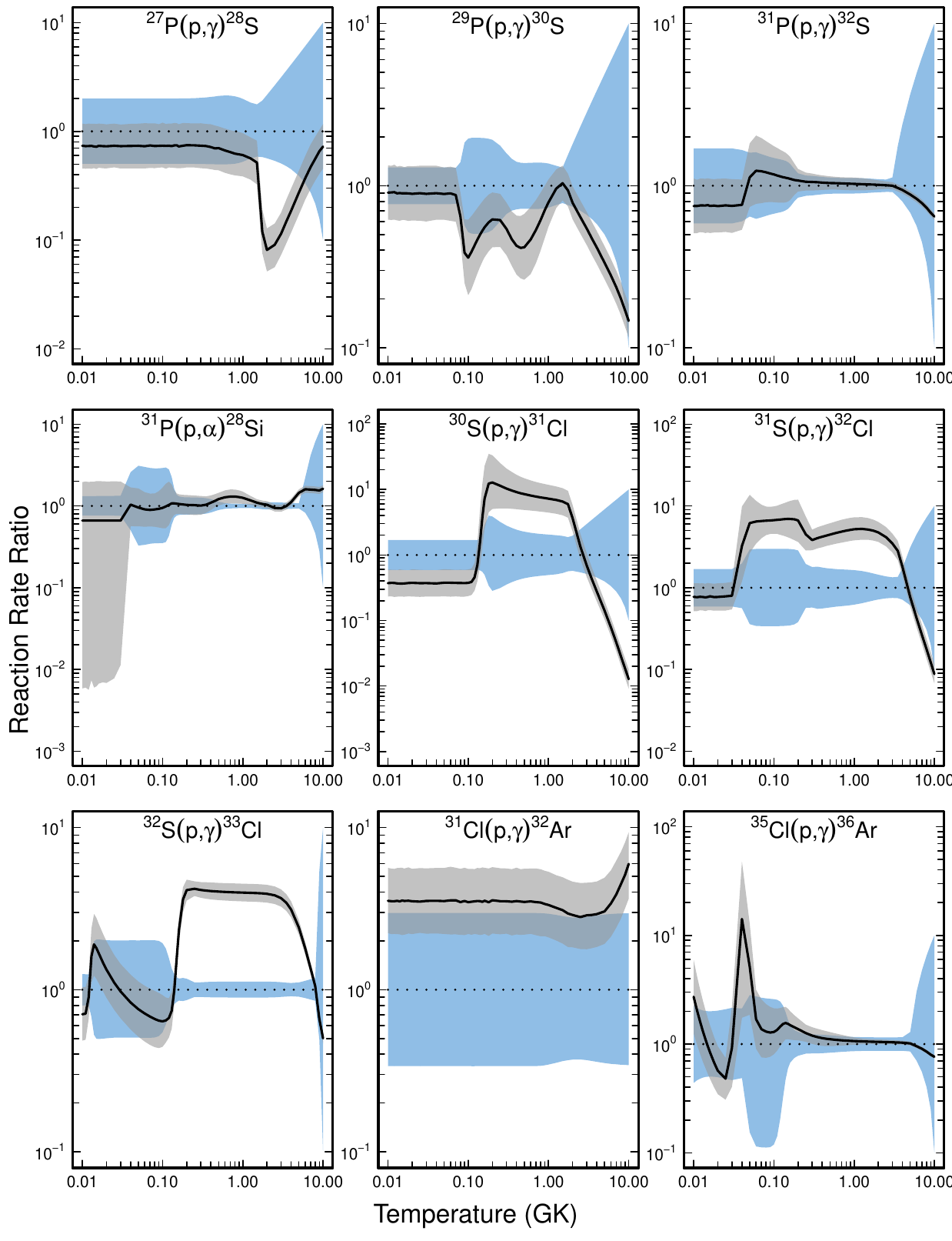}
\caption{
Comparison of our new rates (blue) with previous Monte-Carlo rates (gray; \citealp{ILIADIS2010b}). In each panel, the two bands show 68\% coverage, and rates are normalized to our new recommended values; the solid black line depicts the ratio of previous to present rates.}
\label{fig:comp1}
\end{figure*}
\begin{figure*}[hbt!]
\centering
\includegraphics[width=1\linewidth]{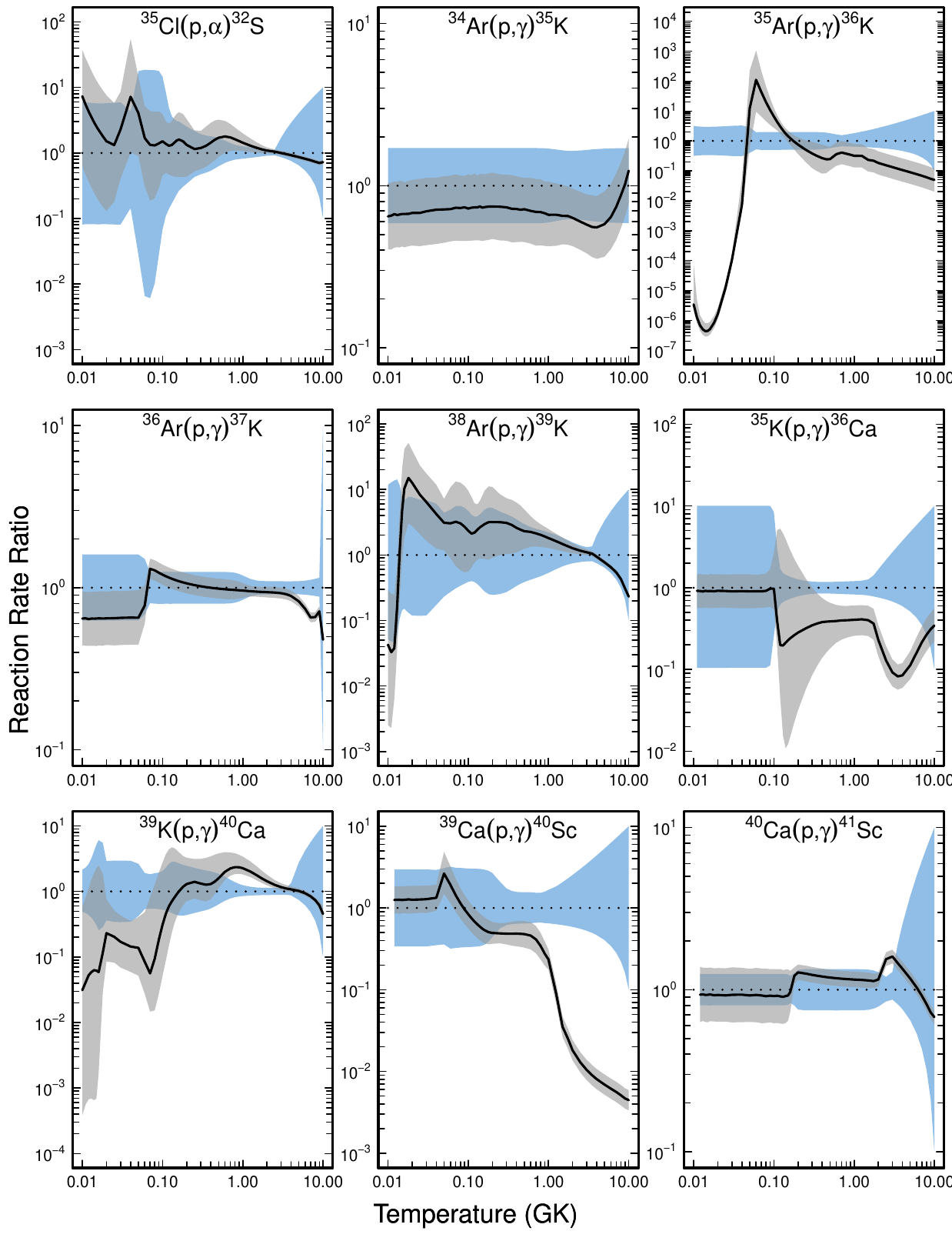}
\caption{
Comparison of our new rates (blue) with previous Monte-Carlo rates (gray; \citealp{ILIADIS2010b,Sallaska,Longland18}). In each panel, the two bands show 68\% coverage, and rates are normalized to our new recommended values; the solid black line depicts the ratio of previous to present rates.}
\label{fig:comp1}
\end{figure*}
%

\clearpage

\bibliography{paper.bib}{}
\bibliographystyle{aasjournal}



\end{document}